\ifdefined\MainOnly
  \def\OPTIONAppendix{0}
\else
  \def\OPTIONAppendix{1}
\fi

\def\OPTIONArxiv{1}%
\def\OPTIONConf{1}% 0 = purple, 1 = ACM, 2 = Springer

\ifnum\OPTIONConf=1%
  \ifnum\OPTIONArxiv=0
\documentclass[acmsmall,review,anonymous]{acmart}
\def~{\nobreakspace{}}

 \usepackage{booktabs}   %% For formal tables:
\setcopyright{none}             %% For review submission

\author{Matthew A. Hammer}
\affiliation{
  \institution{University of Colorado Boulder}
  \department{Department of Computer Science}}
\author{Joshua Dunfield}
\affiliation{
  \institution{Queen's University}
  \department{School of Computing}
}
\author{Kyle Headley}
\affiliation{
  \institution{University of Colorado Boulder}
  \department{Department of Computer Science}
}
\author{Monal Narasimhamurthy}
\affiliation{
  \institution{University of Colorado Boulder}
  \department{Department of Computer Science}
}
\author{Dimitrios J. Economou}
\affiliation{
  \institution{University of Colorado Boulder}
  \department{Department of Computer Science}
}

  \else
         \documentclass[acmsmall]{acmart}
\def~{\nobreakspace{}}
\usepackage{booktabs}   %% For formal tables:

\author{Matthew A. Hammer}
\affiliation{
  \institution{University of Colorado Boulder}
  \department{Department of Computer Science}}
\author{Joshua Dunfield}
\affiliation{
  \institution{Queen's University}
  \department{School of Computing}
}
\author{Kyle Headley}
\affiliation{
  \institution{University of Colorado Boulder}
  \department{Department of Computer Science}
}
\author{Monal Narasimhamurthy}
\affiliation{
  \institution{University of Colorado Boulder}
  \department{Department of Computer Science}
}
\author{Dimitrios J. Economou}
\affiliation{
  \institution{University of Colorado Boulder}
  \department{Department of Computer Science}
}

  \fi%
  \usepackage{joshuadunfield}
  \usepackage{euler} % XXXXXXXXXXXXXXXXXXXXXXXXXXXXXXXXXXX FIXME

\fi

\ifnum\OPTIONConf=2% Springer (LNCS)
   \documentclass[a4paper]{llncs}
   \usepackage[breaklinks]{hyperref}
   \usepackage{breakurl}
   \usepackage{joshuadunfield}

   \usepackage{amsthm}
   \usepackage{amsmath,amssymb,amsthm}
   \usepackage{thmtools,thm-restate}

   \usepackage{euler}
   \author{Matthew A.\ Hammer\INST{1}
     \and Joshua Dunfield\INST{2} \and
     \\  Kyle Headley\INST{1}
     \\  Dimitrios J.\ Economou\INST{1}
     \and Monal Narasimhamurthy\INST{1}
   }
   \INSTitute{%
     University of Colorado Boulder
     \and
     Queen's University (Canada)
   }
\fi%

\ifnum\OPTIONConf=0%
  \documentclass{purple}
  \usepackage{joshuadunfield}
  \usepackage{euler}
\fi

\usepackage{pst-node,graphicx,eepic}

\ifnum\OPTIONConf=0%
    \usepackage{fullpage}
\fi

\usepackage{listings}

\ifnum\OPTIONConf=1%
  \setcitestyle{nosort}    % even if I wanted to sort, the ACM .bst fails to do it correctly
\else
  \usepackage[authoryear]{natbib}
  \bibpunct{(}{)}{;}{a}{}{,}
\fi

\usepackage{pgf,tikz}
\usetikzlibrary{chains,shapes,scopes,arrows,calc,decorations.pathmorphing,snakes}
\ifnum\OPTIONConf=0%
    \usepackage{charter}
\fi
\ifnum\OPTIONConf=1%
\fi
\ifnum\OPTIONConf=2%
\fi

\ifnum\OPTIONConf=2%
  \usepackage{latexsym}
\else
  \usepackage{amsmath,amssymb,amsthm}
  \usepackage{thmtools,thm-restate}
\fi

\usepackage{multirow}
\usepackage{enumerate}
\usepackage{comment}
\usepackage{mathpartir} \usepackage{proof}
\usepackage{framed}
\usepackage{rotating}
\usepackage{relsize}
\usepackage{wrapfig}
\usepackage{llproof}
\usepackage{rulelinks}
\usepackage{soul}

\usepackage{latexsym}
\newcommand{\bindnasrepma}{\mathrel{\rotatebox[origin=c]{180}{\textsf{\&}}}}
        
\usepackage{amsmath,amsthm,amssymb}
\usepackage{thmtools,thm-restate}

\usepackage{fancybox}

\def\MathparLineskip{\lineskiplimit=0.8em\lineskip=0.6em plus 0.1em}

\ifnum\OPTIONConf=0%
    \declaretheoremstyle[
      bodyfont=\sl
    ]{mytheoremstyle}
\fi

\usepackage{semantic}   % Tools for typesetting PL semantics
\usepackage{braket}     % Easy angle-bracket notation
\usepackage{mathpartir} % Used to typeset blocks of inference rules
\usepackage{rotating} % for sidewaysfigure
\usepackage{proof}
\usepackage{pdflscape}
\usepackage{fancyvrb} % to use \verb in footnotes
\usepackage{amsmath,amsthm,amssymb}
\usepackage{thmtools,thm-restate}
\usepackage{mathtools}
\usepackage{color}

\let\MathRightArrow\Rightarrow % save original definition of \Rightarrow
\usepackage{marvosym} % feloniously overrides \Rightarrow
\def\Rightarrow{\MathRightArrow}

\ifnum\OPTIONConf=2
 \makeatletter

 \let\c@lemma\undefined

 \let\c@property\undefined

 \let\c@example\undefined

 \let\c@proposition\undefined

 \let\c@remark\undefined

 \let\c@definition\undefined

 \makeatother
\fi
\declaretheoremstyle[
  bodyfont=\sl
]{mytheoremstyle}

\ifnum\OPTIONConf=1

\else
    \declaretheorem[name=Theorem, style=mytheoremstyle]{theorem}
    \declaretheorem[name=Lemma, style=mytheoremstyle]{lemma}
    \declaretheorem[name=Corollary, style=mytheoremstyle, sibling=theorem]{corollary}
    \declaretheorem[name=Conjecture, style=mytheoremstyle, sibling=lemma]{conjecture}
    \declaretheorem[name=Example]{example}
    \makeatletter
    \declaretheorem[name=Proposition,style=mytheoremstyle,
        sibling=lemma,
      ]{proposition}
    \makeatother
    \declaretheorem[name=Definition,style=mytheoremstyle]{definition}
\fi
\declaretheorem[name=Property, style=mytheoremstyle]{property}

\declaretheorem[name=Remark]{remark}
\newcommand{\Label}[1]{\label{#1}} %
\newcommand{\FLabel}[1]{\label{#1}} %
\newcommand{\D}{\mathcal{D}}
\newcommand{\Ss}{\mathcal{S}}

\newcommand{\derives}{::}

\newtheorem{thm}{Theorem}[section]

\ifnum\OPTIONConf=0
\newtheorem*{thm*}{Theorem}

\fi

\ifnum\OPTIONConf=2

\else
  \newtheorem{defn}[thm]{Definition}
\fi

\newcommand{\such}{\mathrel{|}}

\newcommand{\Type}{\textsf{type}} % Introducing this version; will phase-out \type version
\newcommand{\type}{\Type}

\newcommand{\St}{S} % Store meta variable
\newcommand{\StoreType}[1]{{#1}\,\normalfont\textsf{store-type}} % Store meta variable

\newcommand{\sort}{\gamma} % sort meta variable
\newcommand{\namesort}{\boldsf{Nm}}
\newcommand{\namesetsort}{\boldsf{NmSet}}
\newcommand{\unitsort}{\boldsf{1}}
\newcommand{\Nm}{\namesort}
\newcommand{\NmSet}{\boldsf{NmSet}}

\newcommand{\effsubsym}{{\preceq}}
\newcommand{\effsub}{\mathrel{\effsubsym}}

\newcommand{\xwfeff}{\text{wf-effects}}
\newcommand{\wfeff}{\;\xwfeff}

\newcommand{\xctype}{\text{ctype}}
\newcommand{\ctype}{~\xctype}
\newcommand{\xtefftype}{\text{efftype}}
\newcommand{\tefftype}{~\xtefftype}

\newcommand{\xwfprop}{\text{prop}}
\newcommand{\wfprop}{~\xwfprop}

\newcommand{\Unit}{\tyname{unit}}
\newcommand{\unitty}{\Unit}
\newcommand{\unit}{\textvtt{()}}
\newcommand{\unitexp}{\unit}

\newcommand{\unitindex}{\unit}

\newcommand{\IdxNmFnApp}[2]{#1[[#2]]}

\newcommand{\runonboldsf}{\sffamily\bfseries\selectfont}
\newcommand{\boldsf}[1]{\text{\normalfont\runonboldsf #1}}

\newcommand{\xF}{\boldsf{F}}
\newcommand{\F}{\xF\,}
\newcommand{\xU}{\boldsf{U}}

\newcommand{\disj}{\mathrel{\bot}}

\newcommand{\Value}{{~\normalfont\textsf{val}}}
\newcommand{\mapset}[3]{#1 [[ #2 ]] \leadsto #3 }

\newcommand{\trueprop}{\boldsf{tt}}
\newcommand{\andpropsym}{\boldsf{and}}
\newcommand{\andprop}{\mathrel{\andpropsym}}
\newcommand{\impty}{\supset}
\newcommand{\convsym}{=_{\beta}}
\newcommand{\conv}{\mathrel{\convsym}}
\newcommand{\convPf}[3]{\Pf{#1}{\conv}{#2}{#3}}
\newcommand{\continueconvPf}[2]{\Pf{}{\conv}{#1}{#2}}

\newcommand{\Allsym}{\forall}

\newcommand{\All}[1]{\Allsym{#1}.\,}

\newcommand{\DAll}[2]{\Allsym{{#1} \such {#2}}.\,}

\newcommand{\DExists}[2]{\exists{{#1} \such {#2}}.\,}

\newcommand{\idxapp}[2]{{#1}({#2})}

\newcommand{\Split}[4]{\keyword{split}\Lparen{#1}, {#2}.{#3}.{#4}\Rparen}
 \newcommand{\Proj}[2]{\keyword{prj}_{#1}{#2}}
\newcommand{\Case}[5]{\keyword{case}\Lparen{#1}, {#2}.{#3}, {#4}.{#5}\Rparen}

\newcommand{\VPack}[2]{\keyword{pack}\Lparen{#1}.{#2}\Rparen}
\newcommand{\VUnpack}[4]{\keyword{vunpack}\Lparen{#1}, {#2}.{#3}.{#4}\Rparen}

\newcommand{\Scope}[2]{\keyword{scope}\Lparen{#1}, {#2}\Rparen}

\mathlig{::=}{\bnfas}
\mathlig{||-}{\Vdash}
\mathlig{|}{\;|\;}
\mathlig{[[}{\texttt{\upshape[}}
      \mathlig{]]}{\texttt{\upshape]}}
\mathlig{@@}{{\cdot}}
\mathlig{**}{\times}
\mathlig{|>}{\rhd}
\mathlig{->}{\arr}
\mathlig{-->}{\impty}
\mathlig{=>}{\Rightarrow}
\mathlig{<=}{\Leftarrow}

\mathlig{@>}{\stackrel{{}_{{}_{{}_\namesort}}}{\arr}}
\mathlig{@@>}{\stackrel{{}_{{}_{{}_\textsf{idx}}}}{\arr}}
\mathlig{*>}{\curvearrowright}

\newcommand{\effseqsym}{\mathsf{then}}
\newcommand{\effseq}{\,\effseqsym\,}

\newcommand{\effcoalsym}{\mathsf{after}}
\newcommand{\effcoal}{\,\effcoalsym\,}

\newcommand{\Lparen}{\texttt{(}}
\newcommand{\Rparen}{\texttt{)}}
\newcommand{\NmBin}[2]{\textvtt{$\left<\!\left<\hspace{-1pt}\right.\right.\!$}#1\textvtt{,}\;#2\textvtt{$\!\left.\left.\hspace{-1pt}\right>\!\right>$}}

\newcommand{\NmSetBinSym}{\ensuremath{\textcolor{dRed}{\bullet}}}
\newcommand{\NmSetBin}[2]{\ensuremath{#1\NmSetBinSym#2}}

\newcommand{\RefNr}[1]{\textsf{Ref}(#1)}
\newcommand{\Line}[1]{\textsf{Line}[[#1]]}

\newcommand{\SetNat}[1]{\textsf{Set}[[#1]]}
\newcommand{\SeqNat}[1]{\textsf{Seq}[[#1]]}
\newcommand{\ListNat}[1]{\textsf{List}[[#1]]}
\newcommand{\VecNat}[0]{\textsf{Vec}}
\newcommand{\Level}[0]{\textsf{Lev}}

\newcommand{\SeqBinCons}[0]{\texttt{SeqBin}}
\newcommand{\SeqLeafCons}[0]{\texttt{SeqLf}}

\newcommand{\SetBinCons}[0]{\texttt{SetBin}}
\newcommand{\SetLeafCons}[0]{\texttt{SetLf}}


\newcommand{\chkcolor}{dBlue}
\newcommand{\syncolor}{dRed}

\newcommand{\chk}{\mathrel{\mathcolor{\chkcolor}{\Leftarrow}}}
\newcommand{\uncoloredsyn}{{\Rightarrow}}
\newcommand{\syn}{\mathrel{\mathcolor{\syncolor}{\uncoloredsyn}}}

\mathlig{<<}{\langle}
\mathlig{>>}{\rangle}
\mathlig{==>}{\syn}
\mathlig{<==}{\chk}
\mathlig{||}{\Downarrow}
\mathlig{!!}{\Downarrow}
\mathlig{-->}{\rightarrow_{\beta}}
\mathlig{-C>}{\rightarrow_{\text{\upshape\selectfont CC}}}
\mathlig{->>}{\twoheadrightarrow}
\newcommand{\e}{\epsilon}

\newcommand{\ambns}{M}

\newcommand{\disjoint}{\mathrel{\bot}}
\newcommand{\ApartSym}{\textcolor{dRed}{\mathrel{\bot}}}

\newcommand{\tbrack}[1]{\texttt{\upshape[}{#1}\texttt{\upshape]}}

\newcommand{\leafname}{\textvtt{leaf}}

\newcommand{\Mv}{V}
\newcommand{\ntevalsym}{\Downarrow_{\textit{M}}}
\newcommand{\nteval}{\mathrel{\ntevalsym}}

\newcommand{\xrefv}{\keyword{ref}}
\newcommand{\xthunk}{\keyword{thunk}}

\newcommand{\xname}{\keyword{name}}
\newcommand{\xName}{\tyname{Nm}}

\newcommand{\xRef}{\tyname{Ref}}
\newcommand{\xThunk}{\tyname{Thk}}

\newcommand{\refv}[1]{\xrefv\;{#1}}
\newcommand{\thunk}[1]{\xthunk\;{#1}}

\newcommand{\name}[1]{\xname\;{#1}}
\newcommand{\Name}[1]{\xName\tbrack{#1}}

\newcommand{\xnamefn}{\keyword{nmfn}}

\newcommand{\namefn}[1]{\xnamefn\;{#1}}
\newcommand{\nametm}[1]{\namefn{#1}}

\newcommand{\Ref}[1]{\xRef\tbrack{#1}\,}
\newcommand{\Thk}[1]{\xThunk\tbrack{#1}\,}

\newcommand{\Thunk}[2]{\keyword{thunk}\Lparen{#1},{#2}\Rparen}

\newcommand{\Refe}[2]{\keyword{ref}\Lparen{#1},{#2}\Rparen} % Ref expression form

\newcommand{\List}[3]{\tyname{List}\tbrack{#1;#2}\,{#3}}
\newcommand{\ListDiff}[3]{\tyname{List}\diffbox{\tbrack{#1;#2}}\,{#3}}

\newcommand{\Int}[0]{\tyname{Int}}
\newcommand{\Nat}[0]{\tyname{Nat}}
\newcommand{\Bool}[0]{\tyname{Bool}}

\let\Force\undefined
\newcommand{\Force}[1]{\keyword{force}\Lparen{#1}\Rparen}
\newcommand{\Get}[1]{\keyword{get}\Lparen{#1}\Rparen}

\newcommand{\Ret}[1]{\keyword{ret}\Lparen{#1}\Rparen}
\newcommand{\Let}[3]{\keyword{let}\Lparen{#1},{#2}.{#3}\Rparen}
\newcommand{\Archivist}[1]{\keyword{archivist}\Lparen{#1}\Rparen}

\newcommand{\PreSt}[3]{{#1}\vdash^{#2}_{#3}}

\newcommand{\emptystore}{\cdot}

\newcommand{\DHasEffects}[3]{#1\,\textsf{\upshape reads}\,#2\,\textsf{\upshape writes}\,#3}
\newcommand{\DByHasEffects}[4]{#1\,\textsf{by}\,#2\,\textsf{\upshape reads}\,#3\,\textsf{\upshape writes}\,#4}
\newcommand{\mergeRds}{\mathrel{\cup}}

\newcommand{\dom}[1]{\textsf{\upshape\selectfont dom}(#1)}

\newcommand{\te}{t}

\newcommand{\NotInScope}[1]{}
\newcommand{\REVISEME}[1]{{\color{blue}{(Text to revise goes here; See latex source)}}}

\definecolor{light-gray}{gray}{0.95}
\definecolor{white}{gray}{1}

\newcommand{\diffbox}[1]{\colorbox{dLighterBlue}{\!\ensuremath{#1}\!}}
\newcommand{\samebox}[1]{\colorbox{white}{\!\ensuremath{#1}\!}}

\newcommand{\ByStoreTypingRuleApp}{By rule (\Figref{fig:store-typing})}

\newcommand{\ByRWSetRule}{By \Defnref{def:rw}}

\let\inst\undefined
\newcommand{\inst}[2]{{#1}\tbrack{#2}}

\newcommand{\extractsym}{\textit{extract}}
\newcommand{\extract}[1]{\extractsym({#1})}

\newcommand{\extractassnssym}{\textit{extract-assns}}
\newcommand{\extractassns}[1]{\extractassnssym({#1})}
\newcommand{\extractctxsym}{\textit{extract-ctx}}
\newcommand{\extractctx}[1]{\extractctxsym({#1})}

\newcommand{\flipsym}{\textit{flip}}
\newcommand{\flip}[1]{\flipsym(#1)}

\newcommand{\vtyperule}[1]{vtype-{#1}\xspace}

\newrulecommand{vtypeAllIntro}{\vtyperule{$\Allsym$IndexIntro}}
\newrulecommand{vtypeAllElim}{\vtyperule{$\Allsym$IndexElim}}
\newrulecommand{vtypeExistsIntro}{\vtyperule{$\exists$IndexIntro}}
\newrulecommand{vtypeExistsElim}{\vtyperule{$\exists$IndexElim-NONEXISTENT-RULE}}
\let\vtypeAllIndexIntro\vtypeAllIntro
\let\vtypeAllIndexElim\vtypeAllElim
\let\vtypeExistsIndexIntro\vtypeExistsIntro

\newrulecommand{vtypeSub}{\vtyperule{sub}}

\newcommand{\etyperule}[1]{etype-{#1}\xspace}
\newrulecommand{etypeSub}{\etyperule{sub}}
\newrulecommand{etypeExistsElim}{\etyperule{$\exists$IndexElim}}
\newrulecommand{etypeAllIntro}{\etyperule{$\forall$Intro}}
\newrulecommand{etypeAllIndexIntro}{\etyperule{$\forall$IndexIntro}}
\newrulecommand{etypeAllElim}{\etyperule{$\forall$Elim}}
\newrulecommand{etypeAllIndexElim}{\etyperule{$\forall$IndexElim}}
\let\etypeExistsIndexElim\etypeExistsElim

\newcommand{\vsynrule}[1]{vsyn-{#1}\xspace}
\newrulecommand{vsynvar}{\vsynrule{var}}
\newrulecommand{vsynanno}{\vsynrule{anno}}
\newrulecommand{vsynAllIndexElim}{\vsynrule{$\forall$IndexElim}}
\newcommand{\vchkrule}[1]{vchk-{#1}\xspace}
\newrulecommand{vchkunit}{\vchkrule{unit}}
\newrulecommand{vchkpair}{\vchkrule{pair}}
\newrulecommand{vchksub}{\vchkrule{sub}}
\newrulecommand{vchkname}{\vchkrule{name}}
\newrulecommand{vchknamefn}{\vchkrule{namefn}}
\newrulecommand{vchkref}{\vchkrule{ref}}
\newrulecommand{vchkthunk}{\vchkrule{thunk}}
\newrulecommand{vchkinjOne}{\vchkrule{inj1}}
\newrulecommand{vchkinjTwo}{\vchkrule{inj2}}
\newrulecommand{vchknameapp}{\vchkrule{name-app}}
\newrulecommand{vchkAllIntro}{\vchkrule{$\forall$Intro}}
\newrulecommand{vchkAllIndexIntro}{\vchkrule{$\forall$IndexIntro}}
\newrulecommand{vchkExistsIndexIntro}{\vchkrule{$\exists$IndexIntro}}

\newcommand{\esynrule}[1]{esyn-{#1}\xspace}
\newrulecommand{esynanno}{\esynrule{anno}}
\newrulecommand{esynapp}{\esynrule{app}}
\newrulecommand{esynforce}{\esynrule{force}}
\newrulecommand{esynget}{\esynrule{get}}
\newrulecommand{esynnameapp}{\esynrule{name-app}}
\newrulecommand{esynAllIndexElim}{\esynrule{$\forall$IndexElim}}
\newrulecommand{esynAllElim}{\esynrule{$\forall$Elim}}

\newcommand{\echkrule}[1]{echk-{#1}\xspace}
\newrulecommand{echksub}{\echkrule{sub}}
\newrulecommand{echksplit}{\echkrule{split}}
\newrulecommand{echkcase}{\echkrule{case}}
\newrulecommand{echkret}{\echkrule{ret}}
\newrulecommand{echklam}{\echkrule{lam}}
\newrulecommand{echklet}{\echkrule{let}}
\newrulecommand{echkthunk}{\echkrule{thunk}}
\newrulecommand{echkref}{\echkrule{ref}}
\newrulecommand{echkscope}{\echkrule{scope}}
\newrulecommand{echkAllIndexIntro}{\echkrule{$\forall$IndexIntro}}
\newrulecommand{echkAllIntro}{\echkrule{$\forall$Intro}}
\newrulecommand{echkExistsIndexElim}{\echkrule{$\exists$IndexElim}}

\newcommand{\vsubtypesym}{\subtypesym_{\textsf{\itshape\selectfont V}}}
\newcommand{\vsubtype}{\mathrel{\vsubtypesym}}

\newcommand{\vsubrule}[1]{$\vsubtypesym$-{#1}\xspace}
\newrulecommand{vsubrefl}{\vsubrule{refl}}
\newrulecommand{vsubproduct}{\vsubrule{$**$}}
\newrulecommand{vsubsum}{\vsubrule{$+$}}
  \newrulecommand{vsubname}{\vsubrule{name}}
  \newrulecommand{vsubref}{\vsubrule{ref}}
  \newrulecommand{vsubthk}{\vsubrule{thk}}
  \newrulecommand{vsubnamefn}{\vsubrule{namefn}}
\newrulecommand{vsubAllL}{\vsubrule{$\forall$L}}
\newrulecommand{vsubAllR}{\vsubrule{$\forall$R}}
  \newrulecommand{vsubExists}{\vsubrule{$\exists$}}

\newcommand{\esubtypesym}{\subtypesym_{\textsf{\itshape\selectfont E}}}
\newcommand{\esubtype}{\mathrel{\esubtypesym}}
\newcommand{\esubrule}[1]{$\esubtypesym$-{#1}\xspace}
\newrulecommand{esubeff}{\esubrule{eff}}
\newrulecommand{esubAllType}{\esubrule{all-type}}
\newrulecommand{esubAllIndexL}{\esubrule{all-index-L}}
\newrulecommand{esubAllIndexR}{\esubrule{all-index-R}}

\newcommand{\csubtypesym}{\subtypesym_{\textsf{\itshape\selectfont C}}}
\newcommand{\csubtype}{\mathrel{\csubtypesym}}
\newcommand{\csubrule}[1]{$\csubtypesym$-{#1}\xspace}
\newrulecommand{csublift}{\csubrule{lift}}
\newrulecommand{csubarr}{\csubrule{arr}}

\newcommand{\isavaluePf}[2]{\Pf{{#1}}{\text{is~a}}{\text{value}}{#2}}

\newcommand{\tevalrule}[1]{\text{teval-{#1}}\xspace}

\newrulecommand{tevalvalue}{\tevalrule{value}}
\newrulecommand{tevalbin}{\tevalrule{bin}}
\newrulecommand{tevalapp}{\tevalrule{app}}

\newcommand{\Sortrule}[1]{\text{sort-{#1}}\xspace}

\newrulecommand{Sortunit}{\Sortrule{unit}}
\newrulecommand{Sortpair}{\Sortrule{pair}}
\newrulecommand{Sortproj}{\Sortrule{proj}}
\newrulecommand{Sortempty}{\Sortrule{empty}}
\newrulecommand{Sortsingleton}{\Sortrule{singleton}}
\newrulecommand{Sortunion}{\Sortrule{union}}
\newrulecommand{Sortsepunion}{\Sortrule{sep-union}}
\newrulecommand{Sortabs}{\Sortrule{abs}}
\newrulecommand{Sortapply}{\Sortrule{apply}}
\newrulecommand{Sortmap}{\Sortrule{map}}
\newrulecommand{Sortbuild}{\Sortrule{build}}
\newrulecommand{Sortstar}{\Sortrule{star}}
\newrulecommand{Sortvar}{\Sortrule{var}}

\newcommand{\dynrule}[1]{\text{$!!$-{#1}}\xspace}

\newrulecommand{dynsplit}{\dynrule{split}}
\newrulecommand{dyncase}{\dynrule{case}}
\newrulecommand{dynunpack}{\dynrule{unpack}}
\newrulecommand{dynlet}{\dynrule{let}}
\newrulecommand{dynapp}{\dynrule{app}}
\newrulecommand{dynscope}{\dynrule{scope}}
\newrulecommand{dynnameapp}{\dynrule{name-app}}
\newrulecommand{dynthunk}{\dynrule{thunk}}
\newrulecommand{dynref}{\dynrule{ref}}
\newrulecommand{dynforce}{\dynrule{force}}
\newrulecommand{dynget}{\dynrule{get}}
\newrulecommand{dynterm}{\dynrule{term}}

\newcommand{\lesscaptionspace}{\vspace*{-1.0ex}}

\pgfdeclarelayer{dcgbackground}
\pgfdeclarelayer{dcgeffects}
\pgfdeclarelayer{background}
\pgfdeclarelayer{foreground}
\pgfdeclarelayer{tiles}
\pgfdeclarelayer{nodes}
\pgfdeclarelayer{staticdynamic}

\pgfsetlayers{tiles,dcgeffects,dcgbackground,background,nodes,main,foreground,staticdynamic}

\newcommand{\LstVar}[1]{\large \texttt{#1}}
\newcommand{\Num}[1]{\large #1}
\newcommand{\NmStr}[1]{\textbf{\textsf{#1}}}
\newcommand{\NmNum}[1]{\textbf{\textsf{#1}}}
\newcommand{\NmBinOp}[2]{#1{\cdot}#2}
\newcommand{\False}[0]{\textbf{\texttt{False}}}

\newcommand{\RefNil}[2]{
  \path (#1.north)+(0,0)            node (#1n) {};
  \path (#1.east)+(0,0)             node (#1e) {};
  \path (#1.north |- #1.west)+(-0.42,-0.38) node (a2x) {};
  \path (#1.south |- #1.east)+(+0.42,+0.38) node (a2y) {};
  \path[fill=violet!10,rounded corners, line width=1.5pt, draw=black]
  (a2x) rectangle (a2y);
  \draw (#1.north west)+(-10pt,0) node {$a_5$};
}
\newcommand{\RefCons}[2]{
  \path (#1.north)+(0,0)            node (#1n) {};
  \path (#1.east)+(0,0)             node (#1e) {};
  \path (#1.west)+(0,0)             node (#1w) {};
  \path (#1.north |- #1.west)+(-0.45,-0.85) node (#1x) {};
  \path (#1.south |- #1.east)+(+0.45,+0.85) node (#1y) {};
  \path[fill=violet!10,rounded corners, line width=1.5pt, draw=black]
  (#1x) rectangle (#1y);
  \draw (#1.north west)+(-12pt,0) node (#1nw) {#2};
}
\newcommand{\RefConsOut}[2]{
  \path (#1.north)+(0,0)            node (#1n) {};
  \path (#1.east)+(0,0)             node (#1e) {};
  \path (#1.north |- #1.west)+(-0.45,-0.85) node (#1x) {};
  \path (#1.south |- #1.east)+(+0.45,+0.85) node (#1y) {};
  \path[fill=violet!10,rounded corners, line width=1.5pt, draw=black]
  (#1x) rectangle (#1y);
  \draw (#1.north west)+(-16pt,0) node {#2};
}
\newcommand{\RefConsC}[2]{
  \path (#1.north)+(0,0)            node (#1n) {};
  \path (#1.east)+(0,0)             node (#1e) {};
  \path (#1.north |- #1.west)+(-0.45,-0.85) node (#1x) {};
  \path (#1.south |- #1.east)+(+0.45,+0.85) node (#1y) {};
  \path[fill=violet!10,rounded corners, line width=1.5pt, draw=black]
  (#1x) rectangle (#1y);
  \draw (#1.north west)+(-18pt,0) node {#2}; %% <-- 'C' means change; so, extra space on this line
}
\newcommand{\RefTrie}[2]{
  \path (#1.north)+(0,0.45) node {#2};
  \path (#1.south)+(0,0) node (#1s) {};
  \path (#1.west)+(0,0) node (#1w) {};
  \path (#1.east)+(0,0) node (#1e) {};
  \path (#1.south west)+(0,0) node (#1sw) {};
  \path (#1.north west)+(0,0) node (#1nw) {};
  \path (#1.south |- #1.west)+(-0.35,-0.55) node (#1x) {};
  \path (#1.north |- #1.east)+(+0.35,+0.55) node (#1y) {};         
  \path[reftrie]
  (#1x) rectangle (#1y);
}
\newcommand{\RefLeaf}[2]{
  \path (#1.north)+(0,0.45) node {#2};
  \path (#1.west)+(0,0) node (#1w) {};
  \path (#1.south west)+(0,0) node (#1sw) {};
  \path (#1.south |- #1.west)+(-0.65,-0.40) node (#1x) {};
  \path (#1.north |- #1.east)+(+0.65,+0.40) node (#1y) {};         
  \path[refleaf]
  (#1x) rectangle (#1y);
}
\newcommand{\RefLeafC}[2]{
  \path (#1.north)+(0,0.55) node {#2};
  \path (#1.west)+(0,0) node (#1w) {};
  \path (#1.south west)+(0,0) node (#1sw) {};
  \path (#1.south |- #1.west)+(-0.65,-0.40) node (#1x) {};
  \path (#1.north |- #1.east)+(+0.65,+0.40) node (#1y) {};         
  \path[refleaf]
  (#1x) rectangle (#1y);
}
\newcommand{\ThunkReEval}[1]{
    \path (#1.south |- #1.west)+(-0.85,-0.80) node (#1x) {};
    \path (#1.north |- #1.east)+(+0.85,+0.80) node (#1y) {};
    \path[thunkreeval] (#1x) rectangle (#1y);
}
\newcommand{\RefConsChange}[1]{
    \path (#1.south |- #1.west)+(-0.60,-0.95) node (#1x) {};
    \path (#1.north |- #1.east)+(+0.60,+0.95) node (#1y) {};
    \path[refchange] (#1x) rectangle (#1y);
}
\newcommand{\RefTrieChange}[1]{
    \path (#1.south |- #1.west)+(-0.45,-0.65) node (#1x) {};
    \path (#1.north |- #1.east)+(+0.45,+0.65) node (#1y) {};
    \path[refchange] (#1x) rectangle (#1y);
}
\newcommand{\RefLeafChange}[1]{
    \path (#1.south |- #1.west)+(-0.75,-0.55) node (#1x) {};
    \path (#1.north |- #1.east)+(+0.75,+0.55) node (#1y) {};
    \path[refchange]
    (#1x) rectangle (#1y);
}
\newcommand{\ConsTailPtr}[2]{
         \path[refptr] (#1.third)+(0.091,0.1) circle (1.3pt) edge[->] (#2) {};
}
\newcommand{\ConsTailPtrBend}[3]{
         \path[refptr,#3] (#1.third)+(0.091,0.1) circle (1.3pt) edge[->] (#2) {};
}
\newcommand{\LeftNilPtr}[1]{
        \draw (#1)+(0.021,.18) node[nilcross] {};
}
\newcommand{\RightNilPtr}[1]{
        \draw (#1.second)+(0.091,.071) node[nilcross] {};
}
\newcommand{\LeftRefPtr}[2]{
        \path[bend left,draw=black,fill=black,>=stealth,thick, line width=1.2pt] (#1)+(0.021,.18) circle (1.3pt) edge[->] (#2) {};
}
\newcommand{\LeftRefPtrBend}[3]{
        \path[draw=black,fill=black,>=stealth,thick, line width=1.2pt,#3] (#1)+(0.021,.18) circle (1.3pt) edge[->] (#2) {};
}
\newcommand{\RightRefPtr}[2]{
        \path[bend right,draw=black,fill=black,>=stealth,thick, line width=1.2pt] (#1.second)+(0.091,.071) circle (1.3pt) edge[->] (#2) {};
}
\newcommand{\MysRefPtr}[2]{
        \path[draw=black,fill=black,>=stealth,thick, line width=1.2pt] (#1) circle (0pt) edge[->] (#2) {};
}
\newcommand{\RightRefPtrBend}[3]{
        \path[draw=black,fill=black,>=stealth,thick, line width=1.2pt,#3] (#1.second)+(0.091,.071) circle (1.3pt) edge[->] (#2) {};
}
\tikzset{
  lstvarptr/.style={
     dotted, fill=gray!70!white, draw=gray!70!white,line width=1pt,
  },
  reftrie/.style={
    fill=violet!10,rounded corners, line width=1.5pt, draw=black
  },
  refleaf/.style={
    fill=violet!10,rounded corners, line width=1.5pt, draw=black
  },
  nilcross/.style={strike out, draw=black, minimum size=3.6mm, inner sep=0pt, outer sep=0pt, line width=1.2pt},
  dcgrefget/.style={
    dotted, fill=blue!30!white, draw=blue!30!white,line width=4pt
  },
  dcgthunkforce/.style={
    dotted, fill=blue!30!white, draw=blue!30!white,line width=6pt
  },
  REDOdcgrefget/.style={
    dotted, fill=blue!90!white, draw=blue!90!white,line width=4pt
  },
  REDOdcgthunkforce/.style={
    dotted, fill=blue!90!white, draw=blue!90!white,line width=6pt
  },
  refptr/.style={
        fill=black, draw=black,>=stealth,thick, line width=1.2pt
  },           
  ptrsnd/.style={
    bend left,
    draw=black,
    fill=black,
    >=stealth,
    thick
  },
  lstvar/.style={
      dotted,
      rectangle, 
      very thick, 
      draw=gray!80, 
      fill=gray!05,
      node distance=10pt,
    },
    pointer/.style={    
      draw=black,
      text centered,
      fill=black,
      circle,
    },
    nil/.style={
      fill=yellow!05,
      rectangle split,
      rectangle split parts=1,
      draw,
      text centered,
    },
    cons/.style={
      baseline=-.5ex,
      fill=yellow!05,
      rectangle split,
      rectangle split parts=2,
      rectangle split horizontal,
      draw,
      text centered,
      minimum size=5mm,
    },
    leaf/.style={
      baseline=-.5ex,
      fill=orange!05,
      rectangle split,
      rectangle split parts=2,
      rectangle split horizontal,
      draw,
      text centered,
      minimum size=5mm,
    },
    thunk/.style={
      text width=15mm,
      fill=blue!05,
      rectangle,
      rounded corners=4mm,
      draw,
      line width=1.5pt,
      text centered,
      minimum size=15mm,
    },
    thunkreeval/.style={
      snake=bumps, segment length=12pt, segment amplitude=1.5mm, fill=blue!40, line width=0.5pt, draw=blue
    },
    refchange/.style={
      snake=zigzag, segment length=5pt, segment amplitude=0.5mm, fill=red!40, line width=1pt, draw=red
    },
    pad/.style={
      rectangle,
      minimum size=10mm,
    },
    xpad/.style={
      circle=0mm,
    },
    xpadt/.style={
      circle=0mm,
      minimum size=15mm,
    },
    xpadConsV/.style={
      circle=0mm,
      minimum size=9mm,
    },
    xpadTrie/.style={
      circle=0mm,
      minimum size=5mm,
    },
    xpadLeaf/.style={
      circle=0mm,
      minimum size=13mm,
    },
    apad/.style={
      circle=0mm, node distance=2mm,
    },
    smallpad/.style={
      node distance=0mm, circle=0mm,
    },
    prepad/.style={
      circle=0mm, node distance=0mm,
    },
    postpad/.style={
      circle=0mm, node distance=4mm,
    },
    trie/.style={
      fill=orange!05,
      rectangle split,
      rectangle split parts=2,
      draw,
      text centered,
      minimum size=5mm,
    },
    consV/.style={
      fill=yellow!05,
      rectangle split,
      rectangle split parts=3,
      draw,
      text centered,
      minimum size=5mm,
    },
    dcgedge/.style={
      bend right,
      fill=blue!50!white,
      draw=blue!50!white,
      line width=2pt
    }
    terminal/.style={
      rectangle,minimum size=6mm,rounded corners=3mm,
      very thick,draw=black!30,
      top color=white,bottom color=black!20,
      font=\ttfamily
    },
    node distance=6mm,
    every on chain/.style={},
    every join/.style=
}

\newcommand{\RuleHead}[1]{\text{\raisebox{1em}[0pt]{\ensuremath{\mathsz{\ifnum\OPTIONConf=1 14pt\else 18pt \fi}{#1}}}}~~~~~}

\newcommand{\Fungi}{\textsf{Fungi}\xspace}

\newcommand{\inj}[1]{\keyword{inj}_{#1}\,}
\newcommand{\Inj}[1]{\inj{#1}}

\newcommand{\rulename}[1]{\text{\normalfont\textsf{#1}}}

\newrulecommand{TAbort}{\targettyperulename{Abort}}

\definecolor{darkgreen}{rgb}{0,0.5,0}
\definecolor{darkpurple}{rgb}{0.5,0,0.5}

\newcommand{\code}[1]{\lstinline[basicstyle=\ttfamily]|#1|}
\newcommand{\hlcode}[1]{{\hl{\texttt{#1}}}}

\begin{document}

% %\title{A type system for Adapton}
\ifnum\OPTIONConf=1
    %% \title[Typed Adapton]{Typed Adapton:}
    %% \subtitle{Practical Refinement Types for
    %%   Nominal Allocation and Memoization
    %% }
%    \title[Typed Adapton]{Typed Adapton:}
%    \subtitle{Refinement types for incremental computations with unambiguous names}
%    \subtitle{Refinement types for dynamic dependencies with precise names}
%    \title{Refinement~types~for~computation~dependencies with~precise~names}
%    \title{Refinement~types~for~precisely named cache locations}
    %% \title{\Fungi: A typed, functional language for programs that
    %%   dynamically name their own cached dependency graphs}
%     \title{\Fungi: A type and effect system for incremental computation with names}
%     \title{\Fungi: Types and effects for incremental computation with names}
     \title{\Fungi: Typed incremental computation with names}
%    \subtitle{Incremental computations with precise names}

\else
%    \title{Refinement~types~for~precisely named cache locations}
\fi

% new ACM style:
%
% %% Title information
% \title[Short Title]{Full Title}         %% [Short Title] is optional;
%                                         %% when present, will be used in
%                                         %% header instead of Full Title.
% \titlenote{with title note}             %% \titlenote is optional;
%                                         %% can be repeated if necessary;
%                                         %% contents suppressed with 'anonymous'
% \subtitle{Subtitle}                     %% \subtitle is optional
% \subtitlenote{with subtitle note}       %% \subtitlenote is optional;
%                                         %% can be repeated if necessary;
%                                         %% contents suppressed with 'anonymous'

\iffalse
\newcommand{\MattSays}[1]{\relax}
\newcommand{\MattSaysTODO}[1]{\relax}
\else
\newcommand{\MattSays}[1]{{\color{blue}{\textbf{Matt:}~#1}} \color{black}}
\newcommand{\MattSaysTODO}[1]{{\color{blue}{\textbf{TODO (Matt:)}~#1}} \color{black}}
\fi

% leave this line here
\begin{abstract}

Incremental computations attempt to exploit input similarities over
time, reusing work that is unaffected by input changes.
To maximize this reuse in a general-purpose programming setting, 
programmers need a mechanism to identify dynamic allocations 
(of data and subcomputations) that correspond over time.

We present \Fungi,
a typed functional language for incremental computation with \emph{names}.
Unlike prior general-purpose languages for incremental computing,
\Fungi's notion of names is formal, general, and statically verifiable.
%
% In particular,
\Fungi's type-and-effect system permits the programmer
to encode (program-specific) local % uniqueness
invariants about names,
and to use these invariants to establish \emph{global uniqueness} for their
composed programs, the property of using names correctly.
We prove that well-typed \Fungi programs respect global uniqueness.

We derive a bidirectional version of the type and effect system, and
we have implemented a prototype of~\Fungi in Rust.  %, as a
% deeply embedded DSL.
%
We apply \Fungi to a library of incremental collections,
showing that it is expressive in practice.

\end{abstract}

\maketitle

\section{Introduction}
\label{sec:intro}

\newcommand{\ParAuthorNote}[1]{
  \noindent
  \begin{tabular}{p\textwidth}
    \hline
    \\[-5mm]
  \end{tabular}
}

\newcommand{\Input}{\ensuremath{\textsf{Inp}}}
\newcommand{\Output}{\ensuremath{\textsf{Out}}}

In many software systems,
a fixed algorithm runs repeatedly
over a series of \emph{incrementally changing} inputs
($\Input_1, \Input_2, \ldots$),
producing a series of \emph{incrementally changing}
outputs ($\Output_1, \Output_2, \ldots$).
For example, programmers often change only a single line of source code
and recompile, so $\Input_{t}$ is often similar to $\Input_{t-1}$.

The goal of incremental \emph{computation}
is to exploit input similarity by reusing work from previous runs.
If the source code $\Input_{t}$ is almost the same as $\Input_{t-1}$,
much of the work done to compile $\Input_{t}$
and produce the target $\Output_{t}$
can be reused.
In many settings, this reuse leads to asymptotic improvements in running time.

Such improvements are possible when the recomputation is \emph{stable}:
when the work done by run $t - 1$, %
producing output $\Output_{t-1}$ from input $\Input_{t-1}$, %
is similar to the work needed for run $t$ to produce output $\Output_{t}$ from $\Input_{t}$.
In some cases, such as total replacement of the source program being compiled,
stability is impossible.
Thus, a central design question is how to maximize stability.

Consider a simple program
that applies a binary operation $g$ to two parts ($x$, $y$) of the input, 
and then applies another binary operation $f$ to the result of $g$
and a third part ($z$) of the input.
This program has three inputs, one output, and one \emph{intermediate result}
(the result of $g$ on $x$ and $y$).
Assuming efficient equality tests for $x$, $y$ and the result of $g$,
we can save this intermediate result and, potentially, reuse it across runs.

%\clearpage
\begin{figure}
%% \begin{minipage}{\textwidth}
%% \tiny
%% \begin{verbatim}
%% [
%%   try to arrange side by side (3 minipage environments?): 
%%   First run    Second run    Third run
%% ]

%% f(g(x, y), z)

%% First run

%%   input          output

%%     x---
%%         \
%%          g-----f-
%%         /     /
%%     y--'     /
%%             /
%%            /
%%     z-----'

%% Second run     [show thunk of g as being reused, thunk of f as recomputed]

%%   input          output

%%     x---
%%         \
%%          g-----f-
%%         /     /
%%     y--'     /
%%             /
%%            /
%%    z' ----'

%% Third run     [show thunks of g and f as recomputed]

%%   input          output

%%    x' --
%%         \
%%          g-----f-
%%         /     /
%%     y--'     /
%%             /
%%            /
%%    z' ----'

%% [arrow directions to match dedup figures]
%% \end{verbatim}
%% \end{minipage}
\hspace{-0.9in}~\includegraphics[width=1.15\textwidth]{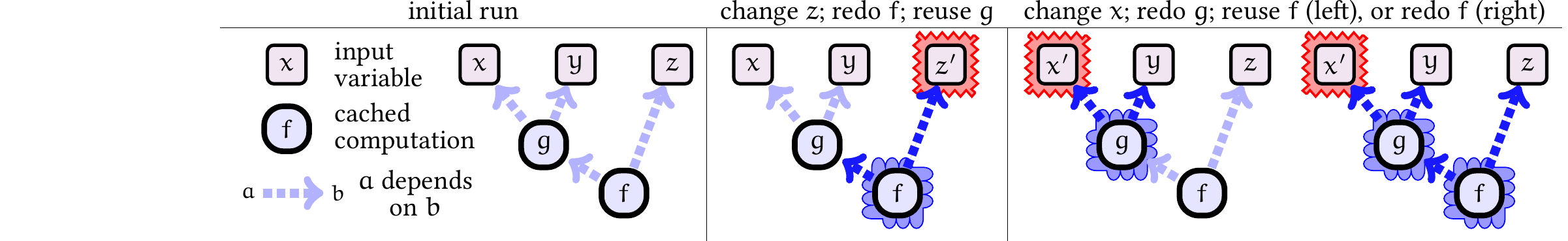}
\caption{Reuse across several program runs}
\label{fig:example}
\end{figure}
% XXX explain conventions, \eg arrow direction

\Figref{fig:example} shows some example runs.
In the first run, we have stored $g(x, y)$.
In the second run, the user has changed the input $z$ to $z'$---%
but since the inputs $x$ and $y$ have not changed,
we can reuse the result $g(x, y)$ and perform only the operation $f$.
In the third run, the user has changed $x$ to $x'$,
which requires doing the operation $g$ again.

%XXX TODO Talk about the input of g changing, but the result of g not changing, and reusing the result of f.

Thus, between the first and second runs we had to recompute only $f$;
between the second and third runs, we had to recompute $g$.
Depending on whether $g$'s result changes, we might recompute $f$ as well.

At this low level of complexity, it may seem straightforward to ensure that
the incremental program is both \emph{consistent} and \emph{efficient}:

\begin{itemize}
\item An incremental program is \emph{from-scratch consistent}
  if its output matches the output that would be produced by running
  the program from scratch (that is, without using saved intermediate results).

  As long as we reuse the result of $g$ only when $x$ and $y$ have not changed,
  and reuse $f$ only when $g$ and $z$ have not changed,
  this simple program is from-scratch consistent.
  
\item An incremental program is \emph{incrementally efficient}
  (or \emph{achieves incremental efficiency})
  if it does only the \emph{necessary} new work.

  As long as we \emph{always} reuse the result of $g$ when $x$ and $y$ have not changed,
  and \emph{always} reuse $f$ when $g$ and $z$ have not changed,
  this simple program is incrementally efficient.
\end{itemize}

For nontrivial programs, however, achieving both incremental consistency and incremental efficiency
can be extremely difficult.
Consider GNU \texttt{make}, a relatively simple build system:
it achieves consistency (at least in principle) only by working at a very
coarse level of granularity---entire programs (\texttt{cc}, \texttt{ld}, etc.) and entire files.
Opportunities to reuse work \emph{within} a 5,000-line input to \texttt{cc} are missed,
and understandably so:
compilers are large systems that use complex data structures and clever algorithms.
Merely comparing file modification times (or even file contents)
cannot utilize, say, the fact that the result of a liveness analysis has not changed.
(Or, that the analysis has changed \emph{slightly}, which creates many subtle
dependencies.)
% Since changing 1 line of a 10,000-line C file results in a different object file,

The gold standard for incremental programs is to painstakingly design an incremental
algorithm that explicitly saves results and reuses work,
perhaps in very clever ways.  % Could cite that thesis that Umut always cites
In many development settings, it is not feasible to expend that kind of effort.
Rather than giving up on incremental software (by not attempting to reuse work at all)
or using simplistic approaches (along the lines of \texttt{make})
that miss many opportunities for reuse,
we should offer incremental programming \emph{languages}
that
% is to
allow programmers to easily build incremental programs
that are correct and efficient, \emph{at scale}.
Thus, an incremental programming language should enable programmers
(1) to store and reuse intermediate results, without drastically changing their source program;
(2) to exploit similarities (between inputs, and between stored results),
  including for highly structured input data
  and nontrivial data structures;
(3) to easily combine smaller incremental programs into incremental systems.
Moreover, the language should make it as easy as possible to obtain
both correctness and efficiency.

Incremental languages can be categorized by their breadth of applicability,
with domain-specific languages at one end of the spectrum
and general-purpose languages at the other;
the language in this paper is general-purpose.
The central advance we make is in \emph{statically} verifying
an important aspect of incremental programs:
that subcomputations are named \emph{uniquely} within each run.

The tiny program shown above is not adequate to illustrate the need
for unique naming:
the program's input has no interesting structure,
and there is only one intermediate result.
We argue the need for names themselves here;
we will discuss a concrete example, illustrating the need
for unique names, in \Secref{sec:overview}.

To reuse a unit of work,
we must observe that the newer result \emph{corresponds to} the older result.
The program $f(g(x,y),z)$ uses no control structures
and performs the operations $f$ and $g$ exactly once,
so it is immediate that $g(x,y)$ in the second run
corresponds to $g(x,y)$ in the first run.
Moreover, we say that $g(x',y)$ in the \emph{third} run corresponds to 
$g(x,y)$ in the second run,
even though $x'$ is (probably) not equal to $x$
and hence $g(x',y)$ is (probably) not equal to $g(x,y)$:
Correspondence is not equality; instead, correspondence is the idea that
two uses of $g$ happen ``in the same place''.

The correspondence of $x'$ to $x$, and $z'$ to $z$, is even more immediate.
But what if, instead of giving three discrete inputs ($x$, $y$, $z$),
we gave a list of integers as input?
If the change in input across runs is confined to specific list elements,
say replacing the second element $22$ with $23$,
we could say that the $k$th element of the previous input corresponds to
the $k$th element of the current input.
However, if the change is to \emph{insert} an element in the input list,
identifying the $k$th element at time $t - 1$
with the $k$th element at time $t$ won't work:
the small change of inserting a single element will look like
the complete replacement.
We need some notion of \emph{identity} to realize that, if we insert an element
at (say) the head of the list, the 1st element at time $t - 1$ corresponds
to the 2nd element at time $t$, the 2nd element at time $t - 1$ corresponds
to the 3rd element at time $t$, and so forth.

% Use the term ``name'' below?
% One idea: Switch from using ``notion (of identity)'' everywhere below,
% to calling this needed thing a naming strategy,
% which augments an existing algorithm with a notion of identity.

In our setting of a general-purpose language,
there is no one-size-fits-all notion of identity.
Instead, we need to enable programmers to choose a notion of identity
that is appropriate for each program---%
a notion that exposes appropriate correspondences,
and hence enables reuse.
We call this notion of identity a \emph{naming strategy}.
Choosing a naming strategy that actually enables reuse is often difficult;
the study of incremental cost semantics, which describe the potential for reuse,
is a research area in itself.
Our contribution is to make it easier for programmers to experiment with
different naming strategies: the Fungi type system rules out a large class of
\emph{naming errors} that, in earlier languages such as Nominal Adapton \citep{Hammer15},
could only be caught at run time.

In \Tableref{tab:related},
we compare \Fungi to some related approaches.
The first two rows list work on incremental languages
for substantially different programming models;
those systems' answers to the question of
how to identify corresponding subcomputations
do not apply in our setting (nor would our answer apply in theirs).
We briefly discuss these two systems, and other work in substantially
different settings, in \Sectionref{sec:related}.
The remaining rows in the table---starting with AFL \citep{Acar02}---%
list general-purpose incremental programming languages
that endeavor to provide a standard programming model
with (relatively) \emph{lightweight} incrementality;
as we noted above, we want to support incrementality
without requiring programmers to drastically change their source programs.
Within this broad setting, we can observe an evolution from no mechanism
to identify corresponding subcomputations (AFL in 2002)
to informal or specialized mechanisms (several papers through 2012
and Adapton in 2014), and then to formal mechanisms.

%%%%%
%%%%% \input{table.tex}
%%%%%

\begin{table}[th]
\def\nope{--}
\def\Yup{\checkmark}
\def\BRK{\\[3pt]}
\let\NOTE\undefined
\newcommand{\NOTE}[1]{\textbf{(#1)}\xspace}
\newcommand{\NOTEDEF}[1]{\NOTE{#1}}

\newcommand{\aflkeysnote}{\NOTE{2}}
\newcommand{\aflkeysnotedef}{\NOTEDEF{2}}

\newcommand{\LeftColTwo}[2]{%
    \tabularenvr{%
      #1%
      \\
      #2%
      \BRK%
    }%
}
\newcommand{\ModelTwo}[2]{\hspace*{-0.57ex}\tabularenvl{#1\\#2 \BRK}}
  
\centering

% correspondence
% mechanism
  \runonfontsz{8pt}
  \begin{tabular}[t]{r | lccc}
    \multicolumn{1}{l}{\tabularenvl{
    ~\\
    ~\\
    ~\\
    approach}}
    &
      \tabularenvl{
      ~\\
      ~\\
        programming \\
        model
      }
    &
      \tabularenvl{
         mechanism \\
         to identify \\
         corresponding \\
         subcomputations
%        names
      }
    &
      \tabularenvl{
      ~\\
      ~\\
        detection of \\
          naming errors
      }
    \\ \cmidrule{2-4}
\iffalse
        %
        %  nope
        %    & \nope & \nope & none & \nope
        % \BRK
            DITTO \citep{Shankar07}
            &
            \ModelTwo{pure subset}{of Java \NOTE{1a}}%
        \BRK
            Incoop \citep{Bhatotia11}
            &
            \ModelTwo{
              restricted functional
            }{
              programming (MapReduce)
            }
            &
            time stamps
        \BRK
            i3QL
              \citep{Mitschke14}
            &
              query language \NOTE{2a}
            % &
            %   (2b)
            &
              ?
            % &
            %   ~
        \BRK
            % http://www.informatik.uni-marburg.de/~pgiarrusso/ILC/
            \LeftColTwo{
                incremental $\lambda$-calculus
            }{
                \citep{Cai14}
            }
           &
             \ModelTwo{%
               restricted functional
             }{%
               language
               \NOTE{3a}%
             }
        %
        %   &  \NOTE{3b}
        %  Reactive Imperative Programming with Dataflow Constraints  
        \BRK
\fi
  \citet{Demetrescu11}
  &
    reactive/imperative
  &
    memory address
  &
    n/a
% \\  general purpose: yes.
% \\  general recursion: yes (double check).
% \\  corresp mechanism: ?
% \\  detection: ?
\BRK
%  https://www.microsoft.com/en-us/research/publication/two-for-the-price-of-one-a-model-for-parallel-and-incremental-computation/
    \LeftColTwo{
      Concurrent revisions
    }{
      \citep{Burckhardt11:oopsla}
    }
    &
      \ModelTwo{
        revision-based
      }{
        imperative programming
      }
    &
      call graphs \NOTE{1}
    &
      n/a
% \\  general purpose: yes.
% \\  general recursion: yes (double check).
% \\  corresp mechanism: same global call graph position; does not tolerate small edits to the call graph structure without sacrificing stability/reuse.
% \\  detection: n/a
\BRK
      AFL \citep{Acar02}
    & functional language
    & none   & n/a    
\BRK
      \citet{Carlsson02}
    & functional language
    & none   & n/a    
\BRK
% 1. ML Workshop 2005: http://people.cs.uchicago.edu/~blume/papers/ml05-sal.pdf, (In particular, see Fig 2, page 5; mkLift and mkLiftCC can be misused.)
% published in Elsevier ENTCS, 2006: BibTeX key Acar06 in adapton.bib
%    Programming model: Functional.  Reuse via informal keys.  Dynamic detection (??).
      \citet{Acar06,Acar06Experimental}
      &
      functional language
      & keys (informal)
      & \aflkeysnote
\BRK
%    \LeftColTwo{
      DeltaML \citep{AcarLeyWild08}%;
%    }{
%      [\citeauthor{Hammer08} \citeyear{Hammer08};
%      \\
%      \citeauthor{Hammer09} \citeyear{Hammer09}, \citeyear{Hammer11};
%      \\
%      \citeauthor{HammerThesis} \citeyear{HammerThesis}%
%      ]
%    }
    & functional language
    % & \Yup
    & keys (informal)
    & \aflkeysnote%
      % "sac style"
\BRK
      CEAL \citep{Hammer09}
    & imperative language
    % & \Yup
    & keys (informal)
    & \aflkeysnote %
      % "sac style"
\BRK
      implicit SAC \citep{Chen11,Chen12}
    & functional language
    & keys (informal)
    & \aflkeysnote
\BRK
%    \LeftColTwo{
      Adapton
%    }{
      \citep{Hammer14}
%    }
    & functional language
    & \tabularenvc{structural \\ (hash-consing) \vspace*{0.8ex}} % informal  % or "structural", or "hash-consing"
    & n/a
\BRK
%    \LeftColTwo{
      Nominal Adapton
%    }{
      \citep{Hammer15}
%    }
    & functional language
    % & \Yup
    & names (formal) %${}^\dagger$)
    & dynamic
                             % first-class names
\BRK
%    \LeftColTwo{
      Fungi
%    }{
      (this paper)
%    }
    & functional language
    & names (formal) % ${}^\ddagger$)
    & static
                             % first-class names
  \end{tabular}

  \medskip
\raggedright

\iffalse
    $\dagger$ each global pointer name formally consists of a list of simpler names (bitstrings), 
    each constructed indirectly (via \code{fork}).  Bitstrings identify namespaces.

    $\ddagger$ each name formally consists of a binary tree, constructed directly
    (via direct binary composition), 
    and indirectly (via name function application, and by \code{scope}).  
    Name functions identify namespaces.
\fi

\runonfontsz{8pt}
%    \NOTEDEF{0a} n/a / static dependency graph / trivial correspondence
%    \\
  \NOTEDEF{1}
      position in global call graph;
      small changes in call graph structure prevent reuse
% \\    \NOTEDEF{1a} DITTO is restricted to run-time invariant checking
% \\ \NOTEDEF{2a} i3QL is an incremental query language embedded within Scala
% \\  \NOTEDEF{3a} $\lambda$-calculus with only tables (multiset-like structures); no general recursion
% \\  \NOTEDEF{3b} static differentiation implies static det.\ of correspondences
%      \\  (TODO: double-check this characterization)
\\  \aflkeysnotedef  % ("hint fail";
      fall back to a global counter---preventing reuse now, in the future, or both

  \bigskip

  \caption{Some approaches to incremental computation}
  \label{tab:related}
\end{table}

\paragraph{Contributions}
We make the following contributions:

\begin{itemize}
%%% core calc, with type and effect system
\item
  We develop a type-and-effect system for a general-purpose incremental programming language
  (Sections \ref{sec:progsyntax} and \ref{sec:typesystem}).
  Using refinement types, the system statically relates names
  to allocated data (references)
  and computations (thunks);
  it supports a set of type-level operations on names that is large enough to describe
  sophisticated uses of names, but small enough for decidable type checking.
  
%%% meta theory: type safety; “effect safety” => unique naming
\item
  In \Sectionref{sec:metatheory},
  we prove that the effects tracked by our system are sound with respect to
  our dynamic semantics (\Sectionref{sec:dynamics}).
  As a consequence, our type system ensures, \emph{statically},
  that names are unique within each run of the program---%
  a property that, previously, could only be checked dynamically \citep{Hammer15}.
  In nontrivial programs, this \emph{global uniqueness} property
  is a consequence of \emph{local uniqueness} properties
  that are specific to particular algorithms and data structures;
  see \Sectionref{sec:overview}.

% remark: why big-step

%%% implemented; can type and run examples
\item
  We implement the type system, and demonstrate its applicability to
  a variety of examples (\Sectionref{sec:impl}).
\end{itemize}

% 2018.07.06

\section{Overview}
\label{sec:overview}

In this section, we use an example program to give an overview of
\Fungi as a typed language for incremental computation with names.
Specifically, we consider the from-scratch semantics, typing, and
incremental semantics of \code{dedup}, a list-processing % \Fungi
function that removes duplicates:
the output list retains only the first occurrence of each input list element.

% As we demonstrate,
The implementation of \texttt{dedup} uses names to
create correspondences between similar inputs,
leading to incremental reuse via an efficient application of
a (general-purpose) change propagation algorithm.
The correctness of change propagation relies on the global uniqueness of allocation names, explained below.

% Meanwhile,
The \Fungi type system ensures that % the author of  %%% how would a human observe global uniqueness?
\code{dedup} satisfies global uniqueness; to do so,
the \Fungi programmer uses types
to express several local uniqueness invariants.
Before discussing this example, we briefly discuss these naming
properties, which are each fundamental to the novel design of \Fungi as a language for typed incremental computation with names.

\subsection{Naming properties}
\label{sec:properties}

%\subsubsection{Principles of unique names}
%\subsubsection{Naming properties}
%\label{sec:properties}
%
% The type system of \Fungi permits the \Fungi programmer to encode and
% observe the following \emph{principles of unique names}:
Our \Fungi type system enforces the \emph{global uniqueness} of names.
For nontrivial programs,
global uniqueness requires \emph{local uniqueness} of names;
our type system also checks local uniqueness properties
as stated by the programmer.

%\item \emph{Unique \textbf{pointer names} for cache locations}: 
%
\paragraph{Global uniqueness of allocation names:}
For every allocated reference cell or thunk,
the name used to identify the allocated reference
(or thunk) is unique.
%  When a subcomputation uses a sequence of names to allocate pointers
%  for data (references) and computations (thunks), each allocated name
%  in the sequence is unique.

% \item \emph{Unique \textbf{logical names} for places}: 
%
\paragraph{Local uniqueness properties:}
The data structures in an incremental program may contain names.
For example, if we map over a list, we may need to associate the
third element of the input list with the third element of the output list.
The name used to represent ``being the third element'' may then occur
within related \emph{pointer} names, such as the pointer names of the
third element of the input and the third element of the output.
The name that represents ``being the third element'' may be stored
in several different lists, but it should not occur more than once
within each list: the input list cannot have two third elements.
Since the appropriate local uniqueness properties depend on the details of each program,
they cannot be given \emph{a priori}.
Instead, the programmer or library designer expresses the appropriate properties,
using the \Fungi type system.

In general, local uniqueness---in the form appropriate to each program---%
is needed to ensure global uniqueness.
Our type system rules out, statically, violations of global uniqueness \emph{and}
violations of local uniqueness.
While previous systems such as Nominal Adapton
included constant-time dynamic checks to catch violations of global uniqueness,
most local uniqueness properties cannot be checked in constant time.

Since local uniqueness violations can lead to \emph{subsequent}
global uniqueness violations, being able to statically ensure local uniqueness
rules out a large class of subtle errors---much like the advanced type system
of the Rust language rules out dangling pointers.  
As we show below (\Secref{sec:errors}), some violations in
these principles are only triggered by \emph{certain} inputs, which
may be unlikely, and thus unlikely to show up in randomized dynamic
tests.

% Without a static check, a violation of the first principle (unique
% logical names) often leads to a violation in the second principle
% (unique pointer names), since logical place names are used, sometimes
% in combination with other (dynamic) nominal data, to identify
% data and subcomputations.
%
%
By enforcing these principles of unique names statically, \Fungi
programs enjoy the guarantees they afford, e.g., that change
propagation will work correctly.

These principles about names, which are fundamental to general-purpose
incremental computation, have been applied in some incremental
computing systems of the past, but until now, have not been codified
formally, or statically verified (see \Tableref{tab:related} for details).

In some past systems (based on self-adjusting computation), the
runtime dynamically detects and tolerates violations of these uniqueness properties---%
the names are called ``keys'', and are viewed as hints
that can be wrong, or non-unique.
In the cases that they are not unique, the caching/allocation mechanism falls back to
using a global counter.
%%% > What about non-monotonic SAC?
%%% 
%%% I'm not sure this was ever implemented.
%%%
%%% If it was, I still think there is a total order data structure
%%% (for a single, monolithic trace), that keys act as hints, and that
%%% the system falls back to a global counter.
%
In turn, this cache location choice is not based on the current input,
is not functional, and consequently, it will generally not be reusable
as a ``replayed'' allocation in subsequent invocations of change
propagation on similar inputs.

In other systems (Nominal Adapton), the runtime system simply triggers
a dynamic error for violations of global uniqueness.

No prior system of which we are aware permits programmers to
systematically encode or check local uniqueness, either statically, or
dynamically (which would be expensive).

Next, to make these ideas concrete, we consider an example.

\subsection{The program listing and dynamic semantics of \code{dedup}}
\label{sec:dedup-tile}

\begin{figure}
\begin{tabular}{c@{\hspace{10pt}}|@{\hspace{10pt}}c}
\begin{lstlisting}
idxtm Dedup : NmSet idx-> NmSet
  = lam~X.<~@t~>~@@~(Insert X) ^ <~@dd~>~@@~X ^ <~@r~>~@@~X

dedup : forall~X1^X2:NmSet.
  List[[X1]] -> Trie[[X2]] -> List[[X1]] 
  writeset~[[Dedup X1]]

dedup l t =
  match (get l) with
  Nil => l
  Cons x y ys =>
    let (tx,b) = scope<[@t]> insert x y t
    let ddys = thunk<[@dd~@~x]>[<dedup ys tx>]
    if b then force ddys else
      ref<[@r~@~x]>(Cons x y (force ddys))
\end{lstlisting}
&
\begin{lstlisting}
type List[[X]] = Ref(ListNode[[X]])

type ListNode : NmSet idx=> Type
  Nil  : forall~X:NmSet. ListNode[[X]]
  Cons : forall~X1^X2:NmSet.
    Nm[[X1]] -> Nat -> List[[X2]] -> 
    ListNode[[X1^X2]]

type Trie[[X]] = Ref(TrieNode[[X]])

idxtm Insert : NmSet idx-> NmSet
insert : forall~X1^X2:NmSet. 
  Nm[[X1]] -> Nat -> Trie[[X2]] -> 
   (Trie[[X1^X2]], Bool) 
   writeset~[[Insert X1]]
\end{lstlisting}
\end{tabular}
\caption{The effect, type and code listing for \code{dedup} (left), and definitions for linked lists and hash tries (right).}
\label{fig:dedup}
%
%\end{figure}
%\begin{figure}
% 
  \vspace{10pt}
  \hspace{-0.9in}
  \includegraphics[width=\textwidth]{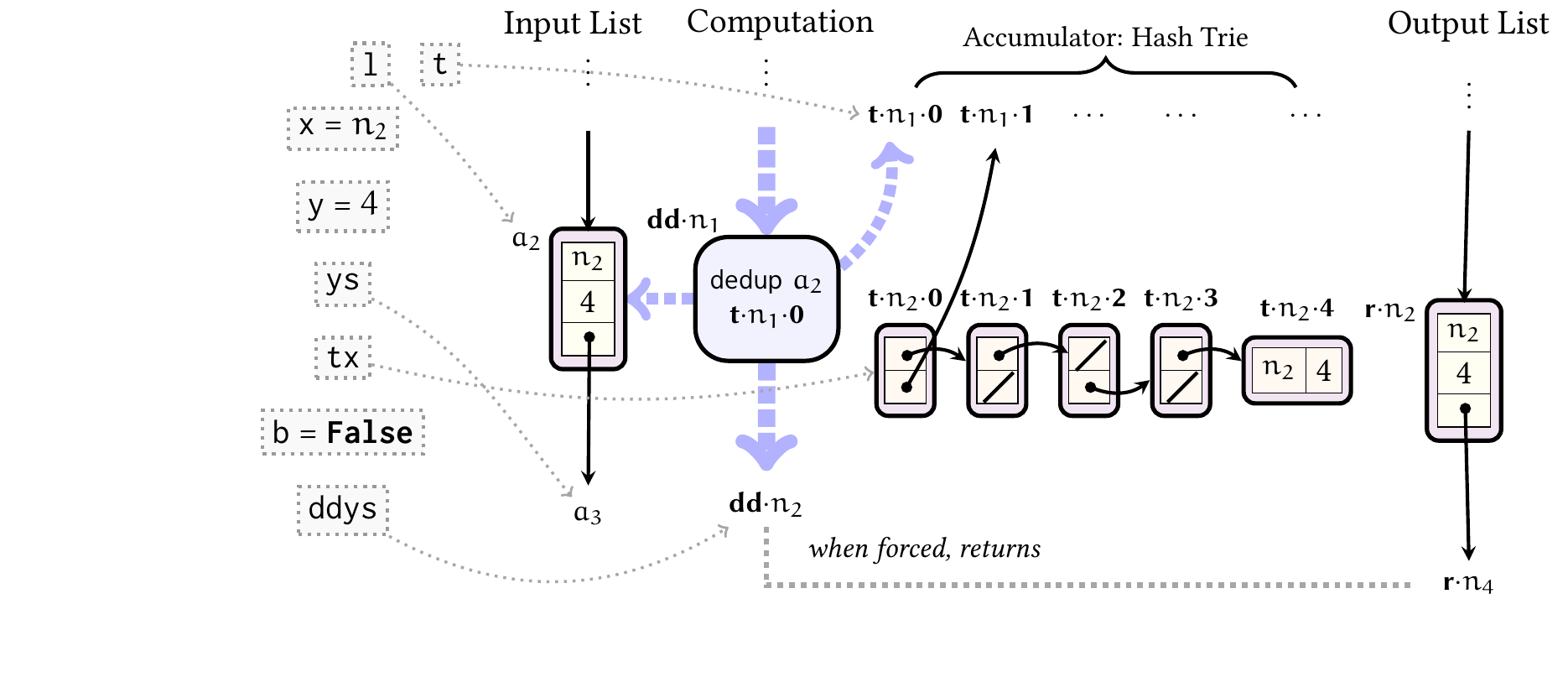}
  \vspace{-30pt}
  \caption{The static variables of~\code{dedup} (left column), and their dynamic values and structure (to the right),
    \\
    for the second iteration of \code{dedup} on the input list $[3,4,3,9]$, where the list element~\code{y} is~$4$.}
\label{fig:dedup-x4x}
\end{figure}

\Figref{fig:dedup} gives the program listing for \code{dedup},
including type declarations.
The right-hand column of the figure shows additional type declarations,
explained further below (\Secref{sec:dedup-types}).

First, let's consider an approximation of the declared type and code for
\code{dedup}, ignoring the \emph{index term} declaration
(\code{idxtm Dedup})
and % the appearance of
other type indices and effects.
The type declaration of \code{dedup} says that it accepts two arguments,
a list of type \code{List[[X1]]}
and a hash trie of type \code{Trie[[X2]]},
and returns a list of type \code{List[[X1]]}.
Before examining the type structure of \code{dedup} in more detail,
we consider the code, and its dynamic semantics.

Consider the initial run of \code{dedup} on the input list~$[3,4,3,9]$,
stored at the sequence of pointer addresses
$\langle a_1, a_2, a_3, a_4, a_5 \rangle$,
which store \texttt{Cons} cells and a terminal \texttt{Nil} value.
In addition to the elements $[3,4,3,9]$,
the \texttt{Cons} cells also contain a sequence of names (as values)
$\langle n_1, n_2, n_3, n_4 \rangle$,
with one name per \code{Cons} cell.

The \code{dedup} function uses these names to % functionally
determine its allocation names---the identities of allocated data and thunks.
Moreover, it stores these names (as values) within the allocated data.
Intuitively, these names identify the logical places of the
\code{Cons} cells in the input list,
and by copying these names into these other allocated values,
they permit the \code{dedup} program to create correspondences % not "correspondances"
with other logical places in its data. % structures that it allocates.
%\footnote{
%\MattSays{XX call back to specifically to ``in the same place'' description from intro section?}
%}
Further below, we will look at a full picture of this entire execution.

\Figref{fig:dedup-x4x} shows a single ``tile'' of this
execution, for the iteration on input cell~$a_2$
containing name~$n_2$ and input element~$4$.
The left side of the figure gives the rest of \texttt{dedup}'s static variables,
and relates them to their dynamic values and structure,
including the input list, accumulator and output list.
%%% There are no "columns" in that figure.
%
We consider the remainder of \Figref{fig:dedup-x4x} in the context of
the \code{dedup} algorithm (\Figref{fig:dedup}).   %, lower left
The \code{dedup} algorithm processes the input list~\code{l} using
structural recursion, retaining the first occurrence of each element
and filtering out subsequent occurrences.  % of these elements.
%
% Specifically,
We use a hash trie argument~\code{t} to
efficiently % maintain and check   % ``maintain whether''?
represent the set of input elements already processed.  % "has been seen earlier" is redundant
% seen in the prefix of input list~\code{l}.
%%% "the prefix of input list l": there is no one prefix of a list
%
The final \code{if}-\code{else} expression branches on this case:
if \code{b} is true, the element has already been seen
so we recur (\code{force ddys}) without building an output \code{Cons} cell;
if \code{b} is false (\code{else} branch),
the element has not been seen
and we construct an output \code{Cons} cell.
% , otherwise, it returns the result from its (\code{thunk}ed, and
% \code{force}d) recursive call.
%%% "else"..."otherwise" was confusing

In this tile (\Figref{fig:dedup-x4x}), the \code{Cons} case
of~\code{dedup} consists of handling the given named list position
($\texttt{x}=n_2$), holding a list element ($\texttt{y}=4$), and
recursively processing the elements in the tail of the \code{Cons}
cell ($\texttt{ys}=a_3$).
To do so efficiently, \code{dedup} inserts element~$\texttt{y}=4$ into
the current trie~($\texttt{t}$) using a helper function~\code{insert}.

The call to \code{insert} uses a \emph{write scope} (written \code{scope}),
which permits the call site to specialize the names
written by this function:
%%% "permits...to" feels clunky when we're talking about a specific thing
%%% that's happening
%
%By placing the call to \code{insert} in a write \code{scope} named by the
%constant name value~\code{@t},
%
the names written by function~\code{insert} are each prefixed
by the given name constant~\code{@t}.
(In listings, we include the symbol~\code{@} to distinguish the
constant name~\NmStr{t} from the static program variable~\code{t};
to save space, in figures we use the face~\NmStr{t} for this same name
constant \code{@t}, and omit the \code{@} symbol.)

As \Figref{fig:dedup-x4x} shows,
the value of variable~\code{t} is (pointer)
name~$\NmBinOp{\NmStr{t}}{\NmBinOp{n_1}{\NmNum{0}}}$, which contains the
prefix (sub-name)~\NmStr{t}, and suffix (sub-name)~\NmNum{0}.
Similarly, the value of variable~\code{tx} is (pointer)
name~$\NmBinOp{\NmStr{t}}{\NmBinOp{n_2}{\NmNum{0}}}$, which contains the
same prefix~\NmStr{t} and suffix~\NmNum{0}, but uses name~$n_2$ in place of name~$n_1$.
For reasons we explain below (\Secref{sec:errors}), omitting this write scope
name~\NmStr{t} would constitute a static typing error for this
program, as it would allow \emph{some} input lists to violate
global uniqueness (\Secref{sec:properties}).%
%

%%% no hyphen in explicitly-*
Next, \code{dedup} allocates an \emph{explicitly named} thunk
to identify its recursive call, giving the subexpression~\code{@dd~@~x},
which prepends the name constant \NmStr{dd}
to the name value of variable~\code{x},
in this case~$n_2$,
resulting in the dynamic name $\NmBinOp{\NmStr{dd}}{n_2}$.
As \Figref{fig:dedup-x4x} shows, our ``current thunk'' for this
iteration is~$\NmBinOp{\NmStr{dd}}{n_1}$, and the next iteration
(statically named \code{ddys}) is dynamically named
$\NmBinOp{\NmStr{dd}}{n_2}$, using the ``current name''~$n_2$.
As we explain in \Secref{sec:errors}, omitting this prefix~\NmStr{dd}
constitutes a static naming error,
since it violates global uniqueness (unique pointer names).

Finally, the \code{dedup} code executes the \code{else} case, since
input element~$4$ is not present in the input trie~$\texttt{t} = \NmBinOp{\NmStr{t}}{\NmBinOp{n_1}{\NmNum{0}}}$, and is making its first
occurrence with this \code{Cons} cell.
Similarly with \code{thunk} above, 
the \code{dedup} code allocates an \emph{explicitly named} reference cell
to identify its resulting list cell, giving the
subexpression~\code{@r~@~x}, which prepends the name constant
$\NmStr{r}$ to $\texttt{x}=n_2$,
resulting in the dynamic name~$\NmBinOp{\NmStr{r}}{n_2}$.
As \Figref{fig:dedup-x4x} shows, the ``next thunk'' that this
iteration forces, $\NmBinOp{\NmStr{dd}}{n_2}$, returns the list pointer
$\NmBinOp{\NmStr{r}}{n_4}$, which we store as the tail of the output
\code{Cons} cell~$\NmBinOp{\NmStr{r}}{n_2}$.
As above, omitting this prefix~\NmStr{r} violates global uniqueness.  %the principle of
%unique pointer names.
%
Other naming mistakes are possible as well; see \Secref{sec:errors}.

In the remainder of the overview,
we define the internal structure of names and
discuss reasoning about uniqueness (\Secref{sec:apartness}),
discuss the details of the \code{insert} function used by \code{dedup},
and consider static reasoning for \code{dedup}'s use of names (\Secref{sec:dedup-types}).

%%%%%%%%%%%%%%%%%%%%%%%%%%%%%%%%%%%%%%%%%%%%%%%%%%%%%%%%%
\subsection{Apartness for names, name sets and name functions}
\label{sec:apartness}

We briefly describe the structure of names, and discuss a notion that underpins
\Fungi's design: \emph{apartness} of names, name sets and name functions.

Our core calculus defines names as binary trees,
$n \bnfas \leafname \bnfalt \NmBin{n}{n}$.
In practice, we augment this definition in two small ways.
First, we extend the $\leafname$ production
with other terminal productions for
numbers and symbolic constants,
written $\NmNum{0}$ and $\NmStr{t}$ (respectively) in the example above.
For the purposes of reasoning formally, we assume (unspecified)
encodings of these terminal productions into the simple formal grammar above.
Second, we use a more lightweight notation for binary name composition:
% right-associative, infix
$\NmBinOp{n_1}{\NmBinOp{n_2}{n_3}}$
denotes
%%% using := to mean "notation for" is confusing; when a word works just as well, use it
$\NmBin{n_1}{\NmBin{n_2}{n_3}}$.
This is only a convenient notation; names are still trees, so (unlike string concatenation)
binary name composition is not associative:
%%% "never associative" is too strong; associativity is a general property;
%%% "division is never commutative" is technically wrong because 1/1, 2/2 are commutative
%
$\NmBinOp{n_1}{\NmBinOp{n_2}{n_3}} \ne \NmBinOp{(\NmBinOp{n_1}{n_2})}{n_3}$,
since
$\NmBin{n_1}{\NmBin{n_2}{n_3}} \ne \NmBin{\NmBin{n_1}{n_2}}{n_3}$.

To respect the principles of unique names, \Fungi encodes ``uniqueness''
through \emph{apartness}.
Apartness plays a central role in our type-and-effects system
and metatheory.
In \Fungi code, we read the connective \code{^} as ``apart'', a
notion that (1) generalizes the operation of (\emph{disjoint}) set
union and (2) asserts that the left- and right-hand operands
are indeed disjoint, with no common names.
%%% this is problematic but possibly okay at this point...
%%% apartness as a predicate *generalizes* disjointness (e.g. to name functions);
%%% for name sets, it asserts that the operands are disjoint;
%%% for name functions f and g, f \apart g isn't defined.

Unlike disjoint set union, which is only defined for sets, our type system % of \Fungi
defines apartness over (pairs of) % for every pair of well-sorted
\emph{name terms} and \emph{index terms}.
Name terms include functions over names, as well as literal names; 
index terms include name sets.
Informally,
we say that
(1) two \emph{names} $n_1$ and $n_2$ are apart if they are not
equal (if $n_1$ and $n_2$ are distinct binary trees),
(2) two \emph{sets of names} are apart if they are disjoint,
and
(3) two \emph{functions} are apart if the functions' images are apart.
For example, two functions from names to names are apart
if their images (name sets) are apart.
% %
% % More generally, we say that two \emph{functions} over names, name sets
% % or other functions are apart if the images are apart.

% Below, we give more examples of apartness in the context of \code{dedup}.

%%%%%%%%%%%%%%%%%%%%%%%%%%%%%%%%%%%%%%%%%%%%%%%%%%%%%%%%%
\subsection{Static effects and types for \texttt{dedup}}
\label{sec:dedup-types}

Having seen part of a dynamic execution, we consider a static view of \code{dedup}, 
how \Fungi enforces global uniqueness for it, 
and how \Fungi permits the programmer to express and enforce
the local uniqueness invariants that support global uniqueness.

\paragraph{Global uniqueness: Static effects for \texttt{dedup}.}
Returning to \Figref{fig:dedup},
the index term declaration \code{idxtm} \code{Dedup} defines a function
from name sets to name sets, of sort \code{NmSet} \code{idx->} \code{NmSet}.
Given the names in \code{dedup}'s input list,
the name set function \code{Dedup} gives an overapproximation of
\code{dedup}'s \emph{write set}---the set of names written by executing
\code{dedup} on an input list associated with the given name set.
This name set function \code{Dedup} appears in the type of \code{dedup},
defining the write set in terms of \code{X1} as \code{writeset~[[Dedup X1]]}.%
\footnote{Our full type system also tracks \emph{read sets},
and checks that the read and write sets are in harmony:
it is not possible to read a location before it has been allocated.}
This annotation says that % we can view the body of
\code{Dedup} is a
static abstraction of the dynamic allocation effects in the body of \code{dedup}.

As explained in detail above,
\code{dedup} uses each input
name \code{x} (drawn from name set \code{X1}) three times.
However, in each of these uses \code{x} is composed with other name constants,
resulting in unique global names.
% global uniqueness is satisfied.
% but to satisfy global uniqueness, each use involves
% (or more) other name constants.
%
The three uses are as follows.

%
% Punctuation was broken; if each list item is a sentence, the first word must be capitalized.
% If the whole list is (part of) a sentence, the items can't end with a period.
%
\begin{enumerate}
\item \emph{Allocate a new path in the trie.}
  % "prepended by" sounds wrong
  In aggregate, these allocations write names in the set \code{Insert} \code{X1},
  but with the name constant \code{@t} prepended.

\item
  \emph{Allocate a recursive thunk.}
  In aggregate, these allocations write names in \code{X1}, but
  with the name constant \code{@dd} prepended.

\item
  \emph{Allocate an output list cell.}
  In aggregate, these allocations write \emph{some} names in \code{X1}
  (for names of non-duplicated input list elements),
  but with the name constant \code{@r} prepended.
\end{enumerate}

These three terms appear in   %%% not "as"
the body of \code{Dedup}.
To describe pointwise binary name combination over pairs of name
\emph{sets}, \Fungi uses the notation $\NmSetBinSym$. %,
% which ``lifts'' ordinary binary name combination to name sets.
%%% that's what "pointwise" means
%
%
(Above, we write ``${\cdot}$'' and ``$\textcolor{dPurple}{\bullet}$''
for binary combination of name constants and name values,
respectively.)
Using the apart name set operator $\ApartSym$, the body of
\code{Dedup} combines these three (disjoint) subsets, simultaneously
asserting that they stand apart.

To see why these terms indeed stand apart, consider the following
expansion, where we expand the definition of $\texttt{Dedup}$ over $\{ n_2 \}$, to
account for the write set of the $n_2$ tile only (\Figref{fig:dedup-x4x}):
\[
\begin{array}{@{}r@{~~~}c@{~~~}cccccccc}
\texttt{Dedup}~\{n_2\} 
&=&
\NmSetBin{\{ \NmStr{t} \}}{(\texttt{Insert}~\{n_2\})} 
&\ApartSym&
\NmSetBin{\{ \NmStr{dd} \}}{\{n_2\}}
&\ApartSym&
\NmSetBin{\{ \NmStr{r} \}}{\{n_2\}}
%\\
%&=
%\NmSetBin{\NmStr{t}}{(\texttt{Insert}~\{n_2\})} 
%~\ApartSym~
%\{ \NmBinOp{\NmStr{dd}}{n_2} \}
%~\ApartSym~
%\{ \NmBinOp{\NmStr{r}}{n_2} \}
\\
&=&
\NmSetBin{\{ \NmStr{t} \}}{ \NmSetBin{\{n_2\}}{\texttt{Nat}} }
&\ApartSym&
\{ \NmBinOp{\NmStr{dd}}{n_2} \}
&\ApartSym&
\{ \NmBinOp{\NmStr{r}}{n_2} \}
\\
&=&
\NmSetBin{\{ \NmStr{t} \}}{ \NmSetBin{\{n_2\}}{\texttt{Succ}^{\ast}\{ \texttt{Zero} \}} }
&\ApartSym&
\multicolumn{3}{c}{
\{ 
\NmBinOp{\NmStr{dd}}{n_2},
\NmBinOp{\NmStr{r}}{n_2} 
\}
}
\\
&=&
\NmSetBin{\{ \NmStr{t} \}}{ \NmSetBin{\{n_2\}}{
  \left(
    \texttt{Succ}^{\ast}\{ \texttt{Succ}( \texttt{Zero} ) \}} 
  \ApartSym
  \{
  \texttt{Zero}
  \}
  \right)

}
&\ApartSym&
\multicolumn{3}{c}{
\{ 
\NmBinOp{\NmStr{dd}}{n_2},
\NmBinOp{\NmStr{r}}{n_2} 
\}
}
\\
&=&
\NmSetBin{\{ \NmStr{t} \}}{ \NmSetBin{\{n_2\}}{
    \texttt{Succ}^{\ast}\{ \texttt{Succ}( \texttt{Zero} ) \}} 
}
&\ApartSym&
\multicolumn{3}{c}{
\{ 
\NmBinOp{\NmStr{t}}{
  \NmBinOp{n_2}{\texttt{Zero}}},
\NmBinOp{\NmStr{dd}}{n_2},
\NmBinOp{\NmStr{r}}{n_2} 
\}
}
\\
&=&
\multicolumn{5}{c}{
  \NmSetBin{\{ \NmStr{t} \}}{ \NmSetBin{\{n_2\}}{
      \texttt{Succ}^{\ast}\{ \texttt{Succ}( \textbf{4} ) \}} 
  }
  ~
  \ApartSym
  ~
  \{ 
\NmBinOp{\NmStr{t}}{\NmBinOp{n_2}{\textbf{4}}},
\NmBinOp{\NmStr{t}}{\NmBinOp{n_2}{\textbf{3}}},
\NmBinOp{\NmStr{t}}{\NmBinOp{n_2}{\textbf{2}}},
\NmBinOp{\NmStr{t}}{\NmBinOp{n_2}{\textbf{1}}},
\NmBinOp{\NmStr{t}}{\NmBinOp{n_2}{\textbf{0}}},
\NmBinOp{\NmStr{dd}}{n_2},
\NmBinOp{\NmStr{r}}{n_2} 
\}
}
\end{array}
\]
%
% As \Secref{sec:insert} explains in further detail, %%% redundant with reference in next sentence
The definition of
\texttt{Insert} uses \texttt{Nat}, an infinite set defined by
Kleene closure: $\texttt{Succ}^{\ast}\{ \texttt{Zero} \}$.
\Secref{sec:insert} explains this definition and the corresponding
implementation of \texttt{insert},
but note that the ``unrolled'' set includes
the five names that appear in \Figref{fig:dedup-x4x}
that are each based on $n_2$,
with $\NmStr{t}$ prepended
and $\NmNum{0}$--$\NmNum{4}$ appended.
We use decimal notation in place of the actual unary
\texttt{Zero} and \texttt{Succ}.   % ; of course, other encodings of naturals are possible too.

As this expansion shows, the names in the image of
$\texttt{Dedup}~\{n_2\}$ are pairwise distinct: we can
distinguish them by their prefixes (\NmStr{t}, \NmStr{dd} and
\NmStr{r}), or---for those with the common prefix~\NmStr{t}---%
by their distinct suffix $\NmNum{0}$--$\NmNum{4}$.

What about the other tiles, for input positions $n_1$, $n_3$ and $n_4$?
Global uniqueness for the entire execution of
\texttt{dedup} rests on the assumptions of local uniqueness for the
input list and input trie, e.g., that $n_2$ is distinct from all other
names, which are also pairwise distinct.  %%% not "pair-wise"
Next, we explain how the \Fungi programmer establishes and maintains
the local uniqueness invariants.  %, which is each localized to the definition of a data
%structure or function.

\paragraph{Local uniqueness: Type indices for \texttt{dedup}.}

The \Fungi programmer encodes local uniqueness invariants by
attaching \emph{apartness constraints} % (as logical propositions)
to the type indices used in the definitions of data structures and functions.
Consider the type indices for the two (user-defined)   % \Fungi
data structures used by \texttt{dedup}, linked lists and hash tries. %,
% and the local uniqueness invariants that they each encode.
%
% While \texttt{dedup} makes use of these invariants,
The invariants expressed in the types are also useful for many other functional algorithms.

The programmer defines \code{List} and \code{ListNode} recursively,
giving a reference cell at the head of each list and recursive sub-list (\Figref{fig:dedup}, right).
Though not shown, \code{TrieNode} is defined similarly.
The type indices enforce that, in each structure, each name appears at most once;
but names may be shared across different structures.
%%% in general, *any* statement "animal X does thing Y" could be changed to
%%% "the programmer uses animal X to do thing Y", 
%%% but that barely adds any information the first time, and definitely
%%% doesn't need to be repeated within one paragraph

In the type for \code{Cons}, the quantifier for name sets \code{X1} and \code{X2} 
%%% while I'm complaining: nonbreaking spaces are almost never a good idea,
%%% and here you weren't even doing it consistently.
includes the % apartness
constraint % (a proposition)
\code{X1^X2}, which
says that \code{X1} and \code{X2} are apart (disjoint).
%
% Meanwhile, %%% these aren't two events or processes that are happening concurrently,
%%% they're two things you're mentioning
These indices appear in the types of the \code{Cons} cell's
name (\code{X1}), and its list tail (\code{X2}).
Consequently, to form lists inductively,
% the \Fungi programmer must %%% this makes it sound like the programmer has to manually prove it
% show that...holds
the constraint \code{X1^X2} must hold,
showing that each additional \code{Cons} cell name is distinct from the others already in its tail.

The type indices for \code{insert} are similar to those of
\code{Cons}.
% Though not a primitive constructor like \code{Cons},  %%% sure, but why does that matter?
They express a similar function in terms of name sets:
stating that the resulting structure (a trie) contains an additional name (in \code{X1})
not present in the input structure (name set \code{X2}).
% % to augment  %%% "augment" is overblown when we just mean it's being added
% % the final structure (a trie) with an additional name that the input
% % structure does not already contain.
%
The type for \code{Nil} allows any name set (a safe overapproximation),
since \code{Nil} contains no concrete names at runtime.
%%% Question: why not define it to be empty?
%
Similarly, the type for an empty trie (not shown) allows any name set.

Turning to the type signature of \code{dedup},
it includes the apartness constraint 
\code{X1^X2},
encoding the invariant that the type indices for
\code{dedup}'s input structures (name sets \code{X1} and \code{X2})
are apart.
The codomain of the type, \code{List[[X1]]},
says that the resulting list contains the same names as the input list.
%
% % encode the invariant that the names of the input list and input trie are apart,
% % and that the resulting list's names coincide with that of the input list.
% %
% %
% % introduced above, in the context of
% %\code{Cons} and \code{insert}.
% %
% %Due to the use of \code{X1} and \code{X2} and their apartness
% %constraint, we have that the names in the input list (with name
% %set~\code{X1}) also appear in the output list, and no names from the
% %trie (with name set~\code{X2}) appear in either.
% %
The type system uses the apartness constraints within
the types of \code{Cons} and \code{insert}
% type~\code{List} (with constructor \code{Cons}) and function~\code{insert},
% as discussed above,
%
to show that \code{dedup}'s apartness constraint holds
for the recursive invocation of \code{dedup}.

Intuitively, that invocation moves name \code{x}
($\texttt{x}=n_2$ in \Figref{fig:dedup-x4x})
from the head of input list \code{l} to the accumulated trie \code{tx},
maintaining the pairwise apartness of names in each of the two structures.  
In terms of \Figref{fig:dedup-x4x}, the inductive
reasoning about \code{dedup}'s invariants goes as follows.
%
%%% First, 
By assumption, the name sets of \code{l} and \code{t} are apart.
(In \Figref{fig:dedup-x4x}, 
the name set of \code{l} is $\{ n_2, n_3, n_4 \}$, 
% and \code{t} contains $\{ n_1 \}$,   %%% this says that \code{t} is a set (it's not), and specifically a set of sets, one of whose members is the singleton $\{n_1\}$
and the name set of \code{t} is $\{ n_1 \}$.)
% respectively). %%% unnecessary when saying "A1 is B1 and A2 is B2"
%
In the \code{Cons} branch,     %%% Question: when *what* is pattern-matched?
the apartness constraint in the type signature
for \code{Cons} provides that the name \code{x} at the \code{Cons}
cell is apart from the names of the list tail \code{ys}, if any.
(In \Figref{fig:dedup-x4x}, recall that \code{x} = $n_2$, and the names in \code{ys} consist of $n_3$ and $n_4$.)
The type signature for \code{insert} provides  %%% the fact
that the names
of the output trie consist of the existing names from the existing trie,
along with the new name for the inserted element.
In \Figref{fig:dedup-x4x}, the inserted trie \code{tx} contains the names $n_1$ and $n_2$.

Putting these facts together, in the recursive invocation of thunk \code{ddys}, 
we have that the names of the list tail \code{ys} ($n_3$ and $n_4$)
and those of the updated trie~\code{tx} ($n_1$ and $n_2$) are apart. %,
%as were \code{l} and \code{t}, by assumption.

\paragraph{Static reasoning:}  %{based on decision procedures.}
To statically enforce both global and local uniqueness,
\Fungi uses decision procedures to determine whether (static
approximations of) name sets are apart.
When it needs to prove such an assertion, but decides otherwise, it
tells the programmer that the name sets in question---%
describing either global effects or local type indices---%
cannot be proven to be apart.  
% but have been decided not to be so.
%
For instance, if the programmer mistakenly passed \texttt{l}
instead of \texttt{ys} in the recursive call, % instead of the list tail \texttt{ys},
the inductive invariant would not hold:
the names of \texttt{l} and \texttt{tx} overlap at name \texttt{x}.  
As a result, \Fungi would report that it cannot show the invariant for the recursive call.
%%% "if the programmer had mistakenly written...will not be able...reports": mismatched
%%% "had...written" (= "wrote") needs "would not be able" and "would report"

Below, we consider \texttt{insert} in more depth (\Secref{sec:insert})
before exploring other possible uniqueness errors within \code{dedup}
(\Secref{sec:errors}),

%%%%%%%%%%%%%%%%%%%%%%%%%%%%%%%%%%%%%%%%%%%%%%%%%%%%%%%%%

\begin{figure}
\begin{tabular}{c@{\hspace{10pt}}|@{\hspace{10pt}}c}
\begin{lstlisting}    
type TrieNode : NmSet => Type
  Emp :forall~X:NmSet. TrieNode[[X]]
  Leaf:forall~X:NmSet. 
    Nm[[X]] -> Nat -> TrieNode[[X]]
  Bin :forall~X1^X2:NmSet. 
    Trie[[X1]] -> Trie[[X2]] -> 
    TrieNode[[X1^X2]]

nmtm Zero : Nm = leaf
nmtm Succ : Nm nm-> Nm = lam~x.<<leaf,,x>>
idxtm Gte : Nm idx-> NmSet = lam~x.Succ~*<~x~>
idxtm Nat~ : NmSet = Gte Zero
idxtm Insert X = X~@@~Nat~

insert x y t = insrec x y t 0 Zero
\end{lstlisting}
&
\begin{lstlisting}
insrec : forall~X1^X2:NmSet. forall~m:Nm.
  Nm[[X]] -> Nat -> Trie[[X2]] ->
  Nat -> Nm[[<~m~>]] -> Trie[[X1^X2]]
  writeset~[[X~@@~Gte m]]

insrec x y t i ni = if i = 4 then
  ref <[x~@~ni]> (Leaf x y)
else
 let (j, nj) = (i+1, Succ ni)
 let (txl, txr) = 
   if hash_bit y i then
    (insrec x y (left t) j nj, right t)
   else
    (left t, insrec x y (right t) j nj)
 ref <[x~@~ni]> (Bin txl txr)
\end{lstlisting}
\end{tabular}
\caption{Types and implementation of hash tries; \code{insrec} illustrates general recursion with named effects. }
\label{fig:insert}
\end{figure}

\subsection{Helper function \code{insert}}
\label{sec:insert}

\Figref{fig:insert} shows the \Fungi programmer's implementation of
\code{insert}, in terms of a recursive function \code{insrec} (right column).
The left column gives the type definition of \code{TrieNode}, whose
use of indices is similar to \code{ListNode} from \Figref{fig:dedup}.
Below this definition, the programmer defines various name and index
terms, culminating in the definition of \code{Insert}, which gives the
write set for \code{insert}, just as \code{Dedup} did for
\code{dedup} in \Figref{fig:dedup}.

Recall that \code{dedup} used structural recursion over a list with a
name at each recursive position (\code{Cons} cell).
Here, \code{insrec} illustrates a pattern of naming allocations within
\emph{general recursion}.
The \code{insert} function takes a name, an element (natural number)
and a trie; it returns the hash trie obtained by hashing the
given element and inserting it into the given trie.
In addition to the updated trie, \code{insert} returns a boolean
indicating whether the element was already present in the trie
(but with a distinct name).
For clarity, we discuss a simpler variant that only returns
the updated trie; the other variant is similar, with similar
allocation effects.  %, as described below.

%
%% Point:
%% structural recursion (\code{dedup} example) 
%% versus 
%% general recursion (\code{insrec} implementation of \code{insert}).

To allocate a new trie path, the \Fungi programmer uses names to identify each allocation.
Rather than use names from an input structure, as with the structural recursion of \code{dedup},
\code{insert} generates name sets (statically) with Kleene closure:
repeated application of a name function \code{Succ~*<~x~>} to an initial set (\Figref{fig:insert}).
By defining write sets in this way,
we % the \Fungi programmer
can name any allocations within general recursion 
based on each allocation's (complete) path in the recursive call graph.
Since \code{insert} recurs only once, there is a single chain of calls
and a natural number suffices to name each call.

The implementation of \code{insrec} (dynamically) computes the sequence of names, % as the loop executes,
starting from \code{Zero}.
(As \Figref{fig:dedup-x4x} illustrates, the last computed name corresponds to $4$.)
%%% (Sentence). is wrong.  (Sentence.) is right.
%
In the final iteration, \code{insert} creates a leaf node holding the
inserted element~\code{y}, and its associated name~\code{x}.
In a more complex structure, we would handle hash collisions by
creating linked lists at these leaf positions; for simplicity, we
assume here that hash collisions are impossible.

The index function \code{Gte} gives the inductive invariant for % of the recursion in
\code{insrec}: 
Every numeral suffix written by the recursive call to \code{insrec}
is greater than the one written by the current iteration of the loop,~\code{ni}.
Recursive iterations will use \code{nj}, or some larger numeral.
While natural numbers are not built-in to \Fungi's type
index system, the programmer encodes the ``greater than''
constraint using \Fungi's notion of apartness.  % (as in \Figref{fig:insert}).

\begin{comment}

As \Figref{fig:insert} shows, \code{insert} essentially performs a
functional ``for'' loop (as a recursive function), counting up to a
maximum iteration number, with one iteration per bit in the hash trie
(in this case, four bits total).
%
In each iteration, \code{insert} uniquely names a trie node by adding
the current iteration count to the (fixed) name parameter~\code{x}.
%
In the final iteration, \code{insert} creates a leaf node holding the
inserted element~\code{y}, and its associated name~\code{x}.

%
In a more complex structure, we would handle hash collisions by
creating linked lists at these leaf positions; for simplicity, we
assume here that hash collisions are impossible.

To give a write set for \code{insert} (via name set function
\code{Insert}), we create a name set for the natural numbers (as
names), \code{Succ~*<~Zero~>}.
%
This definition uses the Kleene-star name set construction form (via
\code{Succ~*}), over the initial set containing only the name constant
representing zero (\code{Zero}).

\end{comment}

%%%%%%%%%%%%%%%%%%%%%%%%%%%%%%%%%%%%%%%%%%%%%%%%%%%%%%%%%%%%%%%%%%%%%%%%%%%%%%%%%%%%

\subsection{Apartness failures violate global uniqueness}
\label{sec:errors}

%%% Replaced "fallacies" with "failures".  "Fallacies" could (maybe) be reasonably used as the
%%% "mistake" heading (I changed it to "class of mistake"), but no one would say
%%% "I fixed twelve type fallacies today".

% possible violations "in the wild": 
% - maybe? https://github.com/plum-umd/adapton.ocaml/blob/master/src/collections/trie.ml#L623
% - maybe? https://github.com/plum-umd/adapton.ocaml/blob/master/src/collections/trie.ml#L633
% - yes?   https://github.com/Adapton/adapton.rust/blob/dev/src/catalog/collections.rs#L647

As explained above, \Fungi enforces global uniqueness and local uniqueness  % invariants
by statically reasoning about apartness.
In writing \code{dedup} without the \Fungi type system, 
it is easy to make naming mistakes that violate an apartness constraint---%
breaking local uniqueness, global uniqueness, or both.

In \Tableref{tab:mistakes}, we list three classes of naming mistakes,
showing concrete examples in the context of \code{dedup} and
the apartness constraint that does not hold ($\not\entails$).
We explain each mistake in more detail, to see why it
violates apartness, sometimes on very specific (and unlikely) inputs.
It is easy to overlook these mistakes; % by the \Fungi programmer
%%% without the help of a static check.
%
the authors have made all of them.
%
%The \Fungi type system reports

\begin{table}
\begin{tabular}{@{}p{.23\textwidth}|p{.30\textwidth}|p{.41\textwidth}@{}}
  class of mistake 
  & the \code{dedup} programmer\ldots
  & apartness failure
  \\ 
  \hline
  missing tag 
  & forgets \code{@dd} and/or \code{@r}
  &   $\not\vdash \lambda x.(\NmBinOp{\NmStr{dd}}{x})~\bot~\texttt{id}$
  and $\not\vdash \lambda x.(\NmBinOp{\NmStr{r}}{x})~\bot~\texttt{id}$
  \\
  prefix/suffix mismatch
  & uses \code{@r} as suffix, or forgets \code{@t}
  & {
      $\not\vdash \lambda x.(\NmBinOp{\NmStr{dd}}{x})~\bot~\lambda x.(\NmBinOp{x}{\NmStr{r}})$
      
      $\not\vdash \lambda x.(\NmBinOp{\NmStr{dd}}{x})~\bot~\lambda x.(\NmBinOp{x}{\NmStr{0}})$
    }
  \\
  ``overlapping primes''
  & defines \code{Insert X = Succ~*X}
  & $X:\NmSet \not\vdash \texttt{Succ}(\texttt{Succ}(X)) \bot \texttt{Succ}(X)$
\end{tabular}

\medskip

\caption{Naming mistakes, with examples from \code{dedup} and their associated apartness failure}
\label{tab:mistakes}
\end{table}

To see the problem with a missing tag, 
consider mapping the name set $\{ \NmNum{0}, \NmBinOp{\NmNum{0}}{\NmNum{1}} \}$ 
by \texttt{id} (the identity function) and 
by $\lambda x. \NmBinOp{x}{\NmNum{1}}$:
The images overlap on name $\NmBinOp{\NmNum{0}}{\NmNum{1}}$, since the
two names in the input set are already related by the second function~$\lambda x. \NmBinOp{x}{\NmNum{1}}$, 
which \texttt{id} fails to distinguish by adding any tag of its own.
By contrast, consider mapping the same name set by two \emph{apart} name functions, 
$\lambda x. \NmBinOp{x}{\NmNum{0}}$ 
and 
$\lambda x. \NmBinOp{x}{\NmNum{1}}$; the two images are disjoint 
%%% every time you write "viz.", the darkness grows
($\{ \NmBinOp{\NmNum{0}}{\NmNum{0}}, \NmBinOp{(\NmBinOp{\NmNum{0}}{\NmNum{1}})}{\NmNum{0}} \}~
~\bot~
\{ \NmBinOp{\NmNum{0}}{\NmNum{1}}, \NmBinOp{(\NmBinOp{\NmNum{0}}{\NmNum{1}})}{\NmNum{1}} \}$).

To see the problem with prefix/suffix mismatches, 
consider mapping the name set $\{ \NmNum{0}, \NmNum{1} \}$ 
by 
$\lambda x. \NmBinOp{\NmNum{0}}{x}$
and 
$\lambda x. \NmBinOp{x}{\NmNum{1}}$;
the two images overlap on name $\NmBinOp{\NmNum{0}}{\NmNum{1}}$, since the
two functions disagree about how they distinguish names in the input set.

Finally, to see the problem with what we call ``overlapping primes'',
consider mapping the name set $\{ \NmNum{0}, \NmNum{1} \}$ 
by \texttt{Succ} and $\texttt{Succ} \circ \texttt{Succ}$: the images overlap at name $\NmNum{2}$.
The problem is similar to the missing tag problem, but at the level of
name sets.
Overlapping primes may involve Kleene closure, e.g.,
$X:\NmSet \not\vdash \texttt{Succ}^{\ast}( \texttt{Succ}(X) ) \bot X$.
There is a critical difference between this apartness violation
and the apartness \emph{property} used by \texttt{insert} and \texttt{insrec} in \Secref{sec:insert},
$x:\Nm \vdash \texttt{Succ}^{\ast}(\{ \texttt{Succ}(x) \}) \bot \{ x \}$.
Like the violation above, this valid apartness property also involves Kleene closure,
but the ``seed'' set used by \texttt{insrec} is a \emph{single} unknown name.
We call this kind of mistake ``overlapping primes'' because an analogous issue
arises when manually ``freshening'' meta-variables in a proof: %% from existing ones:
adding a prime to each existing variable does not work when an
existing variable is the prime of another.
For example, priming the variable $A$ in the set $\{A, B, A'\}$ clashes with
the existing $A'$.

%%%%%%%%%%%%%%%%%%%%%%%%%%%%%%%%%%%%%%%%%%%%%%%%%%%%%%%%%

\subsection{Global uniqueness implies correct change propagation}

The metatheory of the \Fungi type system considers one run
at a time.
Within each run, it ensures global uniqueness, 
the prerequisite for change propagation to ensure from-scratch consistency:
%%%%%% on all possible inputs and input changes.
%
When global uniqueness holds, the outcome of change propagation for any iteration
is always consistent with the outcome of a from-scratch run on the current input.

\begin{figure}
  \begin{tabular}{c@{\hspace{15pt}}c}
    \hspace{-0.80in}
    \includegraphics[width=0.62\textwidth]{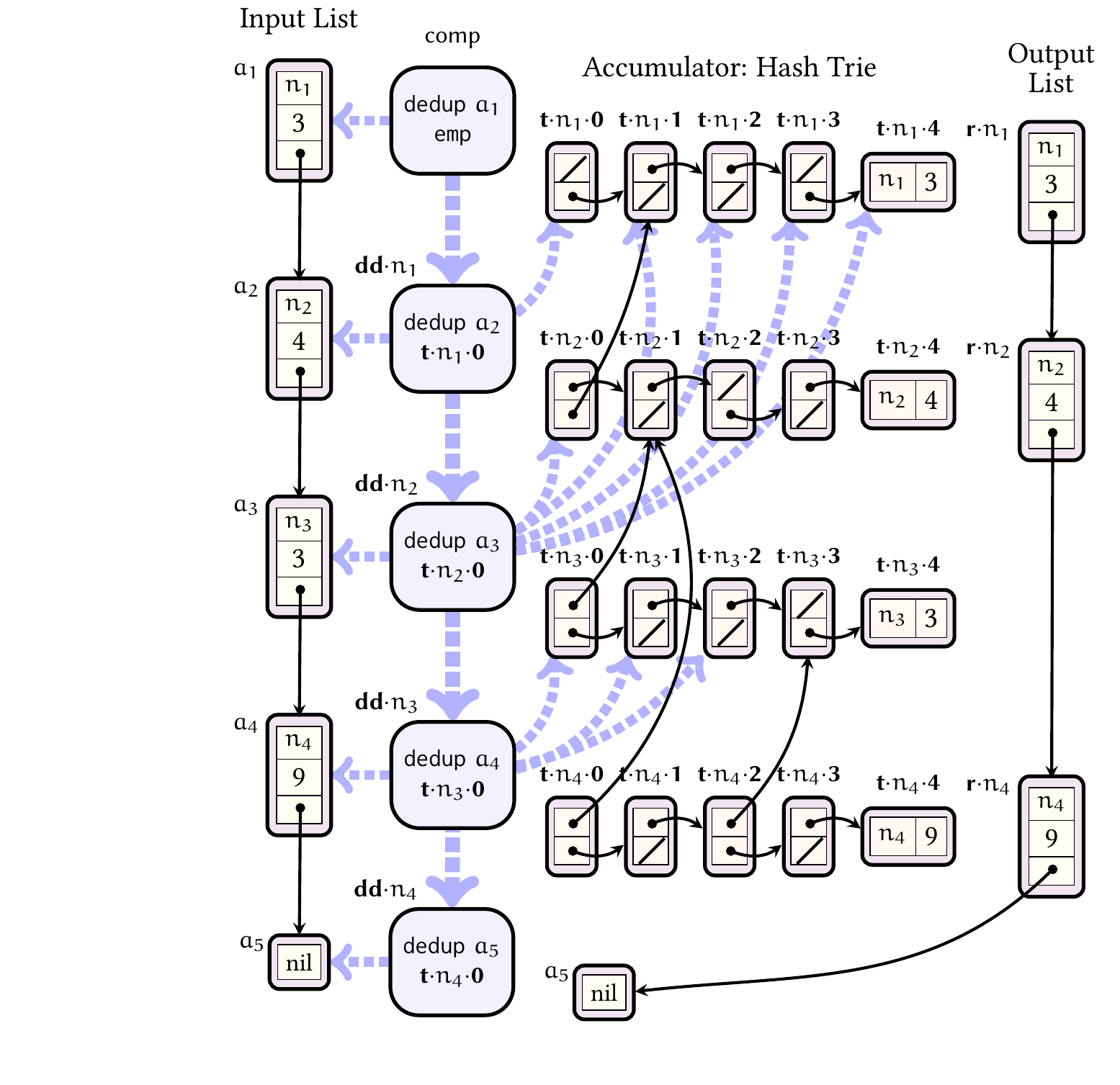}
    &
    \hspace{-0.80in}
    \includegraphics[width=0.62\textwidth]{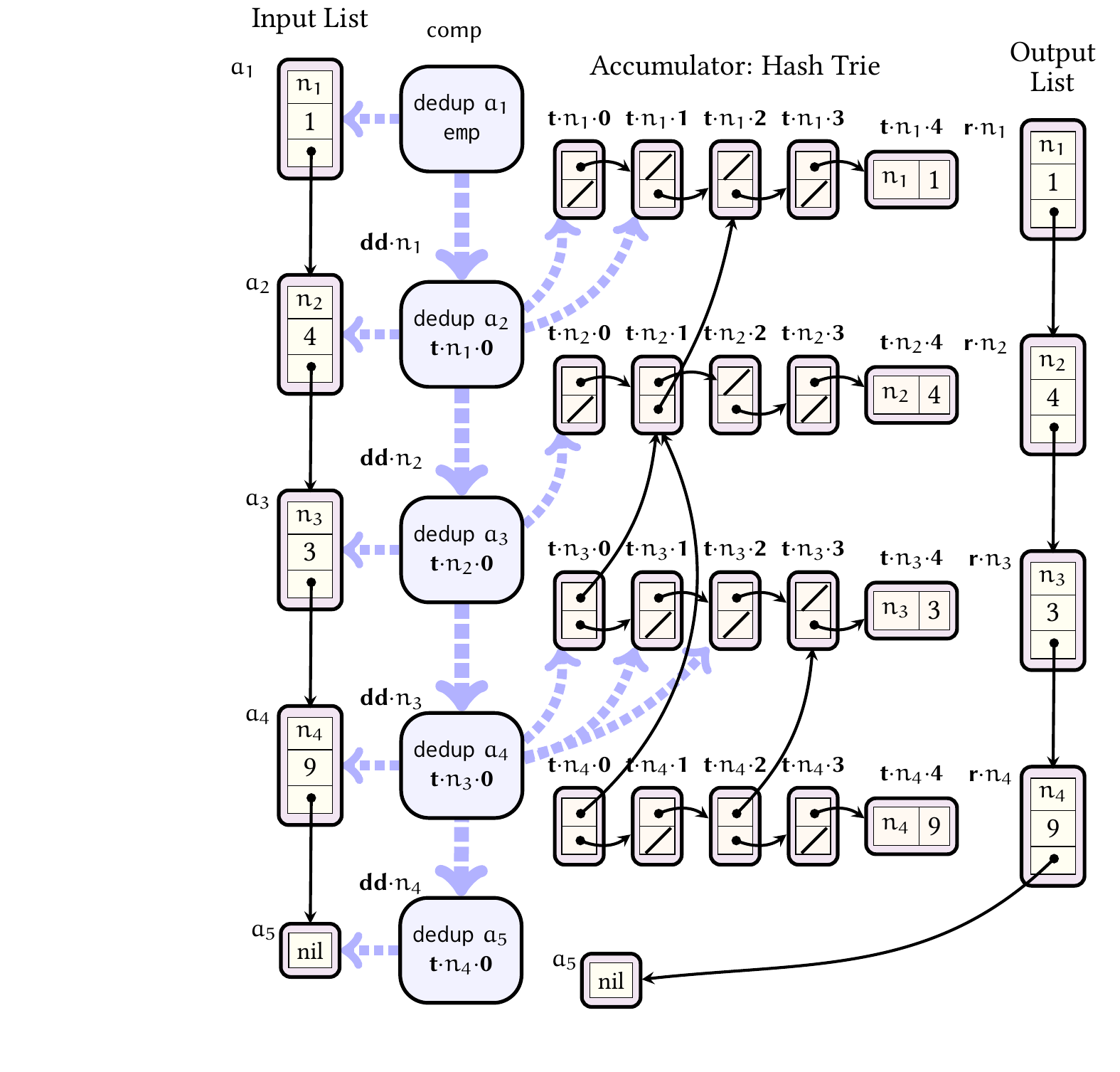}
  \end{tabular}
  \vspace{-20pt}
  \caption{Distinct runs of \code{dedup} on 
    (similar) input lists $[3,4,3,9]$ (left side) and~$[1,4,3,9]$ (right side).
  }
\label{fig:dedup-two-runs}
\end{figure}

\paragraph{Using names to compare similar from-scratch runs.}
Because change propagation is from-scratch consistent, 
we can predict its time complexity (and other dynamic behavior) by comparing two from-scratch runs and seeing where they differ.
Due to their use of names, the two runs similarities and differences can be identified precisely, by name.
Change propagation, described in detail below, attempts to exploit these similarities to reuse past work wherever possible.

\Figref{fig:dedup-two-runs} shows two full runs of \code{dedup} on
(similar) input lists $[3,4,3,9]$ and~$[1,4,3,9]$, 
stored in identical list cells ($a_1$--$a_5$) and initial recursive thunk (\textbf{comp}).
The tile in \Figref{fig:dedup-x4x} is consistent with the left run.
The left run consists of two occurrences of element~$3$, at logical positions~$n_1$ and $n_3$; 
in the right run, logical position~$n_1$ instead contains the (unique) element~$1$.

The later recursive calls depend on the trie paths allocated in earlier calls.
For instance, by carefully comparing the left and right runs' allocated trie paths 
rooted at $\NmBinOp{\NmStr{t}}{\NmBinOp{n_1}{\NmNum{0}}}$, 
we see that the hashes of $3$ and $1$ consist of inverted bits: 
all of the pointers have ``flipped'' between left and right.
Moving downward in the figure,
the allocated trie paths rooted at $\NmBinOp{\NmStr{t}}{\NmBinOp{n_2}{\NmNum{0}}}$ 
differ at that name, and at $\NmBinOp{\NmStr{t}}{\NmBinOp{n_2}{\NmNum{1}}}$, 
but then ``sync up'' at $\NmBinOp{\NmStr{t}}{\NmBinOp{n_2}{\NmNum{2}}}$---$\NmBinOp{\NmStr{t}}{\NmBinOp{n_2}{\NmNum{4}}}$.
The allocated trie paths rooted at
$\NmBinOp{\NmStr{t}}{\NmBinOp{n_3}{\NmNum{0}}}$ and
$\NmBinOp{\NmStr{t}}{\NmBinOp{n_4}{\NmNum{0}}}$ are the same in the
two runs.

Because of the input list's logical position names ($n_1$---$n_4$),
the output list uses identical addresses in the left- and right-hand runs, where they overlap.
The right-hand run's list contents are similar, with three (necessary) exceptions:
(a) the element at logical position $n_1$ is changed to $1$;
(b) $3$ appears at logical position $n_3$ (and pointer name $\NmBinOp{\NmStr{r}}{n_3}$),
whereas the left-hand run had a duplicate $3$ at position $n_3$;
(c) the tail pointer in the output \code{Cons} cell $\NmBinOp{\NmStr{r}}{n_2}$
differs, since position $n_3$ was absent in the left-hand run.
%%%
%%%
%%%
% (a) The elements $3$ versus $1$ at logical position $n_1$, 
% (b) $3$ appears at logical position $n_3$ (and pointer name $\NmBinOp{\NmStr{r}}{n_3}$) in the right run, but not the left run, where this position holds a duplicate $3$, and 
% (c) the preceding tail pointer (from the \code{Cons} cell at $\NmBinOp{\NmStr{r}}{n_2}$)
% differs, due to position $n_3$ being present on the right, and absent
% on the left.

\begin{figure}
  \begin{tabular}{cc}
    %\hspace{-0.85in}
    %\includegraphics[width=0.65\textwidth]{dedup_3439}
    %&
    \hspace{-0.9in}
    \includegraphics[width=0.75\textwidth]{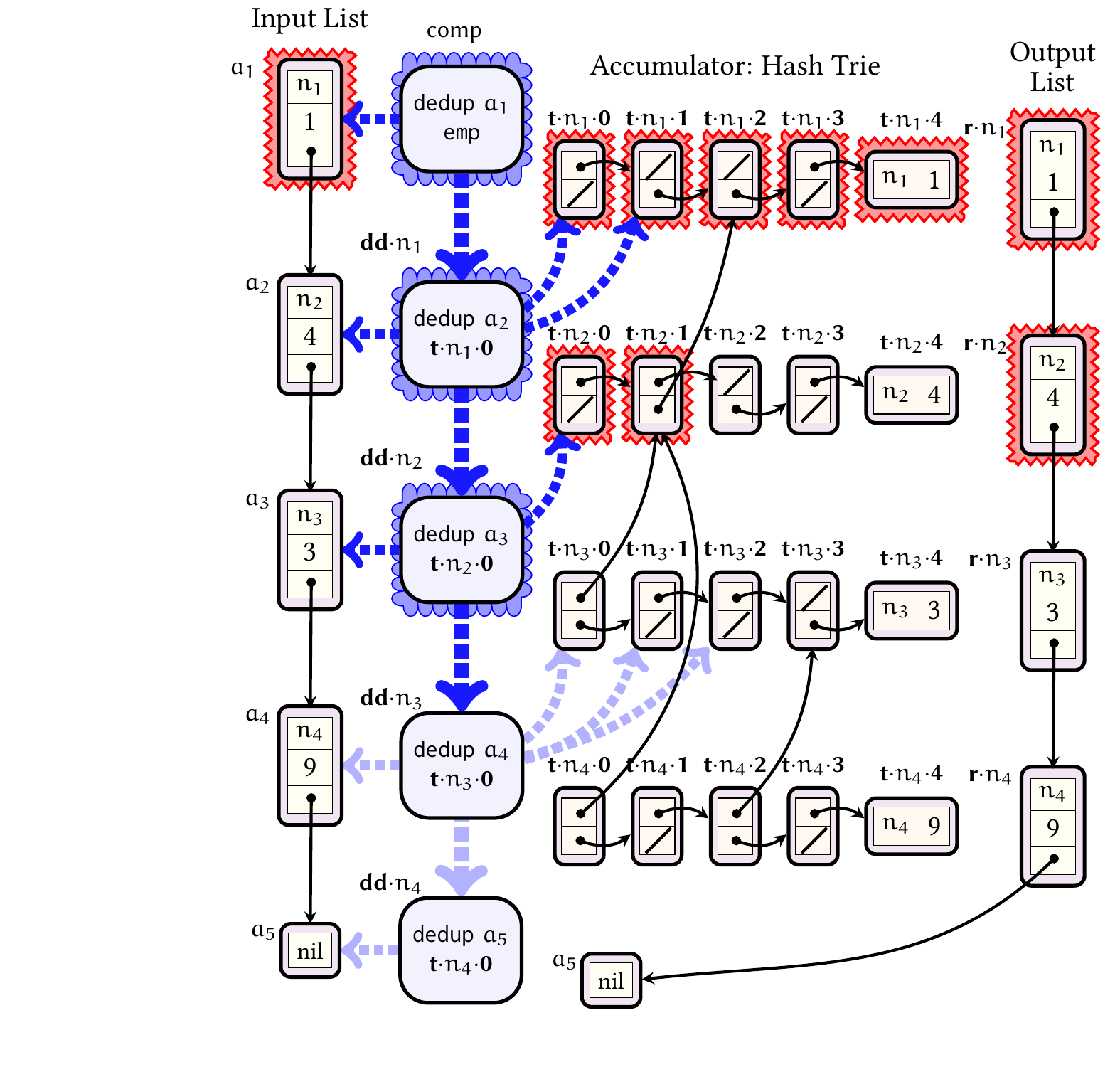}
  \end{tabular}
  \vspace{-20pt}
  \caption{Change propagation uses names to identify the correspondences between subsequent runs of \code{dedup} and efficiently exploit the similarities between these inputs, outputs and intermediate structures (hash tries).}
\label{fig:dedup-changeprop}
\end{figure}

\paragraph{Change propagation}
\Figref{fig:dedup-changeprop} considers the behavior of using change
propagation where the left run of \Figref{fig:dedup-two-runs} happens
first, followed by an input change (at $a_1$), that precipitates
change propagation updating this dependence graph to be from-scratch
consistent with the right run.
As explained above, the two runs in \Figref{fig:dedup-two-runs} differ
at certain allocated names; change propagation selectively re-executes
thunks in the dependence graph in an order that is consistent with a
from-scratch run on the \emph{current} input (in this case, the right run of
\Figref{fig:dedup-two-runs}).
We indicate the re-executed thunks with an additional (blue) border pattern.

Change propagation re-executes thunk $\NmStr{comp}$ first, since it
observes the changed input list cell~$a_1$ that replaces the first $3$ with $1$.
As described above, the new element~$1$ hashes differently, resulting
in a different pattern of pointers in this trie path rooted at 
$\NmBinOp{\NmStr{t}}{\NmBinOp{n_1}{\NmNum{0}}}$.

We indicate overwritten (and changed) reference cells with an
additional (red) border pattern.
\Fungi dynamically records thunks that depend on changed
reference cells, and avoids reusing their results without first
applying the change propagation algorithm to their dependence graphs.
For this reason, change propagation next re-executes
$\NmBinOp{\NmStr{dd}}{n_1}$.

When re-executed, $\NmBinOp{\NmStr{dd}}{n_1}$ changes \emph{some} trie
names with its overwrites, but not all of them (as described above).
Next, it re-forces $\NmBinOp{\NmStr{dd}}{n_2}$, which re-executes (due to the trie overwrites).

When a re-execution results in behavior that is the same as the prior
run, the frontier of change propagation may end, as with
thunk~$\NmBinOp{\NmStr{dd}}{n_2}$.
Its allocations overwrite the prior reference cells with \emph{identical} values.
It (re-)forces
$\NmBinOp{\NmStr{dd}}{n_3}$, whose local effects are unaffected by the
input change (either directly or indirectly).
In this case, \Fungi reuses the cached result of this (unaffected) thunk.

Next, control returns to 
$\NmBinOp{\NmStr{t}}{\NmBinOp{n_1}{\NmNum{0}}}$, which overwrites its output cell's tail pointer with the inserted (and new) cell at $\NmBinOp{\NmStr{r}}{n_3}$.
Finally, control returns to $\NmStr{comp}$, which overwrites $\NmBinOp{\NmStr{r}}{n_1}$ with the changed input value~$1$, but returns the same result, $\NmBinOp{\NmStr{r}}{n_1}$.
If $\NmStr{comp}$ were occuring in the context of more recursive calls in a longer input list, these earlier calls would be unaffected, and not re-executed.

In summary, the change propagation behavior described above critically
relies on (unique) names to bring the initial and updated runs into a
correspondence that it can efficiently exploit.
Unique names are generally \emph{necessary} for efficient (stable) change
propagation, but not alone sufficient.
In particular, the \Fungi type-and-effect system enforces global
uniqueness by reasoning about a single from-scratch execution, not the
relationship between two (or more) similar executions on similar inputs.
In \Secref{sec:related}, we discuss connections to (relational) cost semantics for incremental computation.

\section{Program Syntax}
\Label{sec:progsyntax}

The examples from the prior section use an informally defined variant
of ML, enriched with a (slightly simplified) variant of our proposed
type system.
In this section and the next, we focus on a core calculus for
programs and types, and on making these definitions precise.

\begin{figure}[t]
  \centering
 
\begin{grammar}
  Values
  & $v$
      &$\bnfas$&
      $x
      \bnfalt \unitexp
      \bnfalt \Pair{v_1}{v_2}
      \bnfalt \inj{i}{v}
      \bnfalt \name{n}
      \bnfalt \namefn{M}
      \bnfalt \refv{n}
      \bnfalt \thunk{n}
%      \bnfalt \unthunk{n}
      \bnfalt \VPack{a}{v}
      $
  \\[1ex]
  Terminal exprs.\!\!\!\!
  & $t$
      &$\bnfas$&
              $\Ret{v}
      \bnfalt \lam{x} e
      $
  \\[1ex]
  Expressions
  & $e$
      &$\bnfas$&
           $t
      \bnfalt \Split{v}{x_1}{x_2}{e}
      \bnfalt \Case{v}{x_1}{e_1}{x_2}{e_2}
%      \bnfalt \Ret{v}
%      \bnfalt \lam{x} e
      \bnfaltBRK e\;v
      \bnfalt \Let{e_1}{x}{e_2}
      \bnfalt \Thunk{v} e
%      \bnfalt \Unthunk{v}
      \bnfalt \Force{v}
      \bnfalt \Refe{v} v
      \bnfalt \Get{v}
%      \bnfalt \Ns{v}{e}
      \bnfaltBRK \Scope{v}{e}
      \bnfalt v_M\;v
      \bnfalt%BRK
      \VUnpack{v}{a}{x}{e}
      $
%      \bnfalt \namearrintro{a}{e}
%      \bnfalt \namearrelim{e}{t}
\lesscaptionspace
\end{grammar}

  \caption{Syntax of expressions}
  \label{fig:expr}
\end{figure}

% Local Variables:
% mode: latex
% TeX-master: "types"
% End:

\subsection{Values and Expressions}

\Figref{fig:expr} gives the grammar of
values $v$
and expressions $e$.
We use call-by-push-value (CBPV) conventions
 in this syntax, and in
the type system that follows.
There are several reasons for this.
First,
CBPV can be interpreted as a ``neutral'' evaluation order that
includes both call-by-value or call-by-name, but prefers neither in
its design.
Second, since we make the unit of memoization a thunk,
and CBPV
% "explicate" does not mean "make explicit" -j.
makes explicit the creation of thunks and closures,
it exposes exactly the structure that we extend to a general-purpose
abstraction for incremental computation.
In particular, thunks are the means by which we cache results and track
dynamic dependencies.
%
% Finally, focusing on thunks helps us model demand-driven incremental
% computations, e.g., incremental, demand-driven sorting algorithms.

Values $v$ consist of variables, the unit value, pairs, sums, and several
special forms (described below).

We separate values from expressions, rather than considering values to be a subset of 
expressions.  Instead, \emph{terminal expressions} $t$ are a subset of expressions.
A terminal expression $t$ is either $\Ret{v}$---the expression
that returns the value $v$---or a $\lambda$.
Expressions $e$ include terminal expressions,
elimination forms for pairs, sums, and functions
(\keyword{split}, \keyword{case} and $e\,v$, respectively);
% intro and
% elimination forms for producing values (\keyword{ret} and
% \keyword{let}, respectively);
% intro and elimination forms for function
% types (lambda and application, respectively);
let-binding (which evaluates $e_1$ to $\Ret{v}$
and substitutes $v$ for $x$ in $e_2$);
introduction (\keyword{thunk}) and elimination (\keyword{force})
forms for thunks; and introduction (\keyword{ref})
and elimination (\keyword{get}) forms for pointers (reference
cells that hold values).
% a scoping construct for names;
% and name-level function application $M\;v$.
% % [described below]

The special forms of values are
names $\name{n}$,
name-level functions $\namefn{M}$,
references (pointers), and thunks.
References and thunks include a name $n$,
which is the name of the reference or thunk,
\emph{not} the contents of the reference or thunk.

This syntax is similar to Adapton \citep{Hammer15};
we add the notion of a \emph{name function},
which captures the idea of a namespace and other % simple
transformations on names. %functions  as well as operations
%such as ``forking'' names by composing them.
% XX make sure the previous fork construct is explained somewhere
%
The $\Scope{v}{e}$ construct controls monadic state for the current
name function, composing it with a name function $v$
within the dynamic extent of its subexpression $e$.
Name function application $M\;v$ permits programs to
compute with names and name functions that reside within the type indices.
Since these name functions always terminate, they do not affect a
program's termination behavior.

% The syntax for values consists of the unit value, pairs of values,
% tagged injected values (one of the two ``halves'' of a sum type),
% first-class names, value pointers and thunk pointers.

We do not distinguish syntactically between value pointers (for
reference cells) and thunk pointers (for suspended expressions); the
store maps pointers to either of these.

%% Note: We don't use this sugar any longer; we now use a desugared
%% version (something like the RHS) in Section 2.
%
%% \paragraph{Desugaring \keyword{memo\_ref}.}
%% %
%% We can desugar the syntax \keyword{memo\_ref} into the following construction of primitive operations:
%% the creation and elimination of a named thunk that allocates a named reference cell.
%% %
%% We distinguish the two name uses with the tags 1 and 2:
%% %
%% \[
%% \keyword{memo\_ref}~n~[e] = 
%% \keyword{force}~\big(
%%   \keyword{thunk $n.1$ (let\,$x$\,=\,$e$\,\keyword{in}\,\keyword{ref}($n.2$,x))}
%% \big)
%% \]

\subsection{Names}
\label{sec:names}

\begin{figure}[t]
\begin{grammar}
  Names
  & $m, n$
     &$\bnfas$&
           $\leafname$  & leaf name
   \\ 
   ~~(binary trees)
&&& $\bnfaltbrk \NmBin{n_1}{n_2}$ & binary name composition
%   \\ &&& \Cover{$\bnfaltbrk n.1 \bnfalt n.2$}  & append bit (special cases of binary comp)
 \\[0.7ex]
  Name terms
 &\!\!\!\!\!\!\!$M, N$
  &$\bnfas$& $n \bnfalt \NmBin{M_1}{M_2}$ & literal names, binary name composition
 \\ 
 \multicolumn{2}{l}{(STLC+names)\!\!\!}
 && $\bnfaltbrk a \bnfalt \lam{a} M \bnfalt \idxapp{M}{N}$
       & variable, abstraction, application
  \\[0.7ex]
      Name term values\!\!\!\!\!\!\!
     &$\Mv$
          &$\bnfas$&
          \multicolumn{2}{@{}l@{}}{%
          $n
          \bnfalt \lam{a} M
%          \bnfalt \unitexp
%         \bnfalt \Pair{\Mv}{\Mv}
          $
          }
\\[0.7ex]
  Name term sorts
  & $\sort$
      &$\bnfas$&
      $\namesort$  &\!\!\!\!\! name; inhabitants $n$
%   \\ &&& $\bnfaltbrk \unitsort$     &\!\!\!\!\! unit index sort; inhabitant $\unitexp$
%   \\ &&& $\bnfaltbrk \sort * \sort$  &\!\!\!\!\! product index sort; inhabitants $\Pair{M_1}{M_2}$
   \\ &&& $\bnfaltbrk \sort @> \sort$  &\!\!\!\!\! name term function; inhabitants $\lam{a} M$
%   \\ &&& \Cover{$\bnfaltbrk \namesetsort$}   & name set sort (only used to sort indices)
\\[0.7ex]
  Typing contexts
  & $\Gamma$
     &$\bnfas$&
           $
           \cdot
           \bnfalt
           \Gamma, a : \sort
           \bnfalt
           \cdots
           $  
           &\!\!\!\!\! full definition in \Figureref{fig:syntax-types}
\end{grammar}

\lesscaptionspace

  \caption{Syntax of name terms: a $\lambda$-calculus over names, as binary trees}
  \label{fig:syntax-name-terms}
\end{figure}

% Local Variables:
% TeX-master: "typed-adapton"
% End:

\begin{figure}[t]
  \centering

  \judgbox{\Gamma |- M : \sort}
          {Under $\Gamma$,
            name term $M$ has sort $\sort$
          }
  \vspace{-2.5ex}
  \begin{mathpar}
    %% \Infer{\!M-unit}
    %%     {\strut}
    %%     {\Gamma |- \unitexp : \unitsort}
    %% ~~~
    \Infer{\!M-const}
          %{\Gamma |- n \wfname}
        {\strut}
        {\Gamma |- n : \namesort}
    \and
    \Infer{\!M-var}
        {(a : \sort) \in \Gamma}
        {\Gamma |- a : \sort}
    \and
    \Infer{\!M-bin}
        {
          \Gamma |- M_1 : \namesort
          \\\\
          \Gamma |- M_2 : \namesort
        }
        {\Gamma |- \NmBin{M_1}{M_2} : \namesort}
   \\
    \Infer{\!M-abs}
        {\Gamma, a : \sort' |- M : \sort}
        {
          \arrayenvbl{
          \Gamma |- (\lam{a} M) : (\sort' @> \sort)
          }
        }
  \and
  \Infer{\!M-app}
        {
          %\arrayenvbl{
          \Gamma |- M : (\sort' @> \sort)
          \\
          \Gamma |- N : \sort'
          %}
        }
        {\Gamma |- M(N) : \sort}
    %% ~~~
    %% \Infer{\!M-pair}
    %%     {
    %%       \Gamma |- M_1 : \sort_1
    %%       \\\\
    %%       \Gamma |- M_2 : \sort_2
    %%     }
    %%     {\Gamma |- \Pair{M_1}{M_2} : (\sort_1 * \sort_2)}
    %% ~~~
%    \and
%% \Cover{
%%     \Infer{M-append}
%%         {
%%           \Gamma |- M : \namesort
%%         }
%%         {%\arrayenvbl{
%%            \Gamma |- M.1 : \namesort
%%            \\
%%            \Gamma |- M.2 : \namesort
%%          %}
%%          }
%% }
  \end{mathpar}

  %% \judgbox{\Gamma |- i : \sort}
  %%         {Under $\Gamma$,
  %%           index $i$ has sort $\sort$
  %%         }
  %% \begin{mathpar}
  %%   \Infer{sort-var}
  %%       {
  %%         (a : \sort) \in \Gamma
  %%       }
  %%       {
  %%         \Gamma |- a : \sort
  %%       }
  %%   \and
  %%   \Infer{sort-singleton}
  %%       {
  %%        %  \Gamma |- n \wfname
  %%         \Gamma |- n : \namesort
  %%       }
  %%       {
  %%         \Gamma |- \{n\} : \namesetsort
  %%       }
  %%   \and
  %%   \Infer{sort-sep-union}
  %%       {
  %%         \Gamma |- X : \namesetsort
  %%         \\
  %%         \Gamma |- Y : \namesetsort
  %%       }
  %%       {
  %%         \Gamma |- (X \disj Y) : \namesetsort
  %%       }
  %%   \and
  %%   \Infer{sort-unit}
  %%       {}
  %%       {
  %%         \Gamma |- \unitindex : \unitsort
  %%       }
  %%   \and
  %%   \Infer{sort-pair}
  %%       {
  %%         \Gamma |- i_1 : \sort_1
  %%         \\
  %%         \Gamma |- i_2 : \sort_2
  %%       }
  %%       {
  %%         \Gamma |- \Pair{i_1}{i_2} : (\sort_1 * \sort_2)
  %%       }
  %%   ~~~~
  %%   \Infer{sort-apply}
  %%       {
  %%         \Gamma |- M : \sort_1 @> \sort_2
  %%         \\
  %%         \Gamma |- i : \sort_1
  %%       }
  %%       {
  %%         \Gamma |- M[[i]] : \sort_2
  %%       }
  %%   \and
  %%   \Infer{sort-apply-pointwise}
  %%       {
  %%         \Gamma |- M : \namesort @> \namesort
  %%         \\
  %%         \Gamma |- i : \namesetsort
  %%       }
  %%       {
  %%         \Gamma |- M[[i]] : \namesetsort
  %%       }
  %% \vspace*{-2.5ex}
  %% \end{mathpar}

  \runonfontsz{9.5pt}
  \judgbox{M \nteval \Mv}
          {Name term $M$ evaluates to name term value $\Mv$
          }
  \vspace*{-3.5ex}
  \begin{mathpar}
    \Infer{\tevalvalue}  % name, abstraction, [variable], unit
       {}
       {\Mv \nteval \Mv}
    ~~~~
    \Infer{\tevalbin}
        {
          M_1 \nteval n_1
          \\
          M_2 \nteval n_2
        }
        {
          \NmBin{M_1}{M_2} \nteval \NmBin{n_1}{n_2}
        }
    ~~~~
    \Infer{\tevalapp}
          {
         \arrayenvbl{
          M \nteval \lam{a} M'
          \\
          N \nteval \Mv
          \\{}
          [V / a]M' \nteval \Mv'
         }
        }
       {M(N) \nteval \Mv'}
  \end{mathpar}

  \lesscaptionspace

  \caption{Sorting and evaluation rules for name terms $M$
    %, and the name indices $i$ that index types
  }
  \label{fig:sorting-name-terms}
  \label{fig:eval-name-terms}
\end{figure}

% Local Variables:
% TeX-master: "typed-adapton"
% End:

% see fig-sorting-name-terms.tex

% \begin{figure}[t]
%  \centering

%  \caption{Evaluation rules for name terms}

% \end{figure}

% Local Variables:
% TeX-master: "fungi-lang"
% End:

\Figureref{fig:syntax-name-terms}
shows the syntax of literal names, name terms, name term values, and name term sorts.
Literal names $m$, $n$ are simply binary trees: either an empty leaf $\leafname$
or a branch node $\NmBin{n_1}{n_2}$.
Name terms $M$, $N$ consist of
%the unit name $\unitexp$, a pair of names $\Pair{M_1}{M_2}$,
literal names $n$ and branch nodes $\NmBin{M_1}{M_2}$,
abstraction $\lam{a} M$ and application $\idxapp{M}{N}$.

Name terms are classified by sorts $\sort$: sort $\namesort$
for names $n$,
%unit $\unitsort$ for $\unitexp$,
%product $\sort * \sort$ for pairs $\Pair{M_1}{M_2}$
and $\sort @> \sort$ for (name term) functions.
%Note that $\NmBin{n_1}{n_2}$ is a name $n$---a tree with subtrees
%$n_1$ and $n_2$---and has sort
%$\namesort$, while $\Pair{n_1}{n_2}$ is a pair of names
%and has sort $\Nm * \Nm$.

The rules for name sorting $\Gamma |- M : \sort$
are straightforward  (\Figureref{fig:sorting-name-terms}),
as are the rules for name term evaluation $M \nteval \Mv$
(\Figureref{fig:eval-name-terms}). 
We write $M \conv M'$ when name terms $M$ and $M'$
are convertible, that is, applying any series of $\beta$-reductions
and/or $\beta$-expansions changes one term into the other.

% Local Variables: 
% mode: latex
% TeX-master: "typed-adapton"
% End: 

\section{Type System}
\label{sec:typesystem}

The structure of our type system is inspired by Dependent ML \citep{Xi99popl,Xi07}.
Unlike full dependent typing, DML is separated into
a \emph{program level} and a less-powerful \emph{index level}.
The classic DML index domain is integers with linear inequalities,
making type-checking decidable.
Our index domain includes names, sets of names, and functions over names.
Such functions constitute a tiny domain-specific language that is
powerful enough to express useful transformations of names,
but preserves decidability of type-checking.

Indices in DML have no direct computational content.
For example, when applying a function on vectors that is indexed by
vector length, the length index is not directly manipulated at run time.
However, indices can indirectly reflect properties of run-time values.
The simplest case is that of an indexed \emph{singleton type}, such as $\Int[[k]]$.
Here, the ordinary type $\Int$ and the index domain of integers are in one-to-one
correspondence; the type $\Int[[3]]$ has one value, the integer 3.

While indexed singletons work well for the classic index domain of integers,
they are less suited to names---at least for our purposes.
Unlike integer constraints, where integer literals are common in types---%
for example, the length of the empty list is $0$---%
literal names are rare in types.
Many of the name constraints we need to express look like
``given a value of type $A$ whose name in the set $X$,
this function produces a value of type $B$ whose name is in the set $f(X)$''.
A DML-style system can express such constraints, but the types become verbose:
%
%\begin{mathdispl}
$
   \arrayenvl{
     \forall a : \namesort.\; \forall X : \namesetsort.\;
%     \\
     (a \in X) \impty
     \big(
          A[[a]] -> B[[f(a)]] % \conjty (b \in f(X)))
     \big)
    }
$.
%\end{mathdispl}
%
The notation is taken from one of DML's descendants, Stardust \citep{DunfieldThesis}.
The type is read ``for all names $a$ and name sets $X$,
such that $a \in X$, given some $A[[a]]$ the function returns
$B[[f(a)]]$''.

We avoid such locutions by indexing single values by name sets, rather than names.
For types of the shape given above, this removes half the quantifiers
and obviates the $\in$-constraint attached via $\impty$:
%
%\begin{mathdispl}
$
   \arrayenvl{
     \forall X : \namesetsort.\;
%     \\ ~~
         A[[X]] -> B[[f(X)]]
    }
$.
%\end{mathdispl}
%
This type says the same thing as the earlier one, but
now the approximations
are expressed within the indexing of $A$ and $B$.
Note that $f$, a function on names, is interpreted pointwise:
$f(X) = \{ f(N) \such N \in X \}$.
Standard singletons are handy for index functions on names,
where one usually needs to know the specific function.

For aggregate data structures such as lists, indexing by a name set denotes
\emph{overapproximation} of names:
the proper DML type
%
%\begin{mathdispl}
$
   \arrayenvl{
     \forall Y : \namesetsort.\; \forall X : \namesetsort.\;
%     \\
     %~~
     (Y \subseteq X) \impty
     \big(
          A[[Y]] -> B[[f(Y)]] % \conjty (b \in f(X)))
     \big)
    }
$
%\end{mathdispl}
%
can be expressed by
$%\begin{mathdispl}
   \arrayenvl{
     \forall X : \namesetsort.\;
%     \\
          A[[X]] -> B[[f(X)]] % \conjty (b \in f(X)))
    }
$.%\end{mathdispl}

% XXX
% While making disjointness constraints implicit is convenient,
% it leads to some new subleties.
% For example, if $\Gamma$ contains the typing assumption
% $x : \Nm[\{n_1\} \disj \{n_2\}]$,
% then $\Gamma$ \emph{implicitly} expresses the constraint that
% $n_1 \neq n_2$.

% \semiflaming{Question: For aggregate data structures, like lists,
% do we usually need to know exactly what set of names we have?
% Is it potentially useful to express overapproximation directly within
% indexing, like $\textsf{List}[[Y]]$ meaning
% ``this list has names that are a subset of $Y$''?
% Matt's answer:  No and yes.  That is, exact knowledge is not necessary, and overapproximation is useful.}
%
Following call-by-push-value \citep{Levy99,LevyThesis},
we distinguish \emph{value types} from \emph{computation types}.
Our computation types will also model effects, such as the allocation of a thunk with
a particular name.

% moved to progsyntax.tex:
% [
% \input{fig-syntax-name-terms.tex}
% \input{fig-sorting-name-terms.tex}
% \input{fig-eval-name-terms.tex}
% ]

\begin{figure}[t]
\begin{grammar}
%  Index variables
%  & $a$
%  \\
  Index exprs.\ 
  & $i, j,$
      &$\bnfas$& $a$ & index variable
 \\ 
& $X, Y, Z,$
&& $\bnfaltbrk \{ N \}$ & singleton name set
 \\ 
& $R, W$
&& $\bnfaltbrk \emptyset ~|~ X \disj Y$ & empty set, separating union
\\ &&& $\bnfaltbrk X \union Y$ & union (not necessarily disjoint)
 \\ &&& $\bnfaltbrk \unitindex \bnfalt \Pair{i}{i} \bnfalt \Proj{1}{i} \bnfalt \Proj{2}{i}$
 & unit, pairing, and projection
 \\ &&& $\bnfaltbrk \lam{a} i \bnfalt i(j)$ & function abstraction and application
 \\ &&& $\bnfaltbrk M [[ i ]] \bnfalt i [[ j ]] \bnfalt i^\ast [[ j ]] $ & name set mapping and set building
%   \\[1ex]
%   Propositions
%   & $P$
%       &$\bnfas$&
%       $
%       \trueprop
%       \bnfaltBRK P \andprop P
%       \bnfaltBRK X \disj Y
%       \bnfaltBRK X \subseteq Y
%       $
 \\[1ex]
  Index sorts
  & $\sort$ 
     &$\bnfas$& 
  $\cdots~\bnfaltbrk \namesetsort$  & name set sort
  \\
  &&& $\bnfaltbrk \unitsort$      &unit index sort; inhabitant $\unitexp$
  \\
  &&& $\bnfaltbrk \sort * \sort$  &product index sort; inhabitants $\Pair{i}{j}$
  \\
& & & $\bnfaltbrk \sort_1 @@> \sort_2$ & index functions over name sets % (functions over name sets)
\end{grammar}

\lesscaptionspace

  \caption{Syntax of indices, name set sort}
  \label{fig:syntax-indices}
\end{figure}

% Local Variables:
% TeX-master: "typed-adapton"
% End:

\begin{figure}[t]
  \centering
\runonfontsz{9pt}

\begin{grammar}
  Kinds
  & $K$
      &$\bnfas$&
      $\type$ & kind of value types
      \\ &&& $\bnfaltbrk \type => K$ & type argument (binder space)
      \\ &&& $\bnfaltbrk \sort => K$ & index argument (binder space)
  \\[1ex]
  Propositions
  & $P$
     &$\bnfas$&
     $\trueprop \bnfalt P \andprop P$ & truth and conjunction
     \\ &&& $\bnfaltbrk i \disj j : \gamma$ & index apartness
     \\ &&& $\bnfaltbrk i \equiv j : \gamma$ & index equivalence
  \\[1ex]
  Effects
  & $\e$
      &$\bnfas$&
               $<<W; R>>$
%   \\[1ex]
%   \multicolumn{4}{@{}l}{Type variables~~$\alpha$}
%   \\
%   \multicolumn{4}{@{}l}{Type constructors~~~$d$}
 \\[1ex]
  Value types
  & $A, B$
      &$\bnfas$&
            $\alpha \bnfalt d \bnfalt \Unit$ & type variables, type constructors, unit
      \\ &&& $\bnfaltbrk A + B \bnfalt A ** B$ & sum, product
      \\ &&& $\bnfaltbrk \Ref{i}{A}$  & named reference cell
      \\ &&& $\bnfaltbrk \Thk{i}{E}$  & named thunk (with effects) % thunk
%      \\ &&& $\bnfaltbrk \Unthk{i}{E}$  & named ``unthunk'' (no effects) % unthunk
      \\ &&& $\bnfaltbrk  A[[i]]$  & application of type to index
      \\ &&& $\bnfaltbrk  A\;B$  & application of type constructor to type
      \\ &&& $\bnfaltbrk  \Name{i}$  & name type (name in name set $i$)
      \\ &&& $\bnfaltbrk  (\namesort @> \namesort)[[M]]$  & name function type (singleton)
      \\ &&& $\bnfaltbrk  \DAll{a : \sort}{P} A$  & universal index quantifier
      \\ &&& $\bnfaltbrk  \DExists{a : \sort}{P} A$  & existential index quantifier
%\vspace*{-3ex}
\end{grammar}

\begin{grammar}
  Computation types
  & $C, D$
      &$\bnfas$&
      $\F A \bnfalt A -> E$  & li$\xF$t, functions
  \\[1ex]
  \dots with effects
  & $E$
      &$\bnfas$&
      $C |> \e$  & effects
 \\ &&& $\bnfaltbrk \All{\alpha : K} E  $   & type polymorphism
% % \\ &&& $\bnfaltbrk \Univ{a : \sort} E  $   & index polymorphism
 \\ &&& $\bnfaltbrk (\DAll{a : \sort}{P} E)  $   & index polymorphism
% % \\ &&& $\bnfaltbrk \Namearr{a}{\sort} E$   & name function polymorphism
%  \\ &&& $\bnfaltbrk P --> E  $   & guarded type
 \\[1ex]
  Typing contexts
  & $\Gamma$
    &$\bnfas$&
    $\cdot$
  \\ &&& $\bnfaltbrk \Gamma, a : \sort$ & index variable sorting
  \\ &&& $\bnfaltbrk \Gamma, \alpha : K$ & type variable kinding
  \\ &&& $\bnfaltbrk \Gamma, d : K$ & type constructor kinding
  \\ &&& $\bnfaltbrk \Gamma, N : A$ & ref pointer
  \\ &&& $\bnfaltbrk \Gamma, N : E$ & thunk pointer
  \\ &&& $\bnfaltbrk \Gamma, x : A$ & value variable
  \\ &&& $\bnfaltbrk \Gamma, P$ & proposition $P$ holds
\end{grammar}
% XX need index-level assumptions $P$? equivalence/disequivalence?
% answer: no and no;
% the former are embedded in indices appearing in \Gamma;
% the latter only appear in the relational contexts
\lesscaptionspace

  \caption{Syntax of kinds, effects, and types}
  \label{fig:syntax-types}
\end{figure}

% Local Variables:
% mode: latex
% TeX-master: "typed-adapton"
% End:

\begin{figure}
  \judgbox{\Gamma |- i : \sort}
          {Under $\Gamma$,
            index $i$ has sort $\sort$
          }
  \begin{mathpar}
    \Infer{\Sortvar}
        {
          (a : \sort) \in \Gamma
        }
        {
          \Gamma |- a : \sort
        }
    ~~~~
    \Infer{\Sortunit}
        {}
        {
          \Gamma |- \unitindex : \unitsort
        }
    ~~~~
    \Infer{\Sortpair}
        {
          \Gamma |- i_1 : \sort_1
          \\
          \Gamma |- i_2 : \sort_2
        }
        {
          \Gamma |- \Pair{i_1}{i_2} : (\sort_1 * \sort_2)
        }
    \and
    \Infer{\Sortproj}
        {
          \Gamma |- i : \sort_1 * \sort_2
        }
        {
          \Gamma |- \Proj{b}{i} : \sort_b
        }
%     \and
%     \Infer{\Sortproj2}
%         {
%           \Gamma |- i : \sort_1 * \sort_2
%         }
%         {
%           \Gamma |- \Proj{2}{i} : \sort_2
%         }
    ~~~~
    \Infer{\Sortempty}
        { }
        {
          \Gamma |- \emptyset : \namesetsort
        }
    ~~~~
    \Infer{\Sortsingleton}
        {
         %  \Gamma |- n \wfname
          \Gamma |- N : \namesort
        }
        {
          \Gamma |- \{ N \} : \namesetsort
        }
    \and
    \Infer{\!\Sortunion}
        {
          \arrayenvbl{
            \Gamma |- X : \namesetsort
            \\
            \Gamma |- Y : \namesetsort
          }
        }
        {
          \Gamma |- (X \union Y) : \namesetsort
        }
    ~~
    \Infer{\!\Sortsepunion}
        {
          \arrayenvbl{
            \Gamma |- X : \namesetsort
            \\
            \Gamma |- Y : \namesetsort
          }
          ~~~
          \extract{\Gamma} ||- X \disj Y : \namesetsort
        }
        {
          \Gamma |- (X \disj Y) : \namesetsort
        }
   \and
    \Infer{\Sortabs}
    {
      \Gamma, a : \sort_1 |- i : \sort_2
    }
    {
      \Gamma |- (\Lam{a}{i}) : (\sort_1 @@> \sort_2)
    }
   ~~~~
     \Infer{\Sortapply}
        {
          \Gamma |- i : \sort_1 @@> \sort_2
          \\
          \Gamma |- j : \sort_1
        }
        {
          \Gamma |- i(j) : \sort_2
       }        
     \\
     \Infer{\!\Sortmap}
          {
          \arrayenvbl{
          \Gamma |- M : \namesort @> \namesort
          \\
          \Gamma |- j : \namesetsort
          }
        }
        {
          \Gamma |- M[[j]] : \namesetsort
        }
    ~~
    \Infer{\!\Sortbuild}
          {
          \arrayenvbl{
          \Gamma |- i : \namesort @@> \namesetsort
          \\
          \Gamma |- j : \namesetsort
          }\!\!\!\!\!\!
        }
        {
          \Gamma |- i[[j]] : \namesetsort
        }
    ~~
    \Infer{\!\Sortstar}
          {
          \arrayenvbl{
          \Gamma |- i : \namesort @@> \namesetsort
          \\
          \Gamma |- j : \namesetsort
          }\!\!\!\!
        }
        {
          \Gamma |- i^\ast[[j]] : \namesetsort
        }
  \lesscaptionspace
  \end{mathpar}

  \caption{Sorts statically classify name terms $M$, and the name indices $i$ that index types}
  \label{fig:sorting}
\end{figure}

% Local Variables: 
% mode: latex
% TeX-master: "typed-adapton"
% End: 

% \input{fig-syntax-types-compact.tex}

\subsection{Index Level}
\label{sec:idxtm}

\Figureref{fig:syntax-indices} gives the syntax of
% names, name terms (which include simple total computations over names),
index expressions
and
index sorts (which classify indices).
We use several meta-variables for index expressions;
by convention, we use $X$, $Y$, $Z$, $R$ and $W$ only for sets of names---index
expressions of sort $\NmSet$.
%and $i$ is any index (perhaps a pair of indices).

%
% \subsubsection{Names}
% Names are the value form of name terms.
% A name $n$ is either
% a symbol $s$,
% the root name $\rootname$,
% the ``halves'' $n.1$ and $n.2$,
% or a pair of two names $\Pair{n_1}{n_2}$.
%
% \subsubsection{Name terms}
%
% Name terms model names and functions over names;
% they correspond to terms in a $\lambda$-calculus extended
% with a type of names.
% A name term $M$ is either a name (value) $n$,
% the unit name $\unitexp$,
% a function $\lam{a} M$ or argument $a$,
% application $M_1\,M_2$,
% the tuple $\Pair{M_1}{M_2}$,
% or a ``half'' ($M.1$ or $M.2$).
% Name terms do \emph{not} allow recursion or
% destruction: a name function cannot case-analyze its argument.
%
% We write $M \nteval \Mv$ for big-step evaluation of name terms;
% the rules are given in \Figureref{fig:namestep}.
%
% (out of date comment) In the grammar,
% $M.1$ and $M.2$ are called ``append'' under ``Name terms'',
% but ``descend'' under ``Names''---where something else is called ``append''.

\paragraph{Name sets.}
If we give a name to each element of a list,
then the entire list should carry the set of those names.
We write $\{N\}$ for the singleton name set,
$\emptyset$ for the empty name set,
and $X \disjoint Y$ for a union of two sets $X$ and $Y$
that requires $X$ and $Y$ to be disjoint;
this is inspired by the separating conjunction
of separation logic \citep{Reynolds02}.
While disjoint union is common in the types that we believe programmers need,
our effects discipline requires non-disjoint union $X \union Y$,
so we include it as well.

\paragraph{Variables, pairing, functions.}
An index $i$ (also written $X$, $Y$, \dots when the index is a set of names)
is either an index-level variable $a$,
a name set (described above: $\{N\}$, $X \disjoint Y$ or $X \union Y$),
the unit index $\unitindex$,
a pair of indices $\Pair{i_1}{i_2}$,
pair projection $\Proj{b}{i}$ for $b \in \{1,2\}$,
an abstraction $\lam{a} i$, application $i(j)$,
or name term application $M[[i]]$.

Name terms $M$ are \emph{not} a syntactic subset of indices $i$,
though name terms can appear inside indices (for example, singleton name sets $\{M\}$).
%However, name terms and indices overlap---for example,
%$\lam{a} \unitexp$
%is both a name term function and an index-level function.
%On the other hand, $\lam{a} \Proj{2}{a}$ is an index-level function
%but not a name term function, because projection is not included
%in the syntax of name terms (\Figureref{fig:syntax-name-terms}).
%
Because name terms are not a syntactic subset of indices (and name sets
are not name terms), the application form $i(j)$ does not allow us to
apply a name term function to a name set.  Thus, we also need name term
application $M[[i]]$, which applies the name function $M$ to each
element of the name set $i$.
The index-level map form $i[[j]]$ collects the
output sets of function~$i$ on the elements of the input set~$j$.
The Kleene star variation $i^\ast[[j]]$ applies the function~$i$ zero or more
times to each input element in set~$j$.

% For example, if $M = (\lam{a} a.1.2)$ then $M[[\rootname]] = \rootname.1.2$.
% $\idxapp{M}{M'}$
%
% (Have we settled on the notation for this?  Are we using ``$M\;i$''?)

\paragraph{Sorts.}

We use the meta-variable $\sort$ to classify indices as well as name terms.
% The sort $\namesort$ classifies indices that are single names,
We inherit the function space ${@>}$ from the name term sorts (\Figureref{fig:syntax-name-terms}).
The sort $\namesetsort$ (\Figureref{fig:syntax-indices}) classifies indices that are name sets.
The function space ${@@>}$ classifies functions over \emph{indices} (e.g., tuples of name sets), not merely name terms.
The unit sort and product sort classify tuples of index expressions.

%while the sort $@>$ classifies index-level functions:
% $(\namesort * \namesort) @> \namesort$   allowed? we don't have projection
%$\namesort @> \namesort$ takes a name and returns a name.

Most of the sorting rules in \Figureref{fig:sorting} are straightforward,
but rule `sort-sep-union' includes a premise
$\extract{\Gamma} ||- X \disj Y : \namesetsort$,
which says that $X$ and $Y$ are \emph{apart} (disjoint).

\paragraph{Propositions and extraction.}

Propositions $P$ are conjunctions of atomic propositions
$i \equiv j : \sort$
and
$i \disj j : \sort$, which express equivalence and apartness
of indices $i$ and $j$.  For example, $\{n_1\} \disj \{n_2\} : \NmSet$
implies that $n_1 \neq n_2$.
Propositions are introduced into $\Gamma$ via index polymorphism
$\DAll{a:\sort}{P} E$, discussed below.

The function $\extract{\Gamma}$
(\Figureref{fig:extract} in the appendix)
looks for propositions $P$, which become equivalence and apartness assumptions.
It also translates $\Gamma$ into the relational context used in the definition
of apartness.
We give semantic definitions of equivalence and apartness in the appendix
(Definitions \ref{def:super-semantic-equivalence-i}
and \ref{def:super-semantic-apartness-i}).

% \subsubsection{Entailment}
% We assume an entailment relation
% $\Gamma |- P$,
% where $P$ is a set-theoretic proposition such as
% $n \in X$ or $X \subseteq Y$ or $X \disjoint Y$;
% the latter is interpreted as $(X \sect Y) = \emptyset$.
%
% We assume that entailment is closed under conversion:
% for example, if $M(n) \conv n'$,
% then $\Gamma |- M(n) \in N$ iff $\Gamma |- n' \in N$.
% We also assume that weakening holds:
% if $\Gamma_1, \Gamma_3 |- P$
% then $\Gamma_1, \Gamma_2, \Gamma_3 |- P$.

% \subsubsection{Propositions}
% 
% Propositions $P$ consist of the true constant 
% $\trueprop$;
% conjunction
% $P \andprop P$;
% set inclusion $X \subseteq Y$;
% and
% disjointness
% $X \disj Y$, which asserts that $X \sect Y = \emptyset$.

\subsection{Kinds}

We use a simple system of \emph{kinds} $K$ (\Figureref{fig:kinding} in the appendix).
Kind $\type$ classifies value types, such as $\unitty$ and $(\Thk{i}{E})$.
% %  Definitely value types (and probably \emph{only} value types for $\type => K$);
% %  check if we need it for computation types (any premises of rules that are kinding
% %  with a computation type?).
% \semiflaming{I think I get it now.  Kinding is needed for value types.
% For computation types, we can write down (indeed, we did) rules for a judgment
% ``computation type $C$ (or type-with-effects $E$) has kind $K$'',
% but the kind is always $\type$.  So those judgments could equally well look like
% ``$\Gamma |- C \textsf{\;ctype}$'' (or \textsf{\;efftype}).}

Kind $\type => K$ classifies type expressions that are parametrized by a type.
Such types are called \emph{type constructors} in some languages.
% for example, $\tyname{List}$ has kind $\type => \type$,
% and $\tyname{List}\;\unitty$ has kind $\type$.
  
Kind $\sort => K$ classifies type expressions parametrized by an index.
For example, the $\tyname{List}$ type constructor from \Sectionref{sec:overview}
takes a name set: $\tyname{List}[[X]]$, so $\tyname{List}$ has kind $\namesetsort => \type$.
% For example, the $\tyname{Seq}$ type constructor from \Sectionref{sec:examples}
% takes a name set, \eg $\tyname{Seq}[[X]]$.
% Therefore, this (simplified variant of) $\tyname{Seq}$ has kind $\namesetsort => \type$.
%
A more general $\tyname{Seq}$ type would also track its pointers (not just its names), and permit any element type, and would thus have kind
$\namesetsort
=>
\big(
  \namesetsort => (\type => \type)
\big)$.

% (The inner kind $(\type => \type)$ is an instance of
% kind $\type => K$, not kind $\sort => K$.)

\subsection{Effects}

Effects are described by $<<W; R>>$,
meaning that the associated code may write names in $W$,
and read names in $R$.
(To simplify the example in the overview, we omitted the read set.)

Effect sequencing (\Figureref{fig:comp-typing})
is a (meta-level) partial function over
a pair of effects:
the judgment
$\Gamma |- \e_1 \effseq \e_2 = \e$,
means that $\e$ describes the combination
of having effects $\e_1$ followed by effects $\e_2$.
Sequencing is a partial function because the effects are only
valid when
(1) the writes of $\e_1$ are disjoint from the writes of $\e_2$,
and
(2) the reads of $\e_1$ are disjoint from the writes of $\e_2$.
Condition (1) holds when each cell or thunk is not written more than once
(and therefore has a unique value).
Condition (2) holds when each cell or thunk is written before it is read.
% XX explain why (1) and (2) are necessary (or refer to explanation elsewhere in paper)

Effect coalescing,
``$E \effcoal \e$'',
combines ``clusters'' of effects:
%\[
$
\big(C |> <<\{n_2\}; \emptyset>>\big)
\effcoal
<<\{n_1\}; \emptyset>>
=
C |> (<<\{n_1\}; \emptyset>> \effseq <<\{n_2\}; \emptyset>>)
=
C |> <<\{n_1, n_2\}; \emptyset>>
$.
%\]
%
% Coalescing goes under quantifiers:
% \[
%       \big(
%       \All{\alpha : \type} C |> <<\emptyset; \emptyset>>
%       \big) \effcoal <<\emptyset; \{n_3\}>>
%       ~=~
%       \All{\alpha :\type}
%          C |> \big(<<\emptyset; \emptyset>> \effcoal <<\emptyset; \{n_3\}>>\big)
%        ~=~
%       \All{\alpha :\type}
%          (C |> <<\emptyset; \{n_3\}>>)
% \]
%
Effect subsumption $\e_1 \effsub \e_2$ holds when
the write and read sets of $\e_1$ are subsets of the respective sets of $\e_2$.

\subsection{Types}

\begin{figure}[t]
  \centering

  \judgbox{\Gamma |- v : A}{Under assumptions $\Gamma$, value $v$ has type $A$}
  \vspace*{-0.8ex}
  \begin{mathpar}
    \Infer{\!var}
        {(x : A) \in \Gamma}
        {\Gamma |- x : A}
    \and
    \Infer{\vtypeSub}
         {
           \Gamma |- v : A_1
           \\
           \Gamma |- A_1 \vsubtype A_2
         }
         {\Gamma |- v : A_2}
    \and
    \Infer{\!unit}
        {}
        {\Gamma |- \unitexp : \Unit}
    ~~~~
    \Infer{\!pair}
        {
          \Gamma |- v_1 : A_1
          \\
          \Gamma |- v_2 : A_2
        }
        {\Gamma |- \Pair{v_1}{v_2} : (A_1 ** A_2)}
    ~~~~
    \Infer{\!inj}
        {
          \Gamma |- v_i : A_i
        }
        {\Gamma |- \Inj{i}{v_i} : (A_1 + A_2)}
    \\
    \Infer{name}
       {
         \Gamma |- n \in X
       }
       {
         \Gamma |- (\name{n}) : \Name{X}
       }
    ~~~~~
    \Infer{namefn}
       {
         \Gamma |- M_v : \namesort @> \namesort
         \\
         M_v \conv M
       }
       {
         \Gamma |- (\namefn{M_v}) : (\namesort @> \namesort)[[M]]
       }
   \\
    \Infer{ref}
       {
         \Gamma |- n \in X
         \\
         \Gamma(n) = A
       }
       {
         \Gamma |- (\refv{n}) : (\Ref{X}{A})
       }
    ~~~~~
    \Infer{thunk}
       {
         \Gamma |- n \in X
         \\
         \Gamma(n) = E
       }
       {
         \Gamma
         |-
         (\thunk{n})
         :
        (\Thk{X}{E})
       }
    %%\and
    %% \Infer{unthunk}
    %%    {
    %%      \Gamma |- n \storetype (C |> \e)
    %%    }
    %%    {
    %%      \Gamma
    %%      |-
    %%      \unthunk{n}
    %%      :
    %%      \big(
    %%          \Unthk{n}{(C |> <<\emptyset; \emptyset>>)}
    %%      \big)
    %%    }
   \\
    \Infer{\vtypeAllIndexIntro}
       {
         \Gamma, a:\sort, P |- v : A
       }
       {
         \Gamma |- v : (\DAll{a : \sort}{P} A)
       }
    ~~~
    \Infer{\vtypeAllIndexElim}
       {
         \Gamma |- i : \sort
         \\
         \arrayenvbl{
         \extract{\Gamma} ||- [i/a]P
         \\  
         \Gamma |- v : (\DAll{a : \sort}{P} A)
         }
       }
       {
         \Gamma |-
         v % \idxapp{e}{i}
         : [i/a]A
       }
   \\
    \Infer{\vtypeExistsIndexIntro}
       {
         \Gamma |- i : \sort
         \\
%         \arrayenvbl{
         \extract{\Gamma} ||- [i/a]P
         \\  
         \Gamma |- v : [i/a]A
%         }
       }
       {
         \Gamma |- \VPack{a}{v} : (\DExists{a : \sort}{P} A)
       }
  \lesscaptionspace
  \end{mathpar}
  
  \caption{Value typing}
  \label{fig:value-typing}
\end{figure}

\begin{figure}[htbp]
\runonfontsz{9.0pt}
  \centering

  $~$\hspace*{-2.0ex}\begin{minipage}[t]{0.64\linewidth}
  \judgbox{\Gamma |- (\e_1 \effseq \e_2) = \e}{Effect sequencing}
%  \vspace*{-4.5ex}
  \begin{mathpar}
  \hspace*{-9.0ex}
    \Infer{}
       {
       \arrayenvbl{
        \extract \Gamma |- W_1 \disj W_2
         \\
         \extract \Gamma |- R_1 \disj W_2
       }
      ~~~~~~
       \arrayenvbl{
         \extract \Gamma |- W_1 \cup W_2 \equiv W_3\!\!\!\!\!
         \\
        \extract \Gamma |- R_1 \cup R_2 \equiv R_3
       }
       }
       {
         \Gamma |-
         % \big(
         <<W_1; R_1>> \effseq <<W_2; R_2>>
         % \big)
         =
         <<
         W_3;
         R_3
         >>
      }    
  \end{mathpar}
%   \begin{align*}
%         &\text{if $\Gamma |- W_1 \disj W_2$ and $\Gamma |- R_1 \disj W_2$}
%   \end{align*}
  \end{minipage}
  \hspace*{-1ex}
\begin{minipage}[t]{0.34\linewidth}
  \hspace*{-6.0ex}\judgbox{\Gamma |- \e_1 \effsub \e_2}{Effect subsumption}
%  \vspace*{-4.5ex}
  \begin{mathpar}
   \hspace*{-3.0ex}\Infer{}
       {
         \extract \Gamma |- (X_1 \disj Z_1) \equiv Y_1 : \NmSet
         \\
         \extract \Gamma |- (X_2 \disj Z_2) \equiv Y_2 : \NmSet
       }
       {
         \Gamma
         |-
         <<X_1; X_2>> \effsub <<Y_1; Y_2>>
      }    
  \end{mathpar}
%   \begin{align*}
%         &\text{if $\Gamma |- W_1 \disj W_2$ and $\Gamma |- R_1 \disj W_2$}
%   \end{align*}
  \end{minipage}
%  \vspace*{-0.5ex}

  \medskip

  \judgbox{\Gamma |- (E \effcoal \e) = E'}{\small Effect coalescing}
  \vspace*{-3.5ex}
  \begin{mathpar} \hfill
    \Infer{}
       {
         \Gamma |- (\e_1 \effseq \e_2) = \e
       }
       {
         \Gamma |-
         \big(
           (C |> \e_2) \effcoal \e_1
         \big)
         =
         (C |> \e)
       }    
    ~~
      \Infer{}
         {
           \Gamma |- (E \effcoal \e) = E'
         }
         {
           \Gamma |- %      \big(
           (\All{\alpha : K} E) \effcoal \e
           % \big)
           \,=\,
           (\All{\alpha : K} E')
           \\\\
           \Gamma |- %      \big(
           (\DAll{a : \sort}{P} E) \effcoal \e
           % \big)
           \,=\,
           (\DAll{a : \sort}{P} E')
         }
%     ~~~
%       \Infer{}
%          {
%            \Gamma |- (E \effcoal \e) = E'
%          }
%         {
%         }
%       \\[0.3ex]
% \Gamma |- 
% %      \big(
%         (P \impty E) \effcoal \e
% %      \big)
%           &=&
%          P \impty (E \effcoal \e)
  \vspace*{-0.8ex}
  \end{mathpar}

  \judgbox{\Gamma |-^\ambns e : E}
          {Under $\Gamma$, within namespace~$\ambns$,
            %after history $\hist$,
            computation $e$ has type-with-effects $E$
            % and effects $\e$
          }
  \lesscaptionspace
  \begin{mathpar}
    \Infer{\etypeSub}
        {
          \Gamma |-^\ambns e : E_1
          \\
          \Gamma |- E_1 \esubtype E_2
        }
        {
          \Gamma |-^\ambns e : E_2
        }
%     \Infer{\etypeEffSubsume}
%         {
%           \Gamma |-^\ambns e : (C |> \e_1)
%           \\
%           \Gamma |- \e_1 \effsub \e_2
%         }
%         {
%           \Gamma |-^\ambns e : (C |> \e_2)
%         }
%     \and
%     \Infer{eff-sequence}
%         {
%           \Gamma |-\ambns e : (C |> \e_1) |> \e_2
%           \\
%           (\e_1 \effseq \e_2) = \e_3
%         }
%         {
%           \Gamma |-\ambns e : (C |> \e_3)
%         }
    \vspace*{-2.0ex}
    \\
    \Infer{split}
        {
          \arrayenvbl{
          \Gamma |-%^\ambns
          v : (A_1 ** A_2)
          \\
          \Gamma, x_1:A_1, x_2:A_2 |-^\ambns e : E
          }
        }
        {
          \Gamma |-^\ambns \Split{v}{x_1}{x_2}{e} : E
        }
    \and
    \Infer{case}
        {
          \Gamma |-%^\ambns
          v : (A_1 + A_2)
          \\
          \arrayenvbl{
            \Gamma, x_1:A_1 |-^\ambns e_1 : E
            \\
            \Gamma, x_2:A_2 |-^\ambns e_2 : E
          }
        }
       {
          \Gamma |-^\ambns \Case{v}{x_1}{e_1}{x_2}{e_2} : E
       }
    \and
    \Infer{ret}
        {
          \Gamma |- v : A
        }
        {
          \Gamma |-^\ambns
          \Ret{v}
          :
              (\F A)  |>  <<\emptyset; \emptyset>>
        }
    \and
    \Infer{let}
        {
          \arrayenvbl{
            \Gamma |-^M e_1 : (\F A) |> \e_1
            \\
            \Gamma, x:A |-^M e_2 : (C |> \e_2)
          }
          \\
          \Gamma |- (\e_1 \effseq \e_2) = \e
        }
        {
          \Gamma |-^M \Let{e_1}{x}{e_2}
          :
          (
             C |> \e
          )
        }
    \and
    \Infer{lam}
        {
          \Gamma,x:A |-^\ambns e : E
        }
        {
          \Gamma |-^\ambns
          (\lam{x} e)
          :
          \big(
              (A -> E) |> <<\emptyset; \emptyset>>
          \big)
       }
    \and
    \Infer{app}
        {
          \arrayenvbl{
          \Gamma |- (E \effcoal \e_1) = E_1
          \\
          \Gamma |-^\ambns e
              : \big(
                 (A -> E) |> \e_1
                \big)
          }
          \\
          \Gamma |- v : A
        }
        {
          \Gamma |-^\ambns (e\;v) : E_1
        }
    \and
    \Infer{thunk}
        {
          \Gamma |- v : \Name{X}
          \\
          \Gamma |-^\ambns e : E
        }
        {
          \Gamma |-^\ambns
          \Thunk{v}{e}
          :
          \big(
             \F (\Thk{M[[X]]}{E})
          \big)
          |>
          <<M[[X]]; \emptyset>>
        }
    %% \and
    %% \Infer{unthunk}
    %%     {
    %%       \Gamma |- v : \Thk{X}{(C |> \e)}
    %%     }
    %%     {
    %%       \Gamma |-^\ambns
    %%       \Unthunk{v}
    %%       :
    %%       \big(
    %%          \F (\Unthk{X}{(C |> <<\emptyset;\emptyset>>)})
    %%       \big)
    %%       |>
    %%       (<<\emptyset; X>> \effseq \e)
    %%     }
    \and
    \Infer{force}
        {
%          \Gamma |- v : \Thk{n}{E}
          \Gamma |- v : \Thk{X}{(C |> \e)}
%          \Gamma |- v : \Unthk{X}{ \big(C |> <<\emptyset;\emptyset>>\big) }
          \\
          \Gamma |- (<<\emptyset; X>> \effseq \e) = \e'
        }
        {
          \Gamma |-^\ambns
          \Force{v}
          :
          (
            C |> \e'
          )
        }
    \and
    \Infer{ref}
        {
          \Gamma |- v_1 : \Name{X}
          \\
          \Gamma |- v_2 : A
        }
        {
          \Gamma |-^\ambns
          \Refe{v_1}{v_2}
          :
%          \big(
             \F (\Ref{M[[X]]}{A})
%          \big)
          |>
          <<M[[X]]; \emptyset>>
        }
    ~~~
    \Infer{get}
        {
          \Gamma |- v : \Ref{X}{A}
        }
        {
          \Gamma |-^\ambns
          \Get{v}
          :
          (
             \F A
          )
          |>
          <<\emptyset; X>>
        }
    \and
    \Infer{name-app}
        {
          \arrayenvbl{
            \Gamma |- v_M : (\namesort @> \namesort)[[M]]
            \\
            \Gamma |- v : \Name{i}
          }
        }
        {
          \Gamma |-^N (v_M\;v) : \F (\Name{M[[i]]}) |> <<\emptyset; \emptyset>>
        }
    ~~~
    \Infer{scope}
        {
          \arrayenvbl{
          \Gamma |- v : (\namesort @> \namesort)[[N']]
          \\
          \Gamma |-^{N \circ N'} e : C |> <<W; R>>
          }
       }
       {
         \Gamma |-^N \Scope{v}{e} : C |> <<W; R>>
       }
    \\
    \Infer{\etypeAllIntro}
       {
         \Gamma, \alpha : K |-^\ambns t : E
       }
       {
         \Gamma |-^\ambns
         t
         : (\All{\alpha : K} E)
       }
    \and
    \Infer{\etypeAllElim}
       {
         \Gamma |-^\ambns e : (\All{\alpha : K} E)
         \\
         \Gamma |- A : K
       }
       {
         \Gamma |-^\ambns
         e % \tyapp{e}{A}
         : [A/\alpha]E
       }
   \\
    \Infer{\etypeAllIndexIntro}
       {
         \Gamma, a:\sort, P |-^\ambns t : E
       }
       {
         \Gamma |-^\ambns t : (\DAll{a : \sort}{P} E)
       }
    ~~~
    \Infer{\etypeAllIndexElim}
       {
         \Gamma |- i : \sort
         \\
         \arrayenvbl{
         \extract{\Gamma} ||- [i/a]P
         \\  
         \Gamma |-^\ambns e : (\DAll{a : \sort}{P} E)
         }
       }
       {
         \Gamma |-^\ambns
         e % \idxapp{e}{i}
         : [i/a]E
       }
    \\
    \Infer{\etypeExistsIndexElim}
        {
            \Gamma |- v : (\DExists{a : \sort}{P} A)
            \\
            \Gamma, a : \sort, P, x : A |-^\ambns e : E
        }
        {
          \Gamma |-^\ambns \VUnpack{v}{a}{x}{e} : E
        }
    %% \Infer{NamearrIntro}
    %%     {
    %%       \Gamma, a:\sort |-^\ambns e : E
    %%     }
    %%     {
    %%       \Gamma |-^\ambns
    %%       (\namearrintro{a}{e})
    %%       : (\Namearr{a}{\sort} E)
    %%     }
    %% \and
    %% \Infer{NamearrElim}
    %%     {
    %%       \Gamma |-^\ambns e : (\Namearr{a}{\sort} E)
    %%       \\
    %%       \Gamma |- t : \sort
    %%     }
    %%     {
    %%       \Gamma |-^\ambns
    %%       (\namearrelim{e}{t}) % \tyapp{e}{A}
    %%       : [t/a]E
    %%     }
%     \and
%     \Infer{GuardIntro}
%        {
%          \Gamma, P |-^\ambns \te : E
%        }
%        {
%          \Gamma |-^\ambns
%          \te
%          : (P \impty E)
%        }
%     \and
%     \Infer{GuardElim}
%        {
%          \Gamma |-^\ambns e : (P \impty E)
%          \\
%          \Gamma |- P  % XXX define judgment!
%        }
%        {
%          \Gamma |-^\ambns
%          e % \tyapp{e}{A}
%          : E
%        }
  \lesscaptionspace
  \end{mathpar}
  
  \caption{Computation typing}
  \label{fig:comp-typing}
\end{figure}

% Local Variables: 
% mode: latex
% TeX-master: "fungi-lang"
% End: 

The value types (\Figureref{fig:syntax-types}), written $A$, $B$, include
standard sums $+$ and products $**$;
a unit type;
the type $\Ref{i}{A}$ of references named $i$ containing a value of type $A$;
the type $\Thk{i}{E}$ of thunks named $i$ whose contents have type $E$ (see below);
the application $A[[i]]$ of a type to an index;
the application $A\;B$ of a type $A$ (\eg a type constructor $d$) to a type $B$;
the type $\Name{i}$;
and a singleton type $(\namesort @> \namesort)[[M]]$ where $M$ is a function on names.

As usual in call-by-push-value, computation types $C$ and $D$ include
a connective $\xF$, which ``li$\xF$ts'' value types to computation types:
$\F A$ is the type of computations that, when run,
return a value of type $A$.
(Call-by-push-value usually has a connective dual to $\xF$, written $\xU$,
that ``th\textbf{U}nks'' a computation type into a value type;
in our system, $\mathsf{Thk}$ plays the role of $\xU$.)

Computation types also include functions, written $A -> E$.
In standard CBPV, this would be $A -> C$, not $A -> E$.
We separate computation types alone, written $C$, from computation types with
effects, written $E$; this decision is explained in \Appendixref{apx:remarks}.

Computation types-with-effects $E$ consist of $C |> \e$, which is the bare computation type $C$
with effects $\e$, as well as universal quantifiers (polymorphism) over types
($\All{\alpha : K} E$)
and indices
($\DAll{a : \sort}{P} E$).
In the latter quantifier, the proposition $P$ lets us express quantification over
disjoint sets of names.

\paragraph{Value typing rules.}
The typing rules for values (\Figureref{fig:value-typing})
for unit, variables and pairs are standard.
Rule `name' uses index-level entailment to check that the name $n$ is in the name set $X$.
Rule `namefn' checks that $M_v$ is well-sorted, and that $M_v$ is convertible to $M$.
Rule `ref' checks that $n$ is in $X$, and that $\Gamma(n) = A$, that is,
the typing $n : A$ appears somewhere in $\Gamma$;
rule `thunk' is similar. % to `ref'.
 
\paragraph{Computation typing rules.}
Many of the rules that assign computation types (\Figureref{fig:comp-typing})
are standard---for call-by-push-value---with the addition of effects and the namespace $M$.
The rules `split' and `case' have nothing to do with namespaces or effects, so they
pass $M$ up to their premises, and leave the type $E$ unchanged.
Empty effects are added by rules `ret' and `lam', since both \textkw{ret} and
$\lambda$ do not read or write anything.
The rule `let' uses effect sequencing to combine the effects of $e_1$ and the let-body $e_2$.
The rule `force' also uses effect sequencing, to combine the effect of forcing the thunk
with the read effect $<<\emptyset; X>>$.

The only rule that modifies the namespace is `scope', which composes the given
namespace $N$ (in the conclusion) with the user's $v = \namefn{N'}$ in the second premise (typing $e$).

% We distinguish computation types from types-with-effects for reasons
% discussed in \Appendixref{apx:remarks}.

\subsection{Subtyping}

As discussed above, our type system can overapproximate names.
The type $\Name{X}$ means that the name is contained in the set of $X$;
unless $X$ is a singleton, the type system does not guarantee the specific name.
Approximation induces subtyping: we want to allow a program to
pass $\Name{X_1}$ to a function expecting $\Name{X_1 \disj X_2}$.

For space reasons, the subtyping rules are given and explained
in the appendix (\Secref{sec:subtyping}).

\subsection{Bidirectional Version}

The typing rules in Figures \ref{fig:value-typing}
and \ref{fig:comp-typing}
are declarative:
they define what typings are valid, but not how to derive those typings.
% For example, the rule `namefn' says that
% $\namefn{M_v}$ has a type indexed by $M$
% for some $M$ equivalent to $M_v$, which would require
% guessing $M$. 
The rules' use of names and effects annotations
means that standard unification-based techniques,
like Damas--Milner inference, are not readily applicable.
For example, it is not obvious when to apply \etypeAllIntro,
or how to solve unification constraints over names and name sets.

Bidirectional typing \citep{Pierce98popl}
alternates between checking an expression against a known type (\eg from a type annotation)
and synthesizing a type from an expression.
Since checking rules utilize the given type,
bidirectional typing is decidable for a wide range of rich type systems;
see the citations in \citet{Dunfield13}.
Therefore, we formulate bidirectional typing rules that are decidable
and directly give rise to an algorithm.

For space reasons, this system is presented
in the supplementary material (\Appendixref{apx:bidir}).
We prove in \Appendixref{apx:bidirectional-proofs} that
our bidirectional rules are sound and complete with respect
to the type assignment rules in this section:

Soundness (Thms.\ \ref{thmrefsoundbival}, \ref{thmrefsoundbicomp}):
Given a bidirectional derivation for an annotated expression $e$,
there exists a type assignment derivation for $e$ without annotations.

Completeness (Thms.\ \ref{thmrefcomplbival}, \ref{thmrefcomplbicomp}):
Given a type assignment derivation for $e$ without annotations,
  there exist two annotated versions of $e$: one that synthesizes,
  and one that checks.  (This result is sometimes called \emph{annotatability}.)

% We could add a veneer of isolation by writing
% $C |> <<\emptyset; \emptyset>>$ as just $C$,
% making the `ret' and `lam' rules look simpler, but at the cost of confusion.

% Local Variables: 
% mode: latex
% TeX-master: "fungi-lang"
% End: 

\section{Dynamic Semantics}
\label{sec:dynamics}

% OBSOLETE; see fig-eval-name-terms.tex
%     \input{fig-namestep.tex}

\def\DYNSEMFIGMODE{1}
\ifnum\DYNSEMFIGMODE=1
\begin{figure}[t]
\else
\begin{figure}[h]
\fi

\begin{grammar}
Pointers & $p,q$ & $\bnfas$ & $n$ & name constants
\\[0.2ex]
Stores
& $\St$
    & $\bnfas$ &
    $\emptystore$ & empty store
\\ &&& $\bnfaltbrk$ $\St, p {:} v$ & $p$ points to value $v$
\\ &&& $\bnfaltbrk$ $\St, p {:} e\,{@}\,M$
         & $p$ points to thunk $e$, run in scope~$M$
\end{grammar}
%\vspace{-1.3ex}

  \textbf{Notation}:
  \begin{tabular}[t]{l@{~~}l@{~~}l@{~~}l@{~~}l@{}}
  $\St\{p{\mapsto}v\}$
  &
  and
  &
  $\St\{p{\mapsto}e@M\}$
  &
  \emph{extend} $\St$ at $p$
  &
  when $p \notin \textsf{dom}(\St)$
  \\
%   $\St\{p{\mapsto}e@M\}$ 
   $\St\{p{\mapsto}v\}$ 
  &
  and
  &
  $\St\{p{\mapsto}e@M\}$
  &
  \emph{overwrite} $\St$ at $p$
  &
  when $p \in \textsf{dom}(\St)$
   \end{tabular}
\vspace{1ex}

\judgbox{\PreSt{\St_1}{M}{m} e !! \St_2 ; t}{
  Under store $\St$ in namespace $M$ at current node $m$, \\
  expression
  $e$ produces new store $\St_2$ and result $t$}
\begin{mathpar}
  \runonfontsz{9.0pt}
  \noindent
\ifnum\DYNSEMFIGMODE=2
    \Infer{\!\dynsplit}
        {\PreSt{\St_1}{M}{m} [v_2/x_2][v_1/x_1]e             !! \St_2 ; e' }
        {\PreSt{\St_1}{M}{m} \Split{(v_1, v_2)}{x_1}{x_2}{e} !! \St_2 ; e' }
    ~~~~~
    \Infer{\!\dyncase}
        {\PreSt{\St_1}{M}{m} [v_i/x_i]e_i                          !! \St_2 ; e' }
        {\PreSt{\St_1}{M}{m} \Case{\Inj{i}{v}}{x_1}{e_1}{x_2}{e_2} !! \St_2 ; e' }
    \\
    \Infer{\dynunpack}
         {
           \PreSt{\St_1} {M}{m} [v/x]e         !! \St_2 ; e'
         }
         {
           \PreSt{\St_1}{M}{m}
           \VUnpack{\VPack{a}{v}}{b}{x}{e}
           !!
           \St_2 ; e'
         }
    \and
    \Infer{\dynlet}
      {
%        \arrayenvbl{
          \PreSt{\St_1} {M}{m} e_1              !! \St_1' ; \Ret{v} 
          \\
          \PreSt{\St_1'} {M}{m} [v/x]e_2         !! \St_2' ; e_2'
%        }
      }
      {\PreSt{\St_1'}{M}{m} \Let{e_1}{x}{e_2} !! \St_2' ; e_2' }
    \and
    \Infer{\dynapp}
        {
%          \arrayenvbl{
            \PreSt{\St_1} {M}{m} e_1      !! \St_1' ; \lam{x} e_2
            \\
            \PreSt{\St_1} {M}{m} [v/x]e_2 !! \St_2' ; e_2'
%          }
        }
        {\PreSt{\St_1'}{M}{m} e_1\;v    !! \St_2' ; e_2' }
   \\
\fi
    \Infer{\!\dynscope}
        {
          \PreSt{\St_1} {M_1 \circ M_2}{m} e !! \St_2 ; e'
        }
        { \PreSt{\St_1} {M_1}{m} \Scope{M_2}{e} !! \St_2 ; e' }
    ~~~
    \Infer{\!\dynnameapp}
        {
            M_1 \nteval \lam{a} M_2
            \\
            [n/a]M_2 \nteval p
        }
        {
          \PreSt{\St}{M}{m}
              M_1\;(\name{n})
          !! \St ; \Ret{\name{p}}
        }
\vspace*{-2.0ex}
\\
    \Infer{\!\dynthunk}
      {
        (M\,n) \nteval p
        \\
        \St_1\{p{\mapsto}e\,{@}\,M\} = \St_2
      }
      {
        \arrayenvl{
        \PreSt{\St_1} {M}{m} \Thunk{\name{n}} e !! \St_2; \Ret{\thunk{p}}
        \\~
        }
      }
    ~~~~
    \Infer{\!\dynref}
        {
          \arrayenvbl{ ~\\
          (M\,n) \nteval p
          }
          ~~~~~~~
          \St_1\{p{\mapsto}v\} = \St_2 
        }
        {
          \PreSt{\St_1} {M}{m} 
          \arrayenvl{
          ~~~\Refe{\name{n}}{v}
          \\
          !! \St_2; \Ret{\refv{p}}
          }
        }
   \vspace*{-2.7ex}
    \\
    \Infer{\!\dynforce}
      {
        S(p) = e~@~M_0
        \\\\
        \PreSt{\St_1}{M_0}{p} e !! \St_2; t
      }
      {
        \PreSt{\St_1} {M}{m} \Force{\thunk{p}} !! \St_2; t
      }
    ~~~
    \Infer{\!\dynget}
      { \St(p) = v }
      { \PreSt{\St} {M}{m} \Get{\refv{p}} !! \St ; \Ret{v} }
    ~~~
  \Infer{\!\dynterm}
        %{ e~\textsf{terminal} }
        {}
        {\PreSt{\St}{M}{m} t !! \St ; t }
 \lesscaptionspace
\end{mathpar}
\ifnum\DYNSEMFIGMODE=1
  \caption{Excerpt from the dynamic semantics (see also \Figureref{fig:dynamics})}
  \label{fig:dynamics-excerpt}
\else
  \caption{Dynamic semantics, complete}
  \label{fig:dynamics}
\fi

\end{figure}

% Local Variables:
% mode: latex
% TeX-master: "typed-adapton"
% End:

%\MattSaysTODO{
%\begin{itemize}
%\item
%Fig 14. How are graphs different than "heaps" in other presentations of
%languages with stores?
%\item
%Is the operation G{p |-> e@M} defined when G already has a mapping for p?
%If so, is the current mapping updated by the new definition?
%\end{itemize}
%}

\paragraph{Name terms.}

Recall \Figref{fig:eval-name-terms} (\Secref{sec:names}), which gives
the dynamics for evaluating name term~$M$ to name term
value~$\Mv$.
Because name terms have no recursion, % and pattern-matching,
evaluating a well-sorted name term
always produces a value~(\Theoremref{thm:norm-name}).

\paragraph{Program expressions (\Figureref{fig:dynamics-excerpt}).}
Stores hold the mutable state that names dynamically identify.
Big-step evaluation
for expressions relates an initial and final store, and the
``current scope'' and ``current node'', to a program and value.
We define this dynamic semantics, which closely mirrors prior work,
to show that well-typed evaluations always allocate 
%precisely.
with unique names.

To make this theorem meaningful, the dynamics permits programs to
\emph{overwrite} prior allocations with later ones: if a name is used
ambiguously, the evaluation will replace the old store content with
the new store content.
The rules \dynref and \dynthunk either extend or
overwrite the store, depending on whether the allocated pointer name
is unique or ambiguous, respectively.
We prove that, in fact, well-typed programs always extend (and never
overwrite) the store in any single derivation.  %
(During change propagation, not
  modeled here, we begin with a store and dependency graph
  from a prior run, and even programs without naming errors overwrite the
  (old) store/graph, as discussed in \Secref{sec:intro}.)

While motivated by incremental computation, we are interested in
the allocation effects of a single run, not change propagation between runs.
Consequently, this semantics is simpler than the dynamics of prior work.
First, the store never caches values from evaluation, that is, it does
not model function caching (memoization).
Next, we do not build the dependency edges required for change
propagation.
Likewise, the ``current node'' is not strictly necessary here, but we
include it for illustration.
Were we modeling change propagation, rules
\dynref, \dynthunk, \dynget and \dynforce
would create
dependency edge structure that we omit here.
(These edges relate the current node with the node being observed.)
%, as illustrated in
%\Figref{fig:listmap}~(\Secref{sec:intro}).

% arrrggghh
% Local Variables: 
% mode: latex
% TeX-master: "typed-adapton"
% End: 

%\clearpage

\section{Metatheory: Type Soundness and Unique Names}
\label{sec:metatheory}

In this section, we prove that our type system is sound with respect
to evaluation: 
Every
well-typed,
terminating program produces a terminal computation of the program's type, 
the set of dynamic allocations match the program's static approximation,
and each allocation is globally unique.
\Defnref{def:rw} defines which evaluation derivations have
\emph{precise effects} matching the requirements above.

We sometimes constrain typing contexts to be \emph{store types},
which type store pointers but not program variables; hence, they only
type \emph{closed} values and programs:
\begin{defn}[Store type]
\Label{def:storetype}
We say that $\Gamma$ is a \emph{store typing},
written $\StoreType{\Gamma}$, when each assumption in $\Gamma$ has the
reference-pointer form $p : A$ or the thunk-pointer form $p : E$.
\end{defn}

% \subsection{Precise effects: Read and write sets}
% \Label{sec:rw-sets}

\NotInScope{
(XXX:
Definition of the read sets for \dynapp and \dynlet had $R_1 \mergeRds (R_2 - W_1)$.
We don't know why we were subtracting $W_1$ out; it means that the definition of a read
set excludes locations written to in the first subderivation, which doesn't match the type system.
So we are removing the subtractions.)
}

\begin{defn}[Precise effects]
\Label{def:rw} %\raggedright
  Given an evaluation derivation $\D$, we write
  $\DHasEffects{\D}{R}{W}$ for its \emph{precise effects}
  (\Figureref{fig:readswrites} in the appendix).
\end{defn}

This is a (partial) function over derivations.
We call these effects ``precise'' since sibling sub-derivations must have
disjoint write sets.  

% \vspace*{-0.7ex}
% \paragraph{Main theorem:}
%
% In the statement below, ``$\derives$'' is read ``derives'',
% so that $\Ss$ and $\Dee$ are the given typing and evaluation derivations.
We write $<<W'; R'>> \effsub <<W; R>>$
to mean that $W' \subseteq W$ and $R' \subseteq R$.
For proofs, see \Appendixref{apx:proofs}.

%\clearpage

\begin{restatable}[Subject Reduction]{theorem}{thmrefsubj}
\Label{thm:refsubj}
%\begin{tabular}{@{\hspace{0.75in}}l}
% ~\\
If
$\StoreType{\Gamma_1}$ 
and
%\begin{minipage}{\textwidth}
$        \Gamma_1 |- M : \namesort\,{@>}\,\namesort$
% \\
and
% $\Ss$ derives
$\Gamma_1 |-^M e : C |> <<W; R>> $
% \\
% \\
and
$                 |- \St_1 : \Gamma_1$
\!and
$\Dee$ \!derives $\PreSt{\St_1}{M}{m} e !! \St_2 ; t$
then
% \\
there exists $\Gamma_2\,{\supseteq}\,\Gamma_1$
s.t.\ % such that
$\StoreType{\Gamma_2}$ and
$|- \!\St_2 : \Gamma_2$
and
$\Gamma_2 |- t : C |> <<\emptyset; \emptyset>> $
% \\
and
$\DHasEffects{\Dee}{R_\Dee}{W_\Dee}$
and
$<<W_\Dee; R_\Dee>> \effsub <<W; R>>$.
%\end{minipage}
%\end{tabular}
\end{restatable}

Our implementation (\Secref{sec:impl}) 
follows the change propagation algorithm of \citet{Hammer15}, 
which has been formalized and proven correct (from-scratch consistent) when 
\Defnref{def:rw} (precise effects) holds 
for every program run under consideration---a guarantee of Fungi's type-and-effect system, as stated above.

% Local Variables: 
% mode: latex
% TeX-master: "fungi-lang"
% End: 

\section{Implementation}
\label{sec:impl}

\subsection{Prototype in Rust}

%
% ``Software arch'' at a high level:
%
Using this on-paper design as a guide, we have implemented a preliminary
prototype of \Fungi in Rust.
In particular, we implement each abstract syntax definition and typing
judgement presented in this paper and appendix as a Rust datatype (a
``deep'' embedding of the language into Rust).
We implement the bidirectional type system (\Secref{apx:bidir}) as a family
of Rust functions that produce judgement data structures (possibly
with nested type or effect errors) from a \Fungi syntax tree.

%
% Program syntax notes
%
By using Rust macros, we implement a concrete syntax and associated
parser that suffices for authoring examples similar to those in
\Secref{sec:overview}.
In two ways, we deviate from the \Fungi program syntax presented here:
(1) Rust macros can only afford certain concrete syntaxes
(2) \Fungi programs use explicit (not implicit) index and type
applications; inferring these arguments is future work.

We implement an \emph{incremental} semantics for \Fungi 
based on Adapton in Rust, as provided by an existing external library~\cite{Adapton}.
This library uses the change propagation algorithm(s) of \citet{Hammer14,Hammer15}.
The implementation of \Fungi is documented and publicly available.
At present, it consists of about 10K~lines of Rust.
%
%, and is complete
%enough to type check many basic examples, including the effects of
%structural recursion (\code{max} and \code{filter},
%\Figref{sec:structuralrec}).
%
%We are actively extending the implementation to encompass the full
%reasoning power of the index and name term sub-languages, to type the
%rest of~\Secref{examples}, and beyond.
%
For the latest version of \Fungi, see \texttt{crates.io} and/or
\texttt{docs.rs}, and search for ``fungi-lang''.
\emph{Note to reviewers: visiting those sites will deanonymize the
  authors; see supplemental material instead.}

\subsection{Ongoing and Future Work}

\Secref{sec:future-work-roles}
discusses a proposal for \emph{imperative} (name) effects
in the context of incremental sub-computations 
that (still) require unique names.
Conceivably, 
future \Fungi-based systems could track \emph{reactive} names and their effects,
potentially encoding reactive aspects of FRP language semantics~\citep{
  Elliott97,WanHu00,Cooper06embeddingdynamic,Krishnaswami11,Krishnaswami13,Czaplicki2013AFR}.
In the long term, we intend \Fungi as a \emph{target} language for higher-level 
incremental programming languages.
%, perhaps with more implicit features.
%
%In the meantime, 
%we are exploring ways to make the \Fungi type system more accessible.

\paragraph{Interactive type derivations.}
To debug the examples' type and effect errors, we load the
(possibly incomplete) typing derivations in an associated interactive, web-based
tool.
The tool makes the output typing derivation \emph{interactive}: using
a pointer, we can inspect the syntactic family/constructor, typing
context, type and effect of each subterm in the input program,
including indices, name terms, sorts, values, expressions, etc.
Compared with getting parsing or type errors out of context (or else, only
with an associated line number), we've found this interactive tool
very helpful for teaching newcomers about \Fungi's abstract syntax
rules and type system, and for debugging examples (and
\Fungi)~ourselves.
This tool, the \emph{Human-Fungi Interface} (HFI), is publicly
available software.

As future work, we will extend HFI into an interactive \emph{program
  editor}, based on our existing bidirectional type system, and the
(typed) structure editor approach developed by \citet{Omar17}.
We speculate that \Fungi \emph{itself} may be useful in the implementation of
this tool, by providing language support for interactive,
\emph{incremental} developer features~\cite{Omar17snapl}.
Current approaches prescribe conversion to a distinct,
``co-contextual'' judgement form that requires transforming 
the desired typing rules and their modes~\cite{Erdweg15,Kuci2017}.
\Fungi's explicit-name programming model may offer an alternative
approach for authoring incremental type checkers, based on their
``ordinary'' judgments (rule and typing context structure).

\section{Related Work}
\label{sec:relatedwork}
\label{sec:related}
\label{related}

DML \citep{Xi99popl,Xi07} is an influential system of limited dependent types
or \emph{indexed} types.  Inspired by \citet{Freeman91},
who created a system in which datasort refinements were clearly separated from ordinary types,
DML separates the ``weak'' index level of typing from ordinary typing;
the dynamic semantics ignores the index level.

Motivated in part by the perceived burden of type annotations in DML,
liquid types \citep{Rondon08,Vazou13} deploy machinery to infer
more types.  These systems also provide more flexibility:
types are not indexed by fixed tuples.

% GADTs?  nah

To our knowledge, \citet{Gifford86} were the first to express effects within (or alongside) types.
Since then, a variety of systems with this power have been developed.
A full accounting of this area is beyond the scope of this paper;
for an overview, see \citet{Henglein05:ATTAPL-Chapter}.
We briefly discuss a type system for regions \citep{Tofte97},
in which allocation is central.
Regions organize subsets of data, so that they can be deallocated together.
The type system tracks each block's region, which in turn requires effects on types:
for example, a function whose effect is to return a block within a given region.
% Functions can be polymorphic in regions and effects.
% XX double-check what is in Tofte97 vs. the ATTAPL chapter (see around p. 116)
Our type system shares region typing's emphasis on allocation,
but we differ in how we treat the names of allocated objects.
First, names in our system are fine-grained, in contrast to giving all the objects in
a region the same designation.
Second, names have structure---for example, the names
$0{\cdot}n = \NmBin{\leafname}{n}$
and
$1{\cdot}n = \NmBin{\NmBin{\leafname}{\leafname}}{n}$
share the right subtree~$n$---which
allows programmers to deterministically compute two distinct names from one.

Substructural type systems \citep{OHearn03,Walker05:ATTAPL-Chapter}
might seem suitable for statically verifying the correct usage of names.
We initially believed that an affine type system would be good
for checking global uniqueness, but we abandoned that route.
First, sharing between data structures can be essential for
efficiency (\eg a \texttt{suffixes} function over a list).
Second, while global uniqueness itself seems within the scope of affine typing,
the justification for global uniqueness rests on local uniqueness properties
that fall outside the scope of affine typing.
It is conceivable that some not-yet-invented substructural type system
could accomplish our goals, but ``off-the-shelf'' affine typing is not viable.

\iffalse
\MattSays{ Around where we state these principles/invariants, we may
  want to also state that one may turn to substructural type systems
  (a common comment made by reviewers).  

  However, as it turns out, these are actually a poor match for the
  intended programming model: 

  First, some \Fungi programs are functional programs, which may not be linear, 
  and may use sharing deliberately.

  Second, for most \emph{idiomatic} \Fungi programs, global
  uniqueness instead rests on local uniqueness holding within one or more data
  structures, and local uniqueness does not itself seem naturally expressed
  with substructural rules,
  since the larger goal of local uniqueness 
  is to \emph{not} be the same as global uniqueness, 
  and to permit sharing of names across distinct structures, 
  to define dynamic input-output correspondences.
  %
  A substructural discipline seems \emph{too} limiting to
  permit locally unique names to be shared and recombined, and thus,
  to express correspondences across multiple structures.  
}
\fi

Type systems for variable binding and fresh name generation,
such as FreshML \citep{Pitts00} and Pure FreshML \citep{Pottier07}, can express that sets
of names are disjoint.
But the names lack internal structure that relates specific names
across disjoint name sets.

% Wadler and Thiemann 2003: no particular relevance, I think  -j.
%
% Gaboardi et al., ICFP 2016: I have no idea what is happening  -j.
%
% 

Compilers have long used alias analysis to support optimization passes.
\citet{Brandauer15}
extend alias analysis with disjointness domains,
which can express local (as well as global) aliasing constraints.
Such local constraints are more fine-grained than classic region systems;
our work differs in having a rich structure on names.

\paragraph{Approaches to incremental computation.}
General-purpose incremental computation techniques use
% general-purpose
\emph{change propagation} algorithms.
Change propagation is a provably sound approach for
recomputing the affected output, as the input changes dynamically
after an initial run of the program
\citep{Acar06,AcarLeyWild08,Hammer14,Hammer15}.

Our type and effect system complements past work on
self-adjusting computation. In particular, we expect that variations
of the proposed type system can express and verify the use of names in
some of the work cited above.

Incremental computation can deliver \emph{asymptotic}
speedups for certain algorithms \citep{AcarIhMeSu07,%HammerAcRaGh07,
%AcarIhMeSu08,
%AcarBlTaTu08,
AcarAhBl08,
Acar09,
SumerAcIhMe11b,
Chen12},
and has even addressed open problems~\citep{AcarCoHuTu10}.
Incremental computing abstractions exist in many settings
\citep{Shankar07,
%  HammerAcRaGh07,
Hammer09, AcarLeyWild08}.
\citet{Cai14} use \emph{derivatives} in an incremental $\lambda$-calculus,
which is more restricted than our setting (for example, their calculus lacks rich datatypes).
Approaches such as
concurrent revisions \citep{Burckhardt11:oopsla},
% provide a programming model that is general-purpose but relatively distant from conventional programming languages.
hybrid reactive/imperative programming \citep{Demetrescu11},
% with reactive features (through designated memory locations),
and embedded incremental query languages \citep{Mitschke14}
constitute alternate approaches to incremental computation,
but diverge more markedly from conventional programming languages.

%%%%%%%%%%%%%%%%%%%
%\citet{Cicek15} -- inplace updates to input/output, and no structural
%changes to the dependency graph (``inplace only'').

%%%%%%%%%%%%%%%%%%%%%%%%
%\citet{Cicek16} -- inplace updates to input/output, but you can
%replace dependency graph sub%-graphs with ``fresh'' from-scratch

\citet{Cicek15} develop cost semantics for a limited class of
incremental programs:
they support only in-place input changes and fixed control flow,
so that the structure of the dynamic dependency graph is fixed.
% dynamic dependency graph with fixed control flow, and thus, fixed structure:
For example, the length of an input list cannot
change across successive incremental runs, nor can the structure of
its dependency graph.
\citet{Cicek16} relax the restriction on control flow (but
not input changes) to permit replacing a dependency subgraph
% which may replace a subcomputation
according to a different, from-scratch execution.
Extending their cost semantics to allow general structural
changes (\eg insertion or removal of list elements),
while describing the cost of change propagation for programs like
\texttt{dedup} from \Secref{sec:overview},
would require integrating a general notion of names.
Without such a notion, constant-sized input changes may cascade,
preventing reuse.
% precipitating needlessly inefficient change propagation behavior.

%%%%%%%%
% Hyphen rule: Composing adj out of several words (needlessly
% inefficient).  If ``needlessly'' were not an adverb, you may want a
% hyphen; otherwise, you do not.

%
% Hence, we feel that our work complements theirs, and combining these
% approaches remains an exciting direction for future work.

\paragraph{Detection of naming errors.}
%\paragraph{Imprecise (ambiguous) names.}  % "in past literature"---it's a related work section, it's all past literature :-P  -j.
%
Some past systems dynamically detect ambiguous names, either forcing
the system to fall back to a non-deterministic name
choice~\citep{Acar06,Hammer08}, or to signal an error and
halt~\citep{Hammer15}.
In scenarios with a non-deterministic fall-back mechanism, a name
ambiguity carries the potential to degrade incremental performance,
making it less responsive and asymptotically unpredictable in
general \citep{AcarThesis}.
To ensure that incremental performance gains are predictable, past
work often merely assumes, without enforcement, that names are
precise~\citep{Ley-Wild09}.  
%
% Fortunately,
These existing approaches are complementary to \Fungi,
whose type and effect system is applicable to each, either
\emph{directly} (in the case of Adapton, and variants), or with some
minor adaptations (as we speculate for the others).

%Other incremental computing techniques go further, \emph{forbidding}
%ambiguous names \emph{and} non-determinism; instead of using a
%non-deterministic fallback mechanism, these systems signal a dynamic
%error when the programmer mistakenly names values or computations
%ambiguously~\citep{Hammer15}.
%
%Ideally, we could prove when these measures are unnecessary, e.g.,
%using the type system we propose.

%% \paragraph{Functional reactive programming.}
%% %
%% Incremental computation and reactive programming, especially
%% functional reactive programming, share common elements: both
%% attempt to respond to outside changes and their implementations often
%% both employ dependence graphs to model dependencies in a program that
%% change over time
%% \citep{Elliott97,WanHu00,Cooper06embeddingdynamic,Krishnaswami11,Krishnaswami13,Czaplicki2013AFR}.
%% %
%% We expect that 

%In \Secref{sec:discussion}, we discuss an extension of our type system
%that marries the reactive feedback that is unique to FRP with the
%incremental data structures and algorithms that are unique to IC.

%% \paragraph{Incremental sequences.}

%% Cite pugh and headley.

%% Unlike other sequence representations, level trees are
%% \emph{history independent}: given a sequence of elements with
%% interposed \emph{levels}, there exists a single canonical tree
%% structure to represent the sequence as a level tree.

% Local Variables: 
% mode: latex
% TeX-master: "fungi-lang"
% End: 

\section{Conclusion}
\label{sec:futurework}
\label{sec:conclusion}

%We define the \emph{precise name problem} for programs that use
%explicit names to identify their dynamic data and sub-computations.
%
%We define a solution in the form of \Fungi, a core calculus for such
%programs, whose type and effect system describes and verifies their
%allocation names.

We present \Fungi, 
a typed functional language for incremental computation with names.
Unlike prior general-purpose languages for incremental computing (\Tableref{tab:related}),
\Fungi's notion of names is formal, general, and statically verified.
In particular, \Fungi's type-and-effect system permits the programmer
to encode (program-specific) local uniqueness invariants about names,
and to use these invariants to establish \emph{global uniqueness} for their
composed programs, the property of using names correctly.
We derive a bidirectional version of the type and effect system, and
we have implemented a prototype of~\Fungi in Rust.  %, as a
%deeply embedded DSL.
%
We apply \Fungi to a library of incremental collections.

Our ongoing and future work on \Fungi builds on initial prototypes
reported here: We are extending \Fungi to settings that \emph{mix}
imperative and functional programming models, and we are creating
richer tools for developing, debugging and visualizing \Fungi
programs in the context of larger systems (e.g., written in Rust).

%%%%%%%%%%%%%%%%%%%%%%%%%%%%%%%%%%%%%%%%%%%%%%%%%%%%%%%%%%%%%%%%%%%%%%%%%%%%%%%%
\begin{acks}   % mandatory to use `acks' environment to omit for double-blind

We thank Ryan L. Vandersmith, who leads the development of the
\emph{Human-Fungi Interface} described in \Secref{sec:impl}; this tool has
been invalable for implementing and testing our \Fungi prototype in
Rust.

  We thank 
  Neelakantan R. Krishnaswami,
  Deepak Garg,
  Roly Perera,
  and
  David Walker
  for insightful discussions about this work, and for their suggestions and comments.
This material is based in part upon work supported by a gift from
Mozilla, a gift from Facebook, and support from the National Science
Foundation under grant number CCF-1619282.
Any opinions, findings, and conclusions or recommendations expressed
in this material are those of the author(s) and do not necessarily
reflect the views of Mozilla, Facebook or the National Science Foundation.
\end{acks}

\addtolength{\bibsep}{1pt}
\runonfontsz{9pt}
\bibliography{j,adapton}
\runonfontsz{10pt}

\ifnum\OPTIONAppendix=1
\clearpage
\appendix

\clearpage
\section{Omitted Definitions, Figures, and Remarks}
\label{apx:omitted}

\subsection{Subtyping}
\label{sec:subtyping}

\begin{figure}[t]
  \centering

\judgbox{\Gamma |- A \vsubtype B}{Value type $A$ is a subtype of $B$}
\vspace*{-2.2ex}
\begin{mathpar}
    \Infer{\vsubrefl}
       {}
       {\Gamma |- A \vsubtype A}
    \and
     \Infer{\vsubname}
        {
           \Gamma ||- X \subseteq Y
        }
        {
          \Gamma |- \Name{X} \vsubtype \Name{Y}
        }
    \\
%     \Infer{\vsubunit}
%        {}
%        {\Gamma |- \unitty \vsubtype \unitty}
%     \and
    \Infer{\vsubproduct}
       {
         \Gamma |- A_1 \vsubtype B_1
         \\
         \Gamma |- A_2 \vsubtype B_2
       }
       {
         \Gamma |- A_1 ** A_2 \vsubtype B_1 ** B_2
       }
    \and
    \Infer{\vsubsum}
       {
         \Gamma |- A_1 \vsubtype B_1
         \\
         \Gamma |- A_2 \vsubtype B_2
       }
       {
         \Gamma |- A_1 + A_2 \vsubtype B_1 + B_2
       }
     \\
     \Infer{\!\vsubref}
        {
          \extract{\Gamma} ||- X \subseteq Y
          \\
          \Gamma |- A \vsubtype B
%          \\  \Gamma |- B \vsubtype A
        }
        {
          \Gamma |- (\Ref{X}{A}) \vsubtype (\Ref{Y}{B})
        }
     ~~~~
     \Infer{\!\vsubthk}
        {
          \extract{\Gamma} ||- X \subseteq Y
          \\
          \Gamma |- E_1 \csubtype E_2
%          \\  \Gamma |- E_2 \csubtype E_1
        }
        {
          \Gamma |- (\Thk{X}{E_1}) \vsubtype (\Thk{Y}{E_2})
        }
     \and
     \Infer{\vsubnamefn}
         {
           \Gamma |- M_1 \conv M_2
         }
         {
           \Gamma |- 
           (\namesort @> \namesort)[[M_1]]
           \vsubtype
           (\namesort @> \namesort)[[M_2]]
         }
    \\
    \Infer{\vsubAllL}
       {
         \Gamma |- i : \sort
         \\
         \arrayenvbl{
           \extract{\Gamma} ||- [i/a]P
           \\
           \Gamma |- [i/a]A \vsubtype B
         }
       }
       {
         \Gamma |- 
         (\DAll{a : \sort}{P} A)
         \vsubtype
         B
       }
    \and
    \Infer{\vsubAllR}
       {
         \Gamma, b : \sort, P |- A \vsubtype B
       }
       {
         \Gamma |-
         A
         \vsubtype
         (\DAll{b : \sort}{P} B)
       }
   \and
   \Infer{\vsubExists}
      {
        \Gamma, a : \sort, P_a |- A \vsubtype [a/b]B
        \\
        \extract{\Gamma, a : \sort, P_a} ||- [a/b]P_b
      }
      {
        \Gamma |-
        (\DExists{a : \sort}{P_a} A)
        \vsubtype
        (\DExists{b : \sort}{P_b} B)
      }
% The original goal of this subtyping system was to simulate the stationary rules of
%  value typing.  Existential introduction is *not* stationary because of the `pack' marker.
%
%     \and
%     \Infer{\vsubExists}
%        {
%          \Gamma, a : \sort, P |- A \vsubtype [a/b]B
%        }
%        {
%          \Gamma |-
%          (\DExists{a : \sort}{P} A)
%          \vsubtype
%          (\DExists{b : \sort}{P} B)
%        }
  \lesscaptionspace
  \end{mathpar}
  
  \caption{Subtyping on value types}
  \label{fig:value-subtyping}
\end{figure}

\begin{figure}[t]
  \centering

\judgbox{\Gamma |- C \csubtype D}{Computation type $C$ is a subtype of $D$}
  \begin{mathpar}
      \Infer{\csublift}
         {\Gamma |- A \vsubtype B}
         {\Gamma |- \F A \csubtype \F B}
      \and
      \Infer{\csubarr}
         {
           \Gamma |- A_2 \vsubtype A_1
           \\
           \Gamma |- E_1 \esubtype E_2
         }
         {\Gamma |- (A_1 -> E_1) \csubtype (A_2 -> E_2)}
  \end{mathpar}

  \judgbox{\Gamma |- E_1 \esubtype E_2}{Type-with-effects $E_1$ is a subtype of $E_2$}
  \begin{mathpar}
      \Infer{\esubeff}
         {
           \Gamma |- C_1 \csubtype C_2
           \\
           \Gamma |- \e_1 \effsub \e_2
         }
         {\Gamma |- (C_1 |> \e_1) \csubtype (C_2 |> \e_2)}
      \and
      \Infer{\esubAllType}
         {
           \Gamma, \alpha : K |- E_1 \esubtype E_2
         }
         {\Gamma |- (\All{\alpha : K} E_1) \esubtype (\All{\alpha : K} E_2)}
      \and
      \Infer{\esubAllIndexL}
         {
             \Gamma |- i : \sort
             \\
             \arrayenvbl{
               \extract{\Gamma} ||- [i/a]P
               \\
               \Gamma |- [i/a]E_1 \esubtype E_2
             }
         }
         {\Gamma |- (\DAll{a : \sort}{P} E_1) \esubtype E_2}
      \and
      \Infer{\esubAllIndexR}
         {
             \Gamma, a : \sort, P |- E_1 \esubtype E_2
         }
         {\Gamma |- E_1 \esubtype (\DAll{a : \sort}{P} E_2)}
  \lesscaptionspace
  \end{mathpar}
  
  \caption{Subtyping on computation types}
  \label{fig:comp-subtyping}
\end{figure}

% Local Variables: 
% mode: latex
% TeX-master: "fungi-lang"
% End: 

To design subtyping rules that are correct and easy to implement,
we turn to the DML descendant Stardust \citep{DunfieldThesis}.
The subtyping rules in Stardust are generally a helpful guide,
with the exception of the rule that compares atomic refinements.
In Dunfield's system, $\tau[[i]] \subtype \tau[[j]]$ if $i = j$ in the
underlying index theory.
For example, a list of length $i$ is a subtype of a list of length $j$
if and only if $i = j$ in the theory of integers.
While approximate in the sense of considering all lists of
length $i$ to have the same type, the length itself is not approximate.

In contrast, our name set indices are approximations.
Thus, our rule \vsubname (\Figureref{fig:value-subtyping})
says that $\Name{X} \vsubtype \Name{Y}$ if $X \subseteq Y$,
rather than $X = Y$.
Similarly, subtyping for references and thunks (\vsubref, \vsubthk)
checks inclusion of the associated name (pointer) set, not strict equality.

Our polymorphic types combine two fundamental typing constructs,
universal quantification and guarded types (requiring that $P$ hold for the
quantified index $a$), so our rule \vsubAllL combines
the Stardust rules $\Pi$L for index-level quantification
and $\supset$L for the guarded type \citep[p.\ 33]{DunfieldThesis}.
Likewise, our \vsubAllR combines Stardust's $\Pi$R and $\supset$R.

Unlike Stardust's $\Sigma$ (and unlike our $\forall$),
our existential types have a term-level pack construct,
so an $\exists$ cannot be a sub- or supertype of a non-existential
type.  Thus, instead of rules analogous to Stardust's $\Sigma$L and
$\Sigma$R, we have a single rule \vsubExists with $\exists$ on both
sides, which specializes $\Sigma$R to the case when
$\Sigma$L derives its premise.  Like $\forall$, our $\exists$
incorporates a constraint $P$ on the quantified variable, so our
\vsubExists also incorporates the Stardust rules for
\emph{asserting types} ($\bindnasrepma$),
checking that $P_a$ entails $P_b$.

For refs and thunks, rules \vsubref and \vsubthk are covariant in
the name set describing the location.
They are also covariant in the type of their contents:
unlike an ordinary ML \texttt{ref} type, our \textkw{Ref} names
a location, but the programs described by our type system cannot mutate that location.
To extend our theory to \emph{editor} programs, we would need
different rules (\Sectionref{sec:future-work-roles} in the appendix).

In our subtyping rules for computation types (\Figureref{fig:comp-subtyping}),
rule \csubarr reflects the usual contravariance of function domains,
rule \esubeff allows subsumption within effects $\e$,
and the rules for computation-level $\forall$ follow our rules for
value-level $\forall$.
Instead of an explicit transitivity rule, which is not trivial to implement,
the transitivity of subtyping is admissible.

\subsection{Dynamic semantics, read and write sets, sorting and kinding}

\begin{figure}[h]
\begin{mathpar}
\Infer{emp}
    { }
    {
     |- \cdot : \Gamma
    }
~~~
\Infer{ref}
    {
      \St |- \Gamma
      ~~~~
%      \arrayenvbl{
        \Gamma |- v : A
      ~~~~
        \Gamma(p) = A
%      }
    }
    {
      |- ( \St, p : v ) : \Gamma
    }
~~~
\Infer{thunk}
    {
      \St |- \Gamma
      ~~~~
%      \arrayenvbl{
        \Gamma |- e : E
      ~~~~
        \Gamma(p) = E
%      }
    }
    {
      |- ( \St, p : e ) : \Gamma
    }
\lesscaptionspace
\end{mathpar}

\caption{Store typing: $\St |- \Gamma$, read ``store $\St$ typed by $\Gamma$''.}
\Label{fig:store-typing}
\end{figure}

\def\DYNSEMFIGMODE{2}

\begin{figure}[h]
  \centering

    \newcommand{\SEP}{\\[1ex]}
    \begin{array}[t]{rl@{~~}lll}
      \DByHasEffects{\D}{{\dynterm}()}{\emptyset}{\emptyset}
    \SEP
      \DByHasEffects{\D}{{\dynapp}(\D_1, \D_2)}{R_1 \mergeRds (R_2 - W_1)}{W_1 \disjoint W_2}
      &\text{if}&
      \DHasEffects{\D_1}{R_1}{W_1}
      \\ &\text{and}&
      \DHasEffects{\D_2}{R_2}{W_2}
    \SEP
      \DByHasEffects{\D}{{\dynlet}(\D_1, \D_2)}{R_1 \mergeRds (R_2 - W_1)}{W_1 \disjoint W_2}
      &\text{if}&
      \DHasEffects{\D_1}{R_1}{W_1}
      \\ &\text{and}&
      \DHasEffects{\D_2}{R_2}{W_2}
    \SEP
      \DByHasEffects{\D}{{\dynscope}(\D_0)}{R}{W}
      &\text{if}&
      \DHasEffects{\D_0}{R}{W}
    \end{array}
    \medskip
    \begin{array}[t]{rl@{~~}lll}
%       \DByHasEffects{\D}{{\dynfix}(\D_0)}{R}{W}
%       &\text{if}& \DHasEffects{\D_0}{R}{W}
%     \\
      \DByHasEffects{\D}{{\dyncase}(\D_0)}{R}{W}
      &\text{if}& \DHasEffects{\D_0}{R}{W}
    \\
      \DByHasEffects{\D}{{\dynsplit}(\D_0)}{R}{W}
      &\text{if}& \DHasEffects{\D_0}{R}{W}
    \SEP
      \DByHasEffects{\D}{{\dynref}()}{\emptyset}{p}
      &\text{where}&
      e = \Refe{\name{n}}{v} \AND p \equiv M\;n 
    \SEP
      \DByHasEffects{\D}{{\dynthunk}()}{\emptyset}{p}
      &\text{where}&
      e = \Thunk{\name{n}}{e_0} \AND p \equiv M\;n 
    \SEP
      \DByHasEffects{\D}{{\dynget}()}{p}{\emptyset}
      &\text{where}&
      e = \Get{\refv{p}}
    \SEP
      \DByHasEffects{\D}{{\dynforce}()}{q, R'}{W'}
      &\text{where}&
      e = \Force{\thunk{q}}
      \\ &\text{and}&
      \DHasEffects{\D'}{R'}{W'}
      \\ &\text{where}&
      \text{$\D'$ is the derivation that computed $t$}
    \end{array}

  \caption{Read- and write-sets of a non-incremental evaluation derivation}
  \FLabel{fig:readswrites}
\end{figure}

% Local Variables:
% TeX-master: "fungi-lang"
% End:

%
In \Figureref{fig:readswrites}, we write
\[
\DByHasEffects{\D}{\textit{Rulename (Dlist)}}{R}{W}
\]
to mean that rule $\textit{Rulename}$ concludes $\D$ and has
subderivations $\textit{Dlist}$.  
For example,
\[
\DByHasEffects{\D}{\dynscope(\D_0)}{R}{W}
\]
provided that
$\DHasEffects{\D}{R}{W}$, where $\D_0$ derives the only premise of \dynscope.

\NotInScope{
In the Eval-forceClean case, we refer back to the derivation that (most recently) computed
the thunk being forced (that is, the first subderivation of Eval-computeDep).  A completely formal
definition would take as input a mapping from pointers $q$ to sets $R'$ and $W'$,
return this mapping as output, and modify the mapping in the Eval-computeDep case.
}

\begin{figure}[h]
\small
  \centering

  \judgbox{\Gamma |- A : K}
          {Under $\Gamma$,
            value type $A$ has kind $K$
          }
  \vspace*{-3.0ex}
  \begin{mathpar}
    \Infer{k-typevar}
        {
          (\alpha : K) \in \Gamma
        }
        {
          \Gamma |- \alpha : K
        }
    ~~~
    \Infer{k-tycon}
        {
          (d : K) \in \Gamma
        }
        {
          \Gamma |- d : K
        }
    ~~~~
    \Infer{k-binop}
        {
          \Gamma |- A_1 : \type
          \\
          \Gamma |- A_2 : \type
        }
        {
          \arrayenvbl{
              \Gamma |- (A_1 + A_2) : \type
              \\
              \Gamma |- (A_1 ** A_2) : \type
          }
        }
    \and
    \Infer{\!k-unit}
        {}
        {
          \Gamma |- \unitty : \type
%          \\
%          \Gamma |- \Nmsp : \type
        }
    \and
    \Infer{\!k-name}
        {
          \Gamma |- i : \namesetsort
        }
        {
          \Gamma |- \Name{i} : \type
        }
    \and
    \Infer{\!k-ref}
        {
          \Gamma |- i : \namesetsort
          \\
          \Gamma |- A : \type
        }
        {
          \Gamma |- (\Ref{i}{A}) : \type
        }
    \and
    \Infer{\!k-thk}
        {
          \arrayenvbl{
          \Gamma |- i : \namesetsort
          \\
          \Gamma |- E \tefftype
        }
        }
        {
          \Gamma |- (\Thk{i}{E}) : \type
        }
    \and
    \Infer{\!k-app-type}
        {\arrayenvbl{
          \Gamma |- A : (\type => K)\!\!\!\!\!
          \\
          \Gamma |- B : \type
        }}
        {
          \Gamma |- (A\;B) : K
        }
    \and
    \Infer{\!k-app-index}
        {\arrayenvbl{
          \Gamma |- A : (\sort => K)\!\!\!\!
          \\
          \Gamma |- i : \sort
        }}
        {
          \Gamma |- A[[i]] : K
        }
    \\
    \Infer{\!k-all}
        {
          \Gamma, a : \sort |- P \wfprop
          \\
          \Gamma, a : \sort |- A : \type
        }
        {
          \Gamma |- (\DAll{a : \sort}{P} A) : \type
        }
    \and
    \Infer{\!k-exists}
        {
          \Gamma, a : \sort |- P \wfprop
          \\
          \Gamma, a : \sort |- A : \type
        }
        {
          \Gamma |- (\DExists{a : \sort}{P} A) : \type
        }
  \end{mathpar}

  \judgbox{\Gamma |- C \ctype}
          {Under $\Gamma$,
            computation type $C$ is well-formed
          }
  \vspace{-0.7ex}
  \begin{mathpar}
    \Infer{ctype-lift}
         {
             \Gamma |- A : \type
         }
         {
             \Gamma |- (\F A) \ctype
         }
    \and
    \Infer{ctype-arr}
         {
             \Gamma |- A : \type
             \\
             \Gamma |- E \tefftype
         }
         {
             \Gamma |- (A \arr E) \ctype
         }
  \vspace{-0.7ex}
  \end{mathpar}

  \judgbox{\Gamma |- \e \wfeff}{
           Under $\Gamma$,
           effects $\e$ are well-formed}
  \vspace{-1.2ex}
  \begin{mathpar}
    \Infer{wf-eff}
         {
           \Gamma |- W : \namesetsort
           \\
           \Gamma |- R : \namesetsort
         }
         {
           \Gamma |-
           <<
             W; R
           >>
           \wfeff
         }
  \vspace{-1.5ex}
  \end{mathpar}
  
  \judgbox{\Gamma |- P \wfprop}
         {Under $\Gamma$,
           proposition $P$ is well-formed
         }
  \vspace{-3.3ex}
  \begin{mathpar}
    ~~~~~~~~~~
    \Infer{}
         {}
         {\Gamma |- tt \wfprop}
    ~~
    \Infer{}
         {
           \Gamma |- P_1 \wfprop
           ~~~~~
           \Gamma |- P_2 \wfprop
         }
         {
           \Gamma |- (P_1 \andprop P_2) \wfprop
         }
   ~~
   \Infer{}
         {
           \Gamma |- i : \gamma
           \\
           \Gamma |- j : \gamma
         }
         {\arrayenvbl{
             \Gamma |- (i \disj j : \gamma) \wfprop
             \\
             \Gamma |- (i \equiv j : \gamma) \wfprop
             }
         }
  \vspace{-1.0ex}
  \end{mathpar}

  \judgbox{\Gamma |- E \tefftype}
          {Under $\Gamma$,
            type-with-effects $E$ is well-formed
          }
  \begin{mathpar}
    \Infer{\!etype-eff}
        {
          \Gamma |- C \ctype
          \\
          \Gamma |- \e \wfeff
        }
        {
          \Gamma |- (C |> \e) \tefftype
        }
    \\
    \Infer{\!etype-poly}
        {
          \Gamma, \alpha : K |- E \tefftype
        }
        {
          \Gamma |- (\All{\alpha : K} E) \tefftype
        }
    \and
    \Infer{\!etype-idx}
        {
          \Gamma, a : \sort |- P  \wfprop
          \\
          \Gamma, a : \sort |- E  \tefftype
        }
        {
          \Gamma |- (\DAll{a : \sort}{P} E) \tefftype
        }
   % \and
   %%  \Infer{e-k-namearr}
   %%      {
   %%        \Gamma, a : \sort |- E : \type
   %%      }
   %%      {
   %%        \Gamma |- (\Namearr{a}{\sort} E) : \type
   %%      }
   %% \and
%    \Infer{e-k-guarded}
%         {
%           \Gamma |- E \tefftype
%         }
%         {
%           \Gamma |- (P \impty E) \tefftype
%         }
  \lesscaptionspace
  \end{mathpar}

  \caption{Kinding and well-formedness for types and effects}
  \label{fig:kinding}
\end{figure}

% Local Variables:
% TeX-master: "typed-adapton"
% End:

\clearpage

\subsection{Remarks}
\Label{apx:remarks}

\paragraph{Why distinguish computation types from types-with-effects?}
% \Figureref{fig:syntax-types} gives syntax for ``types with effects'' $E$,
% in addition to computation types $C$.
Can we unify computation types $C$ and types-with-effects $E$?  Not easily.
We have two computation types, $\xF$ and $\arr$.
For $\xF$, the expression being typed could create a thunk, so we must
put that effect somewhere in the syntax.
For $\arr$, applying a function is (per call-by-push-value) just a ``push'': the function carries
no effects of its own (though its codomain may need to have some).
However, suppose we force a thunked function of type $A_1 \arr (A_2 \arr \cdots)$
and apply the function (the contents of the thunk) to one argument.
In the absence of effects, the result would be a computation
of type $A_2 \arr \cdots$, meaning that the computation is waiting for a second argument
to be pushed.  But, since forcing the thunk has the effect of reading the thunk,
we need to track this effect in the result type.  So we cannot return $A_2 \arr \cdots$,
and must instead put effects around $(A_2 \arr \cdots)$.
Thus, we need to associate effects to both $\xF$ and $\arr$, that is, to both computation types.

Now we are faced with a choice:
we could
(1) extend the syntax of each connective with an effect (written next to the connective), or
(2) introduce a ``wrapper'' that encloses a computation type, either $\xF$ or $\arr$.
These seem more or less equally complicated for the present system,
but if we enriched the language with more connectives,
choice (1) would make the new connectives more complicated,
while under choice (2), the complication would already be rolled into the wrapper.
We choose (2), and write the wrapper as $C |> \e$, where $C$ is a computation type
and $\e$ represents effects.

Where should these wrappers live?
We could add $C |> \e$ to the grammar of computation types $C$.  
But it seems useful to have a clear notion of \emph{the} effect associated with a type.
When the effect on the outside of a type is the only effect in the type,
as in $(A_1 \arr \F A_2) |> \e$, ``the'' effect has to be $\e$.
Alas, types like $(C |> \e_1) |> \e_2$ raise awkward questions:
does this type mean the computation does $\e_2$ and then $\e_1$,
or $\e_1$ and then $\e_2$?

We obtain an unambiguous, singular outer effect by distinguishing
types-with-effects $E$ from computation types $C$.
The meta-variables for computation types appear only in the production
$E \bnfas C |> \e$, making types-with-effects $E$ the ``common case'' in the grammar.
Many of the typing rules follow this pattern, achieving some isolation of effect tracking
in the rules.

%  ---------------- Matt will revised below (noted 2018.03.15)
%
\paragraph{Future work: Editor and Archivist.}
\label{sec:future-work-roles}
To distinguish imperative name allocation from name-precise
computation, future versions of \Fungi will introduce two \emph{incremental computation roles},
which we term the \emph{editor} and the \emph{archivist},
respectively; specifically, we define the syntax for roles as~$r ::=
\textsf{ed}~|~\textsf{ar}$.
The archivist role~(\textsf{ar}) corresponds to computation whose
dependencies we cache, and the editor role~(\textsf{ed}) corresponds
to computation that feeds the archivist with input changes, and
demands any changed output that is relevant; in short, the editor
represents the world outside the cached computation.

While the current type system prototype focuses only on the \emph{archivist}
role, leaving the editor role to the surrounding Rust code, future
work will integrate the editor role into Fungi programs.
For example, consider the following typing rules, which approximate
(and extend) our full type system with a \emph{role}~$r$ in each rule:
%
%Each specifies a \emph{write set} in their
%conclusion:
%
{ \small
\begin{mathpar}
\Infer{}
{ 
%\textsf{t-ref\hfill}
%\\\\
\Gamma |- v_n : \Name{X} 
\\
\Gamma |- v : A 
}
{ \Gamma |- \textsf{ref}( v_n, v ) : \RefNr{A} |> r(X) }
~\hfill~
\Infer{}
{ 
%\textsf{t-let-ar\hfill}
%\\\\
\Gamma |- e_1 : A |> \textsf{ar}(X)
\\\\
\Gamma, x : A |- e_2 : B |> \textsf{ar}(Y)
\\\\
\Gamma |- (X \disj Y) \equiv Z : \namesetsort
}
{\Gamma |- \Let{e_1}{x}{e_2} : B |> \textsf{ar}(Z)}
~\hfill~
\Infer{}
{ 
%\textsf{t-let-ed\hfill}
%\\\\
\Gamma |- e_1 : A |> \textsf{ed}(X)
\\\\
\Gamma, x : A |- e_2 : B |> \textsf{ed}(Y)
\\\\
\Gamma |- (X \cup Y) \equiv Z : \namesetsort
}
{\Gamma |- \Let{e_1}{x}{e_2} : B |> \textsf{ed}(Z)}
\end{mathpar} 
}
These rules are similar to the simplified rules presented in
\Secref{sec:overview}.  In contrast to those rules, these conclude with
the judgement form~$\Gamma |- e : A |> r(X)$, mentioning the written
set with the notation~$|> r(X)$, where the set~$X$ approximates the
set of written names (as in the earlier formulation), and $r$ is the
role (absent from the earlier formulation).

The first rule types a reference cell allocation, as before; in the
rule's conclusion, this name set~$X$ serves as the allocation's
\emph{write set}.  The undetermined role~$r$ means that this rule is
applicable to both the editor and the archivist roles.

What was one \texttt{let}~sequencing rule (in \Secref{sec:overview}) is now two rules here:
The second rule enforces the archivist role, where names are precise.
The third rule permits the editor role, where names allocated
later may \emph{overwrite} names allocated earlier.
Finally, a new syntax form \Archivist{e} permits the editor's
computations to delegate to archivist sub-computations; the
program~\Archivist{e} has role~$\textsf{ed}$ whenever program~$e$
types under role~$\textsf{ar}$ under the same typing context.

Among the future work for mixing these roles, we foresee that
extending the theory of \Fungi, including covariant index
subtyping, to this mixture of imperative-functional execution
semantics requires mixing imperative effects (for the editor) and type
index subtyping (for the archivist) in a disciplined, sound manner.

% Local Variables: 
% mode: latex
% TeX-master: "typed-adapton"
% End: 

\section{Omitted Lemmas and Proofs}
\label{apx:proofs}

% \begin{property}[Weakening of entailment]
% \Label{pty:index-weak}
%   If $\Gamma_1, \Gamma_3 |- P$
%   then $\Gamma_1, \Gamma_2, \Gamma_3 |- P$.
% \end{property}

\begin{lemma}[Index-level weakening]
\Label{lem:index-weak}
~
\begin{enumerate}
\item 
  If   $\Gamma |- M : \sort$
  then $\Gamma, \Gamma' |- M : \sort$.
\item
  If   $\Gamma |- i : \sort$
  then $\Gamma, \Gamma' |- i : \sort$.
\item
  If   $\Gamma |- A : K$
  then $\Gamma, \Gamma' |- A : K$.
\end{enumerate}
\end{lemma}
\begin{proof}
  By induction on the given derivation.
\end{proof}

\begin{lemma}[Weakening]
\Label{type-weak}
~
\begin{enumerate}
\item 
  If   $\Gamma |- e : A$
  then $\Gamma, \Gamma' |- e : A$.
\item 
  If   $\Gamma |-^M e : C$
  then $\Gamma, \Gamma' |-^M e : C$.
\end{enumerate}
\end{lemma}
\begin{proof}
By induction on the given derivation,
using \Lemmaref{lem:weakening-semantic}
(for example, in the case for the value typing rule `name')
and \Lemmaref{lem:index-weak}
(for example, in the case for the computation typing rule `AllIndexElim').
\end{proof}

\begin{lemma}[Substitution]
\Label{lem:subst}
% We consider two statements, proven by mutual induction:
~
\begin{enumerate}
\item
If   $\Gamma |- v : A$
and  $\Gamma, x : A |- e : C$
then $\Gamma |- \big( [v / x] e \big) : C$.
\item
If   $\Gamma |- v : A$
and  $\Gamma, x : A |- v' : B$
then $\Gamma |- \big( [v / x] v' \big) : B$.
\end{enumerate}
\begin{proof}
By mutual induction on the derivation typing $e$ (in part 1)
or $v'$ (in part 2).
% In each of (1) and (2), by (mutual) induction on the typing derivation
% into which the substitution occurs (that is, the typing derivation
% with the free variable $x$).
\end{proof}
\end{lemma}

In the presence of subtyping, canonical forms (value inversion) is not entirely
straightforward.

\begin{lemma}[Subtyping Weakening]
\Label{lem:sub-weakening}
  If $\Gamma |- A \vsubtype B$
  then $\Gamma, \Gamma' |- A \vsubtype B$
  where $\Gamma'$ consists of $a : \sort$ and $P$ assumptions.
\end{lemma}
\begin{proof}
  By induction on the derivation of $\Gamma |- A \vsubtype B$.
  In the \vsubname, \vsubnamefn, \vsubref, \vsubthk, \vsubAllL and \vsubExists cases,
  use weakening for the relations $|-$ and $||-$.
\end{proof}

\begin{lemma}[Subtyping Substitution]
\Label{lem:sub-subst}
~\\
  If $\Gamma, a : \sort, P |- A \vsubtype B$
  and $\Gamma |- i : \sort$
  and $\extract{\Gamma} ||- P$
  then $\Gamma |- [i/a]A \vsubtype [i/a]B$.
\end{lemma}
\begin{proof}
  By induction on the derivation of $\Gamma |- A \vsubtype B$.
  In the \vsubref, \vsubthk, \vsubAllL and \vsubExists cases, use substitution for the relation $||-$.
\end{proof}

\begin{lemma}[Reflexivity of Subtyping]
\Label{lem:sub-refl}
  For all $\Gamma$ and $A$, it is the case that $\Gamma |- A \vsubtype A$.
\end{lemma}
\begin{proof}
  Immediate by rule \vsubrefl.
\end{proof}

\begin{lemma}[Transitivity of Subtyping]
\Label{lem:sub-trans}
~\\
  If $\Gamma |-  A_L \vsubtype B$
  and $\Gamma |-  B \vsubtype A_R$
  then $\Gamma |- A_L \vsubtype A_R$.
\end{lemma}
\begin{proof}
  By simultaneous induction on the two given derivations.

  If either derivation is by \vsubrefl, we already have our result.

  Consider cases of the rule concluding $\Gamma |- A_L \vsubtype B$.

  \begin{itemize}
%  \item \vsubrefl: Then $A_L = B$, and $\Gamma |- B \vsubtype A_R$ is given,
%    so we have $\Gamma |- A_L \vsubtype A_R$.

  \item \vsubproduct:

    The derivation of $\Gamma |- B \vsubtype A_R$
    must be by \vsubrefl (already handled), \vsubproduct or \vsubAllR.

%    If by \vsubrefl, then $B = A_R$ and we already know $\Gamma |- A_L \vsubtype B$,
%    which is $\Gamma |- A_L \vsubtype A_R$.

    If by \vsubproduct, the result follows by using the i.h.\ twice on the respective subderivations,
    then applying \vsubproduct.

    If by \vsubAllR, then:

    \begin{llproof}
      \eqPf{A_R}{(\DAll{b : \sort}{P} A_{R0})}  {\byinv{\vsubAllR}}
      \ePf{\Gamma, b : \sort, P} {B \vsubtype A_{R0}}   {\ditto}
      \ePf{\Gamma} {A_L \vsubtype B}   {Given}
      \ePf{\Gamma, b : \sort, P} {A_L \vsubtype B}   {By \Lemmaref{lem:sub-weakening}}
      \ePf{\Gamma, b : \sort, P} {A_L \vsubtype A_{R0}}   {By i.h.}
      \ePf{\Gamma} {A_L \vsubtype (\DAll{b:\sort}{P} A_{R0})}   {By \vsubAllR}
    \Hand  \ePf{\Gamma} {A_L \vsubtype A_R}   {By above equation}
    \end{llproof}

    \smallskip

  \item \vsubsum: Similar to the \vsubproduct case.

    \smallskip

  \item \vsubname, \vsubref, \vsubthk:
    Similar to the \vsubproduct case, using transitivity of $\subseteq$ at the index level.

    \smallskip

  \item \vsubnamefn:
    Use transitivity of $conv$.

    \smallskip

  \item \vsubAllL:

    By i.h., $\Gamma |- [i/a]A_{L0} \vsubtype A_R$. \\
    By \vsubAllL, $\Gamma |- \DAll{a : \sort}{P} A_{L0} \vsubtype A_R$,
    which was to be shown.

    \smallskip

  \item \vsubAllR:

    The other derivation is by either \vsubrefl (already handled) or \vsubAllL.

    \begin{llproof}
      \eqPf{B}{(\DAll{b : \sort}{P} B_0)}  {\byinv{\vsubAllR}}
      \ePf{\Gamma, b : \sort, P} {A_L \vsubtype B_0}   {\ditto}
      \ePf{\Gamma} {i : \sort}   {\byinv{\vsubAllL}}
      \EPf{\extract{\Gamma}} {[i/b]P}   {\ditto}
      \ePf{\Gamma} {[i/b]B_0 \vsubtype A_R}   {\ditto}
      \ePf{\Gamma} {[i/b]A_L \vsubtype [i/b]B_0}   {By \Lemmaref{lem:sub-subst}}
      \ePf{\Gamma} {[i/b]A_L \vsubtype A_R}  {By i.h.}
      \eqPf{[i/b]A_L}{A_L}    {$b$ not free in $A_L$}
    \Hand \ePf{\Gamma} {A_L \vsubtype A_R}  {By above equation}
    \end{llproof}
  \qedhere
  \end{itemize}
\end{proof}

%%%% obsolete, now that subtyping is in the main system
%
% We need subsumption in the proof of \Lemmaref{lem:canon}, but 
% to avoid complicating the value typing rules, we omit a subsumption rule.
% Instead, we prove---for values with no free $x$, guaranteed by
% the $\StoreType{\Gamma}$ assumption in \Lemmaref{lem:canon}---subsumption
% is admissible.  Since proving admissibility of subtyping requires using \Lemmaref{lem:canon},
% we add admissibility as the 0th part of that lemma.

% \begin{lemma}[Admissibility of Subsumption]
% \Label{lem:sub-admissible}
%   If $\StoreType{\Gamma}$
%   and $\Gamma |- v : A$
%   and $\Gamma |- A \vsubtype B$
%   then $\Gamma |- v : B$.
% \end{lemma}
% \begin{proof}
% \end{proof}

\begin{lemma}[Canonical Forms]
\Label{lem:canon}
Suppose $\StoreType{\Gamma}$ and $\Gamma |- v : A$.
\\
\begin{tabular}{r@{~~}r@{~}c@{~}lll}
% 0.& If $A$ & $\vsubtype$ & $B$
%   & then $\Gamma |- v : B$.
% \\
1.& If $A$ & $\vsubtype$ & $\unitty$ 
  &  then $v = \unitexp$.
 \\
2.& If $A$ & $\vsubtype$ & $(B_1 ** B_2)$ 
  &  then $v = \Pair{v_1}{v_2}$
  &  and $\Gamma |- v_1 : B_1$ and $\Gamma |- v_2 : B_2$.
 \\
3.& If $A$ & $\vsubtype$ & $(B_1 + B_2)$ 
  &  then $v = \Inj{i}{v_i}$
  &  where $i \in \{1, 2\}$
  and $\Gamma |- v_i : B_i$.
 \\
4.& If $A$ & $\vsubtype$ & $(\Name{X})$ 
  &  then $v = \name{n}$
  &  where $\Gamma |- n \in X$.
 \\
5.& If $A$ & $\vsubtype$ & $(\Ref{X}{A_0})$ 
  &  then $v = \refv{n}$
  &  where $\Gamma |- n \in X$.
 \\
6.& If $A$ & $\vsubtype$ & $(\Thk{X}{E})$ 
  &  then $v = \thunk{n}$
  &  where $\Gamma |- n \in X$.
\\
7.& If $A$ & $\vsubtype$ & $(\namesort @> \namesort)[[M]]$ 
  &  then $v = \nametm{M_v}$
  &  where $M \conv (\lam{a} M')$
  \\ &&&&&
  and $\cdot |- (\lam{a} M') : (\namesort @> \namesort)$
  \\ &&&&&
  and $M_v \conv M$.
\end{tabular}
\end{lemma}
\begin{proof}
  By induction on the derivation of $\Gamma |- A \vsubtype B$.
% the derivation of $\Gamma |- v : A$.
% In each part, exactly one value typing rule is applicable,
% so the result follows by inversion.

%  Part 0 is that subtyping is admissible.

  \begin{enumerate}
     \setcounter{enumi}{0}

  \item % Part 1:
    Consider cases of the rule concluding $\Gamma |- v : A$.
    
    \begin{itemize}
    \ProofCaseRule{unit}  By inversion.

    \ProofCaseRule{pair}  Impossible because $\Gamma |- A_1 + A_2 \vsubtype \unitty$ is not derivable.

    \ProofCaseRule{name}  Impossible because $\Gamma |- \Name{X} \vsubtype \unitty$ is not derivable.

    \ProofCaseRule{namefn}  Impossible because $\Gamma |- (\namesort @> \namesort)[[M]] \vsubtype \unitty$
    is not derivable.

    \ProofCaseRule{ref}  Impossible because $\Gamma |- (\Ref{X}{A_0}) \vsubtype \unitty$ is not derivable.

    \ProofCaseRule{thunk}  Impossible because $\Gamma |- (\Thk{X}{E}) \vsubtype \unitty$ is not derivable.

    \ProofCaseRule{\vtypeAllIndexIntro}

        \begin{llproof}
          \ePf{\Gamma}{(\DAll{a : \sort}{P} A_0) \vsubtype \unitty} {Given}
          \ePf{\Gamma, a : \sort, P}{A_0 \vsubtype \unitty} {\byinv{\vsubAllL}}
%          \EPf{\extract{\Gamma}}{[i/a]P} {\ditto}
%          \ePf{\Gamma}{i : \sort} {\ditto}
          \ePf{\Gamma, a : \sort, P} {v : A_0}   {Subderivation}
        \Hand  \eqPf{v}{\unitexp}   {By i.h.\ (part 1)}
        \end{llproof}
    
    \DerivationProofCase{\vtypeAllIndexElim}
       {
         \Gamma |- i : \sort
         \\
         \arrayenvbl{
         \extract{\Gamma} ||- [i/a]P
         \\  
         \Gamma |- v : (\DAll{a : \sort}{P} A_0)
         }
       }
       {
         \Gamma |- v : [i/a]A_0
       }

       \begin{llproof}
         \ePf{\Gamma}{i : \sort}  {Subderivation}
         \EPf{\extract{\Gamma}}{[i/a]P}  {Subderivation}
         \ePf{\Gamma}{(\DAll{a : \sort}{P} A_0) \vsubtype [i/a]A_0}  {By \vsubAllL}
         \proofsep
         \ePf{\Gamma}{[i/a]A_0 \vsubtype A}  {Given}
         \proofsep
         \ePf{\Gamma}{(\DAll{a : \sort}{P} A_0) \vsubtype \unitty}  {By \Lemmaref{lem:sub-trans}}
         \ePf{\Gamma}{ v : (\DAll{a : \sort}{P} A_0)}  {Subderivation}
       \Hand  \eqPf{v}{\unitexp}   {By i.h.}
       \end{llproof}

    \ProofCaseRule{\vtypeExistsIndexIntro}

    Impossible because $\Gamma |- (\DExists{a : \sort}{P} A_0) \vsubtype \unitty$
    is not derivable.
    \end{itemize}

  \medskip

  \item % Part 2,
    $**$:

    Consider cases of the rule concluding $\Gamma |- v : A$.

    \begin{itemize}
    \ProofCaseRule{unit}  Impossible because $\Gamma |- \unitty \vsubtype (B_1 ** B_2)$ is not derivable.

    \ProofCaseRule{pair}

        \begin{llproof}
          \ePf{\Gamma}{A \vsubtype (B_1 ** B_2)}   {Given}
          \eqPf{A}{(A_1 ** A_2)}   {\byinv{pair}}
        \Hand  \eqPf{v}{\Pair{v_1}{v_2}}   {\ditto}
          \ePf{\Gamma}{v_1 : B_1}   {\ditto}
          \ePf{\Gamma}{v_2 : B_2}   {\ditto}
          \proofsep
          \ePf{\Gamma}{A_1 \vsubtype B_1}   {\byinv{\vsubproduct}}
          \ePf{\Gamma}{A_2 \vsubtype B_2}   {\ditto}
          \proofsep
       \Hand  \ePf{\Gamma}{v_1 : B_1}   {By \vtypeSub}
       \Hand  \ePf{\Gamma}{v_2 : B_2}   {By \vtypeSub}
        \end{llproof}

    \ProofCasesRules{name, namefn, ref, thunk}

        Impossible because the assumed subtyping is not derivable.

    \ProofCaseRule{\vtypeAllIndexIntro}

        \begin{llproof}
          \ePf{\Gamma}{(\DAll{a : \sort}{P} A_0) \vsubtype (B_1 ** B_2)} {Given}
          \ePf{\Gamma, a : \sort, P}{A_0 \vsubtype (B_1 ** B_2)} {\byinv{\vsubAllL}}
          \ePf{\Gamma, a : \sort, P} {v : A_0}   {Subderivation}
        \Hand  \eqPf{v}{\Pair{v_1}{v_2}}   {By i.h.\ (part 2)}
        \Hand  \ePf{\Gamma}{v_1 : B_1}  {\ditto}
        \Hand  \ePf{\Gamma}{v_2 : B_2}  {\ditto}
        \end{llproof}
    
    \DerivationProofCase{\vtypeAllIndexElim}
       {
         \Gamma |- i : \sort
         \\
         \arrayenvbl{
         \extract{\Gamma} ||- [i/a]P
         \\  
         \Gamma |- v : (\DAll{a : \sort}{P} A_0)
         }
       }
       {
         \Gamma |- v : [i/a]A_0
       }

       \begin{llproof}
         \ePf{\Gamma}{i : \sort}  {Subderivation}
         \EPf{\extract{\Gamma}}{[i/a]P}  {Subderivation}
         \ePf{\Gamma}{(\DAll{a : \sort}{P} A_0) \vsubtype [i/a]A_0}  {By \vsubAllL}
         \proofsep
         \ePf{\Gamma}{[i/a]A_0 \vsubtype A}  {Given}
         \decolumnizePf
         \ePf{\Gamma}{(\DAll{a : \sort}{P} A_0) \vsubtype (B_1 ** B_2)}  {By \Lemmaref{lem:sub-trans}}
         \ePf{\Gamma}{ v : (\DAll{a : \sort}{P} A_0)}  {Subderivation}
        \Hand  \eqPf{v}{\Pair{v_1}{v_2}}   {By i.h.\ (part 2)}
        \Hand  \ePf{\Gamma}{v_1 : B_1}  {\ditto}
        \Hand  \ePf{\Gamma}{v_2 : B_2}  {\ditto}
       \end{llproof}

    \ProofCaseRule{\vtypeExistsIndexIntro}

    Impossible because $\Gamma |- (\DExists{a : \sort}{P} A_0) \vsubtype (B_1 ** B_2)$ is not derivable.
    \end{itemize}

  \medskip

  \item % Part 3,
    $+$: 
    Similar to Part 2.

  \item % Part 4,
    $\Name{X}$:

    In the \vsubname case, use the fact that $\Gamma ||- X' \subseteq X$
    and $\Gamma |- n \in X'$ implies $\Gamma |- n \in X$.
    Otherwise similar to Part 1.

  \item % Part 5,
    $\Ref{X}{A_0}$:
    Similar to Parts 1 and 4.

  \item % Part 6,
    $\Thk{X}{E}$:
    Similar to Part 5.

  \item % Part 7,
    $(\namesort @> \namesort)[[M]]$:
    Similar to Part 5.
  \qedhere
  \end{enumerate}
\end{proof}

\begin{lemma}[Application and membership commute]
\Label{app-mem-commute}
If $\Gamma |- n \in i$
and $p \conv \idxapp{M}{n}$
then $\Gamma |- p \in \idxapp{M}{i}$.
\end{lemma}
\begin{proof}
The set $\idxapp{M}{i}$ consists of all elements of $i$,
but mapped by function~$M$.
The name~$p$ is convertible to the name~$\idxapp{M}{n}$.
Since $n \in i$, we have that $p$ is in the
$M$-mapping of $i$, which is $\idxapp{M}{i}$.
\end{proof}

In each case, we write ``\Pointinghand'' to the left of each goal,
as we prove it.

\thmrefsubj*
\begin{proof}
By induction on the derivation $\Ss$ of
$\Gamma_1 |-^M e : C |> <<W; R>> $.

\begin{itemize}

\DerivationProofCase{ret}
        {
          \Gamma_1 |- v : A
        }
        {
          \Gamma_1 |-^\ambns
          \Ret{v}
          :
          \big(
              (\F A)  |>  <<\emptyset; \emptyset>>
          \big)
        }

\begin{llproof}
\Pf{}{}{(e=t)~\textrm{and}~(\St_1 = \St_2)}{Given}
\Pf{}{}{(R_{\D} = W_{\D} = R = W = \emptyset)}{\ditto}
\Pf{}{}{(\Gamma_2=\Gamma_1)}{Suppose}
\Hand
\Pf{}{|-}{\St_2 : \Gamma_2}{by above equalities}
\Hand
\Pf{\Gamma_2}{|-}{t : C |> <<\emptyset, \emptyset>>}{\ditto}
\Hand
\Pf{}{}{\Dee~\text{reads}~R_\Dee~\text{writes}~W_\Dee}{\ByRWSetRule}
\Hand
\Pf{}{}{<<W_\Dee; R_\Dee>> \effsub <<W; R>>}{All are empty}
\end{llproof}

\DerivationProofCase{get}
        {
          \Gamma_1 |- v : \Ref{X}{A}
        }
        {
          \Gamma_1 |-^\ambns
          \Get{v}
          :
          \big(
             \F A
          \big)
          |>
          <<\emptyset; X>>
        }

\begin{llproof}
\Pf{}{}{(W = \emptyset)~\textrm{and}~(R = X)}{Given}
\Pf{\Gamma_1}{|-}{v : \Ref{X}{A}}{Given}
\Pf{\exists p.}{}{(v = \refv{p})}{\Lemmaref{lem:canon}}
\Pf{\Gamma_1}{|-}{p \in X}{\ditto}
\Pf{}{}{\Gamma_1(p) = A}{By inversion of value typing}
\Pf{\exists v_p.}{}{\St_1(p) = v_p}{Inversion on $|- \St_1 : \Gamma_1$}
\Pf{\Gamma_1}{|-}{v_p:A}{\ditto}
\decolumnizePf
\Pf{}{}{(\Gamma_2=\Gamma_1)~\textrm{and}~(t = \Ret{v_p})}{Suppose}
\Pf{}{}{(R_\Dee = \{ p \})~\textrm{and}~(W_\Dee = \emptyset = W)}{\ditto}
\Hand
\Pf{}{|-}{\St_2 : \Gamma_2}{By above equalities}
\Hand
\Pf{\Gamma_2}{|-}{t : C |> <<\emptyset, \emptyset>>}{\ditto}
\decolumnizePf
\Hand
\Pf{}{}{\Dee~\text{reads}~R_\Dee~\text{writes}~W_\Dee}{\ByRWSetRule}
%\Pf{}{}{}{}
\Hand
\Pf{}{}{<<W_\Dee; R_\Dee>> \effsub <<W; R>>}{By above equality~$W_\Dee = W = \emptyset$,}
  \Pf{}{}{}{$\ldots$~and inequality for $(R_\Dee = \{ p \}) \subseteq (X = R)$.}
\end{llproof}

\DerivationProofCase{force}
        {
          \Gamma_1 |- v : \Thk{X}{(C |> \e)}
        }
        {
          \Gamma_1 |-^\ambns
          \Force{v}
          :
          \big(
            C |> (<<\emptyset; X>> \effseq \e)
          \big)
        }

\begin{llproof}

\Pf{}{}{\Gamma |- e : \tau}{Given}

\Pf{\Gamma}{|-}{e : \tau}{Given}

\Pf{}{}{(W = \emptyset)~\textrm{and}~(R = X)}{Given}
\Pf{\Gamma_1}{|-}{v : \Thk{X}{(C |> \e)}}{Given}
\Pf{\exists p.}{}{(v = \thunk{p})}{\Lemmaref{lem:canon}}
\Pf{\Gamma_1}{|-}{p \in X}{\ditto}
\Pf{}{}{\Gamma_1(p) = (C |> \e)}  {By inversion of value typing}
\Pf{\exists e_p.}{}{\St_1(p) = e_p}{Inversion on $|- \St_1 : \Gamma_1$}
$\Ss_0 ::~$
\Pf{\Gamma_1}{|-}{e_p : (C |> \e)}{\ditto}
\decolumnizePf
$\D_0 ::~$
\Pf{\St_1}{|-}{{}^M_m~e_p !! \St_2; t}{Inversion of~$\D$}
\Hand
\Pf{}{|-}{\St_2 : \Gamma_2}{By i.h.\ on $\Ss_0$ and $\D_0$}
\Hand
\Pf{\Gamma_2}{|-}{t : C |> <<\emptyset, \emptyset>>}{\ditto}
%\Hand
\Pf{}{}{\Dee_0~\text{reads}~R_\Dee~\text{writes}~W_\Dee}{\ditto}
\Pf{}{}{<<W_{\Dee_0}; R_{\Dee_0}>> \effsub <<W; R>>}{\ditto}
\Hand
\Pf{}{}{\Dee~\text{reads}~R_{\Dee_0}~\text{writes}~W_{\Dee_0}}{\ByRWSetRule}
%\Pf{}{}{}{}
\Hand
\Pf{}{}{<<W_{\Dee_0}; R_{\Dee_0}>> \effsub <<W; R>>}{by above equality~$W_\Dee = W = \emptyset$,}
  \trailingjust{\hspace*{-9ex}$\ldots$~and inequality $(R_\Dee = \{ p \}) \subseteq (X = R)$.}
\end{llproof}

\DerivationProofCase{scope}
{
  \Gamma_1 |- v : (\namesort @> \namesort)[[M']]
  \\
  \Gamma_1 |-^{M \circ M'} {e_0} : C |> <<W; R>>
}
{
  \Gamma_1 |-^M {\Scope{v}{e_0}}
  :
  C |> <<W; R>>
}

\begin{llproof}
  $\Ss_0 \derives$ \Pf{\Gamma_1}{|-^{M \circ M'}} {e_0 : C |> <<W; R>>}  {Subderivation 2 of $\Ss$}
  $\Dee \derives$ \Pf{}{}{\PreSt{\St_1}{M}{m} \Scope{v}{e_0} !! \St_2 ; t}  {Given}
  $\Dee_0 \derives$ \Pf{}{}{\PreSt{\St_1}{M \circ M'}{m} e_0 !! \St_2 ; t}  {\byinv{scope}}
  \proofsep
  \Pf{\Gamma_1 }{|- }{M : \namesort @> \namesort}{Assumption}
  \Pf{\Gamma_1 }{|- }{v : (\namesort @> \namesort)[[M']]}{Subderivation 1 of $\Ss$}
  \Pf{\Gamma_1 }{|- }{M' : \namesort @> \namesort}{By inversion}
  \Pf{\Gamma_1, x: \namesort }{|- }{M'\,x : \namesort}{\byrule{t-app}}
  \Pf{\Gamma_1, x: \namesort }{|-}{M\,(M'\,x) : \namesort}{\byrule{t-app}}
  \Pf{\Gamma_1 }{|- }{\lam{x} M\,(M'\,x) : \namesort @> \namesort}{\byrule{t-abs}}
  \Pf{\Gamma_1 }{|- }{(M \circ M') : \namesort @> \namesort}{By definition of $M \circ M'$}
  \decolumnizePf
\Hand \Pf{}{|- }{\St_2 : \Gamma_2}{By i.h.\ on $\Ss_0$}
\Hand \Pf{\Gamma_2}{|-}{t : C |> <<\emptyset; \emptyset>>}{\ditto}
      \Pf{}{}{\DHasEffects{\D_0}{R_{\D_0}}{W_{\D_0}}}{\ditto}
      \Pf{}{}{<<W_{\D_0}; R_{\D_0}>> \effsub <<W; R>>}{\ditto}
\Hand \Pf{}{}{\DHasEffects{\D}{R_{\D}}{W_{\D}}}{\ByRWSetRule}
      \Pf{}{}{<<W_{\D}; R_{\D}>> = <<W_{\D_0}; R_{\D_0}>>}{\ditto}
\Hand \Pf{}{}{<<W_{\D}; R_{\D}>> \effsub <<W; R>>}{By above equalities}
\end{llproof}

\DerivationProofCase{thunk}{
 \Gamma_1 |-  v : \Name{X}
 \\
 \Gamma_1 |- e : E
}
{
 \Gamma_1
 |-^M
 { \Thunk{v_1}{v_2} }
 :
 { \F (\Thk{M(X)}{E})
 |>
 <<M(X); \emptyset>>
 }
}

\begin{llproof}
\Pf{}{}{C = \F (\Thk{M(X)}{E}) \AND R=\emptyset \AND W=M(X)}{Given from $\Ss$}
\Pf{}{}{ \Gamma_1 |-  v : \Name{X}}  {Subderivation}
\Pf{}{}{(v = \name{n}) \AND (n \in X)}  {By \Lemref{lem:canon}}
\Pf{}{}{M~n !! p \AND R_\D=\emptyset \AND W_\D=\{p\}}{Given from $\D$}
%\Pf{}{}{L = P \cup \{n\}}{(XXX Do we need to say anything about disjointness here?)}
%\Pf{}{}{\nfap{L}{M}{\{n\}} = q}{By \Lemmaref{lem:XXX} XXX $M$ is a deterministic function over names}
\Pf{}{}{\St_2 = (\St_1, p : e)}{\ditto}
% \LetPf{x}{y}{}
\decolumnizePf
\Pf{}{}{\Gamma_2 = (\Gamma_1, p : \Thk{p}{E})}{Suppose}
\Hand 
\Pf{}{}{|- \St_2 : \Gamma_2}{\ByStoreTypingRuleApp}
\Pf{}{}{\Gamma_2(p) = E}{By inversion of value typing}
% \Pf{}{}{\Gamma_2 |- M(X) \storetype \nfap{L}{M}{A}}{Since $M~n !! q$}
\decolumnizePf
\Pf{}{}{\Gamma_2 |- {\refv{p}} : {\Ref{p}{E}} }{\byrule{thunk}}
\Hand
\Pf{}{}{\Gamma_2 |-^M \Ret{\thunk{p}} : \F{(\Thk{p}{A})} |> <<\emptyset;\emptyset>> }{\byrule{ret}}
\Hand
\Pf{}{}
     {\DHasEffects{\D}{R_\D}{W_\D}
       \AND W_\D = \{ p \}
     }
     {\ByRWSetRule}
 \decolumnizePf
\Pf{}{}{ n \in X}  {Above}
\Pf{}{}{M(n) \in M(X)}  {Name term application is pointwise}
\Pf{}{}{M(n) \in W}  {By above equality}
\Pf{}{}{M(n) = p}  {}
\Pf{}{}{\{ p \}  \subseteq { W }}   {By set theory}
\Hand 
\Pf{}{}{<<W_\Dee; R_\Dee>> \effsub {<<W; R >>}}{}
\end{llproof}

\DerivationProofCase{ref}{
 \Gamma_1 |-  v_1 : \Name{X}
 \\
 \Gamma_1 |- v_2 : { A }
}
{
 \Gamma_1
 |-^M
 { \Refe{v_1}{v_2} }
 :
 { \F (\Ref{M(X)}{A}) 
 |>
 <<M(X); \emptyset>>
 }
}

\begin{llproof}
\Pf{}{}{C = \F (\Ref{M(X)}{A}) \AND R=\emptyset \AND W=M(X)}{Given from $\Ss$}
\decolumnizePf
\Pf{}{}{ \Gamma_1 |-  v_1 : \Name{X}}  {Subderivation}
\Pf{}{}{(v_1 = \name{n}) \AND (n \in X)}  {\Lemmaref{lem:canon}}
\Pf{}{}{M~n !! p \AND R_\D=\emptyset \AND W_\D=\{p\}}{Given from $\D$}
%\Pf{}{}{L = P \cup \{n\}}{(XXX Do we need to say anything about disjointness here?)}
%\Pf{}{}{\nfap{L}{M}{\{n\}} = q}{By \Lemmaref{lem:XXX} XXX $M$ is a deterministic function over names}
\Pf{}{}{\St_2 = (\St_1, p : v_2)}{\ditto}
\decolumnizePf
% \LetPf{x}{y}{}
\Pf{}{}{\Gamma_2 = (\Gamma_1, p : \Ref{p}{A})}{Suppose}
\Hand 
\Pf{}{}{|- \St_2 : \Gamma_2}{\ByStoreTypingRuleApp}
\Pf{}{}{\Gamma_2(p) = A}{\hspace*{-9ex}By inversion of value typing}
% \Pf{}{}{\Gamma_2 |- M(X) \storetype \nfap{L}{M}{A}}{Since $M~n !! q$}
\Pf{}{}{\Gamma_2 |- {\refv{p}} : {\Ref{p}{A}} }{\byrule{ref}}
\Hand
\Pf{}{}{\Gamma_2 |-^M {\Ret{\refv{p}}} : {\Ret{\Ref{p}{A}}} |> <<\emptyset;\emptyset>> }{\byrule{ret}}
\Hand
\Pf{}{}
     {\DHasEffects{\D}{R_\D}{W_\D}
       \AND W_\D = \{ p \}
     }
     {\ByRWSetRule}
\decolumnizePf
\Pf{}{}{ n \in X}  {Above}
\Pf{}{}{M(n) \in M(X)}  {Name term application is pointwise}
\Pf{}{}{M(n) \in W}  {By above equality}
\Pf{}{}{M(n) = p}  {}
\Pf{}{}{\{ p \}  \subseteq { W }}   {By set theory}
\Hand 
\Pf{}{}{<<W_\Dee; R_\Dee>> \effsub {<<W; R >>}}{}
\end{llproof}

\DerivationProofCase{let}
{
  \Gamma_1 |-^M e_1 : (\F A |> \e_1)
  \\
  \Gamma_1, x:A |-^M e_2 : {(C |> \e_2)}
}
{
  \Gamma_1 |-^M {\Let{e_1}{x}{e_2}}
  :
  {C |> (\e_1 \effseq \e_2)}
}

\begin{llproof}
%% i.h.\ on first subderivation:
\Pf{}{}{|- \St_1 : \Gamma_1}{Given}
$\Ss_1$ :: \Pf{}{}{\Gamma_1 |-^M {e_1} : {\F A} |> {\e_1}}{Subderivation 1 of $\Ss$}
$\D_1$  :: \Pf{}{}{\PreSt{\St_1}{M}{m} {e_1} !! \St_{12}; {t_1} }{Subderivation 1 of $\D$}
\Pf{}{}{\text{exists}~\Gamma_{12} \supseteq \Gamma_1~\text{such that}~\St_{12} : \Gamma_{12}}{By i.h.\ on $\Ss_1$}
\Pf{}{}{\Gamma_{12} |- { t_1 } : { \F A |> <<\emptyset; \emptyset>> } }{\ditto }
\Pf{}{}{\text{$\D_1$ reads $R_{\D_1}$ writes $W_{\D_1}$}}{\ditto}
\Pf{}{}{<< W_{\D_1}; R_{\D_1} >> \effsub { \e_1 } }{\ditto }
\Pf{}{}{<< W_{\D_1}; R_{\D_1} >> \effsub { <<W_1, R_1>> } }{\ditto }
\decolumnizePf
%% i.h.\ on second subderivation:
\Pf{}{}{\Gamma_{12} |- { v } : { A } }{inversion of typing rule \textsf{ret},}
\trailingjust{for terminal computation ${t_1}$}
$\Ss_2$ :: \Pf{}{}{\Gamma_1, x:{A} |-^M {e_2} : {C |> \e_2} }{Subderivation 2 of $\Ss$}
\Pf{}{}{\Gamma_{12}, x:{A} |-^M {e_2} : {C |> \e_2} }{\Lemmaref{type-weak}}
\Pf{}{}{         \Gamma_{12}      |-^M [{v}/x] {e_2} : {C |> \e_2} }{\Lemmaref{lem:subst}}
$\D_2$ :: \Pf{}{}{\PreSt{\St_{12}}{M}{m} {[v/x]e_2} !! \St_{2}; {t_2}}{Subderivation 2 of $\D$}
\Pf{}{}{\text{exists}~\Gamma_{2} \supseteq \Gamma_{12} \supseteq \Gamma_1~\text{such that}}{By i.h.\ on $\Ss_2$}
\Hand
\Pf{}{}{|- \St_2 : \Gamma_{2}}{\ditto}
\Hand
\Pf{}{}{\Gamma_{2} |-^M {t_2} : { C } |> <<\emptyset;\emptyset>>}{\ditto}
\Pf{}{}{\text{$\D_2$ reads $R_{\D_2}$ writes $W_{\D_2}$}}{\ditto}
\Pf{}{}{<< W_{\D_2}; R_{\D_2} >> \effsub { \e_{2} } }{\ditto }
\Pf{}{}{<< W_{\D_2}; R_{\D_2} >> \effsub { <<W_2, R_2>> } }{\ditto }
\decolumnizePf
\Pf{}{}{W_1 \disjoint W_2 \AND R_1 \disjoint W_2}{Definition of $\e_1 \effseq \e_2$}
\Pf{}{}{W_{\D_1} \disjoint W_{\D_2} \AND R_{\D_1} \disjoint W_{\D_2}}{$W_{\D_1} \subseteq {W_1}$; $W_{\D_2} \subseteq {W_2}$; $R_{\D_1} \subseteq {R_1}$}
\Pf{}{}{W_\D = W_{\D_1} \disjoint W_{\D_2}}{\ByRWSetRule}
\Pf{}{}{R_\D = R_{\D_1} \cup (R_{\D_2} - W_{\D_1})}{\ditto}
\Hand
\Pf{}{}{\text{$\D$ reads $R_\D$ writes $W_\D$}}{\ditto}
%\Pf{}{}{\text{$\D$ reads $(R_{\D_1} \cup (R_{\D_2} - W_{\D_1}))$ writes $W_{\D_1} \disjoint W_{\D_2}$}}{\ByRWSetRule}
\Hand
\Pf{}{}{<<W_\D, R_\D>> \effsub {<<W, R>>}}{Since $W_\D \subseteq {W}$ and $R_\D \subseteq {R}$}
\end{llproof}

\DerivationProofCase{app}
        {
          \Gamma |-^\ambns e
              : \big(
                 (A -> E) |> \e_1
                \big)
          \\
          \Gamma |- v : A
        }
        {
          \Gamma |-^\ambns (e\;v) : (E \effcoal \e_1)
        }

Similar to the case for \keyword{let}.

\DerivationProofCase{split}
        {
          \Gamma |-^\ambns v : (A_1 ** A_2)
          \\
          \Gamma, x_1:A_1, x_2:A_2 |-^\ambns e : E
        }
        {
          \Gamma |-^\ambns \Split{v}{x_1}{x_2}{e} : E
        }

Similar to the case for \keyword{let}, using \Lemmaref{lem:canon}.

\DerivationProofCase{case}
        {
          \Gamma |-^\ambns v : (A_1 + A_2)
          \\
          \arrayenvbl{
            \Gamma, x_1:A_1 |-^\ambns e_1 : E
            \\
            \Gamma, x_2:A_2 |-^\ambns e_2 : E
          }
        }
       {
          \Gamma |-^\ambns \Case{v}{x_1}{e_1}{x_2}{e_2} : E
       }

Similar to the case for \keyword{let}, using \Lemmaref{lem:canon}.

\DerivationProofCase{name-app}
        {
          \arrayenvbl{
            \Gamma_1 |- v_M : (\namesort @> \namesort)[[M]]
            \\
            \Gamma_1 |- v : \Name{i}
          }
        }
        {
          \Gamma_1 |- (v_M\;v) : \F (\Name{\idxapp{M}{i}}) |> <<\emptyset; \emptyset>>
        }

\begin{llproof}
\Pf{\Gamma_1}{|-}{v_M : (\namesort @> \namesort)[[M]]}{Given}
\Pf{}{}{v_M = \nametm{M_v}}{\Lemmaref{lem:canon}}
\Pf{}{}{M \conv (\lam{a} M')}  {\ditto}
\Pf{}{}{\cdot |- \lam{a} M' : (\namesort @> \namesort)}{\ditto}
\Pf{}{}{M_v \conv M}  {\ditto}
\decolumnizePf
\Pf{\Gamma_1}{|-}{v : \Name{i}}{Given}
\Pf{}{}{v = \name{n}}{\Lemmaref{lem:canon}}
\Pf{}{}{\Gamma |- n \in i}{\ditto}
\proofsep
\Pf{}{}{M \nteval (\lam{a} M')}{By inversion on $\D$ (name-app)}
\Pf{}{}{[n/a]M' \nteval p}{\ditto}
\convPf{p}{[n/a]M'}{By a property of $\nteval$}
\continueconvPf{\idxapp{(\lam{a} M')}{n}}{By a property of $\conv$}
\continueconvPf{\idxapp{M}{n}}{By a property of $\conv$}
\proofsep
\Pf{}{}{(\Gamma_2 = \Gamma_1),(\St_2 = \St_1)}{Suppose}
\Hand
\Pf{}{}{|- \St_2 : \Gamma_2}{By above equalities and $\St_1 |- \Gamma_1$}
\proofsep
\Pf{\Gamma_1}{|-}{ n \in i}{Above}
\convPf{p}{M(n)}   {Above}
\Pf{\Gamma_1}{|-}{ p \in \idxapp{M}{i}}{By \Lemref{app-mem-commute}}
\decolumnizePf
\Pf{\Gamma_1}{|-}{ \name{p} : \Name{\idxapp{M}{i}})}{\byrule{name}}
\Hand
\Pf{\Gamma_1}{|-}{ \Ret{\name{p}} : \F(\Name{\idxapp{M}{i}}) |> <<\emptyset; \emptyset>>}{\byrule{ret}}
\Hand
\Pf{}{}{\DByHasEffects{\D}{{\dynnameapp}}{\emptyset}{\emptyset}}{\ByRWSetRule}
\Hand
\Pf{}{}{(R_\D = R = \emptyset), (W_\D = W = \emptyset)}{By above equalities}
\end{llproof}

%Need canonical forms to turn 
%$\Gamma_1 |- v_M : (\namesort @> \namesort)[[M]]$
%into
%$v_M = \name{(\lam{a} M_0)}$
%such that
%$(\lam{a} M_0) : (\namesort @> \namesort)$.

\NotInScope{
\paragraph{TO DO:} Think about whether/where we need to explicitly show that stuff is well-kinded.
}

% \paragraph{POSSIBLY TO DO:} Generalize typing rule in figure.

\DerivationProofCase{AllIndexIntro}
       {
         \Gamma_1, a:\sort, P |-^\ambns t : E
       }
       {
         \Gamma_1 |-^\ambns t : (\DAll{a : \sort}{P} E)
       }

\begin{llproof}
$\Ss_0 ::$ \Pf{\Gamma_1, a:\sort, P}{|-^\ambns}{t : E}{Subderivation}
$\D_0  ::$ \Pf{\St_1}{|-^\ambns_m}{e !! \St_2 ; t}{Subderivation}
\Pf{\exists \Gamma_2 \subseteq \Gamma_1}{}{}{By i.h.}
\Hand
\Pf{}{|-}{\St_2 : \Gamma_2}{\ditto}
\Pf{}{}{\DHasEffects{\D_0}{R_{\D_0}}{W_{\D_0}}}{\ditto}
\Pf{\Gamma_2}{|-}{t : E}{\ditto}
\Pf{}{}{<<R_{\D_0}; W_{\D_0}>> \effsub <<R; W>>}{\ditto}
\Hand
\Pf{\Gamma_2}{|-^\ambns}{t : (\All{a : \sort} E)}{By typing rule}
\Hand
\Pf{}{}{\DHasEffects{\D}{R_{\D}}{W_{\D}}}{\ByRWSetRule}
\Hand
\Pf{}{}{<<R_{\D}; W_{\D}>> \effsub <<R; W>>}{By set theory}
\end{llproof}

%\paragraph{TO DO:}
%Either apply i.h.\ and rule again, or
%``similar to ret case''.

\DerivationProofCase{AllIndexElim}
       {
         \Gamma_1 |-^\ambns e : (\DAll{a : \sort}{P} E)
         \\
         \arrayenvbl{
         \Gamma_1 |- i : \sort
         \\
         \extract{\Gamma_1} ||- [i/a]P
         }
       }
       {
         \Gamma_1 |-^\ambns
         e % \idxapp{e}{i}
         : [i/a]E
       }

\begin{llproof}
$\Ss_0 ::$ \Pf{\Gamma_1}{|-^\ambns}{e : (\All{a : \sort} E)}{Subderivation}
$\D_0  ::$ \Pf{\St_1}{|-^\ambns_m}{e !! \St_2 ; t}{Subderivation}
\Pf{\exists \Gamma_2 \subseteq \Gamma_1}{}{}{By i.h.}
\Hand
\Pf{}{|-}{\St_2 : \Gamma_2}{\ditto}
\Pf{}{}{\DHasEffects{\D_0}{R_{\D_0}}{W_{\D_0}}}{\ditto}
\Pf{\Gamma_2}{|-}{t : (\All{a : \sort} E)}{\ditto}
\Pf{}{}{<<R_{\D_0}; W_{\D_0}>> \effsub <<R; W>>}{\ditto}
\decolumnizePf
\Pf{\Gamma_1}{|-}{i : \sort}{Subderivation}
\Pf{\Gamma_2}{|-}{i : \sort}{By weakening}
\Hand
\Pf{\Gamma_2}{|-^\ambns}{t : [i/a]E}{By typing rule}
\Hand
\Pf{}{}{\DHasEffects{\D}{R_{\D}}{W_{\D}}}{\ByRWSetRule}
\Hand
\Pf{}{}{<<R_{\D}; W_{\D}>> \effsub <<R; W>>}{By set theory}
\end{llproof}

    \DerivationProofCase{AllIntro}
       {
         \Gamma, \alpha : K |-^\ambns \te : E
       }
       {
         \Gamma |-^\ambns
         \te
         : (\All{\alpha : K} E)
       }

    Similar to the AllIndexIntro case.

    \DerivationProofCase{AllElim}
       {
         \Gamma |-^\ambns e : (\All{\alpha : K} E)
         \\
         \Gamma |- A : K
       }
       {
         \Gamma |-^\ambns
         e % \tyapp{e}{A}
         : [A/\alpha]E
       }

    Similar to the AllIndexElim case.
\qedhere
\end{itemize}
\end{proof}

% Local Variables: 
% mode: latex
% TeX-master: "typed-adapton"
% End: 

\section{Bidirectional Typing}
\label{apx:bidir}

\subsection{Syntax}

As discussed below, bidirectional typing requires some annotations,
so we assume that values $v$ and expressions $e$ have been extended
with annotations $(v : A)$ and $(e : A)$.
We also assume that we have explicit syntactic forms
$\inst{e}{i}$ and $\inst{e}{A}$,
which avoid guessing quantifier instantiations.

\begin{figure}[htbp]
  \centering

  \judgbox{\Gamma |- v => A}{Under $\Gamma$, value $v$ synthesizes type $A$}
%  \lesscaptionspace
  \begin{mathpar}
        \Infer{\vsynvar}
                {
                        (x : A) \in \Gamma
                }
                {
                        \Gamma |- x => A
                }
        \and
        \Infer{\vsynanno}
                {
                        \Gamma |- v <= A
                }
                {
                        \Gamma |- (v : A) => A
                }
  \end{mathpar}

  \judgbox{\Gamma |- v <= A}{Under $\Gamma$, value $v$ checks against type $A$}
  \begin{mathpar}
        \Infer{\vchkunit}
        {}
        {
                \Gamma |- \unitexp <= \Unit
        }
    ~~~~
    \Infer{\vchkpair}
        {
                \Gamma |- v_1 <= A_1
                \\
                \Gamma |- v_2 <= A_2
        }
        { 
                \Gamma |- \Pair{v_1}{v_2} <= (A_1 ** A_2) 
        }
   \\
   \Infer{\vchksub}
       {
         \Gamma |- v => A
         \\
         \Gamma |- A \vsubtype B
       }
       {
         \Gamma |- v <= B
       }
   \and
   \Infer{\vchkname}
                {
                        \Gamma |- n \in X
                }
                {
                        \Gamma |- (\name{n}) <= \Name{X}
                }
    \and
    \Infer{\vchknamefn}
                {
                        \Gamma |- M_v => (\namesort @> \namesort)
                        \\
                        M_v \conv M
                }
                {
                        \Gamma |- (\namefn{M_v}) <= (\namesort @> \namesort)[[M]]
                }
        \and
        \Infer{\vchkref}
                {
                        \Gamma |- n \in X
                        \\
                        \Gamma(n) = A
                }
                {
                        \Gamma |- (\refv{n}) <= \Ref{X}{A}
                }
        \and
        \Infer{\vchkthunk}
                {
                        \Gamma |- n \in X
                        \\
                        \Gamma(n) = E
                }
                {
                        \Gamma
                        |-
                        (\thunk{n})
                        <=
                        (
                          \Thk{X}{E}
                        )
                }
%         \and
%         \Infer{\vchkconv}
%         {
%                 \Gamma |- v => A_1 
%                 \\
%                 A_1 = A_2
%         }
%         { 
%                 \Gamma |- v <= A_2 
%         }
        \and
        \Infer{\vchkinjOne}
        {
          \Gamma |- v <= A_1
        }
        {
          \Gamma |- \inj{1}{v} <= A_1 + A_2
        }
        ~~~~
        \Infer{\vchkinjTwo}
        {
          \Gamma |- v <= A_2
        }
        {
          \Gamma |- \inj{2}{v} <= A_1 + A_2
        }
       \and
    \Infer{\vchkAllIndexIntro}
       {
         \Gamma, a:\sort, P |- v <= A
       }
       {
         \Gamma |- v <= (\DAll{a : \sort}{P} A)
       }
    ~~~
    \Infer{\vsynAllIndexElim}
       {
         \Gamma |- i : \sort
         \\
         \arrayenvbl{
         \extract{\Gamma} ||- [i/a]P
         \\  
         \Gamma |- v => (\DAll{a : \sort}{P} A)
         }
       }
       {
         \Gamma |-
         \inst{v}{i} % \idxapp{e}{i}
         => [i/a]A
       }
   \\
    \Infer{\vchkExistsIndexIntro}
       {
         \Gamma |- i : \sort
         \\
         \arrayenvbl{
         \extract{\Gamma} ||- [i/a]P
         \\  
         \Gamma |- v <= [i/a]A
         }
       }
       {
         \Gamma |- \VPack{a}{v} <= (\DExists{a : \sort}{P} A)
       }
  \lesscaptionspace
  \end{mathpar}

  \caption{Bidirectional value typing}
  \label{fig:bidirectional-value-typing}
\end{figure}

\begin{figure}[htbp]
\runonfontsz{8.5pt}
\centering

  \judgbox{\Gamma |-^\ambns e => E}
          {Under $\Gamma$, within namespace~$\ambns$, \\
            %after history $\hist$,
            computation $e$ synthesizes type-with-effects $E$
            % and effects $\e$.
          }
  \vspace*{-1.5ex}
  \begin{mathpar}
    \hfill
    \Infer{\esynanno}
        {
                \Gamma |-^\ambns e <= E
        }
        {
                \Gamma |-^\ambns (e : E) => E
        }
    \and
    \Infer{\esynapp}
        {
          \arrayenvbl{
          \Gamma |-^\ambns e
              => \big(
                 (A -> E) |> \e_1
                \big)
          \\
          \Gamma |- v <= A
          ~~~~~
          \Gamma |- E \effcoal \e_1 \equiv E'
          }
        }
        {
          \Gamma |-^\ambns (e\;v) => E'
        }
    \and
    \Infer{\esynforce}
        {
          \arrayenvbl{
          \Gamma |- v => \Thk{X}{(C |> \e)}
          \\
          \Gamma |- <<\emptyset; X>> \effseq \e \equiv \e'
          }
        }
        {
                \Gamma |-^\ambns
                \Force{v}
                =>
                C |> \e'
        }
~~~~
    \Infer{\esynget}
        {
                \Gamma |- v => \Ref{X}{A}
        }
        {
                \Gamma |-^\ambns
                \Get{v}
                =>
                (
                  \F A
                )
                |>
                <<\emptyset; X>>
        }
    \and
    \Infer{\esynnameapp}
        {
                \Gamma |- v_M => (\namesort @> \namesort)[[M]]
                \\
                \Gamma |- v => \Name{i}
        }
        {
                \Gamma |-^N (v_M\;v) => \F (\Name{M[[i]]}) |> <<\emptyset; \emptyset>>
        }
    \and
    \Infer{\esynAllIndexElim}
        {
                \Gamma |-^\ambns e => (\DAll{a : \sort}{P} E)
                \\
                \Gamma |- i : \sort
                \\
                \extract{\Gamma} ||- [i/a]P
        }
        {
                \Gamma |-^\ambns
                \inst{e}{i} % \idxapp{e}{i}
                => [i/a]E
        }
    \and
    \Infer{\esynAllElim}
        {
                \Gamma |-^\ambns e => (\All{\alpha : K} E)
                \\
                \Gamma |- A : K
        }
        {
                \Gamma |-^\ambns
                \inst{e}{A} % \tyapp{e}{A}
                => [A/\alpha]E
              }
  \lesscaptionspace
  \end{mathpar}

  \judgbox{\Gamma |-^\ambns e <= E}
                  {Under $\Gamma$, within namespace~$\ambns$, \\
           %after history $\hist$,
           computation $e$ checks against type-with-effects $E$
           % and effects $\e$.
          }
  \vspace*{-0.5ex}
  \begin{mathpar}
    \Infer{\echksub}
        {
          \Gamma |-^\ambns e => E_1
          \\
          \Gamma |- E_1 \esubtype E_2
        }
        {
          \Gamma |-^\ambns e <= E_2
        }
%          \Infer{chk-conv}
%                 {
%                         \Gamma |-^\ambns e => E_1 
%                         \\
%                         E_1 = E_2
%                 }
%                 { 
%                         \Gamma |-^\ambns e <= E_2 
%                 }
%          ~~~
%          \Infer{chk-eff-subsume}
%                 {
%                   \arrayenvbl{
%                         \Gamma |-^\ambns e => (C |> \e_1)
%                         \\
%                         \Gamma |- \e_1 \effsub \e_2
%                     }
%                 }
%                 {
%                         \Gamma |-^\ambns e <= (C |> \e_2)
%                 }
         \vspace*{-1ex}
         \\
         \Infer{\echksplit}
                {
                        \arrayenvbl{
                        \Gamma |-%^\ambns
                        v => (A_1 ** A_2)
                        \\
                        \Gamma, x_1:A_1, x_2:A_2 |-^\ambns e <= E
                        }
                }
                {
                        \Gamma |-^\ambns \Split{v}{x_1}{x_2}{e} <= E
                }
         ~~~
         \Infer{\echkcase}
                {
                  \arrayenvbl{
                        \Gamma |-%^\ambns
                        v => (A_1 + A_2)
                        \\
                        \Gamma, x_1:A_1 |-^\ambns e_1 <= E
                        \\
                        \Gamma, x_2:A_2 |-^\ambns e_2 <= E
                        }
                }
                {
                        \Gamma |-^\ambns \Case{v}{x_1}{e_1}{x_2}{e_2} <= E
                }
         \vspace*{-0.3ex}
         \\
         \arrayenvbl{
         \Infer{\echkret}
                {
                        \Gamma |- v <= A
                }
                {
                        \Gamma |-^\ambns
                        \Ret{v}
                        <=
                        \big(
                        (\F A)  |>  <<\emptyset; \emptyset>>
                        \big)
                }
        \\[0.7ex]
        \Infer{\echklam}
                {
                        \Gamma,x:A |-^\ambns e <= E
                }
                {
                        \Gamma |-^\ambns
                        (\lam{x} e)
                        <=
                        \big(
                        (A{->}E) |> <<\emptyset; \emptyset>>
                        \big)
                }
         }
         \and
         \Infer{\echklet}
                {
                  \arrayenvbl{
                    \Gamma |-^M e_1 => (\F A) |> \e_1
                    \\
                    \Gamma, x:A |-^M e_2 <= (C |> \e_2)
                    \\
                    \Gamma |- \e_1 \effseq \e_2 \equiv \e
                    }
                }
                {
                        \Gamma |-^M \Let{e_1}{x}{e_2}
                        <=
                        C |> \e
                }
        \and
        \Infer{\echkthunk}
                {
                        \Gamma |- v <= \Name{X}
                        \\
                        \Gamma |-^\ambns e <= E
                }
                {\small
                        \Gamma |-^\ambns
                        \Thunk{v}{e}
                        <=
                        \big(
                        \F (\Thk{M[[X]]}{E})
                        \big)
                        |>
                        <<M[[X]]; \emptyset>>
                }
    \and
        \Infer{\echkref}
        {
                \Gamma |- v_1 <= \Name{X}
                \\
                \Gamma |- v_2 <= A
        }
        {
                \Gamma |-^\ambns
                \Refe{v_1}{v_2}
                <=
                \big(
                \F (\Ref{M[[X]]}{A})
                \big)
                |>
                <<M[[X]]; \emptyset>>
        }
    \and
    \Infer{\echkscope}
        {
                \Gamma |- v => (\namesort @> \namesort)[[N']]
                \\
                \Gamma |-^{N \circ N'} e <= (C |> <<W; R>>)
        }
        {
                \Gamma |-^N \Scope{v}{e} <= (C |> <<W; R>>)
        }
    \and
    \Infer{\echkAllIndexIntro}
        {
                \Gamma, a:\sort, P |-^\ambns t <= E
        }
        {
                \Gamma |-^\ambns t <= (\DAll{a : \sort}{P} E)
        }
    \and
    \Infer{\echkAllIntro}
        {
                \Gamma, \alpha : K |-^\ambns t <= E
        }
        {
                \Gamma |-^\ambns
                t
                <= (\All{\alpha : K} E)
        }
   \and
    \Infer{\echkExistsIndexElim}
        {
            \Gamma |- v => (\DExists{a : \sort}{P} A)
            \\
            \Gamma, a : \sort, P, x : A |-^\ambns e <= E
        }
        {
          \Gamma |-^\ambns \VUnpack{v}{a}{x}{e} <= E
        }
  \lesscaptionspace
  \end{mathpar}

  \caption{Bidirectional computation typing}
  \label{fig:bidirectional-comp-typing}
\end{figure}

% Local Variables: 
% mode: latex
% TeX-master: "typed-adapton"
% End: 

\subsection{Bidirectional Typing Rules}

The typing rules in Figures \ref{fig:value-typing}
and \ref{fig:comp-typing}
are declarative:
they define what typings are valid, but not how to derive those typings.
% For example, the rule `namefn' says that
% $\namefn{M_v}$ has a type indexed by $M$
% for some $M$ equivalent to $M_v$, which would require
% guessing $M$. 
The rules' use of names and effects annotations
means that standard unification-based techniques,
like Damas--Milner inference, are not readily applicable.
% For example, it is not obvious when to apply \echkAllIntro.

Following the DML tradition,
we obtain an algorithmic version of our typing rules by defining
a bidirectional system \citep{Pierce00}:
we split judgments with a colon into
judgments with an arrow.
Thus, the computation typing judgment
$\cdots e : E$
becomes two judgments.
The first is the \emph{checking} judgment
$\Gamma |-^\ambns e <= E$,
in which the type $E$ is already known---it is an \emph{input} to the algorithm.
The second is the \emph{synthesis} judgment
$\Gamma |-^\ambns e => E$,
in which $E$ is not known---it is an output---and the rules construct $E$
by examining $e$ (and $\Gamma$).

In formulating the bidirectional versions of value and computation typing
(Figures \ref{fig:bidirectional-value-typing}
and \ref{fig:bidirectional-comp-typing}),
we mostly follow the ``recipe'' of \citet{Dunfield04:Tridirectional}:
introduction rules check, and elimination rules synthesize.
More precisely, the \emph{principal judgment}---the judgment, either a premise
or conclusion, that has the connective being introduced or eliminated---is
checking (${<=}$) for introduction rules,
and synthesizing (${=>}$) for elimination rules.
In many cases, once the direction of that premise (or conclusion) is determined,
the direction of the other judgments follows by considering what information
is known (as input, or as the output type of the principal judgment,
if that judgment is synthesizing).
For example, if we commit to checking the conclusion of \echklam,
we should check the premise because its type is a subexpression of the
type in the conclusion.
(Checking is more powerful than synthesis: every expression that synthesizes
also checks, but not all expressions that check can synthesize.)

When a synthesis (elimination) premise attempts to type
an expression that is a checking (introduction) form,
the programmer must write a type annotation $(e : E)$.
Thus, following the recipe means that we have a straightforward
\emph{annotation discipline}:
annotations are needed only on redexes.
While we could reduce the number of annotations by adding
synthesis rules---for example, allowing the unit value
$\unitexp$ to synthesize $\unitty$---this makes the system larger
without changing its essential properties;
for a discussion of the implications of such extensions in a different context,
see \citet{Dunfield13}.

Dually, when an expression synthesizes but we are trying to derive a checking judgment,
we use (1) \vchksub for value typing,
or (2) \echksub for computation typing.
The latter rule includes effect subsumption.
% Subsumption is common in bidirectional systems with subtyping,
% such as \citet{Dunfield04:Tridirectional}.

% Local Variables: 
% mode: latex
% TeX-master: "typed-adapton"
% End: 

\section{Bidirectional Typing Proofs}
\label{apx:bidirectional-proofs}

\begin{restatable}[Soundness of Bidirectional Value Typing]{theorem}{thmrefsoundbival}
        \Label{thmrefsoundbival}
        ~
        \begin{enumerate}
                \item If
                $\Gamma |- v => A$, 
                then there exists a value 
                $v'$
                such that $\Gamma |- v' : A$
                and $|v| = v'$.
                
                \item If
                $\Gamma |- v <= A$, 
                then there exists a value
                $v'$
                such that $\Gamma |- v' : A$
                and $|v| = v'$.
        \end{enumerate}
\end{restatable}

\begin{proof}  By induction on the given derivation.

        Part (1):  Proceed by cases on the rule concluding $\Gamma |- v => A$.
        
        \begin{itemize}
                \DerivationProofCase{\vsynvar}
                        {
                                (x : A) \in \Gamma
                        }
                        {
                                \Gamma |- x => A
                        }

                        \begin{llproof}
                                \Pf{}{}{(x : A) \in \Gamma}{Given}
                                \Pf{\Gamma}{|-}{x : A}{By rule var}
                                \Pf{}{}{|x| = x}{By definition of $|{-}|$}
                                \Hand \Pf{\Gamma}{|-}{v' : A ~\textrm{and}~ |v| = v'}{where $v' = x$ and $v = x$}
                        \end{llproof}

                \DerivationProofCase{\vsynanno}
                        {
                                \Gamma |- v_1 <= A
                        }       
                        {
                                \Gamma |- (v_1 : A) => A
                        }

                        \hspace*{-4ex}\begin{llproof}
                                \Pf{\exists~v_1' ~\textrm{such that}~ \Gamma}{|-}{v_1' : A ~\textrm{and}~ |v_1| = v_1'}{By inductive hypothesis}
                                \Pf{}{|(v_1:A)| =}{|v_1| = v_1'}{By definition of $|{-}|$}\trailingjust{and $|v_1| = v_1'$}
                                \Hand\!\!\!\!\!\!\!\! \Pf{\Gamma}{|-}{v' : A ~\textrm{and}~ |v| = v'}{where $v' = v_1'$ and $v = (v_1:A)$}
                        \end{llproof}

  {\noindent
        Part (2):}  Proceed by cases on the rule concluding $\Gamma |- v <= A$.
                
                \DerivationProofCase{\vchkunit}
                        {}
                        {
                                \Gamma |- \unitexp <= \Unit
                        }
                        
                        \begin{llproof}
                                \Pf{\Gamma}{|-}{\unitexp : \Unit}{By rule unit}
                                \Pf{}{|\unitexp| =}{\unitexp}{By definition of $|{-}|$}
                                \Hand \Pf{\Gamma}{|-}{v' : \unitexp ~\textrm{and}~ |v| = v'}{where $v' = \unitexp$ and $v = \unitexp$}
                        \end{llproof}
                        
                \DerivationProofCase{\vchkpair}
                        {
                                \Gamma |- v_1 <= A_1
                                \\
                                \Gamma |- v_2 <= A_2
                        }
                        {
                                \Gamma |- \Pair{v_1}{v_2} <= (A_1 ** A_2)
                        }
                
                        \begin{llproof}
                                \Pf{\exists~v_1' ~\textrm{such that}~ \Gamma }{|-}{v_1' : A_1 ~\textrm{and}~ |v_1| = v_1'}{By inductive hypothesis}
                                \Pf{\exists~v_2' ~\textrm{such that}~ \Gamma}{|-}{v_2' : A_2 ~\textrm{and}~ |v_2| = v_2'}{By inductive hypothesis}
                                \Pf{\Gamma}{|-}{\Pair{v_1'}{v_2'} : (A_1 ** A_2)}{By rule pair}
                                \Pf{}{|\Pair{v_1}{v_2}| =}{\Pair{|v_1|}{|v_2|}= \Pair{v_1'}{v_2'}}{By definition of $|{-}|$} \trailingjust{$|v_1| = v_1'$ and $|v_2| = v_2'$}
                                \Hand \Pf{\Gamma} {|-} {v' : (A_1 ** A_2) ~\textrm{and}~ |v| = v'}{where $v' = \Pair{v_1'}{v_2'}$} \trailingjust{and $v = \Pair{v_1}{v_2}$}
                        \end{llproof}
                
                \DerivationProofCase{\vchkname}
                        {
                                \Gamma |- n \in X
                        }
                        {
                                \Gamma |- (\name{n}) <= \Name{X}
                        }

                        \begin{llproof}
                                \Pf{\Gamma} {|-} {n \in X}{Given}
                                \Pf{\Gamma} {|-} {(\name{n}) : \Name{X}}{By rule name}
                                \Pf{}{|(\name{n})| =}{ (\name{n})}{By definition of $|{-}|$}
                                \Hand \Pf{\Gamma} {|-} {v' : \Name{X} ~\textrm{and}~ |v| = v'}{where $v' = (\name{n})$}\trailingjust{ and $v = (\name{n})$}
                        \end{llproof}
        
                \DerivationProofCase{\vchknamefn}
                        {
                                \Gamma |- M_v => (\namesort @> \namesort)
                                \\
                                M_v \conv M
                        }
                        {       
                                \Gamma |- (\namefn{M_v}) <= (\namesort @> \namesort)[[M]]
                        }
                        
                        \begin{llproof}
                                \Pf{\exists~M_v' ~\textrm{such that}~ \Gamma} {|-} {M_v' : (\namesort @> \namesort) ~\textrm{and}~ |M_v| = M_v'}{~~By i.h.}
                                \decolumnizePf
                                \Pf{}{}{M_v \conv M}{Given}
                                \Pf{}{}{|M_v| \conv M}{Type erasure does not affect convertibility}
                                \Pf{}{}{M_v' \conv M}{Since $|M_v| = M_v'$}
                                \decolumnizePf
                                \Pf{\Gamma} {|-} {(\namefn{M_v'}) : (\namesort @> \namesort)[[M]]}{By rule namefn}
                                \Pf{}{|(\namefn{M_v})| =}{ (\namefn{|M_v|}) = (\namefn{M_v'})}{By definition of $|{-}|$}\trailingjust{and $|M_v| = M_v'$}
                                \Hand \Pf{\Gamma} {|-} {v' :  (\namesort @> \namesort)[[M]] ~\textrm{and}~ |v| = v'}{where $v' =  (\namefn{M_v'})$}\trailingjust{ and $v = (\namefn{M_v})$}
                        \end{llproof}

                \DerivationProofCase{\vchkref}
                        {
                                \Gamma |- n \in X
                                \\
                                \Gamma(n) = A
                        }
                        {
                                \Gamma |- (\refv{n}) <= \Ref{X}{A}
                        }
                        
                        \begin{llproof}
                                \Pf{\Gamma} {|-} {n \in X}{Given}
                                \Pf{}{}{\Gamma(n) = A}{Given}
                                \Pf{\Gamma} {|-} {(\refv{n}) : \Ref{X}{A}}{By rule ref}
                                \Pf{}{|(\refv{n})| =}{ \refv{n}}{By definition of $|{-}|$}
                                \Hand \Pf{\Gamma} {|-} {v' : \Ref{X}{A} ~\textrm{and}~ |v| = v'}{where $v' =  (\refv{n})$}\trailingjust{ and $v = (\refv{n})$}
                        \end{llproof}
                        
                \DerivationProofCase{\vchkthunk}
                        {
                                \Gamma |- n \in X
                                \\
                                \Gamma(n) = E
                        }
                        {
                                \Gamma
                                |-
                                (\thunk{n})
                                <=
                                \big(
                                \Thk{X}{E}
                                \big)
                        }
                        
                        \begin{llproof}
                                \Pf{\Gamma} {|-} {n \in X}{Given}
                                \Pf{}{}{\Gamma(n) = E}{Given}
                                \Pf{\Gamma} {|-} {(\thunk{n}) : \Thk{X}{E}}{By rule thunk}
                                \Pf{}{|(\thunk{n})| =}{ \thunk{n}}{By definition of $|{-}|$}
                                \Hand \Pf{\Gamma} {|-} {v' : \Thk{X}{E} ~\textrm{and}~ |v| = v'}{where $v' =  (\thunk{n})$}\trailingjust{and $v = (\thunk{n})$}
                        \end{llproof}
                
                \DerivationProofCase{\vchksub}
                        {
                                \Gamma |- v_1 => A_1 
                                \\
                                \Gamma |- A_1 \vsubtype A_2
                        }
                        { 
                                \Gamma |- v_1 <= A_2 
                        }       
                
                        By i.h.\ and \vtypeSub.
%                       \begin{llproof}
%                               \Pf{\exists~v_1' ~\textrm{such that}~ \Gamma} {|-} {v_1' : A_1 ~\textrm{and}~ |v_1| = v_1'}{By inductive hypothesis}
%                               \Pf{}{}{A_1 = A_2}{Given}
%                               \Pf{\Gamma} {|-} {v_1' : A_2}{By rule conv}
%                               \Hand \Pf{\Gamma} {|-} {v' :  A_2 ~\textrm{and}~ |v| = v'}{where $v' =  v_1'$ and $v = v_1$}
%                               \end{llproof}
  \qedhere
  \end{itemize}
\end{proof}

\begin{restatable}[Completeness of Bidirectional Value Typing]{theorem}{thmrefcomplbival}
        \Label{thmrefcomplbival}
        ~\\
        If
        $       \Gamma  |- v : A$
        then there exist values
        $v'$ and $v''$ such that

        \begin{enumerate}
                \item 
                $       \Gamma  |- v' => A $
                and $   |v'| = v        $
                
                \item 
                $       \Gamma  |- v'' <= A $
                and $   |v''| = v       $
        \end{enumerate}
        
\end{restatable}

\begin{proof}
        By induction on the derivation of $     \Gamma |- v : A $.

        \DerivationProofCase{unit}
                {}
                {
                        \Gamma |- \unitexp : \Unit
                }
        
                \begin{llproof}
                        \Pf{\Gamma} {|-} {\unitexp <= \Unit}{By rule vchk-unit}
                        \Pf{}{|\unitexp| =}{ \unitexp}{By definition of $|{-}|$}
                        \Hand \Pf{\Gamma} {|-} {v'' <= \unitexp ~\textrm{and}~ |v''| = v}{where $v'' = \unit$ and $v = \unit$}
                        \Pf{\Gamma} {|-} {(\unitexp : \Unit) => \Unit}{By rule vsyn-anno}
                        \Pf{}{|(\unitexp : \Unit)| =} {\unitexp}{By definition of $|{-}|$}
                        \Hand \Pf{\Gamma} {|-} {v' => \unitexp ~\textrm{and}~ |v'| = v}{where $v' = (\unitexp : \Unit)$ and $v = \unit$}
                \end{llproof}
                
        \DerivationProofCase{var}
                {(x : A) \in \Gamma}
                {\Gamma |- x : A}
                        
                \begin{llproof}
                        \Pf{}{}{(x : A) \in \Gamma}{Given}
                        \Pf{\Gamma} {|-} {x => A}{By rule vsyn-var}
                        \Pf{}{|x| =}{x}{By definition of $|{-}|$}
                        \Hand \Pf{\Gamma} {|-} {v' => A ~\textrm{and}~ |v'| = v}{where $v' = x$ and $v = x$}
                        \Pf{\Gamma} {|-} {x <= A}{By rule vchk-conv}
                        \Hand \Pf{\Gamma} {|-} {v'' <= A ~\textrm{and}~ |v''| = v}{where $v'' = x$ and $v = x$}\trailingjust{and $|x| = x$}
                \end{llproof}

        \DerivationProofCase{pair}
                {
                        \Gamma |- v_1 : A_1
                        \\
                        \Gamma |- v_2 : A_2
                }
                {\Gamma |- \Pair{v_1}{v_2} : (A_1 ** A_2)}
                
                \hspace*{-4ex}\begin{llproof}
                        \Pf{\exists~v_1'' ~\textrm{such that}~ \Gamma} {|-} {v_1'' <= A_1 ~\textrm{and}~ |v_1''| = v_1}{By inductive hypothesis}
                        \Pf{\exists~v_2'' ~\textrm{such that}~ \Gamma} {|-} {v_2'' <= A_2 ~\textrm{and}~ |v_2''| = v_2}{By inductive hypothesis}
                        \Pf{\Gamma} {|-} {\Pair{v_1''}{v_2''} <= (A_1 ** A_2)}{By rule vchk-pair}
                        \Pf{}{|\Pair{v_1''}{v_2''}| =} {\Pair{|v_1''|}{|v_2''|} = \Pair{v_1}{v_2}}{By definition of $|{-}|$; $|v_1''| = v_1$; $|v_2''| = v_2$}
\decolumnizePf
                        \Hand\!\!\! \Pf{\Gamma} {|-} {v'' <= (A_1 ** A_2) ~\textrm{and}~ |v''| = v}{where $v'' = \Pair{v_1''}{v_2''}$ and $v = \Pair{v_1}{v_2}$}
                        \Pf{\Gamma} {|-} {(\Pair{v_1''}{v_2''}:(A_1 ** A_2)) => (A_1 ** A_2)}{By rule vsyn-anno}
\decolumnizePf
                        \Pf{}{|\Pair{v_1''}{v_2''}:(A_1 ** A_2)| = }
                        {|\Pair{v_1''}{v_2''}| = \Pair{v_1}{v_2}}{By definition of $|{-}|$; $|\Pair{v_1''}{v_2''}| = \Pair{v_1}{v_2}$}
\decolumnizePf
                        \Hand\!\!\! \Pf{\Gamma} {|-} {v' => (A_1 ** A_2) ~\textrm{and}~ |v'| = v}{where $v' = \Pair{v_1''}{v_2''} $ and $v = \Pair{v_1}{v_2}$}
                \end{llproof}

        \DerivationProofCase{name}
                {
                        \Gamma |- n \in X
                }
                {
                        \Gamma |- (\name{n}) : \Name{X}
                }
            
            \begin{llproof}
                \Pf{\Gamma} {|-} {n \in X}{Given}
                \Pf{\Gamma} {|-} {(\name{n}) <= \Name{X}}{By rule vchk-name}
                \Pf{}{|(\name{n})| = }{(\name{n})}{By definition of $|{-}|$}
                \Hand \Pf{\Gamma} {|-} {v'' <= \Name{X} ~\textrm{and}~ |v''| = v}{where $v'' = (\name{n})$ and $v = (\name{n})$}
                \decolumnizePf
                \Pf{\Gamma} {|-} {(\name{n}:\Name{X}) => \Name{X}}{By rule vsyn-anno}
                \Pf{}{|(\name{n}:\Name{X})| = }{|(\name{n})| = (\name{n})}{By definition of $|{-}|$}
                \decolumnizePf
                \Hand \Pf{\Gamma} {|-} {v' => \Name{X} ~\textrm{and}~ |v'| = v}{where $v' = (\name{n}:\Name{X}) $ and $v = (\name{n})$}
            \end{llproof}
            
        \DerivationProofCase{namefn}
                {
                        \Gamma |- M_v : (\namesort @> \namesort)
                        \\
                        M_v \conv M
                }
                {
                        \Gamma |- (\namefn{M_v}) : (\namesort @> \namesort)[[M]]
                }

                \begin{llproof}
                        \Pf{}{}{\hspace{-28pt} \exists~M_v' ~\textrm{such that}~ }{}
                        \Pf{\Gamma} {|-} {M_v' => (\namesort @> \namesort) ~\textrm{and}~ |M_v'| = M_v}{By inductive hypothesis}
                        \Pf{}{}{M_v \conv M}{Given}
                        \Pf{}{}{|M_v'| \conv M}{Since $|M_v'| = M_v$}
                        \Pf{}{}{M_v' \conv M}{Type annotation does not affect convertibility}
                        \decolumnizePf
                        \Pf{\Gamma} {|-} {(\namefn{M_v'}) <= (\namesort @> \namesort)[[M]]}{By rule vchk-namefn}
                        \Pf{}{}{|(\namefn{M_v'})| = (\namefn{|M_v'|}) = (\namefn{M_v})}{By definition of $|{-}|$}\trailingjust{and $|M_v'| = M_v$}
                        \Hand \Pf{\Gamma} {|-} {v'' <= (\namesort @> \namesort)[[M]] ~\textrm{and}~ |v''| = v}{where $v'' = (\namefn{M_v'})$ and $v = (\namefn{M_v})$}
                        \decolumnizePf
                        \Pf{\Gamma} {|-} {(\namefn{M_v'} : (\namesort @> \namesort)[[M]]) => (\namesort @> \namesort)[[M]]}{By rule vsyn-anno}
                        \Pf{}{}{|(\namefn{M_v'} : (\namesort @> \namesort)[[M]])| % = |(\namefn{M_v'})| 
                          = (\namefn{M_v})}{By definition of $|{-}|$} %{ and $|(\namefn{M_v'})| = (\namefn{M_v})$}
                        \decolumnizePf
                        \Hand \Pf{\Gamma} {|-} {v' => (\namesort @> \namesort)[[M]] ~\textrm{and}~ |v'| = v}{where $v' = (\namefn{M_v'} : (\namesort @> \namesort)[[M]]) $}
%                       \trailingjust{and $v = (\namefn{M_v})$}
                \end{llproof}
        
        \DerivationProofCase{ref}
                {
                        \Gamma |- n \in X
                        \\
                        \Gamma(n) = A
                }
                {
                        \Gamma |- (\refv{n}) : \Ref{X}{A}
                }
        
                \begin{llproof}
                        \Pf{\Gamma} {|-} {n \in X}{Given}
                        \Pf{}{}{\Gamma(n) = A}{Given}
                        \Pf{\Gamma} {|-} {(\refv{n}) <= \Ref{X}{A}}{By rule vchk-ref}
                        \Pf{}{}{|(\refv{n})| =(\refv{n})}{By definition of $|{-}|$}
                        \Hand \Pf{\Gamma} {|-} {v'' <= \Ref{X}{A} ~\textrm{and}~ |v''| = v}{where $v'' = (\refv{n})$ and $v = (\refv{n})$}
                        \Pf{\Gamma} {|-} {((\refv{n}) : \Ref{X}{A}) => \Ref{X}{A}}{By rule vsyn-anno}
                        \Pf{}{}{|(\refv{n}: \Ref{X}{A})| = |(\refv{n})| = (\refv{n})}{By definition of $|{-}|$} %{}\trailingjust{and $|(\refv{n})| = (\refv{n})$}
                        \Hand \Pf{\Gamma} {|-} {v' => \Ref{X}{A} ~\textrm{and}~ |v'| = v}{where $v' = (\refv{n}: \Ref{X}{A})$}\trailingjust{ and $v = (\refv{n})$}
                \end{llproof}

        \DerivationProofCase{thunk}
                {
                        \Gamma |- n \in X
                        \\
                        \Gamma(n) = E
                }
                {
                        \Gamma
                        |-
                        (\thunk{n})
                        :
                        \big(
                        \Thk{X}{E}
                        \big)
                }
        
                \begin{llproof}
                        \Pf{\Gamma} {|-} {n \in X}{Given}
                        \Pf{}{}{\Gamma(n) = E}{Given}
                        \Pf{\Gamma} {|-} {(\thunk{n}) <= \big(\Thk{X}{E} \big)}{By rule vchk-thunk}
                        \Pf{}{}{|(\thunk{n})| = (\thunk{n})}{By definition of $|{-}|$}
                        \Hand \Pf{\Gamma} {|-} {v'' <= \big(\Thk{X}{E} \big) ~\textrm{and}~ |v''| = v}{where $v'' = (\thunk{n})$ }\trailingjust{and $v = (\thunk{n})$}
                        \Pf{\Gamma} {|-} {(\thunk{n} : \big(\Thk{X}{E} \big)) => \big(\Thk{X}{E} \big)}{By rule vsyn-anno}
                        \Pf{}{}{|(\thunk{n} : \big(\Thk{X}{E} \big))| %=|\thunk{n}|
                          = \thunk{n}}{By definition of $|{-}|$}
                        \decolumnizePf
                        \Hand \Pf{\Gamma} {|-} {v' => \big(\Thk{X}{E} \big) ~\textrm{and}~ |v'| = v}{where $v' = (\thunk{n} : \big(\Thk{X}{E} \big))$}\trailingjust{ and $v = (\thunk{n})$}
                \end{llproof}
\end{proof}

\begin{restatable}[Soundness of Bidirectional Computation Typing]{theorem}{thmrefsoundbicomp}
        \Label{thmrefsoundbicomp}
        ~
        \begin{enumerate}
                \item If
                $       \Gamma  |-^\ambns e => E $, 
                then there exists a value 
                $       e'      $
                such that $     \Gamma  |-^\ambns e' : E $
                and $   |e| = e'        $
                
                \item If
                $       \Gamma  |-^\ambns e <= E $, 
                then there exists a value 
                $       e'      $
                such that $     \Gamma  |-^\ambns e' : E $
                and $   |e| = e'        $
        \end{enumerate}
\end{restatable}

\begin{proof}
  By induction on the given derivation.

  {\noindent Part (1):} Proceed by case analysis on the rule concluding
  $\Gamma |-^\ambns e => E$.
        
  \begin{itemize}
        \DerivationProofCase{\esynapp}  
                {
                        \Gamma |-^\ambns e_1
                                =>  \big(
                                        (A -> E) |> \e_1
                                        \big)
                        \\
                        \Gamma |- v <= A
                }
                {
                        \Gamma |-^\ambns (e_1\;v) => (E \effcoal \e_1)
                }
        
                \begin{llproof}
                        \Pf{%\exists~e_1' ~\textrm{such that}~
                          \Gamma} {|-^\ambns}{ e_1' : \big(
                                (A -> E) |> \e_1
                                \big) ~\textrm{and}~ |e_1| = e_1'}{By inductive hypothesis}
                        \Pf{%\exists~v' ~\textrm{such that}~
                          \Gamma} {|-} {v' : A ~\textrm{and}~ |v| = v'}{By \Thmref{thmrefsoundbival}}
                        \Pf{\Gamma} {|-^\ambns}{ (e_1'\;v') : (E \effcoal \e_1)}{By rule app}
                        \Pf{}{}{|(e_1\;v)| = (|e_1|\;|v|) = (e_1'\;v')}{By definition of $|{-}|$}
                        \Hand \Pf{\Gamma} {|-^\ambns}{ e' : (E \effcoal \e_1) ~\textrm{and}~ |e| = e'}{where $e' = (e_1'\;v')$ and $e = (e_1\;v)$}
                \end{llproof}   
        
        \DerivationProofCase{\esynforce}
                {
                        \Gamma |- v => \Thk{X}{(C |> \e)}
                }
                {
                        \Gamma |-^\ambns
                        \Force{v}
                        =>
                        \big(
                        C |> (<<\emptyset; X>> \effseq \e)
                        \big)
                }
                
                \begin{llproof}
                        \Pf{%\exists~v' ~\textrm{such that}~ 
                          \Gamma} {|-} {v' : \Thk{X}{(C |> \e)} ~\textrm{and}~ |v| = v'}{By \Thmref{thmrefsoundbival}}
                        \Pf{\Gamma} {|-^\ambns}{ \Force{v'} : \big(C |> (<<\emptyset; X>> \effseq \e) \big)}{By rule force}
                        \Pf{}{}{|\Force{v}| = \Force{|v|} = \Force{v'}}{By definition of $|{-}|$}\trailingjust{and $|v| = v'$}
                        \Hand \Pf{\Gamma} {|-^\ambns}{ e' : \big(C |> (<<\emptyset; X>> \effseq \e) \big) ~\textrm{and}~ |e| = e'}{where $e' = \Force{v'}$ and $e = \Force{v}$}
                \end{llproof}
        
        \DerivationProofCase{\esynget}
                {
                        \Gamma |- v => \Ref{X}{A}
                }
                {
                        \Gamma |-^\ambns
                        \Get{v}
                        =>
                        \big(
                        \F A
                        \big)
                        |>
                        <<\emptyset; X>>
                }
                        
                \begin{llproof}
                        \Pf{%\exists~v' ~\textrm{such that} ~
                          \Gamma} {|-} {v' : \Ref{X}{A} ~\textrm{and}~ |v| = v'}{By \Thmref{thmrefsoundbival}}
                        \Pf{\Gamma} {|-^\ambns}{ \Get{v'} : \big(\F A\big) |> <<\emptyset; X>>}{By rule get}
                        \Pf{}{}{|\Get{v}| = \Get{|v|} = \Get{v'}}{By the definition of $|{-}|$}\trailingjust{and $|v| = v'$}
                        \Hand \Pf{\Gamma} {|-^\ambns}{ e' : \big(\F A\big) |> <<\emptyset; X>> ~\textrm{and}~ |e| = e'}{where $e' = \Get{v'}$ and $e = \Get{v}$}
                \end{llproof}
                
        \DerivationProofCase{\esynnameapp}
                {
                        \arrayenvbl{
                                \Gamma |- v_M => (\namesort @> \namesort)[[M]]
                                \\
                                \Gamma |- v => \Name{i}
                        }
                }
                {
                        \Gamma |-^N (v_M\;v) => \F (\Name{\idxapp{M}{i}}) |> <<\emptyset; \emptyset>>
                }

                \begin{llproof}
                        \Pf{%\exists~v_M' ~\textrm{such that} ~
                          \Gamma} {|-} {v_M' : (\namesort @> \namesort)[[M]] ~\textrm{and}~ |v_M| = v_M'}{By \Thmref{thmrefsoundbival}}
                        \Pf{%\exists~v' ~\textrm{such that}~ 
                          \Gamma} {|-} {v' : \Name{i} ~\textrm{and}~ |v| = v'}{By \Thmref{thmrefsoundbival}}
                        \Pf{\Gamma} {|-}{^N (v_M'\;v') : \F (\Name{\idxapp{M}{i}}) |> <<\emptyset; \emptyset>>}{By rule name-app}
                        \Pf{|(v_M\;v)|}{=} {(|v_M|\;|v|) = (v_M'\;v')}{By definition of $|{-}|$; $|v_M| = v_M'$; $|v| = v'$}
                        \Hand\hspace*{-4ex} \Pf{\Gamma} {|-^\ambns}{ e' : \F (\Name{\idxapp{M}{i}}) |> <<\emptyset; \emptyset>> ~\textrm{and}~ |e| = e'}{where $e' = (v_M'\;v')$ and $e = (v_M\;v)$}
                \end{llproof}
                        
        \DerivationProofCase{\esynAllIndexElim}
                {
                        \Gamma |-^\ambns e => (\All{a : \sort} E)
                        \\
                        \Gamma |- i : \sort
                }
                {
                        \Gamma |-^\ambns
                        \inst{e}{i}
                        => [i/a]E
                }
                
                \begin{llproof}
                        \Pf{%\exists~ e' ~\textrm{such that}~ 
                          \Gamma} {|-^\ambns}{ e' : (\All{a : \sort} E) ~\textrm{and}~ |e| = e' }{By inductive hypothesis}
                        \Pf{\Gamma} {|-} {i : \sort}{Given}
                        \Hand \eqPf{\inst{e}{i}}{|e'|}  {By $|e| = e'$}
                        \Hand \Pf{\Gamma} {|-^\ambns}{ e' : [i/a]E}{By rule AllIndexElim}
                \end{llproof}
                        
        \DerivationProofCase{\esynAllElim}
                {
                        \Gamma |-^\ambns e => (\All{\alpha : K} E)
                        \\
                        \Gamma |- A : K
                }
                {
                        \Gamma |-^\ambns
                        \inst{e}{A}
                        => [A/\alpha]E
                }
                
                               Similar to the syn-AllIndexElim case.

        \DerivationProofCase{\esynanno}
                {
                        \Gamma |-^\ambns e_1 <= E
                }
                {
                        \Gamma |-^\ambns (e_1 : E) => E
                }
        
                \begin{llproof}
                        \Pf{%\exists~ e_1' ~\textrm{such that}~
                          \Gamma} {|-^\ambns}{ e_1' : E  ~\textrm{and}~ |e_1| = e_1'}{By inductive hypothesis}
                        \Pf{}{}{|e_1:E| = |e_1| = e_1'}{By the definition of $|{-}|$}\trailingjust{and $|e_1| = e_1'$}
                        \Hand \Pf{\Gamma} {|-^\ambns}{ e' : E ~\textrm{and}~ |e| = e'}{where $e' = e_1'$ and $e= (e_1 : E)$}
                \end{llproof}
          \end{itemize}

  {\noindent Part (2):} Proceed by case analysis on the rule concluding
  $\Gamma |-^\ambns e <= E$.

          \begin{itemize}       
        \DerivationProofCase{\echksub}
                {
                        \Gamma |-^\ambns e => (C |> \e_1)
                        \\
                        \e_1 \effsub \e_2
                }
                {
                        \Gamma |-^\ambns e <= (C |> \e_2)
                }
        
                By i.h.\ and rule \etypeSub.
%               \begin{llproof}
%                       \Hand \Pf{%\exists~ e' ~\textrm{such that}~
%                           \Gamma} {|-^\ambns}{ e' : (C |> \e_1) ~\textrm{and}~ |e| = e'}{By inductive hypothesis}
%                       \Pf{}{}{\e_1 \effsub \e_2}{Given}
%                       \Hand \Pf{\Gamma} {|-^\ambns}{ e' : (C |> \e_2)}{By rule eff-subsume}
%               \end{llproof}
        
        \DerivationProofCase{\echksplit}
                {
                        \arrayenvbl{
                                \Gamma |-%^\ambns
                                v => (A_1 ** A_2)
                                \\
                                \Gamma, x_1:A_1, x_2:A_2 |-^\ambns e_1 <= E
                        }
                }
                {
                        \Gamma |-^\ambns \Split{v}{x_1}{x_2}{e_1} <= E
                }
        
                \begin{llproof}
                        \Pf{%\exists~ v' ~\textrm{such that}~ 
                          \Gamma} {|-} {v' : (A_1 ** A_2) ~\textrm{and}~ |v| = v'}{By \Thmref{thmrefsoundbival}}
                        \Pf{%\exists~ e_1' ~\textrm{such that}~
                          \Gamma, x_1:A_1, x_2:A_2} {|-^\ambns}{ e_1' : E ~\textrm{and}~ |e_1| = e_1'}{By inductive hypothesis}
                        \Pf{\Gamma} {|-^\ambns}{ \Split{v}{x_1}{x_2}{e_1'} : E}{By rule split}
                        \eqPf{|\Split{v}{x_1}{x_2}{e_1}|}{ \Split{|v|}{x_1}{x_2}{|e_1|}}  {}
                        \decolumnizePf
                                                      \continueeqPf{ \Split{v'}{x_1}{x_2}{e_1'}}{By definition of $|{-}|$}\trailingjust{and $|v| = v', |e_1| = e_1'$}
                        \Hand \Pf{\Gamma} {|-^\ambns}{ e' : E ~\textrm{and}~ |e| = e'}{where $e' = \Split{v'}{x_1}{x_2}{e_1'}$}
                        \Pf{}{}{}{\hspace{8pt} and $e= \Split{v}{x_1}{x_2}{e_1}$}
                \end{llproof}
                
        \DerivationProofCase{\echkcase}
                {
                        \Gamma |-%^\ambns
                        v => (A_1 + A_2)
                        \\
                        \arrayenvbl{
                                \Gamma, x_1:A_1 |-^\ambns e_1 <= E
                                \\
                                \Gamma, x_2:A_2 |-^\ambns e_2 <= E
                        }
                }
                {
                        \Gamma |-^\ambns \Case{v}{x_1}{e_1}{x_2}{e_2} <= E
                }
        
                \begin{llproof}
                        \Pf{%\exists~ v' ~\textrm{such that}~
                          \Gamma} {|-} {v' : (A_1 + A_2) ~\textrm{and}~ |v| = v'}{By \Thmref{thmrefsoundbival}}
                        \Pf{%\exists~ e_1' ~\textrm{such that}~
                          \Gamma, x_1:A_1} {|-^\ambns}{ e_1' : E ~\textrm{and}~ |e_1| = e_1'}{By inductive hypothesis}
                        \Pf{%\exists~ e_2' ~\textrm{such that}~ 
                          \Gamma, x_2:A_2} {|-^\ambns}{ e_2' : E ~\textrm{and}~ |e_2| = e_2'}{By inductive hypothesis}
                        \Pf{\Gamma} {|-^\ambns}{ \Case{v'}{x_1}{e_1'}{x_2}{e_2'} : E}{By rule case}
                        \eqPf{|\Case{v}{x_1}{e_1}{x_2}{e_2}|}{\Case{|v|}{x_1}{|e_1|}{x_2}{|e_2|}}{By definition of $|{-}|$}
                        \continueeqPf{\Case{v'}{x_1}{e_1'}{x_2}{e_2'}}{Since $|v| = v', |e_1| = e_1', |e_2| = e_2'$}
                        \decolumnizePf
                        \Hand \Pf{\Gamma} {|-^\ambns}{ e' : E ~\textrm{and}~ |e| = e'}{where $e' = \Case{v'}{x_1}{e_1'}{x_2}{e_2'}$}
                        \Pf{}{}{}{\hspace{8pt} and $e= \Case{v}{x_1}{e_1}{x_2}{e_2}$}
                \end{llproof}
                
        \DerivationProofCase{\echkret}
                {
                        \Gamma |- v <= A
                }
                {
                        \Gamma |-^\ambns
                        \Ret{v}
                        <=
                        \big(
                        (\F A)  |>  <<\emptyset; \emptyset>>
                        \big)
                }
                
                \begin{llproof}
                        \Pf{%\exists~ v' ~\textrm{such that}~
                          \Gamma} {|-} {v' : A ~\textrm{and}~ |v| = v'}{By \Thmref{thmrefsoundbival}}
                        \Pf{\Gamma} {|-^\ambns}{ \Ret{v'} : \big((\F A)  |>  <<\emptyset; \emptyset>>   \big)}{By rule ret}
                        \Pf{}{}{|\Ret{v}| = \Ret{|v|} = \Ret{v'}}{By definition of $|{-}|$}\trailingjust{and $|v| = v'$}
                        \Hand \Pf{\Gamma} {|-^\ambns}{ e' : \big((\F A)  |>  <<\emptyset; \emptyset>> \big) ~\textrm{and}~ |e| = e'}{where $e' = \Ret{v'}$ and $e = \Ret{v}$}
                \end{llproof}
                
        \DerivationProofCase{\echklet}
                {
                        \Gamma |-^M e_1 => (\F A) |> \e_1
                        \\
                        \Gamma, x:A |-^M e_2 <= (C |> \e_2)
                }
                {
                        \Gamma |-^M \Let{e_1}{x}{e_2}
                        <=
                        \big(
                        C |> (\e_1 \effseq \e_2)
                        \big)
                }
                
                \begin{llproof}
                        \Pf{%\exists~ e_1' ~\textrm{such that}~
                          \Gamma} {|-}{^M e_1' : (\F A) |> \e_1 ~\textrm{and}~ |e_1| = e_1'}{By inductive hypothesis}
                        \Pf{%\exists~ e_2' ~\textrm{such that}~
                          \Gamma, x:A} {|-}{^M e_2' : (C |> \e_2) ~\textrm{and}~ |e_2| = e_2'}{By inductive hypothesis}
                        \Pf{\Gamma} {|-}{^M \Let{e_1'}{x}{e_2'} : \big(C |> (\e_1 \effseq \e_2)\big)}{By rule let}
                        \eqPf{|\Let{e_1}{x}{e_2}|}{\Let{|e_1|}{x}{|e_2|}}{By definition of $|{-}|$}
                        \continueeqPf{\Let{e_1'}{x}{e_2'}}{Since $|e_1| = e_1', |e_2| = e_2'$}
                        \decolumnizePf
                        \Hand \Pf{\Gamma} {|-^\ambns}{ e' : \big(C |> (\e_1 \effseq \e_2)\big) ~\textrm{and}~ |e| = e'}{where $e' = \Let{e_1'}{x}{e_2'}$} \trailingjust{and $e = \Let{e_1}{x}{e_2}$}
                \end{llproof}
        
        \DerivationProofCase{\echklam}
                {
                        \Gamma,x:A |-^\ambns e_1 <= E
                }
                {
                        \Gamma |-^\ambns
                        (\lam{x} e_1)
                        <=
                        \big(
                        (A -> E) |> <<\emptyset; \emptyset>>
                        \big)
                }

                \begin{llproof}
                        \Pf{%\exists~ e_1' ~\textrm{such that}~
                          \Gamma,x:A} {|-^\ambns}{ e_1' : E ~\textrm{and}~ |e_1| = e_1'}{By inductive hypothesis}
                        \Pf{\Gamma} {|-^\ambns}{ (\lam{x} e_1') : \big((A -> E) |> <<\emptyset; \emptyset>> \big)}{By rule lam}
                        \Pf{}{}{|(\lam{x} e_1)| = (\lam{x} |e_1|) = (\lam{x} e_1')}{By definition of $|{-}|$}\trailingjust{and $|e_1| = e_1'$}
                        \decolumnizePf
                        \Hand \Pf{\Gamma} {|-^\ambns}{ e' : \big((A -> E) |> <<\emptyset; \emptyset>> \big) ~\textrm{and}~ |e| = e'}{where $e' = (\lam{x} e_1')$ and $e = (\lam{x} e_1)$}
                \end{llproof}
                
        \DerivationProofCase{\echkthunk}
                {
                        \Gamma |- v <= \Name{X}
                        \\
                        \Gamma |-^\ambns e_1 <= E_1
                }
                {
                        \Gamma |-^\ambns
                        \Thunk{v}{e_1}
                        <=
                        \big(
                        \F (\Thk{M(X)}{E_1})
                        \big)
                        |>
                        <<M(X); \emptyset>>
                }
                
                \begin{llproof}
                        \Pf{\text{Let~}E = }{}{\big(\F (\Thk{M(X)}{E_1}) \big) |> <<M(X); \emptyset>>}{Assumption}
                        \Pf{\exists~ v' ~\textrm{such that}~ \Gamma} {|-} {v' : \Name{X} ~\textrm{and}~ |v| = v'}{By \Thmref{thmrefsoundbival}}
                        \Pf{\exists~ e_1' ~\textrm{such that}~ \Gamma} {|-^\ambns}{ e_1' : E ~\textrm{and}~ |e_1| = e_1'}{By inductive hypothesis}
                        \Pf{\Gamma} {|-^\ambns}{ \Thunk{v'}{e_1'} : E}{By rule thunk}
                        \Pf{}{|\Thunk{v}{e_1}| =}{\Thunk{|v|}{|e_1|} = \Thunk{v'}{e_1'}}{By definition of $|{-}|$}\trailingjust{and $|v| = v'$, $|e_1| = e_1'$}
\decolumnizePf
                        \Hand \Pf{\Gamma} {|-^\ambns}{ e' : E  ~\textrm{and}~ |e| = e'}{where $e' = \Thunk{v'}{e_1'}$ and $e = \Thunk{v}{e_1}$}
                \end{llproof}

        \DerivationProofCase{\echkref}
                {
                        \Gamma |- v_1 <= \Name{X}
                        \\
                        \Gamma |- v_2 <= A
                }
                {
                        \Gamma |-^\ambns
                        \Refe{v_1}{v_2}
                        <=
                        \big(
                        \F (\Ref{M(X)}{A})
                        \big)
                        |>
                        %          <<\{M(X)\}; \emptyset>>
                        <<M(X); \emptyset>>
                }

                \begin{llproof}
                        \Pf{\exists~ v_1' ~\textrm{such that}~ \Gamma} {|-} {v_1' : \Name{X} ~\textrm{and}~ |v_1| = v_1'}{By \Thmref{thmrefsoundbival}}
                        \Pf{\exists~ v_2' ~\textrm{such that}~ \Gamma} {|-} {v_2' : A ~\textrm{and}~ |v_2| = v_2'}{By \Thmref{thmrefsoundbival}}
                        \Pf{\Gamma} {|-^\ambns}{ \Refe{v_1'}{v_2'} : \big( \F (\Ref{M(X)}{A})   \big) |> <<M(X); \emptyset>>}{By rule ref}
                        \Pf{}{|\Refe{v_1}{v_2}| =}{\Refe{|v_1|}{|v_2|} = \Refe{v_1'}{v_2'}}{By definition of $|{-}|$} %{; $|v_1| = v_1'$; $|v_2| = v_2'$}
                        \decolumnizePf
                        \Hand \Pf{\Gamma} {|-^\ambns} { e' :  \big( \F (\Ref{M(X)}{A})  \big) |> <<M(X); \emptyset>>} {}
                        \Pf{}{}{~\textrm{and}~ |e| = e'\text{~where~}e' = \Refe{v_1'}{v_2'}\text{~and~}e = \Refe{v_1}{v_2}} {}
                \end{llproof}
                
                \DerivationProofCase{\echkscope}
                {
                  \Gamma |- v => (\namesort @> \namesort)[[N']]
                  \\
                  \Gamma |-^{N \circ N'} e_1 <= C |> <<W; R>>
                }
                {
                  \Gamma |-^N \Scope{v}{e_1} <= C |> <<W; R>>
                }
        
                \begin{llproof}
                        \Pf{%\exists~ v' ~\textrm{such that}~ 
                          \Gamma} {|-} {v' : \namesort @> \namesort  ~\textrm{and}~ |N'| = v'}{By inductive hypothesis}
                        \Pf{%\exists~ e_1' ~\textrm{such that}~
                          \Gamma} {|-^{N \circ N'}} {e_1' : C |> <<W; R>>  ~\textrm{and}~ |e_1| = e_1'}{By inductive hypothesis}
                        \Pf{\Gamma} {|-^N} {\Scope{v'}{e_1'} : C |> <<W; R>>}{By rule scope}
                        \Pf{}{|\Scope{v}{e_1}| =}{ \Scope{|v|}{|e_1|} = \Scope{v'}{e_1'}}{By definition of $|{-}|$; $|v| = v'$; $|e_1'| = e_1$}
                        \decolumnizePf
                        \Hand \Pf{\Gamma} {|-^\ambns}{ e' : C |> <<W; R>> ~\textrm{and}~ |e| = e'}{where $e' = \Scope{v'}{e_1'}$ and $e = \Scope{v}{e_1}$}
                \end{llproof}
        
        \DerivationProofCase{\echkAllIndexIntro}
                {
                        \Gamma, a:\sort |-^\ambns t <= E
                }
                {
                        \Gamma |-^\ambns t <= (\All{a : \sort} E)
                }
                
                \begin{llproof}
                        \Pf{%\exists~ t' ~\textrm{such that}~ 
                          \Gamma, a:\sort}{|-^\ambns}{ t' : E  ~\textrm{and}~ |t| = t'}{By inductive hypothesis}
                        \Pf{\Gamma} {|-^\ambns}{ t' : (\All{a : \sort} E)}{By rule AllIndexIntro}
                        \Hand \Pf{\Gamma} {|-^\ambns}{ e' : (\All{a : \sort} E) ~\textrm{and}~ |e| = e'}{where $e' = t'$ and $e = t$}
                \end{llproof}
        
        \DerivationProofCase{\echkAllIntro}
                {
                        \Gamma, a:\sort |-^\ambns t <= E
                }
                {
                        \Gamma |-^\ambns
                        t
                        <= (\All{\alpha : K} E)
                }
        
                \begin{llproof}
                        \Pf{\exists~ t' ~\textrm{such that}~ \Gamma, a:\sort} {|-^\ambns}{ t' : E  ~\textrm{and}~ |t| = t'}{By inductive hypothesis}
                        \Pf{\Gamma} {|-^\ambns}{ t': (\All{\alpha : K} E)}{By rule AllIntro}
                        \Hand \Pf{\Gamma} {|-^\ambns}{ e' : (\All{\alpha : K} E) ~\textrm{and}~ |e| = e'}{where $e' = t'$ and $e = t$}
                \end{llproof}   

        \DerivationProofCase{\echksub}
                {
                        \Gamma |-^\ambns e => E_1 
                        \\
                        E_1 = E_2
                }
                { 
                        \Gamma |-^\ambns e <= E_2 
                }

                                   By i.h.\ and \etypeSub.
  \qedhere
  \end{itemize}
%               \begin{llproof}
%                       \Hand \Pf{\exists~ e' ~\textrm{such that}~ \Gamma} {|-^\ambns}{ e' : E_1 ~\textrm{and}~ |e| = e'}{By inductive hypothesis}
%                       \Pf{}{}{E_1 = E_2}{Given}
%                       \Pf{\Gamma} {|-^\ambns}{ e' : E_2}{Since $E_1 = E_2$}
%               \end{llproof}
\end{proof}

\begin{restatable}[Completeness of Bidirectional Computation Typing]{theorem}{thmrefcomplbicomp}
        \Label{thmrefcomplbicomp}
        ~\\
        If
        $       \Gamma  |-^\ambns e : E $, 
        then there exist computations
        $       e'      $, $ e''        $ such that
        \begin{enumerate}
                \item 
                $       \Gamma  |-^\ambns e' => E $
                and $   |e'| = e        $
                
                \item 
                $       \Gamma  |-^\ambns e'' <= E $    and $   |e''| = e       $
        \end{enumerate} 
\end{restatable}

\begin{proof}
        By induction on the derivation of $ \Gamma      |-^\ambns e : E $.

        \begin{itemize}
        \ProofCaseRule{\etypeSub}   By i.h.\ and \echksub.
%               {
%                       \Gamma |-^\ambns e : (C |> \e_1)
%                       \\
%                       \e_1 \effsub \e_2
%               }
%               {
%                       \Gamma |-^\ambns e : (C |> \e_2)
%               }
%               
%               \begin{llproof}
%                       \Pf{%\exists~e' ~\textrm{such that}~ 
%                           \Gamma}{|-^\ambns}{ e' => (C |> \e_1) ~\textrm{and}~ e = |e'|}{By inductive hypothesis}
%                       \Pf{}{}{\e_1 \effsub \e_2}{Given}
%                       \Pf{\Gamma}{|-^\ambns}{ e' <= (C |> \e_2)}{By chk-eff-subsume}
%                       \Pf{\Gamma}{|-^\ambns}{ (e': (C |> \e_2)) => (C |> \e_2)}{By syn-anno}
%                       \Pf{}{|(e': (C |> \e_2))| =}{|e'| = e}{By definition of $|{-}|$}%}\trailingjust{and $|e'| = e$}
%                       \Hand \Pf{\Gamma}{|-^\ambns}{ e' => (C |> \e_2) ~\textrm{and}~ |e'| = e}{}
%                       \Hand \Pf{\Gamma}{|-^\ambns}{ e'' <= (C |> \e_2) ~\textrm{and}~ |e''| = e}{where $e'' = (e': (C |> \e_2))$}
%               \end{llproof}
        
        \DerivationProofCase{split}
                {
                        \arrayenvbl{
                                \Gamma |-%^\ambns
                                v : (A_1 ** A_2)
                                \\
                                \Gamma, x_1:A_1, x_2:A_2 |-^\ambns e_1 : E
                        }
                }
                {
                        \Gamma |-^\ambns \Split{v}{x_1}{x_2}{e_1} : E
                }
                
                \begin{llproof}
                        \Pf{%\exists~e_1' ~\textrm{such that}~
                          \Gamma, x_1:A_1, x_2:A_2}{|-}{e_1' <= E ~\textrm{and}~ e_1 = |e_1'|}{By inductive hypothesis}
                        \Pf{%\exists~v' ~\textrm{such that}~
                          \Gamma}{|-}{v' => (A_1 ** A_2) ~\textrm{and}~ v_1 = |v_1'|}{By \Thmref{thmrefcomplbival}}
                        \Pf{\Gamma}{|-^\ambns}{ \Split{v'}{x_1}{x_2}{e_1'} <= E}{By chk-split}
                        \Pf{\Gamma}{|-^\ambns}{ (\Split{v'}{x_1}{x_2}{e_1'} : E) => E}{By syn-anno}
                        \decolumnizePf
                        \Pf{}{|(\Split{v'}{x_1}{x_2}{e_1'} : E)| =}{ |\Split{v'}{x_1}{x_2}{e_1'}|}{By definition of $|{-}|$}
                        \Pf{}{|\Split{v'}{x_1}{x_2}{e_1'}| =}{\Split{|v'|}{x_1}{x_2}{|e_1'|}}{By definition of $|{-}|$}
                        \Pf{}{\Split{|v'|}{x_1}{x_2}{|e_1'|} = }{\Split{v}{x_1}{x_2}{e_1}}{Since $|v'| = v$, $|e_1'| = e_1$}
                        \decolumnizePf
                        \Hand \Pf{}{}{\Gamma |- e' => E ~\textrm{and}~ |e'| = e}{where $e' = \Split{v'}{x_1}{x_2}{e_1'}$}
                        \Pf{}{}{}{\hspace{8pt} and $e = \Split{v}{x_1}{x_2}{e_1}$}
                        \Hand \Pf{}{}{\Gamma |- e'' <= E ~\textrm{and}~ |e''| = e}{where $e'' = (\Split{v'}{x_1}{x_2}{e_1'} : E)$}
                        \Pf{}{}{}{\hspace{8pt} and $e = \Split{v}{x_1}{x_2}{e_1}$}
                \end{llproof}
        
        \DerivationProofCase{case}
                {
                        \Gamma |-
                        v : (A_1 + A_2)
                        \\
                        \arrayenvbl{
                                \Gamma, x_1:A_1 |-^\ambns e_1 : E
                                \\
                                \Gamma, x_2:A_2 |-^\ambns e_2 : E
                        }
                }
                {
                        \Gamma |-^\ambns \Case{v}{x_1}{e_1}{x_2}{e_2} : E
                }
                
                \begin{llproof}
                        \Pf{%\exists~ v' ~\textrm{such that}~
                          \Gamma}{ |- } {v' => (A_1 + A_2) ~\textrm{and}~ |v'| = v}{By \Thmref{thmrefcomplbival}}       
                        \Pf{%\exists~ e_1'' ~\textrm{such that}~
                          \Gamma, x_1:A_1}{|-^\ambns}{ e_1'' <= E ~\textrm{and}~ |e_1''| = e_1}{By inductive hypothesis}
                        \Pf{%\exists~ e_2'' ~\textrm{such that}~
                          \Gamma, x_2:A_2}{|-^\ambns}{ e_2'' <= E ~\textrm{and}~ |e_2''| = |e_2|}{By inductive hypothesis}
                        \Pf{\Gamma}{|-^\ambns} { \Case{v'}{x_1}{e_1''}{x_2}{e_2''} <= E}{By rule chk-case}
                        \Pf{\Gamma}{|-^\ambns} { (\Case{v'}{x_1}{e_1''}{x_2}{e_2''} : E) => E}{By rule chk-conv}
                        \decolumnizePf
                        \Pf{}{|(\Case{v'}{x_1}{e_1''}{x_2}{e_2''} : E)| = }{|\Case{v'}{x_1}{e_1''}{x_2}{e_2''}|}{By definition of $|{-}|$}
                        \Pf{|\Case{v'}{x_1}{e_1''}{x_2}{e_2''}|}{ = }{\Case{|v'|}{x_1}{|e_1''|}{x_2}{|e_2''|}}{By definition of $|{-}|$}
                        \Pf{}{\Case{|v'|}{x_1}{|e_1''|}{x_2}{|e_2''|} =}{ \Case{v}{x_1}{e_1}{x_2}{e_2}}{}%{Since $|v'| = v$, $|e_1''| = e_1$, $|e_2''| = e_2$}
\decolumnizePf
                        \Hand \Pf{\Gamma}{|-^\ambns}{ e' => E ~\textrm{and}~ |e'| = e}{where $e' = (\Case{v'}{x_1}{e_1''}{x_2}{e_2''} : E)$}
                        \Pf{}{}{}{\hspace{8pt} and $e = \Case{v}{x_1}{e_1}{x_2}{e_2}$}
                        \Hand \Pf{\Gamma}{|-^\ambns}{ e'' <=  E ~\textrm{and}~ |e''| = e}{where $e'' = \Case{v'}{x_1}{e_1''}{x_2}{e_2''}$}
                        \Pf{}{}{}{\hspace{8pt} and $e = \Case{v}{x_1}{e_1}{x_2}{e_2}$}
                \end{llproof}
        
        \DerivationProofCase{ret}
                {
                        \Gamma |- v : A
                }
                {
                        \Gamma |-^\ambns
                        \Ret{v}
                        :
                        \big(
                        (\F A)  |>  <<\emptyset; \emptyset>>
                        \big)
                }
        
                \begin{llproof}
                        \Pf{\text{Let~}E}{=}{\big( (\F A)  |>  <<\emptyset; \emptyset>> \big)}{Assumption}
                        \Pf{\exists~ v'' ~\textrm{such that}~ \Gamma}{|-}{v'' <= A ~\textrm{and}~ |v''| = v}{By \Thmref{thmrefcomplbival}}
                        \Pf{\Gamma}{|-^\ambns}{ \Ret{v''} <= E}{By rule chk-ret}
                        \Pf{\Gamma}{|-^\ambns}{ (\Ret{v''} : E) => E}{By syn-anno}
                        \Pf{}{|(\Ret{v''} : E)| =}{|\Ret{v''}|}{By definition of $|{-}|$}
                        \Pf{}{|\Ret{v''}| =}{\Ret{|v''|} = \Ret{v}}{By definition of $|{-}|$}\trailingjust{and $|v''| = v$}
                        \Hand \Pf{\Gamma}{|-^\ambns}{ e' => E ~\textrm{and}~ |e'| = e}{where $e' = (\Ret{v''} : E)$ and $e = \Ret{v}$}
                        \Hand \Pf{\Gamma}{|-^\ambns}{ e'' <=  E ~\textrm{and}~ |e''| = e}{where $e'' = \Ret{v''}$ and $e = \Ret{v}$}
                \end{llproof}
                
        \DerivationProofCase{let}
                {
                        \Gamma |-^M e_1 : (\F A) |> \e_1
                        \\
                        \Gamma, x:A |-^M e_2 : (C |> \e_2)
                }
                {
                        \Gamma |-^M \Let{e_1}{x}{e_2}
                        :
                        \big(
                        C |> (\e_1 \effseq \e_2)
                        \big)
                }
        
                \begin{llproof}
                        \Pf{\text{Let~}E}{=}{\big(C |> (\e_1 \effseq \e_2)\big)}{Assumption}
                        \Pf{%\exists~ e_1' ~\textrm{such that}~
                          \Gamma}{|-}{ ^M e_1' => (\F A) |> \e_1 ~\textrm{and}~ |e_1'| = e_1}{By inductive hypothesis}
                        \Pf{%\exists~ e_2'' ~\textrm{such that}~
                          \Gamma, x:A}{|-}{^M e_2'' <= (C |> \e_2) ~\textrm{and}~ |e_2''| = e_2}{By inductive hypothesis}
                        \Pf{\Gamma}{|-}{^M \Let{e_1'}{x}{e_2''} <= E}{By rule chk-let}
                        \Pf{\Gamma}{|-}{^M (\Let{e_1'}{x}{e_2''} : E) => E}{By rule chk-conv}
                        \Pf{}{|(\Let{e_1'}{x}{e_2''} : E)| =}{|\Let{e_1'}{x}{e_2''}|}{By definition of $|{-}|$}
                        \Pf{}{|\Let{e_1'}{x}{e_2''}| =}{\Let{|e_1'|}{x}{|e_2''|} = \Let{e_1}{x}{e_2}}{By definition of $|{-}|$}\trailingjust{and $|e_1'| = e_1$, $|e_2''| = e_2$}
                        \decolumnizePf
                        \Hand \Pf{\Gamma}{|-^\ambns}{ e' => E ~\textrm{and}~ |e'| = e}{where $e' = (\Let{e_1'}{x}{e_2''} : E)$ and $e = \Let{e_1}{x}{e_2}$}
                        \Hand \Pf{\Gamma}{|-^\ambns}{ e'' <=  E ~\textrm{and}~ |e''| = e}{where $e'' = \Let{e_1'}{x}{e_2''}$ and $e = \Let{e_1}{x}{e_2}$}
                \end{llproof}
        
        \DerivationProofCase{lam}
                {
                        \Gamma,x:A |-^\ambns e_1 : E_1
                }
                {
                        \Gamma |-^\ambns
                        (\lam{x} e_1)
                        :
                        \big(
                        (A -> E_1) |> <<\emptyset; \emptyset>>
                        \big)
                }
                
                \begin{llproof}
                        \Pf{\text{Let~}E}{=}{\big((A -> E_1) |> <<\emptyset; \emptyset>>\big)}{Assumption}
                        \Pf{\exists~ e_1'' ~\textrm{such that}~ \Gamma,x:A}{|-^\ambns}{ e_1'' <= E_1 ~\textrm{and}~ |e_1''| = e_1}{By inductive hypothesis}
                        \Pf{\Gamma}{|-^\ambns}{ (\lam{x} e_1'') <= E}{By rule chk-lam}
                        \Pf{\Gamma}{|-^\ambns}{ ((\lam{x} e_1'') : E) => E}{By syn-anno}
                        \Pf{}{|(\lam{x} e_1'')| =}{(\lam{x} |e_1''|) = (\lam{x} e_1)}{By definition of $|{-}|$}\trailingjust{and $|e_1''| = e_1$}
\decolumnizePf
                        \Hand \Pf{\Gamma}{|-^\ambns}{ e' => E ~\textrm{and}~ |e'| = e}{where $e' = ((\lam{x} e_1'') : E)$ and $e = (\lam{x} e_1)$}
                        \Hand \Pf{\Gamma}{|-^\ambns}{ e'' <=  E ~\textrm{and}~ |e''| = e}{where $e'' = (\lam{x} e_1'')$ and $e = (\lam{x} e_1)$}
                \end{llproof}
        
        \DerivationProofCase{app}
                {
                        \Gamma |-^\ambns e_1
                        : \big(
                        (A -> E) |> \e_1
                        \big)
                        \\
                        \Gamma |- v : A
                }
                {
                        \Gamma |-^\ambns (e_1\;v) : (E \effcoal \e_1)
                }
        
                \begin{llproof}
                        \Pf{\exists~ e_1' ~\textrm{such that}~ \Gamma}{|-^\ambns}{ e_1' => \big((A -> E) |> \e_1 \big) ~\textrm{and}~ |e_1'| = e_1}{By inductive hypothesis}
                        \Pf{\exists~ v'' ~\textrm{such that}~ \Gamma}{|-}{v'' <= A ~\textrm{and}~ |v''| = v}{By \Thmref{thmrefcomplbival}}
                        \Pf{\Gamma}{|-^\ambns}{ (e_1'\;v'') => (E \effcoal \e_1)}{By rule syn-app}
                        \Pf{\Gamma}{|-^\ambns}{ (e_1'\;v'') <= (E \effcoal \e_1)}{By rule chk-conv}
                        \Pf{}{|(e_1'\;v'')| = }{(|e_1'|\;|v''|) = (e_1\;v)}{By the definition of $|{-}|$}\trailingjust{and $|e_1'| = e_1$, $|v''| = v$}
\decolumnizePf
                        \Hand \Pf{\Gamma}{|-^\ambns}{ e' => E ~\textrm{and}~ |e'| = e}{where $e' = (e_1'\;v'')$ and $e = (e_1\;v)$}
                        \Hand \Pf{\Gamma}{|-^\ambns}{ e'' <=  E ~\textrm{and}~ |e''| = e}{where $e'' = (e_1'\;v'')$ and $e = (e_1\;v)$}
                \end{llproof}
                
        \DerivationProofCase{thunk}
                {
                        \Gamma |- v : \Name{X}
                        \\
                        \Gamma |-^\ambns e_1 : E
                }
                {
                        \Gamma |-^\ambns
                        \Thunk{v}{e_1}
                        :
                        \big(
                        \F (\Thk{M(X)}{E})
                        \big)
                        |>
                        <<M(X); \emptyset>>
                }

                \begin{llproof}
                        \Pf{\text{Let~}E}{=}{\big(\F (\Thk{M(X)}{E})    \big) |> <<M(X); \emptyset>>}{Assumption}
                        \Pf{\exists~ v'' ~\textrm{such that}~ \Gamma}{|-}{v'' <= \Name{X} ~\textrm{and}~ |v''| = v}{By \Thmref{thmrefcomplbival}}
                        \Pf{\exists~ e_1'' ~\textrm{such that}~ \Gamma}{|-^\ambns}{ e_1'' <= E ~\textrm{and}~ |e_1''| = e_1}{By inductive hypothesis}
                        \Pf{\Gamma}{|-^\ambns}{ \Thunk{v''}{e_1''} <= E}{By rule chk-thunk}
                        \Pf{\Gamma}{|-^\ambns}{ (\Thunk{v''}{e_1''} : E) => E}{By rule syn-anno}
                        \Pf{}{|(\Thunk{v''}{e_1''} : E)| = }{|\Thunk{v''}{e_1''}|}{By definition of $|{-}|$}
                        \Pf{|\Thunk{v''}{e_1''}|} {=} {\Thunk{|v''|}{|e_1''|} = \Thunk{v}{e_1}}{By definition of $|{-}|$}\trailingjust{and $|v''| = v$, $|e_1''| = e_1$}
                        \decolumnizePf
                        \Hand \Pf{\Gamma}{|-^\ambns}{ e' => E ~\textrm{and}~ |e'| = e}{where $e' = (\Thunk{v''}{e_1''} : E)$ and $e = \Thunk{v}{e_1}$}
                        \Hand \Pf{\Gamma}{|-^\ambns}{ e'' <=  E ~\textrm{and}~ |e''| = e}{where $e'' = \Thunk{v''}{e_1''}$ and $e = \Thunk{v}{e_1}$}
                \end{llproof}
        
        \DerivationProofCase{force}
                {
                        %          \Gamma |- v : \Thk{n}{E}
                        \Gamma |- v : \Thk{X}{(C |> \e)}
                        %          \Gamma |- v : \Unthk{X}{ \big(C |> <<\emptyset;\emptyset>>\big) }
                }
                {
                        \Gamma |-^\ambns
                        \Force{v}
                        :
                        \big(
                        C |> (<<\emptyset; X>> \effseq \e)
                        \big)
                }
                
            \begin{llproof}
                \Pf{\text{Let~}E}{=}{\big(C |> (<<\emptyset; X>> \effseq \e)
                        \big)}{Assumption}
                \Pf{%\exists~ v' ~\textrm{such that}~ 
                  \Gamma}{|-}{v' => \Thk{X}{(C |> \e)} ~\textrm{and}~ |v'| = v}{By \Thmref{thmrefcomplbival}}
                \Pf{\Gamma}{|-^\ambns}{ \Force{v'} => E}{By rule syn-force}
                \Pf{\Gamma}{|-^\ambns}{ \Force{v'} <= E}{By chk-conv}
                \Pf{}{|\Force{v'}| =}{\Force{|v'|} = \Force{v}}{By definition of $|{-}|$}\trailingjust{and $|v'| = v$}
                \decolumnizePf
                \Hand \Pf{\Gamma}{|-^\ambns}{ e' => E ~\textrm{and}~ |e'| = e}{where $e' = \Force{v'}$ and $e = \Force{v}$}
                \Hand \Pf{\Gamma}{|-^\ambns}{ e'' <=  E ~\textrm{and}~ |e''| = e}{where $e'' = \Force{v'}$ and $e = \Force{v}$}
                \end{llproof}
        
        \DerivationProofCase{ref}
                {
                        \Gamma |- v_1 : \Name{X}
                        \\
                        \Gamma |- v_2 : A
                }
                {
                        \Gamma |-^\ambns
                        \Refe{v_1}{v_2}
                        :
                        \big(
                        \F (\Ref{M(X)}{A})
                        \big)
                        |>
                        %          <<\{M(X)\}; \emptyset>>
                        <<M(X); \emptyset>>
                }
                
                \begin{llproof}
                        \Pf{\text{Let~}E}{=}{\big( \F (\Ref{M(X)}{A}) \big) |> <<M(X); \emptyset>>}{Assumption}
                        \Pf{%\exists~ v_1'' ~\textrm{such that}~
                          \Gamma} {|-} {v_1'' <= \Name{X} ~\textrm{and}~ |v_1''| = v_1}{By \Thmref{thmrefcomplbival}}
                        \Pf{%\exists~ v_2'' ~\textrm{such that}~
                          \Gamma} {|-} {v_2'' <= A ~\textrm{and}~ |v_2''| = v_2}{By \Thmref{thmrefcomplbival}}
                        \Pf{\Gamma} {|-^\ambns} { \Refe{v_1''}{v_2''} <= E}{By rule chk-ref}
                        \Pf{\Gamma} {|-^\ambns} { (\Refe{v_1''}{v_2''} : E) => E}{By rule syn-anno}
                        \decolumnizePf
                        \Pf{}{|(\Refe{v_1''}{v_2''}:E)| = }{|\Refe{v_1''}{v_2''}|}{By definition of $|{-}|$}
                        \Pf{}{|\Refe{v_1''}{v_2''}| = }{\Refe{|v_1''|}{|v_2''|} = \Refe{v_1}{v_2}}{By definition of $|{-}|$}\trailingjust{and $|v_1''| = v_1$, $|v_2''| = v_2$}
                        \decolumnizePf
                        \Hand \Pf{\Gamma}{|-^\ambns}{ e' => E ~\textrm{and}~ |e'| = e}{where $e' = (\Refe{v_1''}{v_2''} : E)$ and $e = \Refe{v_1}{v_2}$}
                        \Hand \Pf{\Gamma}{|-^\ambns}{ e'' <=  E ~\textrm{and}~ |e''| = e}{where $e'' = \Refe{v_1''}{v_2''}$ and $e = \Refe{v_1}{v_2}$}
                \end{llproof}
        
        \DerivationProofCase{get}
                {
                        \Gamma |- v : \Ref{X}{A}
                }
                {
                        \Gamma |-^\ambns
                        \Get{v}
                        :
                        \big(
                        \F A
                        \big)
                        |>
                        <<\emptyset; X>>
                }
        
                \begin{llproof}
                        \Pf{%\exists~ v' ~\textrm{such that}~ 
                          \Gamma} {|-} {v' => \Ref{X}{A} ~\textrm{and}~ |v'| = v}{By \Thmref{thmrefcomplbival}}
                        \Pf{\Gamma} {|-^\ambns} { \Get{v'} => \big( \F A \big) |> <<\emptyset; X>>}{By rule syn-get}
                        \Pf{\Gamma} {|-^\ambns} { \Get{v'} <= \big( \F A \big) |> <<\emptyset; X>>}{By rule chk-conv}
                        \Pf{}{|\Get{v'}| = }{\Get{|v'|} = \Get{v}}{By definition of $|{-}|$}\trailingjust{and $|v'| = v$}
                        \decolumnizePf
                        \Hand \Pf{\Gamma}{|-^\ambns}{ e' =>  \big( \F A \big) |> <<\emptyset; X>> ~\textrm{and}~ |e'| = e}{where $e' = \Get{v'}$ and $e = \Get{v}$}
                        \Hand \Pf{\Gamma}{|-^\ambns}{ e'' <=  \big( \F A \big) |> <<\emptyset; X>> ~\textrm{and}~ |e''| = e}{where $e'' = \Get{v'}$ and $e = \Get{v}$}
                \end{llproof}   
        
        \DerivationProofCase{name-app}
                {
                        \arrayenvbl{
                                \Gamma |- v_M : (\namesort @> \namesort)[[M]]
                                \\
                                \Gamma |- v : \Name{i}
                        }
                }
                {
                        \Gamma |-^N (v_M\;v) : \F (\Name{\idxapp{M}{i}}) |> <<\emptyset; \emptyset>>
                }
        
                \begin{llproof}
                        \Pf{%\exists~ v_M' ~\textrm{such that}~ 
                          \Gamma} {|-} {v_M' => (\namesort @> \namesort)[[M]] ~\textrm{and}~ |v_M'| = v_M}{By \Thmref{thmrefcomplbival}}
                        \Pf{%\exists~ v' ~\textrm{such that}~ 
                          \Gamma} {|-} {v' => \Name{i} ~\textrm{and}~ |v'| = v}{By \Thmref{thmrefcomplbival}}
                        \Pf{\Gamma} {|-} {^N (v_M'\;v') => \F (\Name{\idxapp{M}{i}}) |> <<\emptyset; \emptyset>>}{By rule syn-name-app}
                        \Pf{\Gamma} {|-} {^N (v_M'\;v') <= \F (\Name{\idxapp{M}{i}}) |> <<\emptyset; \emptyset>>}{By rule chk-conv}     
                        \Pf{}{|(v_M'\;v')| =}{(|v_M'|\;|v'|) = (v_M\;v)}{By definition of $|{-}|$}\trailingjust{and $|v_M'| = v_M$, $|v'| = v$}
\decolumnizePf
                        \Hand \Pf{\Gamma}{|-^N} {e' => \F (\Name{\idxapp{M}{i}}) |> <<\emptyset; \emptyset>> ~\textrm{and}~ |e'| = e}{where $e' = (v_M'\;v')$ and $e = (v_M\;v)$}
                        \Hand \Pf{\Gamma}{|-^N} {e'' <= \F (\Name{\idxapp{M}{i}}) |> <<\emptyset; \emptyset>> ~\textrm{and}~ |e''| = e}{where $e'' = (v_M'\;v')$ and $e = (v_M\;v)$}
                \end{llproof}
                
        \DerivationProofCase{scope}
                {
                        \Gamma |- v : (\namesort @> \namesort)[[N']]
                        \\
                        \Gamma |-^{N \circ N'} e_1 : C |> <<W; R>>
                }
                {
                        \Gamma |-^N \Scope{v}{e_1} : C |> <<W; R>>
                }
                
                \begin{llproof}
                                \Pf{\text{Let~}E}{=}{C |> <<W; R>>}{Assumption}
                        \Pf{\exists v'' ~\textrm{such that}~, \Gamma}{|-}{v'' => (\namesort @> \namesort)[[N']] ~\textrm{and}~ |v''| = v}{By \Thmref{thmrefcomplbival}}
                        \Pf{\exists e_1' ~\textrm{such that}~, \Gamma}{|-^{N \circ N'}}{e_1' <= E ~\textrm{and}~ |e_1'| = e_1}{By inductive hypothesis}
                        \Pf{\Gamma}{|-^N}{\Scope{v''}{e_1'} <= E}{By rule chk-scope}
                        \Pf{\Gamma}{|-^N}{(\Scope{v''}{e_1'} : E) => E}{By rule syn-anno}
                        \decolumnizePf
                        \Pf{}{|(\Scope{v''}{e_1'} : E)|}{= |(\Scope{v''}{e_1'}|}{By definition of $|{-}|$}
                        \Pf{}{|(\Scope{v''}{e_1'}| = \Scope{|v''|}{|e_1'|}}{=\Scope{v}{e_1}}{By definition of $|{-}|$}
                        \decolumnizePf
                        \Hand \Pf{\Gamma}{|-^\ambns}{ e' => E ~\textrm{and}~ |e'| = e}{where $e' = (\Scope{v''}{e_1'} : E)$ and $e = \Scope{v}{e_1}$}
                        \Hand \Pf{\Gamma}{|-^\ambns}{ e'' <= E ~\textrm{and}~ |e''| = e}{where $e'' =\Scope{v''}{e_1'}$ and $e = \Scope{v}{e_1}$}
                \end{llproof}
        
        \DerivationProofCase{\etypeAllIndexIntro}
                {
                        \Gamma, a:\sort |-^\ambns t : E
                }
                {
                        \Gamma |-^\ambns t : (\All{a : \sort} E)
                }
        
                \begin{llproof}
                        \Pf{\exists~ t'' ~\textrm{such that}~ \Gamma, a:\sort} {|-^\ambns} { t'' <= E ~\textrm{and}~ |t''| = t}{By inductive hypothesis}
                        \Pf{\Gamma}{|-^\ambns}{ t'' <= (\All{a : \sort} E)}{By rule chk-AllIndexIntro}
                        \Pf{\Gamma}{|-^\ambns}{ (t'' : (\All{a : \sort} E)) => (\All{a : \sort} E)}{By rule syn-anno}
                        \Pf{}{|(t'' : (\All{a : \sort} E))| =}{|t''| = t}{By definition of $|{-}|$}\trailingjust{and $|t''| = t$}
                        \decolumnizePf
                        \Hand \Pf{\Gamma}{|-^\ambns}{ e' => (\All{a : \sort} E) ~\textrm{and}~ |e'| = e}{where $e' = (t'' : (\All{a : \sort} E))$ and $e = t$}
                        \Hand \Pf{\Gamma}{|-^\ambns}{ e'' <= (\All{a : \sort} E) ~\textrm{and}~ |e''| = e}{where $e'' = t''$ and $e = t$}
                \end{llproof}   
        
        \DerivationProofCase{\etypeAllIndexElim}
                {
                        \Gamma |-^\ambns e : (\All{a : \sort} E)
                        \\
                        \Gamma |- i : \sort
                }
                {
                        \Gamma |-^\ambns
                        e
                        : [i/a]E
                }
        
                \begin{llproof}
                        \Pf{\exists~ e' ~\textrm{such that}~ \Gamma} {|-^\ambns} { e' => (\All{a : \sort} E) ~\textrm{and}~ |e'| = e}{By inductive hypothesis}
                        \Pf{\Gamma}{|-}{i : \sort}{Given}
                        \Pf{\Gamma}{|-^\ambns}{ e' => [i/a]E}{By rule syn-AllIndexElim}
                        \Pf{\Gamma}{|-^\ambns}{ e' <= [i/a]E}{By rule chk-conv} 
                        \Hand \Pf{\Gamma}{|-^\ambns}{\inst{e'}{i} => [i/a]E ~\textrm{and}~ |\inst{e'}{i}| = e}{}
                        \Hand \Pf{\Gamma}{|-^\ambns}{\inst{e'}{i} <= [i/a]E ~\textrm{and}~ |\inst{e'}{i}| = e}{}
                \end{llproof}
        
        \DerivationProofCase{\etypeAllIntro}
                {
                        \Gamma, a:\sort |-^\ambns t : E
                }
                {
                        \Gamma |-^\ambns
                        t
                        : (\All{\alpha : K} E)
                }

                Similar to the AllIndexIntro case.
        
%               \begin{llproof}
%                       \Pf{\exists~ t'' ~\textrm{such that}~ \Gamma, a:\sort} {|-^\ambns} { t'' <= E ~\textrm{and}~ |t''| = t}{By inductive hypothesis}
%                       \Pf{\Gamma}{|-^\ambns}{ t'' <= (\All{\alpha : K} E)}{By rule chk-AllIntro}
%                       \Pf{\Gamma}{|-^\ambns}{ (t'' : (\All{\alpha : K} E)) => (\All{\alpha : K} E)}{By rule syn-anno}
%                       \Pf{}{|(t'' : (\All{\alpha : K} E))| =}{|t''| = t}{By definition of $|{-}|$}\trailingjust{and $|t''| = t$}
%                       \Hand \Pf{\Gamma}{|-^\ambns}{ e' => (\All{\alpha : K} E) ~\textrm{and}~ |e'| = e}{where $e' = (t'' : (\All{\alpha : K} E)$ and $e = t$}
%                       \Hand \Pf{\Gamma}{|-^\ambns}{ e'' <= (\All{\alpha : K} E) ~\textrm{and}~ |e''| = e}{where $e'' = t''$ and $e = t$}
%               \end{llproof}   
        
        \DerivationProofCase{\etypeAllElim}
                {
                        \Gamma |-^\ambns e : (\All{\alpha : K} E)
                        \\
                        \Gamma |- A : K
                }
                {
                        \Gamma |-^\ambns
                        e % \tyapp{e}{A}
                        : [A/\alpha]E
                }

                Similar to the AllIndexElim case.                 
%               \begin{llproof}
%                       \Pf{\exists~ e' ~\textrm{such that}~ \Gamma}{|-^\ambns}{ e' => (\All{\alpha : K} E) ~\textrm{and}~ |e'| = e}{By inductive hypothesis}
%                       \Pf{\Gamma}{|-}{A : K}{Given}
%                       \Pf{\Gamma}{|-^\ambns}{ e'=> [A/\alpha]E}{By rule syn-AllElim}
%                       \Pf{\Gamma}{|-^\ambns}{ e'<= [A/\alpha]E}{By rule chk-conv}
%                       \Hand \Pf{\Gamma}{|-^\ambns}{ e' => [A/\alpha]E ~\textrm{and}~ |e'| = e}{}
%                       \Hand \Pf{\Gamma}{|-^\ambns}{ e'' <= [A/\alpha]E ~\textrm{and}~ |e''| = e}{where $e'' = e'$}
%               \end{llproof}

        \ProofCaseRule{\etypeExistsIndexElim}
            By i.h.\ and \echkExistsIndexElim.
        \qedhere
        \end{itemize}
\end{proof}

% Local Variables: 
% mode: latex
% TeX-master: "typed-adapton"
% End: 

\section{Name Term Language}
\label{sec:nametermlang}

We define a restricted \emph{name term} language for computing larger names from smaller names.
This language consists of the following:
\begin{itemize}
\item Syntax for \emph{names}, 
\emph{name terms} and \emph{sorts} 
(\Figref{fig:syntax-name-terms} in \Secref{sec:names}).

\item Name term sorting: 
A judgment that assigns sorts to name terms 
(\Figref{fig:sorting-name-terms} in \Secref{sec:names}).

\item Big-step evaluation for name terms: 
A judgment that assigns \emph{name term values} to 
\emph{name terms} 
(\Figref{fig:eval-name-terms} in \Secref{sec:names}).

\item Semantic definition of equivalent and disjoint name terms 
(\Secref{sec:nmtm-semantic-defs}).

\item Logical proof rules for equivalent and disjoint name terms: 
Two judgements that should be sound with respect to the semantic definitions of equivalence and disjointness
 (\Figref{fig:equiv-rules} and \Figref{fig:disj-rules}).
\end{itemize}

The first three items are each described in \Secref{sec:names}.
We define the last two items in this section.

\subsection{Semantic equivalence and disjointness}
\label{sec:nmtm-semantic-defs}

Below, we define semantic equivalence and disjointness of (sorted)
name terms.
We define these semantic properties inductively, based on the common
sort of the name terms.
In this sense, these definitions can be viewed as instances of logical
relations.

We define contexts~$\Gamma$ that relate two variables;
each declaration either asserts that $a$ and $b$ are equivalent, or disjoint.
We write $\Gamma.1$ and $\Gamma.2$ for the projection of a relational $\Gamma$
into an ordinary $\Gamma$ suitable for the left-hand ($\Gamma.1$) or right-hand ($\Gamma.2$)
sides.
Also, we write $\flip{\Gamma}$ for the operation of exchanging $a$ and $b$
in each declaration: $\flip{(a \disj b : \sort)} = (b \disj a : \sort)$,
so that $\flip{\Gamma}.1 = \Gamma.2$ and $\flip{\Gamma}.2 = \Gamma.1$.
\[\begin{array}[t]{lcll}
\textrm{Substitutions}
& \sigma & ::= & \cdot ~|~ \sigma, N / a
\\
\textrm{Relational sorting contexts}
& \Gamma &  ::= & \cdot
\\
\textrm{~~~(Hypothetical variable equivalence)}
&        &  ~|~ & \Gamma, \left( a \equiv b : \sort \right)
\\
\textrm{~~~(Hypothetical variable apartness)}
&        &  ~|~ & \Gamma, \left(  a \disj b  : \sort \right)
\end{array}
\]
\[
\begin{array}[t]{p{3mm}rcl}
&(\cdot).1 & = & \cdot
\\
&(\Gamma, a \equiv b : \sort).1 & = & (\Gamma).1, a : \sort
\\
&(\Gamma, a \disj b : \sort).1 & = & (\Gamma).1, a : \sort
\end{array}
~
\begin{array}[t]{p{3mm}rcl}
&(\cdot).2 & = & \cdot
\\
&(\Gamma, a \equiv b : \sort).2 & = & (\Gamma).2, b : \sort
\\
&(\Gamma, a \disj b : \sort).2 & = & (\Gamma).2, b : \sort
\\[2px]
\end{array}
\]
\begin{defn}[Closing substitutions]~\\
We define closing substitution pairs related by equivalence and
  disjointness assumptions in a context~$\Gamma$.
These definitions use and are used by the definitions
below for equivalence and apartness of open terms.
\begin{itemize}
\item
  $||- \sigma_1 \equiv \sigma_2 : \Gamma$ means that
  $(x \equiv y : \sort) \in \Gamma$
  implies
  \big(%
  $\sigma_1(x) = N$
  and $\sigma_2(y) = M$
  and $\cdot ||- N \equiv M : \sort$%
  \big)

\item
  $||- \sigma_1 \disj \sigma_2 : \Gamma$ means that
  $(x \disj y : \sort) \in \Gamma$
  implies
  \big($\sigma_1(x) = N$
  and $\sigma_2(y) = M$
  and $\cdot ||- N \disj M : \sort$%
  \big)

\item
  $||- \sigma_1 \sim \sigma_2 : \Gamma$ means that
  $||- \sigma_1 \equiv \sigma_2 : \Gamma$
  and
  $||- \sigma_1 \disj \sigma_2 : \Gamma$
\end{itemize}
\end{defn}

\begin{defn}[Semantic equivalence]
\label{def:semantic-equivalence-M}
  We define $\Gamma ||- M_1 \equiv M_2 : \sort$ as follows:
  \\
  $(\Gamma).1 |- M_1 : \sort$
  and $(\Gamma).2 |- M_2 : \sort$
  and,
  \\
  for all $\sigma_1$, $\sigma_2$
  such that $||- \sigma_1 \equiv \sigma_2 : \Gamma$ and $[\sigma_1]M_1 !! V_1$ and $[\sigma_2]M_2 !! V_2$,
  \\
  we have the following about $V_1$ and $V_2$:

  \medskip

 \begin{tabular}{|c|l|}
\hline
 Sort ($\sort$) & Values~$V_1$ and $V_2$ of sort~$\sort$ are equivalent, written $||- V_1 \equiv V_2 : \sort$
\\
\hline
  $\unitsort$            & Always
\\
  $\Nm$ & When $V_1 = n_1$ and $V_2 = n_2$ and $n_1 = n_2$ (identical binary trees)
\\
  $\sort_1 * \sort_2$ & When $V_1 = (V_{11}, V_{12})$ and $V_1 = (V_{21}, V_{22})$
  \\ & ~~and $||- V_{11} \equiv V_{21} : \sort_1$ and $||- V_{21} \equiv V_{22} : \sort_2$
\\
  $\sort_1 @> \sort_2$ &
 When $V_1 = \Lam{a_1}{M_1}$ and $V_2 = \Lam{a_2}{M_2}$,
 \\ & ~~and
 for all name terms $||- N_1 \equiv N_2 : \sort_1$,
\\
& ~~~~$[N_1/a_1] M_1 !! W_1$ and  $[N_2 / a_2] M_2 !! W_2$ implies $||- W_1 \equiv W_2 : \sort_2$
\\
\hline
\end{tabular}
\end{defn}

\medskip

\begin{defn}[Semantic apartness]
\label{def:semantic-apartness-M}
  We define $\Gamma ||- M_1 \disj M_2 : \sort$ as follows:
  \\
  $(\Gamma).1 |- M_1 : \sort$
  and $(\Gamma).2 |- M_2 : \sort$
  and,
  \\
  for all $\sigma_1$, $\sigma_2$ such that
  $||- \sigma_1 \sim \sigma_2 : \Gamma$ and $[\sigma_1]M_1 !! V_1$ and $[\sigma_2]M_2 !! V_2$,
  \\
  we have the following about $V_1$ and $V_2$:

  \medskip

 \begin{tabular}{|c|l|}
\hline
 Sort ($\sort$) & Values~$V_1$ and $V_2$ of sort~$\sort$ are apart, written $||- V_1 \disj V_2 : \sort$
\\
\hline
  $\unitsort$            & Always
\\
  $\Nm$ & When $V_1 = n_1$ and $V_2 = n_2$ and $n_1 \ne n_2$ (distinct binary trees)
\\
  $\sort_1 * \sort_2$
  & When $V_1 = (V_{11}, V_{12})$
   and $V_1 = (V_{21}, V_{22})$
  \\
  & ~~ and $||- V_{11} \disj V_{21} : \sort_1$
  and $||- V_{21} \disj V_{22} : \sort_2$
\\
  $\sort_1 @> \sort_2$ &
 When $V_1 = \Lam{a_1}{M_1}$ and $V_2 = \Lam{a_2}{M_2}$, \\
 & ~~and
 for all name terms $|- N_1 : \sort_1$ and $|- N_2 : \sort_1$,
\\ ~~
%& \hfill$(\Lam{a_1}{M_1})(N_1) !! W_1$ and  $(\Lam{a_2}{M_2})(N_2) !! W_2$ implies $||- W_1 \disj W_2 : \sort_2$
 & ~~~~$[N_1/a_1] M_1 !! W_1$ and  $[N_2 / a_2] M_2 !! W_2$ implies $||- W_1 \disj W_2 : \sort_2$
\\
\hline
\end{tabular}
\end{defn}

\begin{figure}
\centering
  \judgbox{\Gamma |- M \equiv N : \sort}
          {The name terms $M$ and $N$ are \emph{equivalent} at sort $\gamma$}

\begin{mathpar}
\Infer{Eq-Var}
{(M \equiv N : \sort) \in \Gamma}
{\Gamma |- M \equiv N : \sort}
~~~
\Infer{E-Refl}
      { 
        \arrayenvbl{
          (\Gamma).1 |- M : \sort
          \\
          (\Gamma).2 |- M : \sort
        }
      }
      {\Gamma |- M \equiv M : \sort}
~~~
\Infer{E-Sym}
      {\flip{\Gamma} |- N \equiv M : \sort}
      {\Gamma |- M \equiv N : \sort}
\and
\Infer{Eq-Trans}
    {
      \Gamma  |- M_1 \equiv M_2 : \sort
      \\
      \Gamma  |- M_2 \equiv M_3 : \sort
    }
    { \Gamma |- M_1 \equiv M_3 : \sort }
\\
\Infer{Eq-Pair}
    {
      \Gamma |- M_1 \equiv N_1 : \sort_1
      \\
      \Gamma |- M_2 \equiv N_2 : \sort_2
    }
    {\Gamma |- (M_1, M_2) \equiv (N_1, N_2) : \sort_1 * \sort_2}
\and
\Infer{Eq-Bin}
    {\Gamma |- M_1 \equiv N_1 : \namesort
      \\
     \Gamma |- M_2 \equiv N_2 : \namesort
    }
    {\Gamma |- \NmBin{M_1}{M_2} \equiv \NmBin{N_1}{N_2} : \namesort}
\\
\Infer{Eq-Lam}
    {\Gamma, \big(a \equiv b : \sort_1\big) |- M \equiv N                   : \sort_2 }
    {\Gamma                                 |- \Lam{a}{M} \equiv \Lam{b}{N} : \sort_1 @> \sort_2 }
\and
\Infer{Eq-App}
    {
      \arrayenvcl{
       \Gamma |- M_1 \equiv N_1 : \sort_1 @> \sort_2
        \\
       \Gamma |- M_2 \equiv N_2 : \sort_1
      }
    }
    {\Gamma |- M_1 (M_2) \equiv N_1 (N_2) : \sort_2}
\and
\Infer{Eq-$\beta$}
    {
      \Gamma  |- M_2 \equiv M_2' : \sort_1
      \\
      \Gamma, a \equiv a : \sort_1  |- M_1 \equiv M_1' : \sort_2
    }
    {
      \Gamma |- (\Lam{a}M_1)M_2 \equiv [M_2'/a]M_1' : \sort_2
    }
\end{mathpar}

\caption{Deductive rules for showing that two name terms are equivalent}
\label{fig:equiv-rules}

\end{figure}

\begin{figure}
\centering
  \judgbox{\Gamma |- M \disjoint N : \sort}
          {The name terms $M$ and $N$ are \emph{apart} at sort $\gamma$}
%  \vspace*{-1.8ex}

\begin{mathpar}
\Infer{Var}
    {(a \disjoint b : \sort) \in \Gamma}
    {\Gamma |- a \disjoint b : \sort}
\and
\Infer{D-Sym}
    {\flip{\Gamma} |- N \disjoint M : \sort}
    {\Gamma |- M \disjoint N : \sort}
\and
\Infer{D-trans} 
{
  \Gamma  |- M_1 \equiv M_2 : \sort
  \\
  \Gamma  |- M_2 \disj M_3 : \sort
}
{ \Gamma |- M_1 \disj M_3 : \sort }
\and
\Infer{D-Bin$_1$}
{\Gamma |- M_1 \disj N_1 : \namesort
}
{\Gamma |- \NmBin{M_1}{M_2} \disj \NmBin{N_1}{N_2} : \namesort}
\and
\Infer{D-Bin$_2$}
{
  \Gamma |- M_2 \disj N_2 : \namesort
}
{\Gamma |- \NmBin{M_1}{M_2} \disj \NmBin{N_1}{N_2} : \namesort}
\and
\Infer{D-EqTag$_1$}
{
  \Gamma |- M_1 \equiv M_2 : \namesort
}
{\Gamma |- \NmBin{M_2}{N} \disj M_1 : \namesort}
\and
\Infer{D-EqTag$_2$}
{
  \Gamma |- N_1 \equiv N_2 : \namesort
}
{\Gamma |- \NmBin{M}{N_1} \disj N_2 : \namesort}
\and
\Infer{D-Lam}
{\Gamma |- M \disjoint N                   : \sort_2 }
{\Gamma |- \Lam{a}{M} \disjoint \Lam{b}{N} : \sort_1 @> \sort_2 }
\and
\Infer{D-App}
{\Gamma |- M_1 \disjoint N_1 : \sort_1 @> \sort_2
% Removed premise; No longer assumed by lambda rule.
%  \\
% \Gamma |- M_2 \equiv    N_2 : \sort_1
}
{\Gamma |- M_1 (M_2) \disjoint N_1 (N_2) : \sort_2}
\and
\Infer{D-$\beta$}
{
  \Gamma.1 |- M_2 : \gamma_2
  \\
  \Gamma.1, a : \gamma_2 |- M_1 : \gamma
  \\
  \Gamma |- [M_2/a] M_1 \disj N : \gamma
}
{
  \Gamma |- (\Lam{a}{M_1})\;M_2 \disj N : \gamma
}
%\and
%\Infer{D-$\beta$}
%{
%\Gamma  |- [M_2/a]M_1 \disj N : \sort
%\\\\
%(\Gamma).1  |- M_2 : \sort_2
%\\\\
%(\Gamma).1, a : \sort_2  |- M_1 : \sort
%}
%{ \Gamma |- (\Lam{a}M_1)M_2 \disj N : \sort }
\end{mathpar}
\caption{Deductive rules for showing that two name terms are apart}
\label{fig:disj-rules}
\end{figure}

% Local Variables: 
% mode: latex
% TeX-master: "typed-adapton"
% End: 

\subsection{Metatheory of name term language}
\label{sec:nameterm-rules-soundness}
\label{sec:nameterm-rules-completeness}

Some lemmas in this section are missing complete proofs and should be
considered conjectures
(\Lemmaref{lem:syn-eq-proj}--\Lemmaref{lem:sub-eval-sem-eq}).

\begin{lemma}[Projections of syntactic equivalence]
  \label{lem:syn-eq-proj}
~\\
  If $\Gamma |- M_1 \equiv M_2 : \sort$, then $\Gamma.1 |- M_1 : \sort$ and
  $\Gamma.2 |- M_2 : \sort$.
\end{lemma}

% \begin{lemma}[Compatibility of substitution with name term structure]
% \label{lem:compat-sub-ntm}
% ~
% \begin{enumerate}
% \item We have $[\sigma](M_1 , M_2) = ([\sigma]M_1,[\sigma]M_2)$.
% \item We have $[\sigma]\NmBin{M_1}{M_2} = \NmBin{[\sigma]M_1}{[\sigma]M_2}$.
% \item We have $[\sigma](MN) = ([\sigma]M)([\sigma]N)$.
% \end{enumerate}
% \end{lemma}

\begin{lemma}[Determinism of evaluation up to substitution]
  \label{lem:sub-eval-det}
~\\
  If $\Gamma.1 |- M : \sort$
  and $\Gamma.2 |- M : \sort$
  and $||- \sigma_1 \equiv \sigma_2 : \Gamma$
  and $[\sigma_1]M !! V_1$
  and $[\sigma_2]M !! V_2$
  \\ then there
  exists $V$ such that $V_1 = [\sigma_1]V$ and $V_2 = [\sigma_2]V$.
\end{lemma}

\begin{lemma}[Reflexivity of semantic equivalence]
\label{lem:refl-sem}
  ~
\begin{enumerate}
\item If $|- M : \sort$ then $||- M \equiv M : \sort$.
\item If $\Gamma.1 |- V : \sort$
  and $\Gamma.2 |- V : \sort$
  and $||- \sigma_1 \equiv \sigma_2 : \Gamma$
  and $V_1 = [\sigma_1]V$
  and $V_2 = [\sigma_2]V$
  \\
  then $||- V_1 \equiv V_2 : \sort$.
\end{enumerate}
\end{lemma}

\begin{lemma}[Type safety]
\label{lem:type-safety}
  If $\Gamma |- M : \sort$
  and $[\sigma]M !! [\sigma]V$
  then $\Gamma |- V : \sort$.
\end{lemma}

\begin{lemma}[Symmetry of semantic equivalence]
\label{lem:sym-sem}
  ~
  \begin{enumerate}
  \item If $||- \sigma_1 \equiv \sigma_2 : \Gamma$ then $||- \sigma_2 \equiv \sigma_1 : \flip{\Gamma}$.
  \item If $||- V_1 \equiv V_2 : \sort$ then $||- V_2 \equiv V_1 : \sort$.
  \item If $\cdot ||- M_1 \equiv M_2 : \sort$ then $\cdot ||- M_2 \equiv M_1 : \sort$.
  \item If $\Gamma ||- M_1 \equiv M_2 : \sort$
    then $\flip{\Gamma} ||- M_2 \equiv M_1 : \sort$.
  \end{enumerate}
\end{lemma}

\begin{lemma}[Evaluation respects semantic equivalence]
\label{lem:eval-resp-sem-eq}
~\\
  If $\Gamma ||- M \equiv N : \sort$
  and $||- \sigma_1 \equiv \sigma_2 : \Gamma$
  and $[\sigma_1] M !! V_1$
  then there exists $V_2$
  such that $[\sigma_2]N !! V_2$
  and $||- V_1 \equiv V_2 : \sort$.
\end{lemma}

\begin{lemma}[Closing substitutions respect syntactic equivalence]
\label{lem:fund-sub}
~\\
  If $||- \sigma_1 \equiv \sigma_2 : \Gamma$
  and $\Gamma |- M_1 \equiv M_2 : \sort$
  then $\cdot ||- [\sigma_1]M_1 \equiv [\sigma_2]M_2 : \sort$.
\end{lemma}

% Sorry, this is wrong. I just realized I was trying to generalize Barendregt's
% substitution lemma, but failed. https://isabelle.in.tum.de/nominal/example.html
%\begin{lemma}[Closing substitutions respect composition]
%~\\
%  \label{lem:cl-sub-resp-comp}
%  If  $||- \sigma_1 \equiv \sigma_2 : \Gamma,
%  ||- \sigma_1' \equiv \sigma_2' : \Gamma'$, and
%  $\Gamma$ and $\Gamma'$ are disjoint, then
%  $||- [\sigma_1']\sigma_1 \equiv [\sigma_2']\sigma_2 : \Gamma$.
%\end{lemma}

\begin{lemma}[Reflexivity of name term evaluation]
\label{lem:sub-eval-sem-eq}
~\\
  If $\Gamma.1 |- M : \sort$ and $\Gamma.2 |- M : \sort$
  and $||- \sigma_1 \equiv \sigma_2 : \Gamma$
  and $[\sigma_1]M !! V_1$
  and $[\sigma_2]M !! V_2$
  \\
  then $||- V_1 \equiv V_2 : \sort$.
\end{lemma}

\begin{lemma}[Transitivity of value equivalence]
\label{lem:trans-val}
~\\
  If $|- V_1 : \sort$ and $|- V_2 : \sort$
  and $||- V_1 \equiv V_2 : \sort$
  and $||- V_2 \equiv V_3 : \sort$
  then $||- V_1 \equiv V_3 : \sort$.
\end{lemma}
\begin{proof}
  Uses strong normalization.
\end{proof}

\begin{conjecture}[Soundness of deductive equivalence]
~\\
  If
  $\Gamma |- M_1 \equiv M_2 : \sort$
  then
  $\Gamma ||- M_1 \equiv M_2 : \sort$.
\end{conjecture}
\begin{proof} By induction on the given derivation.
\DerivationProofCase{Eq-Var}
{(a \equiv b : \sort) \in \Gamma}
{\Gamma |- a \equiv b : \sort}

By definition of closing substitutions.

\DerivationProofCase{Eq-Refl}
      { 
        (\Gamma).1 |- M : \sort
        \\
        (\Gamma).2 |- M : \sort
      }
      {\Gamma |- M \equiv M : \sort}

By \Lemmaref{lem:sub-eval-det}, \Lemmaref{lem:type-safety}, and \Lemmaref{lem:refl-sem}.
      
\DerivationProofCase{Eq-Sym}
      {\Gamma |- N \equiv M : \sort}
      {\Gamma |- M \equiv N : \sort}

By \Lemmaref{lem:sym-sem}.

\DerivationProofCase{Eq-Trans}
    {
      \Gamma  |- M_1 \equiv M_2 : \sort
      \\
      \Gamma  |- M_2 \equiv M_3 : \sort
    }
    { \Gamma |- M_1 \equiv M_3 : \sort }

By idempotency of flipping relational contexts, \Lemmaref{lem:eval-resp-sem-eq},
inductive hypotheses on the two given subderivations, \Lemmaref{lem:sym-sem}, and \Lemmaref{lem:trans-val}.

\DerivationProofCase{Eq-Pair}
{\Gamma |- M_1 \equiv N_1 : \sort_1
  \\
 \Gamma |- M_2 \equiv N_2 : \sort_2
}
{\Gamma |- (M_1, M_2) \equiv (N_1, N_2) : \sort_1 * \sort_2}

By the definition of substitution and the i.h.

\DerivationProofCase{Eq-Bin}
{\Gamma |- M_1 \equiv N_1 : \namesort
  \\
 \Gamma |- M_2 \equiv N_2 : \namesort
}
{\Gamma |- \NmBin{M_1}{M_2} \equiv \NmBin{N_1}{N_2} : \namesort}

By the definition of substitution and the i.h.

\DerivationProofCase{Eq-Lam}
{\Gamma, \big(a \equiv b : \sort_1\big) |- M \equiv N                   : \sort_2 }
{\Gamma                                 |- \Lam{a}{M} \equiv \Lam{b}{N} : \sort_1 @> \sort_2 }

By transposition of substitutions and the i.h.
% Barendregt's Substitution Lemma
% You can't apply a result from a different system. -j.
% Barendregt's substitution lemma: https://isabelle.in.tum.de/nominal/example.html

\DerivationProofCase{Eq-App}
    {
      \arrayenvcl{
        \Gamma |- M_1 \equiv N_1 : \sort_1 @> \sort_2
        \\
        \Gamma |- M_2 \equiv N_2 : \sort_1
      }
    }
    {\Gamma |- M_1 (M_2) \equiv N_1 (N_2) : \sort_2}

By definition of substitution and inversion (teval-app) of resulting derivations, the inductive hypothesis on
the two given syntactic equivalence subderivations (of Eq-App), and definition
of semantic equivalence of arrow-sorted values, we get the result.

\DerivationProofCase{Eq-$\beta$}
{
  \Gamma  |- M_2 \equiv M_2' : \sort_1
  \\
  \Gamma, a \equiv a : \sort_1  |- M_1 \equiv M_1' : \sort_2
}
{ \Gamma |- (\Lam{a}M_1)M_2 \equiv [M_2'/a]M_1' : \sort_2 }

Fix $||- \sigma_1 \equiv \sigma_2 : \Gamma$. Suppose $[\sigma_1]((\lam{a}
M_1)M_2) !! V_1$ and $[\sigma_2]([M_2'/a]M_1') !! V_2$. We need to show $||- V_1
\equiv V_2 : \sort_2$.
By the definition of substitution and inversion of teval-app,
$[\sigma_1]M_2 !! V$ and $[V/a]([\sigma_1]M_1) !! V_1$ for some $V$.
Hence, because $\Gamma, a \equiv a : \sort_1$,
we have $[\sigma_1, V/a]M_1 !! V_1$.
Rewrite
$[\sigma_2]([M_2'/a]M_1') !! V_2$
as 
$[\sigma_2, [\sigma_2]M_2' / a]M_1' !! V_2$.
By \Lemmaref{lem:fund-sub}, $\cdot |- V \equiv [\sigma_2]M_2' : \sort_1$.
Therefore,
\[
||-
(\sigma_1, V/a) \equiv (\sigma_2, [\sigma_2]M_2' / a)
:
(\Gamma, a \equiv a : \sort_1)
\]
By the inductive hypothesis on
$\Gamma, a \equiv a : \sort_1 |- M_1 \equiv M_1' : \sort_2$,
we get $||- V_1 \equiv V_2 : \sort_2$.
\end{proof}

\begin{conjecture}[Soundness of deductive disjointness]
  If
  $\Gamma |- M_1 \disj M_2 : \sort$
  then
  $\Gamma ||- M_1 \disj M_2 : \sort$.
\end{conjecture}

%\subsection{Completeness of the deductive rules}

\begin{conjecture}[Completeness of deductive equivalence]
  If
  $\Gamma ||- M_1 \equiv M_2 : \sort$
  then
  $\Gamma |- M_1 \equiv M_2 : \sort$.
\end{conjecture}

\begin{conjecture}[Completeness of deductive disjointness]
  If
  $\Gamma ||- M_1 \disj M_2 : \sort$
  then
  $\Gamma |- M_1 \disj M_2 : \sort$.
\end{conjecture}

%% \MattSaysTODO{Dimitrios is working on the conjectures above.  The
%%   order above seems like the right one.  Proving these conjectures
%%   will likely involve reasoning about substitutions too, requiring
%%   lemmas about them.  We should add those lemmas above, as they
%%   arise in his development.}

% Local Variables:
% mode: latex
% TeX-master: "typed-adapton"
% End:

\clearpage

\section{Index term language}
\label{sec:indextermlang}

We define a restricted \emph{index term} language for describing name sets
and functions that relate them.
This language consists of the following:

\begin{itemize}
\item Syntax for \emph{index terms}, 
and (additional) \emph{sorts} (\Figref{fig:syntax-indices} in \Secref{sec:idxtm}).

\item Index term sorting, which assigns sorts to index terms 
(\Figref{fig:sorting} in \Secref{sec:idxtm}).

\item Reduction rules for index terms, deriving the judgment $i --> j$
  (\Figref{fig:index-step}).

\item Semantic definitions of equivalent and disjoint index terms 
(\Secref{sec:idxtm-semantic-def}).

\item Deductive rules for equivalent and disjoint index terms,
which should be sound with respect 
to the semantic definitions of equivalence and disjointness
 (\Figref{fig:idx-equiv-rules} and \Figref{fig:idx-disj-rules}).
\end{itemize}

The first two items are defined in \Secref{sec:idxtm}.
The remaining items are defined here.

\subsection{Index reduction}

We write $i --> j$ for the one-step head reduction of index $i$ to $j$,
defined in \Figref{fig:index-step}.
This is essentially $\beta$-reduction for functions, products,
and mapping name sets (with map $M [[ X ]] $ distributing over union operators).
% (defined pointwise, using an auxiliary judgment
% $\mapset{M}{X}{Y}$).

Based on one-step head reduction, we define one-step reduction
$i -C> j$ (reducing anywhere within an index, not only at the head)
and multi-step reduction $i ->> j$:

\begin{defn}[Reduction for indices]
~
  \begin{enumerate}
  \item Let $-C>$ be the congruence closure of $-->$.
  \item Let $->>$ be the reflexive-transitive closure of $-C>$.
  \end{enumerate}
\end{defn}

\begin{figure}

\judgbox{i --> j}
        {Index~$i$ reduces to index~$j$}
\vspace{-1.5ex}
\begin{mathpar}
\Infer{reduce-proj}
    {}
    { \Proj{b}{ \Pair{j_1}{j_2} } --> j_b}
\and
\Infer{reduce-app}
    {}
    {\idxapp{(\lam{a} i)}{j} --> [j/a] i}
\\
\Infer{reduce-map-empty}
   { }
   { \mapset{M}{ \emptyset }{ \emptyset } }
\and
\Infer{reduce-map-single}
    {
      M ( N ) \nteval V
    }
    {
      M [[ \{ N \} ]] -->  \{ V \}
    } 
\vspace*{-1.2ex}
\\
\Infer{reduce-map-$\disj$}
    {
      M~\textrm{injective}
    }
    { M [[ X_1 \disj X_2 ]] --> M [[ X_1 ]] \disj M [[ X_2 ]] }
\vspace*{-1.6ex}
\\
\Infer{reduce-map-$\union$}
    {
    }
    { M [[ X_1 \union X_2 ]] --> M [[ X_1 ]] \union M [[ X_2 ]] }
% these are too "big-step"
% \and
% \Infer{reduce-map-$\disj$}
%     {
%       \arrayenvbl{
%         M [[ X_1 ]] --> { Y_1 } 
%         \\
%         M [[ X_2 ]] --> { Y_2 } 
%       }
%     }
%     { M [[ X_1 \disj X_2 ]] --> { Y_1 \disj Y_2 } }
% \and
% \Infer{reduce-map-$\union$}
%     {
%       \arrayenvbl{
%         M [[ X_1 ]] --> { Y_1 } 
%         \\
%         M [[ X_2 ]] --> { Y_2 } 
%       }
%     }
%     { M [[ X_1 \union X_2 ]] --> { Y_1 \union Y_2 } }
\\
\Infer{reduce-kleene-outer}
    {  }
    { M^\ast [[ j ]] --> M [[ M^\ast [[ j ]] ]] }
\and
\Infer{reduce-kleene-inner}
    {  }
    { M^\ast [[ j ]] --> M^\ast [[ M [[ j ]] ]] }
\vspace*{-1.2ex}
\end{mathpar}

\iffalse
          \medskip

          \judgbox{ \mapset{M}{X}{Y} }
                  {Name term function $M$, applied to each member of $X$, yields name set $Y$}
          \begin{mathpar}
          \Infer{Empty}
          { \strut }
          { \mapset{M}{ \emptyset }{ \emptyset } }
          \and
          \Infer{Single}
          { M ( N ) \nteval V }
          { \mapset{M}{ \{ N \} }{ \{ V \} } }
          \and
          \Infer{Apart}
          { 
            \mapset{ M }{ X_1 }{ Y_1 } 
            \\
            \mapset{ M }{ X_2 }{ Y_2 } 
          }
          { \mapset{ M }{ X_1 \disj X_2 }{ Y_1 \disj Y_2 } }
          \and
          \Infer{Union}
          { \mapset{ M }{ X_1 }{ Y_1 } 
            \\
            \mapset{ M }{ X_2 }{ Y_2 } 
          }
          { \mapset{ M }{ X_1 \union X_2 }{ Y_1 \union Y_2 } }
          \vspace*{-1.5ex}
          \end{mathpar}
\fi

\caption{Reduction rules for indices}
\label{fig:index-step}
\end{figure}

% Local Variables:
% TeX-master: "fungi-lang"
% End:

\begin{figure}

\judgbox{ \Gamma |- M \in X}
        {Name term~$M$ is a member of name set~$X$, 
          assuming $X \Value$
        }
\begin{mathpar}
\Infer{Membership${}_1$}
{ \Gamma |- (\{ M \} \disj Y) \equiv X : \NmSet }
{ \Gamma |- M \in X }
\and
\Infer{Membership${}_2$}
{ \Gamma |- (\{ M \} \cup Y) \equiv X : \NmSet }
{ \Gamma |- M \in X }
\end{mathpar}

\judgbox{ \Gamma ||- M \in X}
        {Name term~$M$ is a member of name set~$X$, 
          assuming $X \Value$
        }
\begin{mathpar}
\Infer{Membership${}_1$}
{ \Gamma ||- (\{ M \} \disj Y) \equiv X : \NmSet }
{ \Gamma ||- M \in X }
\and
\Infer{Membership${}_2$}
{ \Gamma ||- (\{ M \} \cup Y) \equiv X : \NmSet }
{ \Gamma ||- M \in X }
\end{mathpar}

\judgbox{\Gamma |- M \notin X}
        {The name of name term $M$ is \emph{not} a member of name set $X$,
          assuming $X \Value$}

\begin{mathpar}
\Infer{NonMembership}
{ \Gamma |- (\{ M \} \disj X) \equiv Y : \NmSet }
{ \Gamma |- M \notin X }
\lesscaptionspace
\end{mathpar}

\judgbox{\Gamma ||- M \notin X}
        {The name of name term $M$ is \emph{not} a member of name set $X$,
          assuming $X \Value$}

\begin{mathpar}
\Infer{NonMembership}
{ \Gamma ||- (\{ M \} \disj X) \equiv Y : \NmSet }
{ \Gamma ||- M \notin X }
\lesscaptionspace
\end{mathpar}

\caption{Name term membership}
\label{fig:name-term-membership}
\end{figure}

% Local Variables:
% TeX-master: "typed-adapton"
% End:

\subsection{Semantic equivalence and apartness of index terms}
\label{sec:idxtm-semantic-def}

\begin{defn}[Closing substitutions for index terms] ~\\
We define closing substitution pairs related by equivalence and
  disjointness assumptions in a context~$\Gamma$.
These definitions use and are used by the definitions
below for equivalence and apartness of open terms.

\begin{itemize}
\item
  $||- \sigma_1 \equiv \sigma_2 : \Gamma$ ~iff~
  $(x \equiv y : \sort) \in \Gamma$ 
  implies 
  \big($\sigma_1(x) = i$ 
  and $\sigma_2(y) = j$ 
  and $\cdot ||- i \equiv j : \sort$
  \big)  
  
\item
  $||- \sigma_1 \disj \sigma_2 : \Gamma$ ~iff~
  $(x \disj y : \sort) \in \Gamma$
  implies
  \big($\sigma_1(x) = i$ 
  and $\sigma_2(y) = j$ 
  and $\cdot ||- i \disj j : \sort$
  \big)

\item
  $||- \sigma_1 \sim \sigma_2 : \Gamma$ ~iff~
  \big(
  $||- \sigma_1 \equiv \sigma_2 : \Gamma$
  and
  $||- \sigma_1 \disj \sigma_2 : \Gamma$
  \big)
\end{itemize}
\end{defn}

\begin{defn}[Index normal form]
\label{def:index-normal}
  An index $i$ is in \emph{index normal form}
  iff none of the rules
  \[
    \text{reduce-proj, reduce-app, reduce-map$\{$-empty, -single, -$\disj$, -$\union\}$}
  \]
  can be applied (anywhere within $i$).
\end{defn}

Under this definition, normal forms are not unique:
the rules reduce-kleene-inner and reduce-kleene-outer can reduce normal forms.
In effect, these are normal forms ``modulo Kleene closure''.
  
  % $\union$,

\begin{defn}[Semantic equivalence of index terms]
\label{def:semantic-equivalence-i}
  We define $\Gamma ||- i_1 \equiv i_2 : \sort$ as follows:
  \\
  $(\Gamma).1 |- i_1 : \sort$
  and $(\Gamma).2 |- i_2 : \sort$
  and, 
  \\
  for all $\sigma_1$, $\sigma_2$, and $j_1$
  such that
  $||- \sigma_1 \equiv \sigma_2 : \Gamma$ and $[\sigma_1]i_1 ->> j_1$
  where $j_1$ is in index normal form,
  there exists $j_2$ such that $j_2$ is in index normal form
  and $[\sigma_2]i_2 ->> j_2$,
  \\
  we have
  the following about $j_1$ and $j_2$:
  
  \medskip

 \begin{tabular}{|c|l|}
\hline
 Sort~$\sort$ & Indices~$j_1$ and $j_2$ of sort~$\sort$ are equivalent, written $||- j_1 \equiv j_2 : \sort$
\\
\hline
  $\unitsort$            & Always
\\
  $\NmSet$ & When \big( $||- M \in j_1$  if and only if~~$||- M \in j_2$  \big)
\\
  $\sort_1 * \sort_2$ &
  When $j_1 = (j_{11}, j_{12})$
  and $j_2 = (j_{21}, j_{22})$
  \\
  & ~~and $||- j_{11} \equiv j_{21} : \sort_1$
  \\
  & ~~and $||- j_{12} \equiv j_{22} : \sort_2$
\\
  $\sort_1 @@> \sort_2$ &
 When $j_1 = \Lam{a_1}{X_1}$ and $j_2 = \Lam{a_2}{X_2}$,
 \\ & ~~ and
 for all name terms $||- Y_1 \equiv Y_2 : \sort_1$,
\\
&  $~~~~~~~~(\Lam{a_1}{X_1})(Y_1) ->> Z_1$ and  $(\Lam{a_2}{X_2})(Y_2) ->> Z_2$
\\
& $~~~~~$implies ~$||- Z_1 \equiv Z_2 : \sort_2$
\\
\hline
\end{tabular}
\end{defn}

\medskip

\begin{defn}[Semantic apartness of index terms]
\label{def:semantic-apartness-i}
  We define $\Gamma ||- i_1 \disj i_2 : \sort$ as follows:
  \\
  $(\Gamma).1 |- i_1 : \sort$
  and $(\Gamma).2 |- i_2 : \sort$
  and, 
  \\
  for all $\sigma_1$, $\sigma_2$, and $j_1$
  such that
  $||- \sigma_1 \equiv \sigma_2 : \Gamma$ and $[\sigma_1]i_1 ->> j_1$
  where $j_1$ is in index normal form,
  there exists $j_2$ such that $j_2$ is in index normal form
  and $[\sigma_2]i_2 ->> j_2$,
%
  % for all $\sigma_1$, $\sigma_2$
  % such that
  % $||- \sigma_1 \sim \sigma_2 : \Gamma$ and $[\sigma_1] i_1 !! j_1$ and $[\sigma_2] i_2 !! j_2$,
  \\
  we have the following about $j_1$ and $j_2$:
  
  \medskip

 \begin{tabular}{|c|l|}
\hline
 Sort ($\sort$) & Index values $j_1$ and $j_2$ of sort~$\sort$ are apart, written $||- j_1 \disj j_2 : \sort$
\\
\hline
  $\unitsort$            & Always
\\
  $\NmSet$ &
  When
%  \big( $||- M \in j_1$~if and only if~~$||- M \notin j_2$  \big)
  \tabularenvl{
  \big( $||- M \in j_1$~implies~~$||- M \notin j_2$  \big)
  \\ and
%  \big( $||- M \in j_2$~if and only if~~$||- M \notin j_1$  \big)
  \big( $||- M \in j_2$~implies~~$||- M \notin j_1$  \big)
  }
\\
  $\sort_1 * \sort_2$ &
  When
  \tabularenvl{
  $j_1 = (j_{11}, j_{12})$
  and $j_2 = (j_{21}, j_{22})$
  \\
  and $||- j_{11} \disj j_{21} : \sort_1$
  and $||- j_{12} \disj j_{22} : \sort_2$
  }
\\
  $\sort_1 @@> \sort_2$ &
 When $j_1 = \Lam{a_1}{X_1}$ and $V_2 = \Lam{a_2}{X_2}$,
 \\ & ~~and
 for all name terms $|- Y_1 : \sort_1$ and $|- Y_2 : \sort_1$,
\\
&  ~~~~~~$(\Lam{a_1}{X_1})(Y_1) ->> Z_1$ and  $(\Lam{a_2}{X_2})(Y_2) ->> Z_2$
\\
& ~~~~implies $||- Z_1 \disj Z_2 : \sort_2$
\\
\hline
\end{tabular}
\end{defn}

\medskip

\begin{figure}[htbp]
  \centering
  \small
  \begin{align*}
    \extractassns{\cdot}
      &= \cdot
    \\
    \extractassns{\Gamma, P}
      &= \extractassns{\Gamma}, P
    \\
    \extractassns{\Gamma, \trueprop}
      &= \extractassns{\Gamma}
    \\
    \extractassns{\Gamma, (P_1 \andprop \dots \andprop P_n)}
      &= \extractassns{\Gamma}, P_1, \dots, P_n
\\ &\text{for $n \geq 1$, where each $P_k$}
   \\ &\text{~~~has the  form $i \disj j : \sort$ or $i \equiv j : \sort$}
    \\
    \extractassns{\Gamma, \mathcal{Z}}
      &= \extractassns{\Gamma}~\text{where $\mathcal{Z}$ is not a proposition}
  \end{align*}
  \begin{align*}
    \extractctx{\cdot}
      &= \cdot
    \\
    \extractctx{\Gamma, a : \sort}
      &= \extractctx{\Gamma}, (a \equiv a : \sort)
    \\
    \extractctx{\Gamma, \alpha : \type}
      &= \extractctx{\Gamma}
    \\
    \extractctx{\Gamma, d : K}
      &= \extractctx{\Gamma}
    \\
    \extractctx{\Gamma, p : \cdots}
      &= \extractctx{\Gamma}
    \\
    \extractctx{\Gamma, x : A}
      &= \extractctx{\Gamma}
    \\
    \extractctx{\Gamma, P}
     & = \extractctx{\Gamma}
  \end{align*}
  \[
  \extract{\Gamma} ~=~
  (\extractassns{\Gamma}; \extractctx{\Gamma})
  \]
  
  \caption{Extraction function on typing contexts}
  \label{fig:extract}
\end{figure}

% Local Variables:
% TeX-master: "typed-adapton"
% End:

The next two definitions bridge the gap with the type system,
in which contexts $\Gamma_T$ also include propositions $P$.
It is defined assuming that $\extract{\Gamma_T}$
(defined in \Figureref{fig:extract})
has given us some propositions $P_1, \dots, P_n$ and 
a relational context $\Gamma$.

\begin{defn}[Extended semantic equivalence of index terms]
\label{def:super-semantic-equivalence-i}
~\\
  We define $P_1, \dots, P_n ; \Gamma ||- i \equiv j  : \sort$
  to hold if and only if
  \begin{center}
      $\mathcal{J}(P_1)$
      and $\cdots$
      and $\mathcal{J}(P_n)$
      implies
      $\Gamma ||- i \equiv j : \sort$    
  \end{center}
  where $\mathcal{J}(i \mathrel{\Theta} j : \sort) = (\Gamma ||- i \mathrel{\Theta} j : \sort)$.
\end{defn}

\begin{defn}[Extended semantic apartness of index terms]
\label{def:super-semantic-apartness-i}
~\\
  We define $P_1, \dots, P_n ; \Gamma ||- i \disj j  : \sort$
  to hold if and only if
  \begin{center}
      $\mathcal{J}(P_1)$
      and $\cdots$
      and $\mathcal{J}(P_n)$
      implies
      $\Gamma ||- i \disj j : \sort$    
  \end{center}
  where $\mathcal{J}(i \mathrel{\Theta} j : \sort) = (\Gamma ||- i \mathrel{\Theta} j : \sort)$.
\end{defn}

When a typing context is weakened, semantic equivalence and apartness
under the extracted context continue to hold:

\begin{lemma}[Weakening of semantic equivalence and apartness]
\label{lem:weakening-semantic}
  ~\\
  If
  $\extract{\Gamma_T} ||- i_1 \equiv i_2 : \sort$
  (respectively $i_1 \disj i_2 : \sort$)
  then
  $\extract{\Gamma_T, \Gamma_T'} ||- i_1 \equiv i_2 : \sort$  
  (respectively $i_1 \disj i_2 : \sort$.
\end{lemma}
\begin{proof}  
  By induction on $\Gamma_T'$.

  We prove the $\equiv$ part; the $\disj$ part is similar.

  \begin{itemize}
  \item 
  If $\Gamma_T' = \cdot$,
  we already have the result.

  \item
  If $\Gamma_T' = (\Gamma', P)$
  then:

    By i.h., $\extract{\Gamma_T, \Gamma'} ||- i_1 \equiv i_2 : \sort$.
    
    That is, $\extractassns{\Gamma_T, \Gamma'};
    \extractctx{\Gamma_T, \Gamma'} ||- i_1 \equiv i_2 : \sort$.

    By its definition, $\extractctx{\Gamma_T, \Gamma', P} = 
    \extractctx{\Gamma_T, \Gamma', P}$.
    \\
    Therefore, we have
    $\extractassns{\Gamma_T, \Gamma'};
    \extractctx{\Gamma_T, \Gamma', P} ||- i_1 \equiv i_2 : \sort$.
    \\
    Adding an assumption before the semicolon only supplements
    the antecedent in \Defnref{def:super-semantic-equivalence-i},
    so
    \[
    \extractassns{\Gamma_T, \Gamma', P};
    \extractctx{\Gamma_T, \Gamma', P} ||- i_1 \equiv i_2 : \sort
    \]
    which was to be shown.
    
  \item
    If $\Gamma_T' = (\Gamma', a : \sort)$
    then by i.h.,
    \[
    \extract{\Gamma_T, \Gamma'} ||- i_1 \equiv i_2 : \sort
    \]
    By definition of $\extractctxsym$,
    \[
    \extractctx{\Gamma_T, \Gamma', a : \sort}
    =
    \extractctx{\Gamma_T, \Gamma'}, a \equiv a : \sort
    \]
    By the i.h.\ and \Defnref{def:semantic-equivalence-i},
    \[
    (\extractctx{\Gamma_T, \Gamma'}).1 |- i_1 : \sort
    \]
    We need to show that
    $(\extractctx{\Gamma_T, \Gamma', a : \sort}).1 |- i_1 : \sort$,
    which follows by weakening on sorting.
    The ``$.2$'' part is similar.

    Since $a$ does not occur in $i_1$ and $i_2$,
    applying longer substitutions that include $a$
    to $i_1$ and $i_2$ does not change them; thus, we get the same
    $j_1$ and $j_2$ as for $\Gamma_T, \Gamma'$.

  \item In the remaining cases of $\mathcal{Z}$ for $\Gamma_T' = (\Gamma', \mathcal{Z})$, neither $\extractassnssym$ nor $\extractctxsym$ change,
    and the i.h.\ immediately gives the result. \qedhere
  \end{itemize}
\end{proof}

\clearpage

\subsection{Deductive equivalence and apartness for index terms}
\label{sec:idxtm-deductive}

\begin{figure}[h]
\centering
  \judgbox{\Gamma |-i \equiv j : \sort}
          {The index terms $i$ and $j$ are \emph{equivalent} at sort $\gamma$}

\begin{mathpar}
\Infer{Eq-Var}
{(i \equiv j : \sort) \in \Gamma}
{\Gamma |- i \equiv j : \sort}
\and
\Infer{E-Refl}
      { (\Gamma).1 |-i : \sort
      \\
      (\Gamma).2 |-i : \sort
      }
      {\Gamma |-i \equiv i : \sort}
\and
\Infer{E-Sym}
      {\flip{\Gamma} |- j \equiv i : \sort}
      {\Gamma |-i \equiv j : \sort}
\\
\Infer{Eq-Pair}
{\Gamma |-i_1 \equiv j_1 : \sort_1
  \\
 \Gamma |-i_2 \equiv j_2 : \sort_2
}
{\Gamma |- (i_1,i_2) \equiv (j_1, j_2) : \sort_1 * \sort_2}
%% \and
%% \Infer{Eq-Bin}
%% {\Gamma |-i_1 \equiv j_1 : \namesort
%%   \\
%%  \Gamma |-i_2 \equiv j_2 : \namesort
%% }
%% {\Gamma |- \NmBin{i_1}{i_2} \equiv \NmBin{j_1}{j_2} : \namesort}
\and
\Infer{Eq-Lam}
{\Gamma, \big(a \equiv b : \sort_1\big) |-i \equiv j                   : \sort_2 }
{\Gamma                                 |- \Lam{a}{i} \equiv \Lam{b}{j} : \sort_1 @@> \sort_2 }
\and
\Infer{Eq-App}
{
\arrayenvcl{
 \Gamma |-i_1 \equiv j_1 : \sort_1 @@> \sort_2
  \\
 \Gamma |-i_2 \equiv j_2 : \sort_1
}
}
{\Gamma |-i_1 (i_2) \equiv j_1 (j_2) : \sort_2}
\and
\Infer{Eq-$\beta$}
{
\arrayenvcl{
  (\Gamma).1, a : \sort_2  |-i_1 : \sort
  \\
  (\Gamma).1  |-i_2 : \sort_2
  \\
  \Gamma      |- [i_2/a]i_1 \equiv j : \sort
}
}
{ \Gamma |- (\Lam{a}i_1)i_2 \equiv j : \sort }
\and
\Infer{Eq-Empty}
{ }
{ \Gamma |- \emptyset \equiv \emptyset : \NmSet }
\and
\Infer{Eq-Single}
{ \Gamma |- M \equiv N : \Nm }
{ \Gamma |- \{ M \} \equiv \{ N \} : \NmSet }
\and
\Infer{Eq-Apart}
{ \Gamma |- X_1 \equiv X_2 : \NmSet 
  \\
  \Gamma |- Y_1 \equiv Y_2 : \NmSet
}
{ \Gamma |- (X_1 \disj Y_1) \equiv (X_2 \disj Y_2) : \NmSet }
\and
\Infer{Eq-Perm}
{ \Gamma |- (X_2 \disj X_1) \equiv Y : \NmSet }
{ \Gamma |- (X_1 \disj X_2) \equiv Y : \NmSet }
\and
\Infer{Eq-Map}
{ \Gamma |- M \equiv N : \Nm @> \Nm
  \\
  \Gamma |- X \equiv Y : \NmSet
}
{ \Gamma |- M [[ X ]] \equiv N [[ Y ]] : \NmSet }
\and
\Infer{Eq-FlatMap}
{ \Gamma |- i \equiv j : \Nm @@> \NmSet
  \\
  \Gamma |- X \equiv Y : \NmSet
}
{ \Gamma |- i [[ X ]] \equiv j [[ Y ]] : \NmSet }
\and
\Infer{Eq-Star}
{ \Gamma |- i \equiv j : \Nm @@> \NmSet
  \\
  \Gamma |- X \equiv Y : \NmSet
}
{ \Gamma |- i^\ast[[ X ]] \equiv j^\ast[[ Y ]] : \NmSet }
\lesscaptionspace
\end{mathpar}

\caption{Deductive rules for showing that two index terms are equivalent}
\label{fig:idx-equiv-rules}

\end{figure}

\begin{figure}
\centering
  \judgbox{\Gamma |-i \disjoint j : \sort}
          {The index terms $i$ and $j$ are \emph{apart} at sort $\gamma$}
%  \vspace*{-1.8ex}

\begin{mathpar}
\Infer{Var}
{(a \disjoint b : \sort) \in \Gamma}
{\Gamma |- a \disjoint b : \sort}
\and
\Infer{D-Sym}
{\flip{\Gamma} |- j \disjoint i : \sort}
{\Gamma |-i \disjoint j : \sort}
\and
\Infer{D-Proj$_1$}
{\Gamma |-i_1 \disjoint j_1 : \sort_1}
{\Gamma |- (i_1,i_2) \disjoint (j_1, j_2) : \sort_1 * \sort_2}
\and
\Infer{D-Proj$_2$}
{\Gamma |-i_2 \disjoint j_2 : \sort_2}
{\Gamma |- (i_1,i_2) \disjoint (j_1, j_2) : \sort_1 * \sort_2}
%% \and
%% \Infer{D-Bin$_1$}
%% {\Gamma |-i_1 \disj j_1 : \namesort
%% }
%% {\Gamma |- \NmBin{i_1}{i_2} \disj \NmBin{j_1}{j_2} : \namesort}
%% \and
%% \Infer{D-Bin$_2$}
%% {
%%   \Gamma |-i_2 \disj j_2 : \namesort
%% }
%% {\Gamma |- \NmBin{i_1}{i_2} \disj \NmBin{j_1}{j_2} : \namesort}
%% \and
%% \Infer{D-EqTag$_1$}
%% {
%%   \Gamma |-i_1 \equiv i_2 : \namesort
%% }
%% {\Gamma |- \NmBin{i_2}{j} \disj i_1 : \namesort}
%% \and
%% \Infer{D-EqTag$_2$}
%% {
%%   \Gamma |- j_1 \equiv j_2 : \namesort
%% }
%% {\Gamma |- \NmBin{i}{j_1} \disj j_2 : \namesort}
\and
\Infer{D-Lam}
  {\Gamma |-i \disjoint j                    : \sort_2 }
  {\Gamma |- \Lam{a}{i} \disjoint \Lam{b}{j} : \sort_1 @@> \sort_2 }
\and
\Infer{D-App}
  {
    \Gamma |-i_1 \disjoint j_1 : \sort_1 @@> \sort_2
% No need to show this 2nd premise anymore: it is no longer assumed by the lambda rule:
%    \\
%    \Gamma |-i_2 \equiv    j_2 : \sort_1
  }
  {\Gamma |-i_1 (i_2) \disjoint j_1 (j_2) : \sort_2}
\and
\Infer{D-$\beta$}
  {
    \Gamma  |- [i_2/a]i_1 \disj j : \sort
    \\
    \arrayenvbl{
      (\Gamma).1  |-i_2 : \sort_2
      \\
      (\Gamma).1, a : \sort_2  |-i_1 : \sort
    }
  }
  { \Gamma |- (\Lam{a}i_1)i_2 \disj j : \sort }
\\
\Infer{D-Empty}
  { (\Gamma).2 |- X : \NmSet }
  { \Gamma |- \emptyset \disj X : \NmSet }
\and
\Infer{D-Single}
  { \Gamma |- M \disj N : \Nm }
  { \Gamma |- \{ M \} \disj \{ N \} : \NmSet }
\and
\Infer{\!D-Apart}
  {
    \Gamma |- X_1 \disj Y : \NmSet 
    ~~~~~
    \Gamma |- X_2 \disj Y : \NmSet
  }
  { \Gamma |- (X_1 \disj X_2) \disj Y : \NmSet }
\and
\Infer{\!D-Map}
  { 
    \Gamma |- M \disj N : \Nm @> \Nm
% Don't need this condition:
%    ~~~~~
%    \Gamma |- X \equiv Y : \NmSet
  }
  { \Gamma |- M [[ X ]] \disj N [[ Y ]] : \NmSet }
\and
\Infer{\!D-FlatMap1}
  {
    \Gamma |- i \disj j : \Nm @@> \NmSet
% Don't need this condition
    %% ~~~~~
    %% \Gamma |- X \equiv Y : \NmSet
  }
  { \Gamma |- i [[ X ]] \disj j [[ Y ]] : \NmSet }
\and
\Infer{\!D-FlatMap2}
  {
    \Gamma |- i \equiv j : \Nm @@> \NmSet
    ~~~~~
    \Gamma |- X \disj Y : \NmSet
  }
  { \Gamma |- i [[ X ]] \disj j [[ Y ]] : \NmSet }
\and
% Don't need this rule:
%% \Infer{\!D-FlatMap3}
%%   {
%%     \Gamma |- i \disj j : \Nm @@> \NmSet
%%     ~~~~~
%%     \Gamma |- X \disj Y : \NmSet
%%   }
%%   { \Gamma |- i [[ X ]] \disj j [[ Y ]] : \NmSet }
%% \and
\Infer{\!D-Star}
  {
    \Gamma |- i \disj j : \Nm @@> \NmSet
    ~~~~~
    \Gamma |- X \disj Y : \NmSet
  }
  { \Gamma |- i^\ast [[ X ]] \disj j^\ast [[ Y ]] : \NmSet }
\lesscaptionspace
\end{mathpar}

\caption{Deductive rules for showing that two index terms are apart}
\label{fig:idx-disj-rules}
\end{figure}

% Local Variables: 
% mode: latex
% TeX-master: "typed-adapton"
% End: 

%% \judgbox{\Gamma |- X \equiv Y}{Under $\Gamma$, the name set~$X$ is equivalent to name set~$Y$}

%% \begin{mathpar}
%% \Infer{Dis$_1$}
%% { \Gamma |- X_1 \equiv X_2  \\
%%   \Gamma |- Y_1 \equiv Y_2 }
%% { \Gamma |- (X_1 \disj Y_1) \equiv (X_2 \disj Y_2)  }
%% \and
%% \Infer{Dis$_2$}
%% { \Gamma |- X_1 \equiv Y_2  \\
%%   \Gamma |- Y_1 \equiv X_2 }
%% { \Gamma |- (X_1 \disj Y_1) \equiv (X_2 \disj Y_2)  }
%% \and
%% \Infer{Apply}
%% { \Gamma |- M ( \AsNm{i} ) \equiv N ( \AsNm{ j }) : \namesetsort }
%% { \Gamma |- M [[ i ]] \equiv N [[ j ]] }
%% \end{mathpar}

%% \judgbox{\Gamma |- X \disj Y}{Under $\Gamma$, the name set~$X$ is disjoint from name set~$Y$}

%% \begin{mathpar}
%% \Infer{Dis$_1$}
%% { \Gamma |- X \disj Z  
%%   \\
%%   \Gamma |- Y \disj Z  
%% }
%% { \Gamma ||- (X \disj Y) \disj Z}
%% \and
%% \Infer{Dis$_2$}
%% { \Gamma |- Z \disj X
%%   \\
%%   \Gamma |- Z \disj Y
%% }
%% { \Gamma |- Z \disj (X \disj Y)}
%% \and
%% \Infer{Apply}
%% { \Gamma |- M ( \AsNm{i} ) \disj N ( \AsNm{ j }) : \namesetsort }
%% { \Gamma |- M [[ i ]] \disj N [[ j ]] }
%% \end{mathpar}

% Local Variables:
% mode: latex
% TeX-master: "fungi-lang"
% End:

\clearpage
\section{Normalization for Name Terms}
\label{apx:normproof-nameterm}

We write ``$M$ halts'' when there exists $V$ such that $M \nteval \Mv$.

We write $\sigma : \Gamma$
when, for all $a \in \dom{\Gamma}$, we have $\cdot |- \sigma(a) : \Gamma(a)$.
It follows that $\sigma(a)$ is closed.

\begin{definition}[$R_{\sort}(M)$]
  \label{def:normop-name}
  ~
  \begin{enumerate}
  \item
    $R_{\sort}(M)$ if and only if $\sort \neq (\sort_1 @> \sort_2)$ and $M$ halts.

  \item $R_{(\sort_1 @> \sort_2)}(M)$ if and only if\!
    \tabularenvl{
    (i) $M$ halts and \\
    (ii) for all closed $M'$, if $R_{\sort_1}(M')$ then $R_{\sort_2}(M~M')$.}
  \end{enumerate}
\end{definition}

\begin{lemma}[Substitution]
  \label{lem:subst-name}
  If $\Gamma, a : \sort_a |- M : \sort$
  and $\Gamma |- M_a : \sort_a$
  then $\Gamma |- [M_a/a]M : \sort$.
\end{lemma}
\begin{proof}
  By induction on the derivation of $\Gamma, a : \sort_a |- M : \sort$.
\end{proof}

\begin{lemma}[Closedness]
  \label{lem:closed-name}
  If $M$ is closed
  and $M \nteval \Mv$
  then $\Mv$ is closed.
\end{lemma}
\begin{proof}
  By induction on the derivation of $M \nteval \Mv$.
\end{proof}

\begin{lemma}[Canonical Forms]
  \label{lem:canon-name}
  Suppose $|- \Mv : \sort$.

  \begin{enumerate}
  \item If $\sort = \namesort$ then $\Mv = n$.
  \item If $\sort = (\sort_1 @> \sort_2)$ then $\Mv = (\lam{a} M_0)$
    and $a : \sort_1 |- M_0 : \sort_2$.
  \end{enumerate}
\end{lemma}
\begin{proof}
  By inspection of the given derivation.
\end{proof}

% \begin{lemma}[Termination]
%   \label{lem:norm-lemma-1-name}
%   If $R_{\sort}(M)$ then $M$ halts.
% \end{lemma}
% \begin{proof}
%   By definition of $R_{\sort}(M)$.
% \end{proof}

\begin{lemma}[Multiple Substitution]
  \label{lem:multi-subst-name}
  If $\Gamma |- M : \sort$ and $\sigma : \Gamma$ then $ |- [\sigma]M : \sort$.
\end{lemma}
\begin{proof}
  By induction on the length of $\sigma$, using \Lemmaref{lem:subst-name}.
\end{proof}

\begin{lemma}[Type Preservation]
  \label{lem:type-preservation-name}
  If $|- M : \sort$
  and $M \nteval \Mv$
  then $|- \Mv : \sort$.
\end{lemma}

\begin{lemma}[Preservation]
  \label{lem:preservation-name}
  If $R_{\sort}(M)$
  and $M \nteval \Mv$
  then $R_{\sort}(\Mv)$.
\end{lemma}
\begin{proof}
  By induction on $\sort$.

  If $\sort$ does not have the form $(\sort_1 @> \sort_2)$,
  then the only requirement is to show there exists $\Mv'$ such that $\Mv \nteval \Mv'$.
  Let $\Mv' = \Mv$.  Then $\Mv \nteval \Mv'$ by \tevalvalue.

  Otherwise, $\sort = (\sort_1 @> \sort_2)$,
  and we also have to show that for all closed $M_1$
  such that $R_{\sort_1}(M_1)$,
  it is the case that $R_{\sort_2}(M\;M_1)$.

  By definition of $R$, there exists $\Mv_1$ such that $M_1 \nteval \Mv_1$.
  By i.h., $R_{\sort_1}(\Mv_1)$.

  \begin{llproof}
    \Pf{M}{\nteval}{\Mv}  {Above}
    \eqPf{\Mv}{(\lam{a} M_0)}  {By \Lemmaref{lem:canon-name}}
    \Pf{M_1}{\nteval}{\Mv_1}  {Above}
    \Pf{[\Mv_1/a]M_0}{\nteval}{\Mv_2}  {By i.h.}
    \Pf{M\;M_1}{\nteval}{\Mv_2}  {By \tevalapp}
  \end{llproof}
\end{proof}

\begin{lemma}[Normalization]
  \label{lem:norm-name}
  ~\\
  If $\Gamma |- M : \sort$ and $\sigma : \Gamma$
  and, for all $a \in \dom{\Gamma}$, we have $R_{\Gamma(a)}(\sigma(a))$,
  \\
  then $R_{\sort}([\sigma]M)$.
\end{lemma}
\begin{proof} By induction on the derivation of $\Gamma |- M : \sort$.

  \begin{itemize}
%
%                 \begin{llproof}
%                         \eqPf{[\sigma]\unitexp}{\unitexp}{By definition of $[\sigma](-)$}
%                         \Pf{\unitexp}{\nteval}{\unitexp}{By rule \tevalvalue}
%                         \Pf{[\sigma]\unitexp}{\nteval}{\unitexp}{By above equation}
%                         \isavaluePf{\unitexp} {}
%                         \Hand \Pf{}{}{R_{\unitsort}([\sigma]M)} {$[\sigma]M$ halts}
%                 \end{llproof}

        \DerivationProofCase{M-const}
                {}
                {\Gamma |- n : \namesort}

                \begin{llproof}
                        \eqPf{[\sigma]n}{n}{By definition of $[\sigma](-)$}
                        \Pf{n}{\nteval}{n}{By rule \tevalvalue}
                        \Pf{[\sigma]n}{\nteval}{n}{By above equation}
                        \isavaluePf{n} {}
                        \Hand \Pf{}{}{R_{\namesort}([\sigma]M)} {By definition of $R$}
                \end{llproof}

%         \DerivationProofCase{M-unit}
%                 {}
%                 {\Gamma |- \unitexp : \unitsort}
%                 %
%                 Similar to the M-name case.                 

        \DerivationProofCase{M-var}
                {(a : \sort) \in \Gamma}
                {\Gamma |- a : \sort}

                We have $\Gamma(a) = \sort$.
                It is given that $R_{\Gamma(a)}(\sigma(a))$.
                Since $\sigma(a) = [\sigma]a$,
                we have $R_{\sort}{[\sigma]a}$,
                which was to be shown.
%                 \begin{llproof}
% %                        \Pf{\exists v_i.([\sigma]a}{\nteval}{v_i)}{By definition of $[\sigma](-)$}
%                  
%                         \Pf{}{|-}{[\sigma] a : \sort}{From \Lemmaref{lem:multi-subst-name}}
%                         \Hand \Pf{}{}{R_{\sort}([\sigma] a)}{Since $[\sigma]a \nteval v_i$ and $|- [\sigma] a : \sort$}
%                 \end{llproof}

        \DerivationProofCase{M-abs}
                {\Gamma, a : \sort_1 |- M_0 : \sort_2}
                {
                  \Gamma |- (\lam{a} M_0) : (\sort_1 @> \sort_2)
                }

                Suppose that, for some closed $M'$, we have $R_{\sort_1}(M')$.
                By the definition of $R$, that means there exists $\Mv'$
                such that $M' \nteval \Mv'$.
                
                We need to show $R_{\sort_2}\big([\sigma]((\lam{a} M_0) \, M')\big)$.

                Let $\sigma_a = (\sigma, \Mv'/a)$.

                \begin{llproof}
%                  \ePf{\Gamma}{\Mv' : \sort_1}  {By \Lemmaref{lem:type-preservation-name}}
                  \Pf{}{}{R_{\sort_1}(\Mv')}  {By \Lemmaref{lem:preservation-name}}
                  \ePf{\Gamma, a : \sort_1}{M_0 : \sort_2}  {Subderivation}
                  \Pf{}{}{R_{\sort_2}([\sigma_a]M_0)}  {By i.h.\ with $\sigma_a$ as $\sigma$}
%                  \eqPf{[\sigma_a]M_0}{[\sigma, \Mv'/a]M_0 = [\sigma][ [\sigma]\Mv'/a]M_0}{By def.\ of subst.}
%                  \Pf{}{}{R_{\sort_2}([\sigma][ [\sigma]\Mv'/a]M_0)}  {By above equation}
%                  \Pf{[\sigma_a]M_0}{\nteval}{\Mv}  {By $R_{\sort_2}([\sigma_a]M_0)$}
%                  \Pf{[\sigma_a]M'}{\nteval}{\Mv'}  {By $R_{\sort_1}(M')$}
%                  \eqPf{[\sigma_a]M'}{[\sigma]M'} {$a$ not free in $M'$}
%                  \Pf{[\sigma]M'}{\nteval}{\Mv'}  {By above equation}
                  \proofsep
                  \Pf{(\lam{a} [\sigma]M_0)}{\nteval}{\lam{a} [\sigma]M_0}  {By \tevalvalue}
                  \Pf{M'}{\nteval}{\Mv'}  {Above}
%                  \Pf{}{}{R_{\sort_1}([\sigma]\Mv')}  {By i.h.}
%                  \Pf{}{}{R_{\sort_2}([\sigma_a]M_0)}  {By i.h.}
                  \Pf{[\sigma_a]M_0}{\nteval}{\Mv} {By definition of $R$}
                  \eqPf{[\sigma_a]M_0}{[\Mv'/a][\sigma]M_0}  {By def.\ of subst.}
                  \Pf{[\Mv'/a][\sigma]M_0}{\nteval}{\Mv} {By above equation}
                  \Pf{(\lam{a} [\sigma]M_0)\,M'}{\nteval}{\Mv}  {By \tevalapp}
                  \eqPf{M'}{[\sigma]M'}  {$M'$ closed}
                  \Pf{(\lam{a} [\sigma]M_0)\,[\sigma]M'}{\nteval}{\Mv}  {By above equation}
                  \eqPf{(\lam{a} [\sigma]M_0)\,[\sigma]M'}{([\sigma]\lam{a} M_0)\,[\sigma]M'}  {By def.\ of subst.}
                  \continueeqPf{[\sigma]\big((\lam{a} M_0)\,M'\big)} {By def.\ of subst.}
                  \Pf{[\sigma]\big((\lam{a} M_0)\,M'\big)}{\nteval}{\Mv}  {By above equations}
                  \proofsep
                \Hand  \Pf{}{R_{\sort_2}}{\big([\sigma]((\lam{a} M_0) \, M')\big)}  {By definition of $R$}
                \end{llproof}

%                 \begin{llproof}
% %                        \Pf{[\sigma, N/a]}{:}{\Gamma'}{Assumption, where $\Gamma' = (\Gamma, a : \sort_1)$}
% % can't assume the variable is the same! -j.
%                         \Pf{|-}{[\sigma, N/a](\lam{a} M)}{ : (\sort_1 @> \sort_2)}{By \Lemmaref{lem:multi-subst-name}}
%                         \eqPf{[\sigma, N/a] (\lam{a} M)}{(\lam{a} M)}{By definition of $[\sigma](-)$}
%                         \Hand \Pf{R_{\sort}([\sigma, N/a] (\lam{a} M))}{}{}
%                                {Since $[\sigma, N/a](\lam{a} M) \nteval  (\lam{a} M)$}\trailingjust{and $|- [\sigma, N/a]  (\lam{a} M) : (\sort_1 @> \sort_2)$}
%                 \end{llproof}

        \DerivationProofCase{\!M-app}
                {
                        \Gamma |- M_1 : (\sort' @> \sort)
                        \\
                        \Gamma |- M_2 : \sort'
                }
                {\Gamma |- (M_1\;M_2) : \sort}

                \begin{llproof}
                        \Pf{}{}{R_{(\sort' @> \sort)}([\sigma] M_1)}{By i.h.}
                        \Pf{}{}{R_{\sort'}([\sigma] M_2)}{By i.h.}
                        \Pf{}{}{R_{\sort}(([\sigma]M_1)\;[\sigma]M_2)}{By definition of $R$}
                        \Hand \Pf{}{}{R_{\sort}([\sigma](M_1\;M_2))}{By def.\ of subst.}
                \end{llproof}

%         \DerivationProofCase{\!M-pair}
%                 {
%                         \Gamma |- M_1 : \sort_1
%                         \\
%                         \Gamma |- M_2 : \sort_2
%                 }
%                 {\Gamma |- \Pair{M_1}{M_2} : (\sort_1 ** \sort_2)}
%
%                                 \begin{llproof}
%                         \Pf{}{}{R_{\sort_1}([\sigma] M_1)}{By inductive hypothesis}
%                         \Pf{}{}{R_{\sort_2}([\sigma] M_2)}{By inductive hypothesis}
%                         \Pf{\exists v_1.([\sigma] M_1}{\nteval}{v_1)}{Since $R_{\sort_1}([\sigma] M_1)$}
%                         \Pf{\exists v_2.([\sigma] M_2}{\nteval}{v_2)}{Since $R_{\sort_2}([\sigma] M_2)$}
%                         \Pf{([\sigma] M_1, [\sigma] M_2)}{\nteval}{(v_1, v_2)}{By rule \tevaltuple}
%                         \Pf{[\sigma](M_1, M_2)}{\nteval}{(v_1, v_2)}{By definition of $[\sigma](-)$}
%                         \Pf{}{|-}{[\sigma] (M_1, M_2) : (\sort_1 ** \sort_2)}{From \Lemmaref{lem:multi-subst-name}}
%                         \Hand \Pf{}{}{R_{(\sort_1 ** \sort_2)}([\sigma]\Pair{M_1}{M_2})}{Since $[\sigma]\Pair{M_1}{M_2} \nteval (v_1, v_2)$}\trailingjust{and $|- [\sigma] \Pair{M_1}{M_2} : (\sort_1 ** \sort_2)$}
%                 \end{llproof}

        \DerivationProofCase{\!M-bin}
                {
                        \Gamma |- M_1 : \namesort
                        \\
                        \Gamma |- M_2 : \namesort
                }
                {\Gamma |- \NmBin{M_1}{M_2} : \namesort}

                \begin{llproof}
                        \Pf{}{}{R_{\namesort}([\sigma] M_1)}{By i.h.}
                        \Pf{}{}{R_{\namesort}([\sigma] M_2)}{By i.h.}
                        \Pf{[\sigma] M_1}{\nteval}{\Mv_1}{By $R_{\namesort}([\sigma] M_1)$}
                        \ePf{\Gamma}{M_1 : \namesort}  {Subderivation}
                        \ePf{}{[\sigma]M_1 : \namesort}  {By \Lemmaref{lem:multi-subst-name}}
                        \eqPf{\Mv_1}{n_1}  {By \Lemmaref{lem:type-preservation-name}}
                                           \trailingjust{and \Lemmaref{lem:canon-name}}
                        \Pf{[\sigma] M_2}{\nteval}{n_2}{Similar}
                        \Pf{\NmBin{[\sigma] M_1}{[\sigma] M_2}}{\nteval}{\NmBin{n_1}{n_2}}{By rule \tevalbin}
                        \Pf{[\sigma]\NmBin{M_1}{M_2}}{\nteval}{\NmBin{n_1}{n_2}}{By definition of $[\sigma](-)$}
%                        \Pf{}{|-}{[\sigma]\NmBin{M_1}{M_2}: \namesort}{From \Lemmaref{lem:multi-subst-name}}
                        \Hand \Pf{}{}{R_{\namesort}([\sigma]\NmBin{M_1}{M_2})}{By definition of $R$}
                \end{llproof} \qedhere
  \end{itemize}
\end{proof}

%
% This is not *strong* normalization.
% Strong normalization would be:
%   If $|- M : \sort$ then there exists $\Mv$ such that $M \nteval Mv$
%   *and* for all $N$ such that $M \nteval N$, we have that $N$ is a value.
%
\begin{restatable}[Normalization]{theorem}{thmnormname}
        \label{thm:norm-name}
        If $|- M : \sort$ then there exists $\Mv$ such that $M \nteval \Mv$.
\end{restatable}
\begin{proof}
  By \Lemmaref{lem:norm-name}, $R_{\sort}(M)$.

  By definition of $R$, there exists $\Mv$ such that $M  \nteval \Mv$.
\end{proof}

% Local Variables: 
% mode: latex
% TeX-master: "typed-adapton"
% End: 

%%%%%%%%%%%%%%%%%%%%%%%%%%%%%%%%%%%%%%%%%%%
%
% Strong normalization for indices is not referred to anywhere in the paper,
%  the current proof is not correct,
%  and a correct proof may be harder than the (now possibly correct)
%  proof of normalization for name terms.
%  Therefore, commenting this out.
%  The old proof is in old-strongnormproof-indextermlang.tex.
% \input{strongnormproof-indextermlang}
%
%%%%%%%%%%%%%%%%%%%%%%%%%%%%%%%%%%%%%%%%%%%

\fi
\end{document}